\documentclass[twocolumn,showpacs,preprintnumbers,amsmath,amssymb,
floatfix]{revtex4}

\usepackage{graphicx}
\usepackage{dcolumn}

\usepackage{mathptmx, courier, pifont}
\usepackage[scaled=0.92]{helvet}
\usepackage[T1]{fontenc}
\usepackage{textcomp} 

\usepackage{bm}

\newcommand{\singlefig}[6]{%
\begin{figure}\vspace{#3}%
\includegraphics*[scale=#5]{#2}%
\caption{\label{fig:#1} #6}%
\vspace{#4}%
\end{figure}}

\newcommand{\doublefig}[6]{%
\begin{figure*} \vspace{#3}%
\includegraphics*[scale=#5]{#2}%
\caption{\label{fig:#1} #6}
\vspace{#4}
\end{figure*}}

\newtheorem{conjecture}{Conjecture}
\newcommand{\doTheorem}[1]{\begin{conjecture} #1 \end{conjecture}}

\newcommand{\tsub}[1]{_{\mbox{\scriptsize#1}}}

\newcommand{\bra}[1]{\langle#1|}
\newcommand{\ket}[1]{|#1\rangle}
\newcommand{\ev}[1]{\langle#1\rangle}
\newcommand{\mel}[3]{\bra{#1}#2\ket{#3}}
\newcommand{\atanh}{{\rm atanh}}

\newcommand{\singletPair}{p}
\newcommand{\AF}{Q}

\newcommand{\Qvector}{\vec\AF}
\newcommand{\QdotQ}{\Qvector\cdot\Qvector}
\newcommand{\SdotS}{\vec S\cdot\vec S}
\newcommand{\DdagD}{\singletPair^\dagger\singletPair}
\newcommand{\pidagpi}{\vec\pi^\dagger\cdot\vec\pi}
\newcommand{\pidotpi}{\vec\pi^\dagger\cdot\vec\pi}

\newcommand{\sufourk}{SU(4)$_k$}

\newcommand{\singletGap}{\Delta\tsub s}
\newcommand{\tripletGap}{\Delta\tsub t}

\newcommand{\pratio}{\sigma}

\newcommand{\DeltaZero}{\Delta_0}

\newcommand{\Deltad}{\Delta_d}
\newcommand{\Deltaq}{\Delta_q}
\newcommand{\Deltapi}{\Delta_\pi}

\newcommand{\Deltadsq}{\Delta_d^2}
\newcommand{\Deltaqsq}{\Delta_q^2}
\newcommand{\Deltapisq}{\Delta_\pi^2}
\newcommand{\Deltapm}{\Delta_\pm}
\newcommand{\DeltaMinus}{\Delta_-}
\newcommand{\DeltaPlus}{\Delta_+}
\newcommand{\lambdaPrime}{\lambda^\prime}

\newcommand{\sofour}{{\rm SO(4)}}
\newcommand{\sufour}{{\rm SU(4)}}
\newcommand{\sutwo}{{\rm SU(2)}}
\newcommand{\sofive}{{\rm SO(5)}}
\newcommand{\soeight}{{\rm SO(8)}}
\newcommand{\uone}{{\rm U(1)}}
\newcommand{\ufour}{{\rm U(4)}}

\newcommand{\eq}[1]{Eq.~(\ref{#1})}
\newcommand{\eqs}[1]{Eqs.~(\ref{#1})}
\newcommand{\eqnoeq}[1]{(\ref{#1})}
\newcommand{\fig}[1]{Fig.~\ref{fig:#1}}
\newcommand{\tableref}[1]{Table~\ref{table:#1}}

\newcommand{\runningheads}[2]{\markboth{\hfill #1\hfill}{\hfill #2\hfill}}

\begin{document}

\title{Fermion Dynamical Symmetry and Strongly-Correlated Electrons: A
Comprehensive Model of High-Temperature Superconductivity}

\author{Mike Guidry$^{(1)}$}
\email{guidry@utk.edu}
\author{Yang Sun$^{(2)}$}
\email{sunyang@sjtu.edu.cn}
\author{Lian-Ao Wu$^{(3)}$}
\email{lianaowu@gmail.com}
\author{Cheng-Li Wu$^{(4)}$}
\email{clwu@phys.cts.nthu.edu.tw}

\affiliation{
$^{(1)}$Department of Physics and Astronomy, University of
Tennessee, Knoxville, Tennessee 37996, USA \\
$^{(2)}$School of Physics and Astronomy, Shanghai Jiao Tong
 University, Shanghai 200240, People's Republic of China \\
 $^{(3)}$IKERBASQUE, Basque Foundation for Science, 48011 Bilbao, Spain,
and Department of Theoretical Physics and History of Science,
Basque Country University (EHU/UPV), Post Office Box 644, 48080 Bilbao, Spain\\
$^{(4)}$Department of Physics, Chung-Yuan Christian University,
Chungli, Taiwan 320, ROC
}

\date{\today}

\begin{abstract}
We review application of the SU(4) model of strongly-correlated electrons to 
cuprate and iron-based superconductors. A minimal self-consistent generalization 
of BCS theory to incorporate antiferromagnetism on an equal footing with pairing 
and strong Coulomb repulsion is found to account systematically for the major 
features of high-temperature superconductivity, with microscopic details of the 
parent compounds entering only parametrically. This provides a systematic 
procedure to separate essential from peripheral, suggesting that many features 
exhibited by the high-$T\tsub c$ data set are of interest in their own right but 
are not central to the superconducting mechanism. More generally, we propose 
that the surprisingly broad range of conventional and unconventional 
superconducting and superfluid behavior observed across many fields of physics 
results from the systematic appearance of similar algebraic structures 
for the emergent effective Hamiltonians, even though the  microscopic 
Hamiltonians of the corresponding parent states may differ radically from each 
other.
\end{abstract}


\maketitle

\runningheads{}{{\em Fermion Dynamical Symmetry and Strongly-Correlated
Electrons} --- M. W. Guidry, Y. Sun, L.-A. Wu, and C.-L. Wu}


\clearpage
\begin{small}
\tableofcontents%
\end{small}

\section{\label{intro} Introduction}

High-temperature superconductivity (HTSC) was discovered in 1986 for the
copper oxides \cite{highTc_discovery} but its theoretical interpretation remains
contentious \cite{bonn06}.  In 2008, new high-temperature superconductivity was
discovered in FeAs compounds \cite{FeAsDiscovery,sun09}, and in 2010 in FeSe 
compounds
\cite{FeSeDiscovery}. This new iron-based superconductivity also is not well
understood, with many open questions about the underlying mechanism and whether
it has any relation to the mechanism for cuprate superconductivity
\cite{FeAsOverview,pagl2010,mou2011,oh2011}. Conventional superconductors are
described well by the BCS (Bardeen--Cooper--Schrieffer) theory \cite{BCS57}, and
correspond to spin-singlet condensates of Cooper pairs \cite{coop56} having
phonon pair-binding and orbitally-symmetric ($s$-wave) pairing formfactors. It
is generally thought that the cuprate and iron-based superconductors result from
condensation of spin-singlet Cooper pairs, but that they have non-phonon
pair binding with formfactors that differ from the conventional symmetric
$s$-wave form; superconductors having such unconventional pairing are
commonly called {\em unconventional superconductors.}

More generally, superconductivity (SC) exhibiting many similarities with that of
cuprate and iron superconductors has been found in other condensed-matter
systems, such as organics and heavy-fermion compounds \cite{norm2011,jero2012}.
Superconductivity (or superfluidity) also is known to play a central role in
nuclear structure \cite{SCnuclear}, and is expected to occur for the neutrons 
and protons in
 neutron stars \cite{NS_SC} and for the color degree of freedom in quark matter
\cite{colorSC,colorSC2}. These other forms of superconductivity or 
superfluidity are all
thought to involve condensates of Cooper pairs, even though the underlying
structure and interactions may differ fundamentally among these instances. The 
mechanism is
often suspected to involve unconventional pairing, but is typically not well
understood. There is even less understanding of how such a broad set of physical
systems, ranging from diverse compounds in condensed matter to many different
isotopes of atomic nuclei to a variety of neutron stars, should exhibit a
similar Cooper instability leading to the superconducting or superfluid state.

\subsection{The Adequacy of Theoretical Tools}

The lack of agreement concerning the mechanism for high-temperature
superconductivity, and the limited understanding of how these mechanisms are
connected to the various other occurrences of superconductivity noted above,
have two ready explanations:

\begin{enumerate}
 \item
In each case the issues are complex but most models emphasize only limited
aspects of the overall problem.
\item
Often the superconducting mechanism is obscured by the complex behavior because
the dearth of solvable models with broad physics content makes it difficult to
separate essential features from secondary ones.
\end{enumerate}
Thus, we believe that the primary issue is not that there are key measurements
remaining to be made that will magically unravel the HTSC problem, but rather
that most theoretical tools are inadequate to assess properly the implications
of the quite sophisticated data sets already in existence. Stated concisely,
theoretical models that deploy a sufficiently broad arsenal of physics tend to
not be solvable, while those models that are solvable tend to be so at the
cost of emphasizing certain aspects of the problem relative to others. This
latter feature of solvable models then tends to produce strong either/or
dichotomies around which various camps rally.

We propose here that a solvable model with an adequate range of physics for the
HTSC problem is possible by applying a set of mathematical tools originating in
the theory of Lie algebras and generalized coherent states. The methods that we
shall use to accomplish this might be viewed as unconventional in that they have
not found broad use in condensed matter physics prior to the work reviewed here.
However, their validity is well established in strongly-correlated fermion and
boson systems for fields such as nuclear \cite{FDSM,IBM}, elementary particle
\cite{Bi94}, molecular \cite{Ia95}, and polymer physics \cite{Ia99}, and---truth
be told---they draw substantially on ideas that originated in condensed matter
and related fields.
%
%
The dynamical symmetry methods discussed here may
be viewed as sophisticated generalizations of pseudospin models, which were
introduced in application to the BCS model by Anderson
\cite{ande58}, and later used extensively in both
condensed matter and nuclear physics. Likewise, the method of generalized
coherent states is a sophisticated extension of Glauber coherent states
\cite{gla63}, which have been employed often in condensed matter contexts.

By making use of these powerful  methods, we shall arrive at a physical picture 
that is
surprisingly conventional relative to many proposed explanations of
high-temperature superconductivity.  We shall show that a theory unifying
self-consistently superconductivity built on a BCS-like wavefunction (possibly
with unconventional order) and the N\'eel model of antiferromagnetism in the 
presence of strong Coulomb repulsion leads,
with few further assumptions, to physics very similar to that observed in actual
 high-temperature superconductors.

\subsection{Areas of Some Consensus}

It is useful to begin the discussion with some things that are are not so
contentious. Despite the absence of general agreement on the mechanism for
high-temperature superconductivity in the cuprates and iron-based compounds, we
believe that there {\em is} fairly broad consensus on four issues. 

(1)~High-temperature superconductivity is related intimately to an interplay
among  pairing, antiferromagnetism, and charge degrees of freedom for
strongly-correlated electrons, as illustrated schematically in
\fig{interplay}.%
\singlefig
{interplay}       
{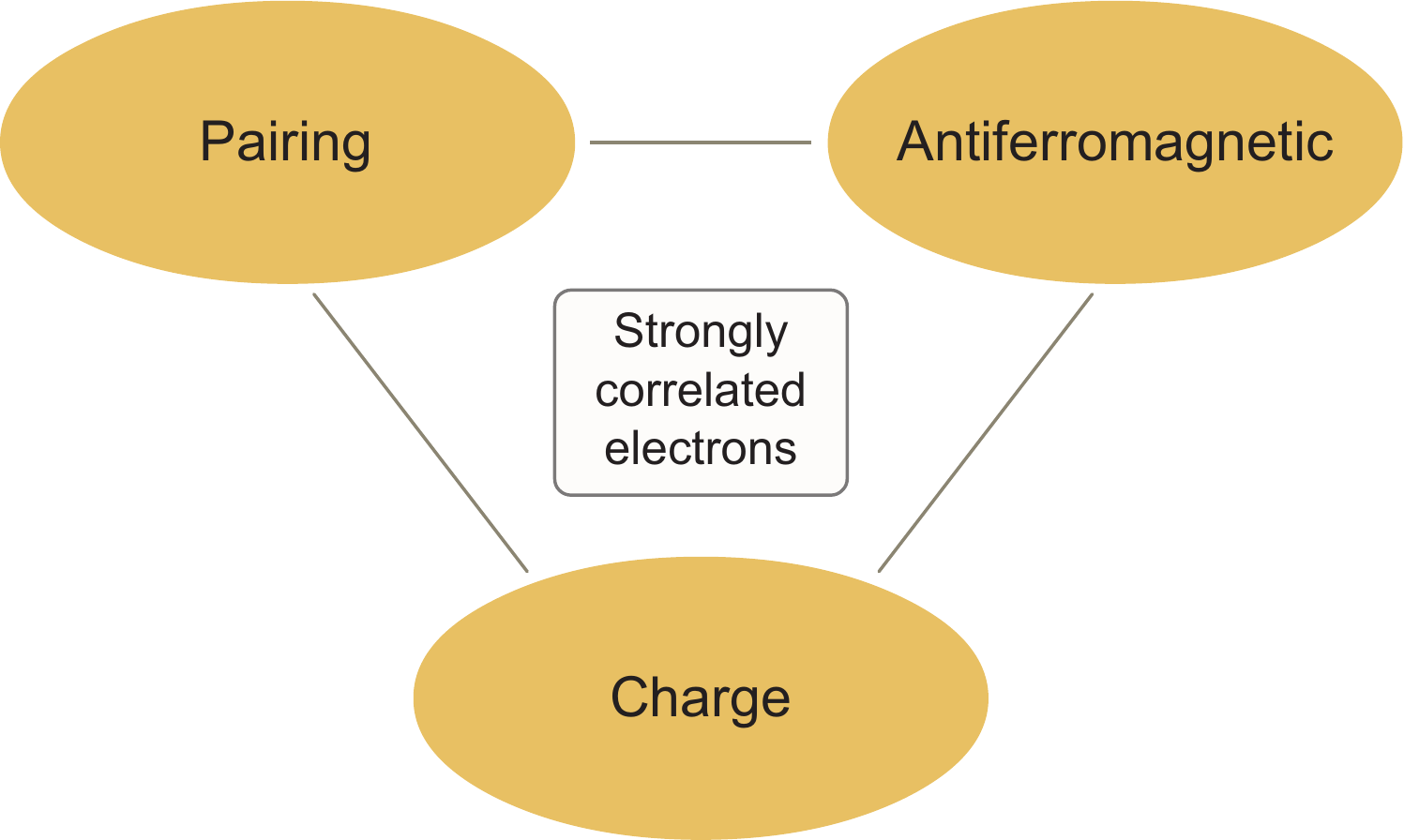}
{0pt}         
{0pt}         
{0.50}         
{Cuprate superconductivity is thought to involve an interplay of pairing,
antiferromagnetism, and charge in a strongly-correlated electron system. }

(2)~The superconducting state is a condensate of spin-singlet Cooper pairs
behaving in many respects as an ordinary BCS superconductor, but with important
differences, particularly in states with low doping. The orbital pairing
formfactor for the cuprate superconductors is of $d_{x^2-y^2}$ form and
dominated by contributions from a single band.  Figure \ref{fig:dwaveBCS}
illustrates.%
\singlefig
{dwaveBCS}       
{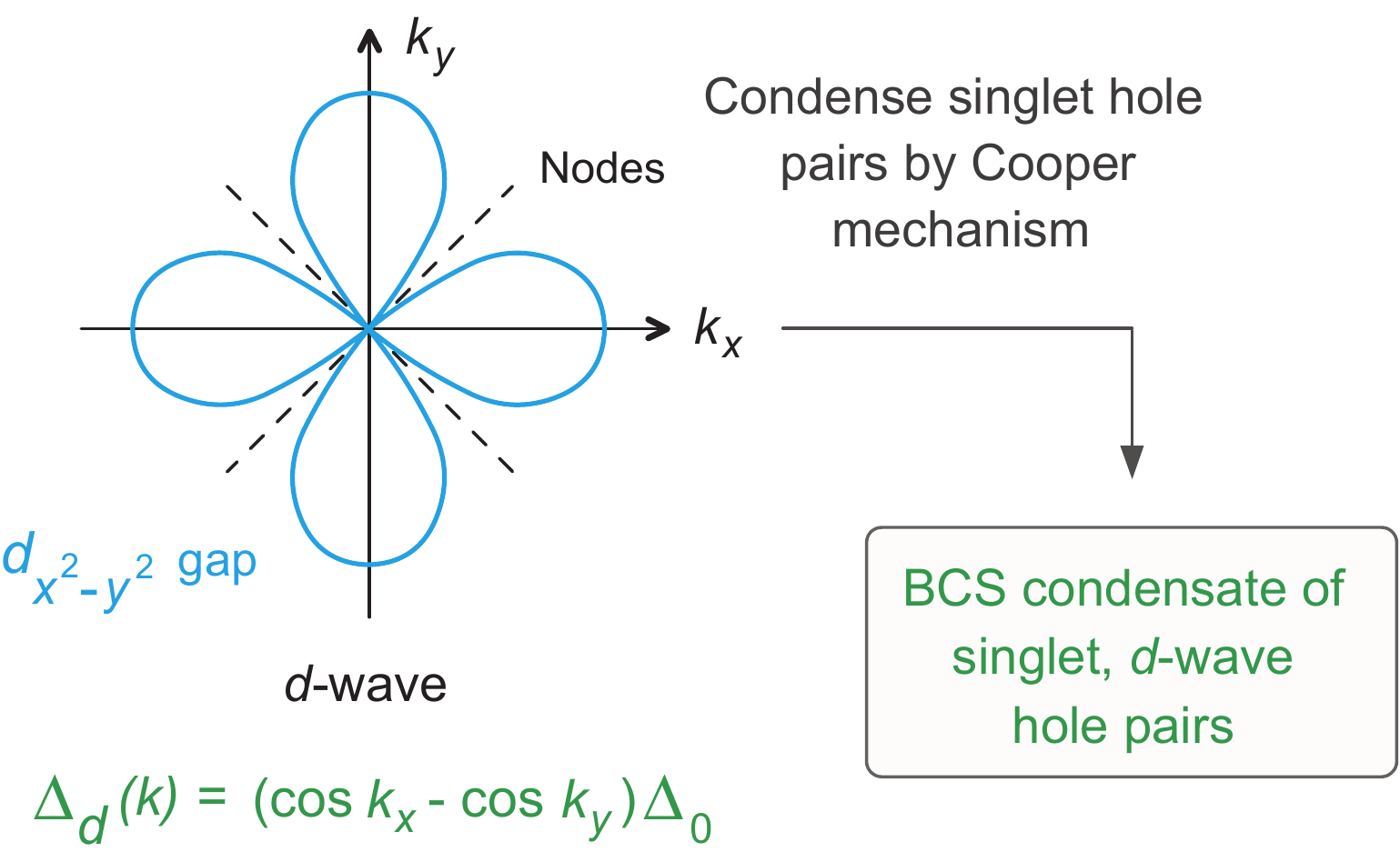}
{0pt}         
{0pt}         
{0.54}         
{Cuprate superconductivity is thought to involve singlet pairing with a
$d$-wave orbital formfactor. }
For the iron-based superconductors there is less certainty but it is generally
believed that multiple bands near the Fermi surface contribute and that there
may be more than one pairing gap. The orbital symmetry of the gap often appears 
to be a
modified $s$-wave form, but it is possible that the symmetry of the gap may vary
with the compound being examined \cite{hirs2011}.

(3)~The precursor to the cuprate superconducting state is an antiferromagnetic
(AF) Mott insulator (an insulating state in which the insulator properties
derive specifically from strong onsite Coulomb repulsion), unlike for ordinary
superconductors where the precursor is a normal Fermi liquid (an interacting
system of fermions having excitations that can be put into one-to-one
correspondence with those of a non-interacting fermion system). Figure
\ref{fig:mottAndFermiLiquid} illustrates.%
\singlefig
{mottAndFermiLiquid}       
{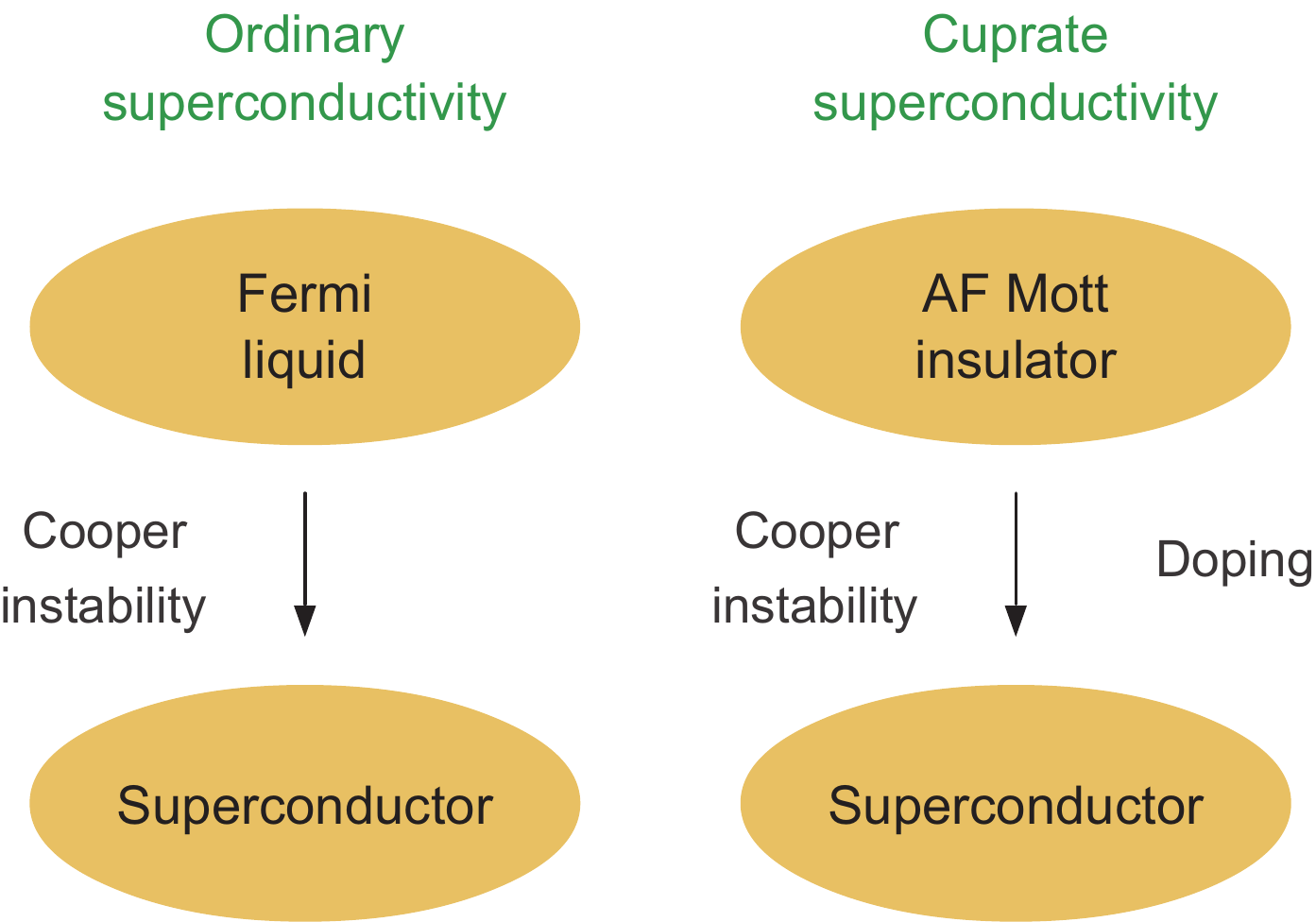}
{0pt}         
{0pt}         
{0.55}         
{Normal superconductivity results from the Cooper instability for a Fermi
liquid.  Cuprate superconductivity also results from Cooper pairing, but it
develops from a Mott insulator state through doping. }
The precursor to the iron-based superconductors is typically a poor AF metal 
that
might (or might not) be near a Mott transition.

(4)~The pseudogap, corresponding to an observed partial gapping of states in the
underdoped region above the superconducting transition temperature
\cite{pseudogap,pseudogap2}, is a well-established enigma for the cuprate 
superconductors
(\fig{PGstate}).%
\singlefig
{PGstate}       
{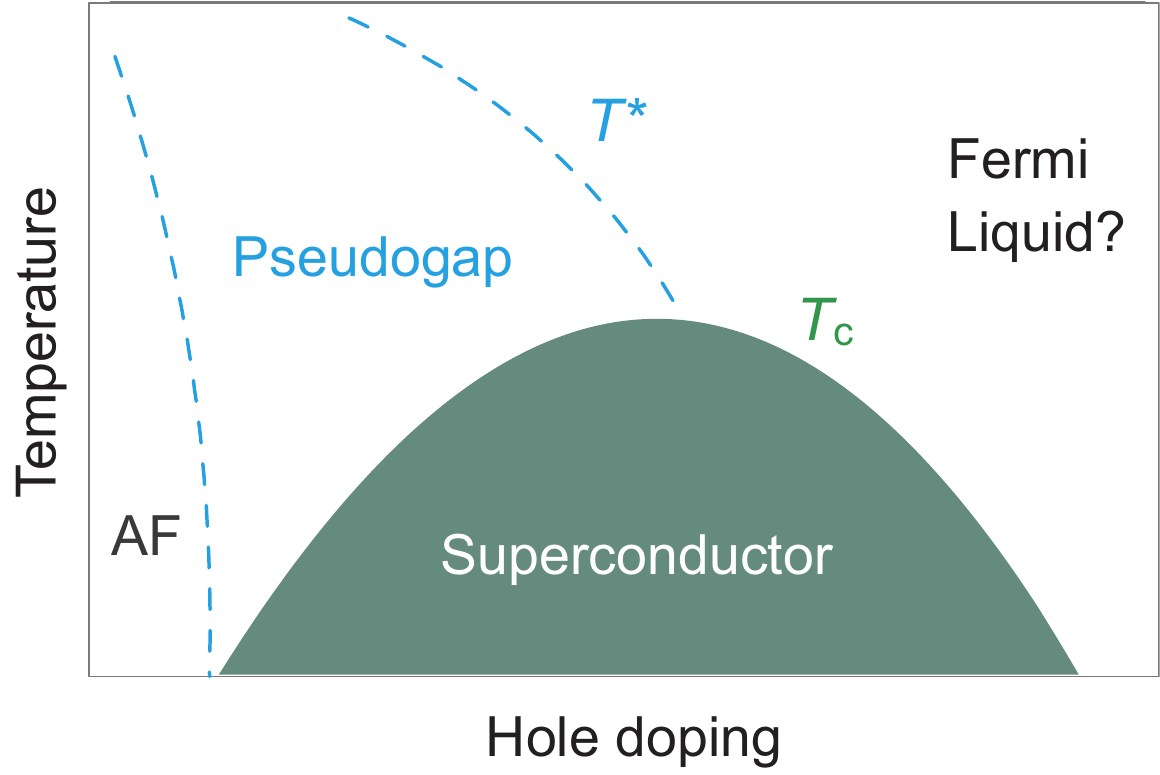}    
{0pt}         
{0pt}         
{0.55}         
{The pseudogap region in the cuprates. }
Whether a pseudogap exists for the iron-based compounds is not settled
(positive evidence for one is given in
Refs.~\cite{shee2010,mert2009,ahil2008,sato2008,ning2008}). The pseudogap state
is not a ``normal'' normal state, and it is not well described by traditional
Fermi liquid concepts.  There is a general feeling that understanding
the pseudogap state is necessary to understanding HTSC, at least for the
cuprates.

\subsection{Fundamental Issues with Little Consensus}

In contrast to the areas enumerated in the previous section where there is
relatively uniform agreement, there are a number of important issues for which
there is little agreement. These share two common features, in our opinion:
(1)~Their resolution may be central to understanding the high-temperature
superconducting mechanism.  (2)~A resolution of these issues requires a
multiphysics approach capable of integrating concepts on a similar footing that
could be treated as approximately independent in many simpler problems.

\subsubsection{Parent States and Rapid Onset of Superconductivity}

Band theory suggests that cuprates at half lattice filling should be metals, but 
they
are in fact insulators with antiferromagnetic (AF) properties.  This behavior is
thought to result from a Mott-insulator normal state, where the insulator 
properties
follow from strong onsite Coulomb repulsion. Doping the normal states with 
electron
holes produces a rapid transition to a superconducting (SC) state, with a 
pairing gap
typically appearing for about 3--5\% hole density per copper site in the
copper--oxygen plane.  Furthermore, at low to intermediate doping a partial 
energy
gap appears at temperatures above the SC transition temperature $T\tsub c$ that 
is
termed a pseudogap (PG), with the SC gap and PG having opposite doping 
dependence at
low doping \cite{pseudogap,pseudogap2}.

Parent states of normal superconductors are Fermi liquids and normal 
superconductors are described by BCS theory \cite{BCS57}, which assumes the 
condensation of zero-spin, zero-momentum fermion pairs into a new collective 
state with long-range coherence of the wavefunction. The key to understanding 
normal superconductivity was the demonstration by Cooper \cite{coop56} that 
normal Fermi liquids possess a fundamental instability:  a zero-momentum 
electron pair above a filled Fermi sea can form a bound state for {\em 
vanishingly small attractive interaction.}  In normal superconductors the 
attraction is provided by interactions with lattice phonons, which bind weakly 
over a limited frequency range because electrons and the lattice have different 
response times. However, it is the Cooper instability, not the source of the 
attractive interaction, that is most fundamental: a weak attraction alone 
cannot 
produce a superconducting state, but the Cooper instability can (in principle) 
produce a superconducting state for {\em any} weakly-attractive interaction 
between a spin-singlet, zero momentum pair.

The rapid onset of HTSC with hole doping in the cuprates (\fig{rapidOnset})%
\singlefig
{rapidOnset}       
{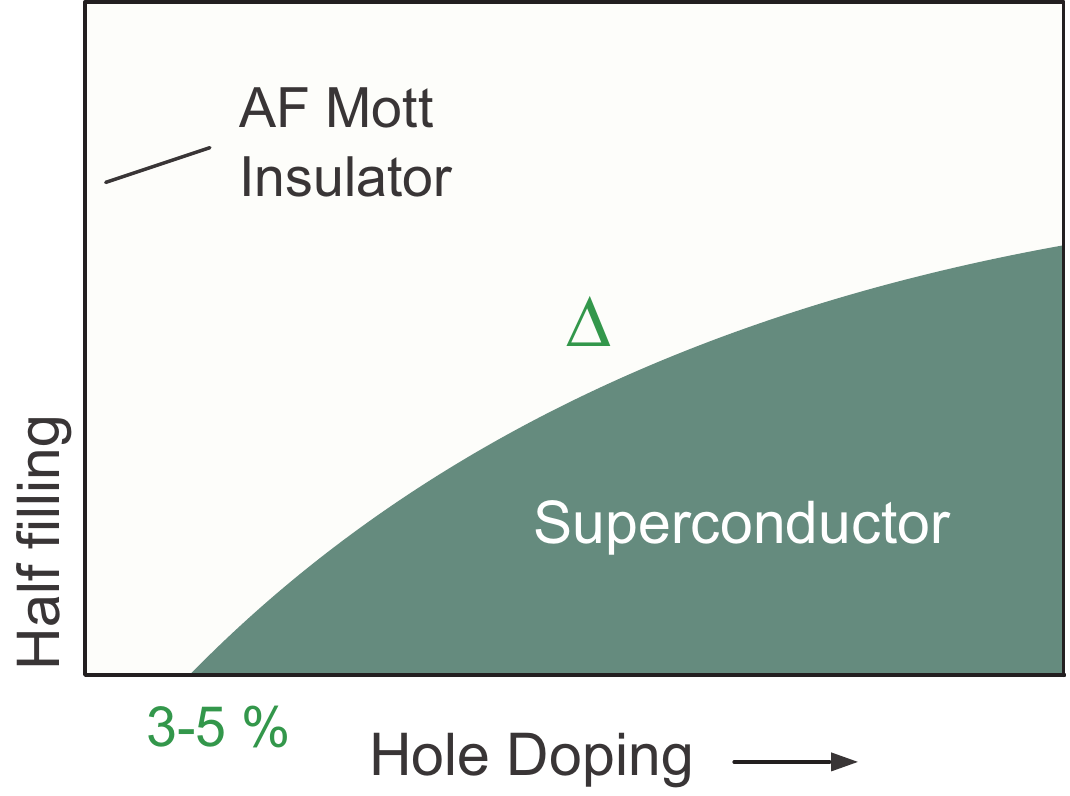}    
{0pt}         
{0pt}         
{0.55}         
{The rapid onset of cuprate superconductivity with doping.
}
suggests that the Mott insulator already contains within it a hidden propensity 
to superconductivity. Such behavior is indicative of a fundamental instability 
with respect to pair condensation, but  this (as well as the origin of 
pseudogap 
states) is difficult to interpret within the standard BCS framework since the 
superconductor appears to derive from a Mott insulator, not a normal Fermi 
liquid.  

As for normal superconductivity, we believe that the key to understanding HTSC 
is not the effective interaction leading to pair binding (a topic important in 
its own right), but rather the nature of the instability that produces the 
superconducting state.  At larger doping the cuprate high-temperature 
superconducting state exhibits many properties of a normal ($d$-wave) BCS 
superconductor, so this instability must reduce to the classical Cooper 
instability at larger doping, but morph into something having more complex 
behavior at lower doping where the normal state approaches a Mott insulator and 
there is a pseudogap lying above the superconducting transition temperature.

\subsubsection{Nature of the Pseudogap State}

As we have noted, there is broad agreement that an explanation of the pseudogap 
(PG) state may be central to understanding the superconductivity.  However, 
there is extensive disagreement over what the appropriate explanation is.  Two 
general ideas (illustrated in \fig{PGpictures})%
\singlefig
{PGpictures}       
{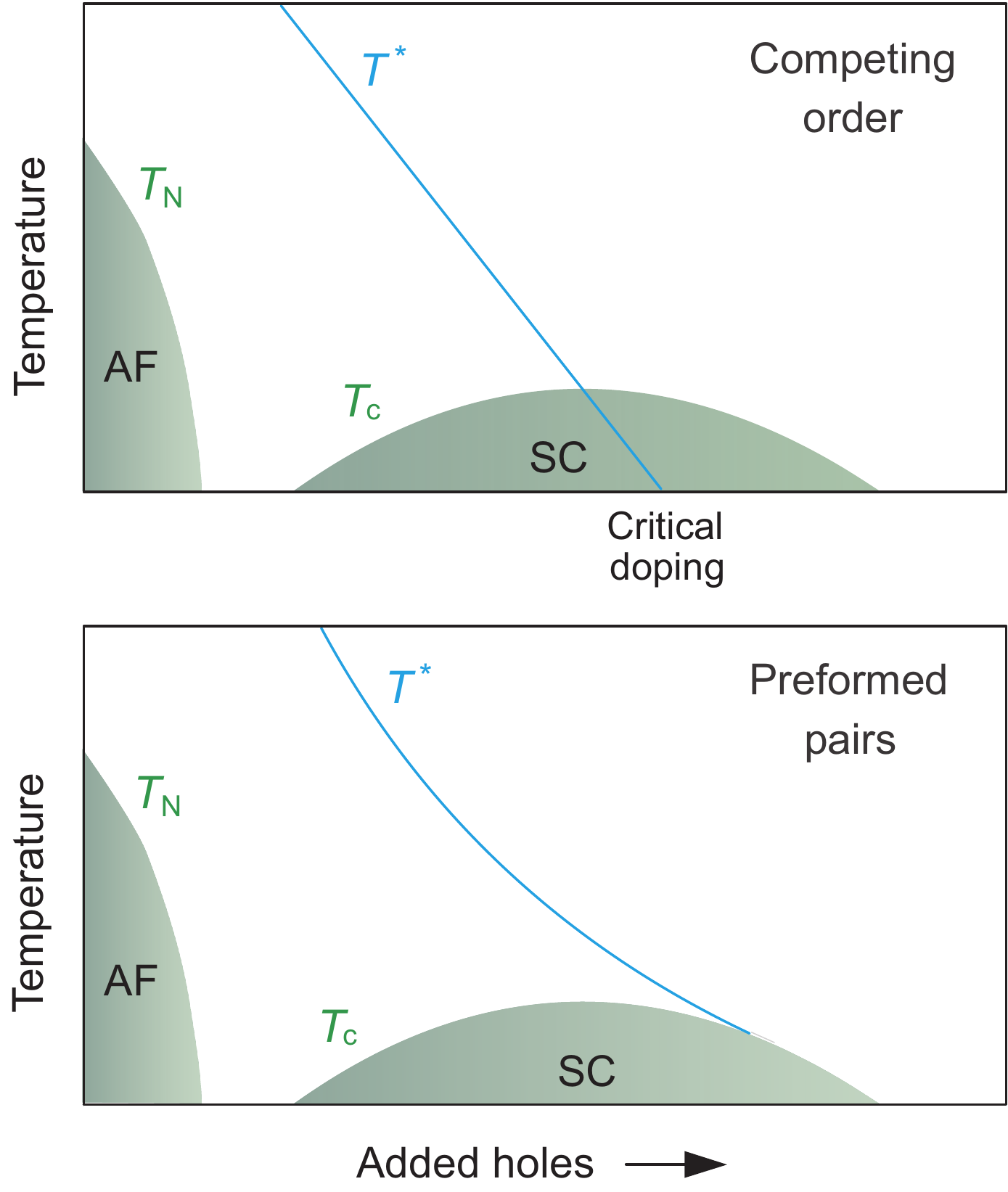}    
{0pt}         
{0pt}         
{0.53}         
{Two views of the nature of the pseudogap. In the {\em preformed pairs} picture,
pairs form on a higher-energy scale than the scale on which they condense into
long-range order.  The preformed pairs {\em aid} the formation of the
superconducting condensate since they are precursors to it.  In the {\em
competing order} picture, some other order {\em competes} with the
superconductivity. It must be {\em suppressed} before superconductivity can set
in fully.
}
have dominated conceptual understanding of the pseudogap.

\begin{enumerate}
 \item 
The {\em preformed pairs picture} assumes that the pseudogap is associated with
formation of correlated pairs at a pseudogap temperature $T^*$, but that these
condense into a state having long-range pairing order---a superconductor---only
at a lower temperature $T\tsub c$ (for example, see Ref.~\cite{EK95}).  
\item
The {\em competing order picture} assumes that pseudogap properties result 
because another form of order competes with superconductivity for $T\tsub c < T 
< T^*$, and the competing order must be suppressed before robust 
superconductivity can appear (for example, see Ref.~\cite{tall01}). 
\end{enumerate}
The order competing with superconductivity often is assumed to involve
antiferromagnetism or charge degrees of freedom, and competing-order approaches
often conjecture a quantum phase transition between states dominated by the two
forms of order, with  quantum-critical scaling that is assumed to
account for many HTSC features, but in a way that is not often clearly
elucidated.

Preformed pairs and competing order are viewed commonly as incompatible
alternatives, and many papers claiming evidence for one or the other view may be
found in the literature.  These seemingly contradictory results raise the issue
of whether the preformed pair and competing order pictures need be mutually
exclusive.  We shall argue that in a solvable theory of adequate complexity,
explanation of the cuprate pseudogap is no longer an either/or choice between
preformed pairs or competing order, but rather is deeply and essentially a
consequence of both.

The question of pseudogap origin is related closely to another. Competing order, 
at
least at a mean-field level, suggests an association of the pseudogap with a 
phase,
but it has proven difficult to find any order parameters that characterize the 
PG
states systematically across all compounds. Thus a common opinion is that the
transition to the pseudogap state may be a crossover and not a true phase 
transition.
Yet there is significant evidence for competing order in the cuprates, and 
mean-field pictures can account for many cuprate properties by treating the PG 
state
as a phase characterized by some order parameter.
 
Hence, an important question that we will address is whether competing order 
is consistent with the
scarcity of evidence for a clear phase associated with the pseudogap state. We 
shall
show that in the pseudogap region the correlated many-body system becomes 
uniquely
susceptible to quantum fluctuations in the pairing and antiferromagnetic degrees 
of
freedom, and that these
fluctuations can reconcile the successes of a mean-field approach to the 
pseudogap
with the elusiveness of well-defined order parameters associated with the 
pseudogap.

\subsubsection{Spatial Inhomogeneity but a Universal Phase Diagram}

There is evidence for a rather universal cuprate phase diagram (at least for
hole-doped compounds), but there is at the same time strong indication of a 
broad
variety of disorder in these same compounds. For example, cuprate 
high-temperature
superconductors exhibit various spatial inhomogeneities such as stripes or
checkerboards, particularly for lower hole doping and near magnetic vortex cores
\cite{hayd04,tran04,hoff02,vers04,hana04}. The behavior of the iron compounds is 
more
varied, but there is a strong suggestion of universal properties in the phase
diagram, and there is evidence for coexistence of AF and SC order both 
homogeneously
(the same patch exhibiting AF and SC order) and inhomogeneously (AF and SC order 
in
separate patches on the nanoscale). How does one reconcile evidence for a 
relatively
universal phase diagram with evidence of a rich variety of inhomogeneity for
individual compounds?

The relationship of such inhomogeneity to the unusual properties of these
systems is not well established.  Does it oppose superconductivity, does it 
enable
superconductivity, or is it a distraction?  Dopant atoms may
favor superconductivity globally by enhancing charge carrier density,
but may suppress superconductivity locally by inducing atomic-scale disorder.
For example, strong disorder was found in atomically-resolved
scanning tunneling microscope images of the superconducting gap for Bi-2212
\cite{mcel05}, and it was concluded that this disorder derives primarily from
dopant impurities. However,  the charge variation between nanoregions was found
to be small, implying that inhomogeneity may be tied to impurities but need not
necessarily couple strongly to charge. 

We shall show that such inhomogeneities follow generically from perturbations on 
the
AF and SC correlations,  largely independent of specifics and not necessarily
coupled to charge variation. Further, we shall show that these properties are
consistent with a global cuprate phase diagram, are directly related to the 
nature of
pseudogap states, and imply a linkage among pseudogaps, inhomogeneity, and 
emergent
behavior.  Thus we shall propose a testable hypothesis for separating primary
features from derivative features in the high-$T\tsub c$ data set.

\subsection{Addressing These Issues within a Unified Framework}

In this paper we use dynamical symmetries implemented in terms of Lie algebras
and generalized coherent state to address all of these issues simultaneously.
Specifically, we demonstrate in a solvable model motivated by cuprate
phenomenology that the ground state of a minimal implementation of competing
singlet pairing, antiferromagnetism, charge, and spin is an antiferromagnetic
Mott insulator that is fundamentally unstable with respect to condensing
electron hole pairs (and thus becoming a superconductor) at any finite hole
doping.  Furthermore, the same solution implies pseudogap states having many
properties that are in quantitative agreement with data.  We shall show that
this ground state is unique near half filling and near optimal doping and 
beyond, but
can become highly degenerate in the underdoped region.  This degeneracy implies 
extreme sensitivity to  perturbations and thus to a
variety of induced inhomogeneity and other emergent behavior, but only in a 
narrow range of doping for underdoped compounds. 

Then, we shall argue that the dynamical symmetry motivated by cuprate
phenomenology is---despite superficial differences---also appropriate for a
description of the Fe-based superconductors, thus providing a unified
description of cuprate and iron superconductors. Indeed we shall argue that
superficial differences between the cuprate and iron superconductors (for 
example,
different orbital pairing symmetry) are not essential to a unified description
at the broad level of understanding emergent superconductivity in these systems.

We shall then discuss the relationship of the theory derived in the present
paper to various other proposals to explain high-temperature superconductors. We
shall show, for example, that SU(4) coherent states contain much of the physics
of resonating valence bond (RVB) models, but without some specifically RVB
assumptions and with broader physics than RVB, that the present dynamical
symmetry methodology provides a microscopic derivation of the Zhang SO(5) model
as one of its approximate solutions, and that Hubbard or $t$-$J$ models and the
present methodology are related by the adoption of fundamentally different (but
equally valid) approaches to truncation of the Hilbert space for the full
problem.

Finally, based on the experience with normal and unconventional superconductors
discussed in this review and in the review of nuclear structure physics
discussed in Ref.~\cite{FDSM}, we shall suggest an even more sweeping
conjecture.  Although superconductivity occurs in many forms in a variety of
disciplines, we believe that all of these forms may represent a single basic
mechanism involving a Cooper pairing instability that may occur in the presence
of other strongly collective modes, and that has a common algebraic description
across diverse systems in terms of fermion dynamical symmetries.

\section{\label{typesTruncation} Truncation of Large Hilbert Spaces}

In complicated many-body systems even the minimal Hilbert space is enormous and 
tractable
theories must reduce this to a more manageable subspace. There are two common
philosophies that may be followed in implementing such a truncation, which we
shall term {\em microscopic-properties truncation} and {\em emergent-symmetry 
truncation.}

\subsection{Truncation Based on Microscopic Properties of the
Weakly-Interacting System}

Microscopic-properties truncation identifies key microscopic physical
features of the idealized weakly-interacting system that are expected to be
valid for the actual correlated many-body system, and uses that as a guide for
truncating the full space. Typically Hubbard or $t$--$J$ model approaches are of
this form. An assumption is made about the microscopic form of the important
physical interactions and this is used to construct a simple Hamiltonian.  In
principle this implies no truncation of the configuration space, but in practice
calculations are possible only if a small-enough subspace is chosen.  The 
choice of this
space is  guided by microscopic physical insight and symmetries of the
interactions (for example, a basis of spin states), often assuming that  only 
states
below some energy cutoff in the non-interacting basis contribute.

However, it is often not feasible to implement microscopic-properties truncation
without drastic assumption. In high-$T\tsub c$ superconductors, the correlations 
may
be so strong that dynamics can no longer be given a meaningful description in 
terms
of individual fundamental particles of the weakly-interacting system (see, for
example, the discussion in Refs.~\cite{An00,pines00,laug00}). The essential 
physics then tends 
to be
governed by a few collective modes that are  {\em emergent.} (An emergent mode 
is a
new collective state that {\em emerges} because of interactions and not because 
it
exists already in the microscopic constituents of the non-interacting system). 
Then a
quite different kind of simplification is possible, based on what we shall term
emergent-symmetry truncation.

\subsection{Emergent-Symmetry Truncation}

Truncation based on emergent symmetries is  tailored for collective modes and
long-range order. It identifies essential forms of the collective modes of
interest and uses that as a guide to remove from the full space all states 
inconsistent with these forms, leaving a small collective subspace that is
highly sympathetic to the relevant collective modes.  In picturesque terms, we
might also call this {\em Michelangelo truncation,} since the famous sculptor is
said to have replied to a query about how he made such beautiful statues that he
looked at the block of stone, envisioned the statue trapped within it, and then
chipped away everything that wasn't statue.

The most powerful systematic method of determining the essential form of
collective modes is to identify a symmetry associated with them.  We shall call
this a symmetry of emergent modes, as opposed to symmetries of the
weakly-interacting system. Note that if a mode is  emergent
there is no reason to expect its dynamical symmetries to have any direct
relationship with the symmetries of the Hamiltonian for the weakly-interacting
system in the full space before truncation.

\subsection{Spontaneously-Broken Symmetries}

This latter point is sufficiently important to merit further attention. A
symmetry that is important at both the microscopic and collective
levels is that associated with angular momentum (isotropy of space), which is
conserved microscopically and also in the collective mode if it is treated
exactly. But if we demand that angular momentum be conserved both
microscopically and in the collective mode, the use of symmetry to truncate the
space is limited to using the Clebsch--Gordan series to decompose
the space into a finite number of subspaces that can be solved independently
because each is labeled by a conserved angular momentum quantum number---in
matrix language, the full matrix is transformed into one with
block-diagonal form by a similarity transformation, permitting each
block-diagonal submatrix to be solved independently.  

A far more spectacular simplification results if we give up the requirement of
angular momentum conservation for the collective state and treat it as
geometrically deformed, thereby breaking rotational invariance. Then the
symmetries of the emergent collective state do not include that of rotational
invariance and the solution fails to conserve angular momentum. We then say that
the emergent state breaks the symmetry of the true Hamiltonian for the system
(which certainly does conserve angular momentum) spontaneously, because the
Hamiltonian is symmetric but the wavefunction is not. Strictly, such a state is
unphysical, because it breaks a symmetry observed by the exact solution.
However, the broken-symmetry approximate solution can often be a very useful
simplification because (1)~the symmetry violation may not be very important for
many physical properties of the system, and (2)~there are established techniques
to restore the broken symmetry of the collective state by the use of projection
integrals when it is important to do so.

\subsection{Examples of Emergent Symmetries}

The Bohr--Mottelson approach to nuclear structure physics utilizes as
approximate solutions to the nuclear many-body problem deformed states that may
break rotational invariance \cite{bohr-mottelson}. In the language employed
here, these imply emergent symmetries. A second example of the emergent
symmetries described above is the ordinary BCS superconducting state, which is
separated by a phase transition from the normal state and cannot be reached by
perturbation of the normal state. If we view the SC state as being described by
a symmetry [pseudospin SU(2) in the case of a simple BCS superconductor], this
symmetry is a symmetry of the collective many-body state, not of the underlying
microscopic system. It is emergent.  In this example, the BCS state may be
viewed as a state that breaks rotational invariance in gauge space; more
prosaically, it fails to conserve particle number
\footnote{
The broken particle number symmetry can be restored by 
particle-number projection, but in practice this procedure may not be necessary 
as we are dealing with a system having a very large number of fermions. 
}.

The SU(4) symmetry described in this review may be understood as a very
sophisticated pseudospin model embodying emergent symmetries associated with the
collective modes of the system. In the symmetry limits it is an exact many-body
solution, but in the coherent state approximation that we shall often employ it
becomes a spontaneously-broken symmetry.

\section{\label{dynamicalSymmetryMethod} The Dynamical Symmetry Method}

This review is about emergent-symmetry truncation of a Hilbert space
corresponding to the strongly-correlated electron problem.  The approach that we
shall use will rely upon the method of dynamical symmetries to identify the
emergent collective subspace and implement the corresponding truncation.  It
begins with the following conjecture \cite{FDSM}:

\doTheorem{Strongly correlated modes in fermion or boson many-body quantum
systems imply a corresponding dynamical symmetry (a symmetry of the Hamiltonian
or Lagrangian dynamics) described by a Lie algebra in the second-quantized
operators representing the physical modes of the system. }

\noindent
This is a conjecture, but there is a large amount of very strong circumstantial
evidence to support its validity from various fields of many-body physics
\cite{FDSM,IBM,Bi94,Ia95,Ia99,guid01}.

\subsection{Solution Algorithm}

Assuming the validity of the preceding conjecture, we may implement the
following algorithm.

\smallskip\noindent
1.~Identify a minimal set of emergent-state degrees of freedom thought to be
physically relevant for the problem at hand, guided by phenomenology and
theory.  

\medskip\noindent
2.~Close a commutation algebra of manageable dimensionality on the
second-quantized operators creating, annihilating, and counting the modes chosen
in the first step, meaning that when all possible bilinear forms of this set of 
operators
are commuted, the result is always a linear combination of the full operator
set. This Lie algebra is termed the {\em highest symmetry} of the problem, and
may be specified completely in terms of the generators for emergent physical 
modes in the
system, expressed as operators in second-quantized form. 

\medskip\noindent
3.~Identify a collective subspace of the full Hilbert space by requiring that 
matrix elements of the operators found in the preceding step 
do not cause transitions out of the collective subspace.  
This (typically dramatic) reduction of the full space is termed {\em 
symmetry-dictated truncation.} The collective (emergent) states in this 
subspace will be of low energy but their wavefunction components  are 
{\em selected by symmetry, not energy,} and may contain a mixture of both low 
and high energy pieces of the basis for the weakly interacting system.

\medskip\noindent
4.~Use standard methods to identify subalgebra chains of the highest algebraic
structure that end in algebras for relevant conservation laws, such as those for
charge and spin. Associated with each Lie algebra and subalgebra will be a
corresponding {\em Lie group.} Each such subalgebra chain or corresponding
subgroup chain defines a {\em dynamical symmetry} of the highest
symmetry. Generally, more than one dynamical symmetry may be
associated with a given highest symmetry.

\medskip\noindent
5.~Construct Hamiltonians that are polynomials in the Casimir invariants ({\em 
dynamical symmetry Hamiltonians}) for each chain. Each symmetry
chain defines a wavefunction basis labeled  by the eigenvalues of chain
invariants (the Casimirs and the elements of the Cartan subalgebras), and a
Hamiltonian that is diagonal in that basis because it is constructed explicitly
from invariants.  Thus, the Sch\"odinger equation is solved analytically for
each chain, by construction.

\medskip\noindent
6.~Calculate the physical implications of each of these dynamical symmetries by
considering the diagonal and transitional matrix elements of physical relevance
for the problem at hand. This is possible because of the eigenvalues and
eigenvectors that were obtained in step 4, and because consistency of the
symmetry
requires that transition operators be related to group generators; otherwise
transitions would mix irreducible multiplets and break the symmetry.

\medskip\noindent
7.~If the results of step (5) agree with experimental observables, 
indicating that a wise choice was made in step (1), construct the most general
Hamiltonian  in the model space, which is  a linear combination of
the terms in all the Hamiltonians for the symmetry group chains. The Casimir
operators of different group chains do not generally commute with each other, so
a Casimir invariant for one group chain may be a symmetry-breaking term for
another group chain.  Thus the competition between different dynamical
symmetries and the corresponding (quantum) phase transitions may be studied.

\medskip\noindent
8.~Symmetry-limit solutions may be used as a starting point for more ambitious
calculations that incorporate symmetry-breaking terms.  Such more realistic
approximations may be solved by (a)~perturbation theory around the symmetry
solutions (which are generally non-perturbative, so this corresponds to
perturbation theory around a non-perturbative vacuum), (b)~by numerical
diagonalization of symmetry-breaking terms, or (c)~by coherent-state or other
approximations to the full Hamiltonian described above in step (6).

\medskip\noindent
Representative application of these ideas for both fermion and boson systems 
may 
be found in  Refs.~\cite{FDSM,IBM,Bi94,Ia95,Ia99,guid01}. Our primary interest 
here is in strongly-correlated electron systems so we shall deal only with 
fermionic applications.  Note also that the central concept employed here that 
symmetry can have dynamical and not just conservation-law implications, and 
that 
non-abelian symmetries imply theories that are at once richer and have fewer 
free parameters than abelian counterparts, are also key ingredients of local 
non-abelian gauge field theories in elementary particle physics, though the 
degrees of freedom and methodology are different there \cite{guid92}.

\subsection{Validity and Utility of the Approach}

The {\em only approximation} to the full quantum-mechanical problem in our
approach is the space truncation.  If all degrees of freedom are
incorporated the resulting theory is microscopic and exact.  Of course,
practically only a few carefully-selected degrees of freedom can be
included and the effect of the excluded space must be incorporated through 
renormalized (effective) interactions operating in the truncated space.  

Thus, the utility of this approach depends on making a wise choice for the
relevant collective degrees of freedom and on the availability of sufficient
phenomenological or theoretical information to specify the effective
interactions that operate within the truncated space. The validity of the
resulting formalism then stands on whether  predicted matrix elements agree
with corresponding physical observables, once a small set of
effective interaction parameters has been fixed  by
comparison with the global data set.

\section{\label{su4Model} Strongly-Correlated SU(4) Electrons}

Let us now introduce a formalism based on the approach described in
\S\ref{dynamicalSymmetryMethod} that is capable of dealing with the issues
outlined in \S\ref{intro}, by virtue of being complex enough to incorporate the
essential physics and yet amenable to solutions that may be compared
with data. We do so through a theory of strongly-correlated
electrons that uses the power of Lie algebras, Lie groups, and generalized
coherent states  to truncate the Hilbert space to a manageable collective
subspace
\cite{guid99,guid01,lawu03,guid04,sun05,sun06,sun07,guid08,sun09,guid09,guid09b,
guid10,guid11,FDSM}.

\subsection{Structure of the Coherent Pair Basis}

Motivated  by  phenomenology of the cuprates, our basic physical
assumption is that the configuration space for a minimal theory of 
high-temperature
superconductivity is built from {\it coherent pairs} representing superpositions 
of
particles or holes centered on different lattice sites (which we shall term {\em
bondwise pairs;} see \fig{bondwise-onsitePairs}b). In the interest of 
constructing a 
minimal theory, we shall neglect pairing between next nearest neighbors. We 
shall also consider an extended
theory including in addition pairs of particles or holes defined on the same 
lattice
sites (which we shall term {\em onsite pairs;} see \fig{bondwise-onsitePairs}a).
However, as will be discussed further in \S\ref{reduction_so8-su4} below, 
such onsite
pairing configurations are important in conventional superconductivity but are 
likely
of less importance for the low-lying states in high-temperature superconductors. 
The
next step is to use phenomenology as a guide to identify a set of operators 
that 
can be
associated with the relevant physical degrees of freedom exhibited by this
collective subspace.
\singlefig
{bondwise-onsitePairs}       
{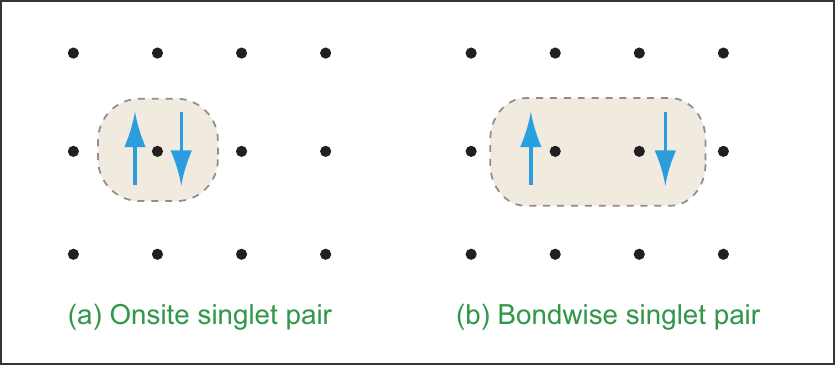}    
{0pt}         
{0pt}         
{1.01}         
{Onsite and nearest-neighbor bondwise singlet pairs. }

\subsection{\label{su4Operators} The Collective Operators}

We propose to solve for the doping and temperature dependence of observables in
a theory that incorporates on an equal footing antiferromagnetism and (possibly
unconventional) superconductivity. To construct a Hamiltonian embodying these
degrees of freedom, and conservation laws for charge and spin, we employ the
concept of a {\em complete set of quantum operators,}  which may be defined in
either physical or mathematical terms.  (1)~Physically, a complete set of
quantum operators represents all degrees of freedom produced if an initial set
of operators is allowed to undergo all possible (those not forbidden by
fundamental principles) interactions among themselves.  (2)~Mathematically, a
complete set of quantum operators corresponds to a set of creation and
annihilation operators closed under the operation of commutation, implying that
the operators form a {\em Lie algebra}.

For the high temperature superconductor problem we require at a minimum three
staggered magnetization operators $\vec Q$ to describe antiferromagnetism,
creation and annihilation operators $\singletPair^\dagger$ and $\singletPair$
for bondwise singlet pairs plus a number operator $\hat n$ to describe
superconductivity, and three spin operators $\vec S$ to describe electron spin. 
However, this set of nine operators is physically incomplete since scattering of
singlet pairs (antiparallel spins on adjacent sites) from the AF particle--hole
degrees of freedom can produce triplet pairs (parallel spins on adjacent sites),
which are not part of the operator set. This is illustrated in
\fig{completeSet}.%
\singlefig
{completeSet}       
{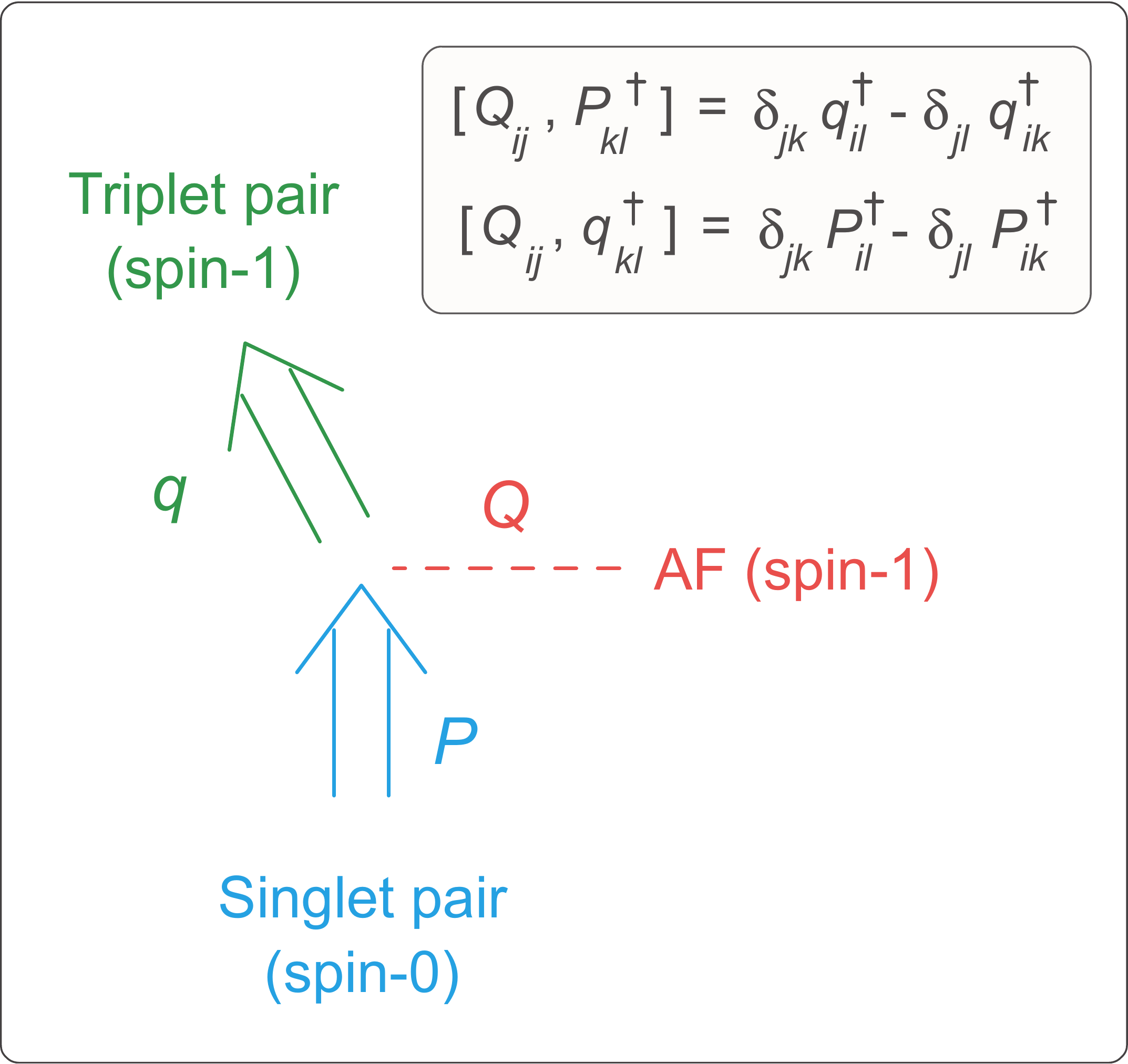}    
{0pt}         
{0pt}         
{0.27}         
{Scattering of singlet pairs by AF
operators will necessarily produce triplet pairs, even if none existed before.
Thus a physically and mathematically consistent Hilbert space must contain both
kinds of pairs in the presence of antiferromagnetism.
 }

The mathematical statement of this incompleteness is that the  set
$\{\vec \AF, \singletPair^\dagger, \singletPair, \hat n, \vec S\}$ does not 
close a
Lie algebra under commutation of set members, because commuting singlet-pair
operators with antiferromagnetic operators produces triplet pair operators and
commuting triplet-pair operators with antiferromagnetic operators produces
singlet pair operators (inset to \fig{completeSet}).  The physical picture of
\fig{completeSet} suggests that a self-consistent Hilbert space containing
singlet pairs must contain triplet pairs also if antiferromagnetic interactions
are present.  As we shall now demonstrate, a minimal physically-complete
operator set results if we add to the original nine operators three creation and
three annihilation operators for bondwise spin-triplet pairs defined on adjacent
lattice sites.

The operators that we shall use are particle--hole symmetric (though our
predictions for observables generally will not be; see the
discussion in note \cite{particle-hole}).
Unless specified explicitly in the following, we shall use ``electrons" to
reference either electrons or electron holes. We begin by introducing the
following 16 operators
\begin{subequations}
\begin{align}
\vec S\ &= \left( \frac{S_{12}+S_{21}}{2}, \ -i \, \frac
{S_{12}-S_{21}}{2}, \ \frac {S_{11}-S_{22}}{2} \right)
\\
\vec \AF \ &= \left(\frac{Q_{12}+
Q_{21}}{2},-i\frac{ Q_{12}- Q_{21}}{2}, \frac{
Q_{11}- Q_{22}}{2} \right)
\\
\vec \pi^\dagger &= \left(\ i\frac {q_{11}^\dagger\ -
q_{22}^\dagger}2, \ \frac{q_{11}^\dagger + q_{22}^\dagger}2, \
-i\frac {q_{12}^\dagger + q_{21}^\dagger}2 \right)
\\
\vec \pi &= \left(\ -i\frac {q_{11} -
q_{22}}{2}, \ \frac{q_{11} + q_{22}}{2}, \
i\frac {q_{12} + q_{21}}{2} \right)
\\
\singletPair^\dagger &= p^\dagger_{12} \qquad \singletPair = p_{12}
\\
\hat{n}\ &= \sum_{k,i}
c_{k,i}^\dagger c_{k,i} =S_{11}+S_{22}+\Omega\
\label{operatorset_n}
\\
Q_+ &= Q_{11}+Q_{22} = \sum_k (c_{k+Q\uparrow}^\dagger 
c_{k\uparrow} + c_{k+Q\downarrow}^\dagger c_{k\downarrow})
\label{operatorsetQ+}
\end{align}
\label{operatorset}
\end{subequations}
in which we define
\begin{subequations}
 \label{E1}
\begin{align}
\singletPair^\dagger &= \sum_{\bm k b b'} g(\bm k) \alpha_{\bm k b}
\alpha_{\bm -k b'}
c_{\bm k b \uparrow}^\dagger
c_{-\bm k b' \downarrow}^\dagger
\quad \singletPair=(\singletPair^\dagger)^\dagger
\label{E1.1}
\\
q_{ij}^\dagger &= \sum_{\bm k b b'} g(\bm k) \alpha_{\bm k + \bm Q, b}
\alpha_{\bm -k b'}
c_{\bm k+\bm Q,b i}^\dagger c_{-\bm k,b' j}^\dagger
\quad q = (q^\dagger)^\dagger
\label{E1.2}
\\
Q_{ij} &= \sum_{\bm k b b'} \alpha_{\bm k + \bm Q, b} \alpha^*_{\bm k
b'}
c_{\bm k+\bm Q,b i}^\dagger c_{\bm k b' j}
\label{E1.3}
\\
 S_{ij} &=
\sum_{\bm k b b'} \alpha_{\bm k b} \alpha_{\bm k b'}^*
c_{\bm k,b i}^\dagger c_{\bm k,b' j} - \tfrac12 \Omega \delta_{ij}
\label{E1.4}
\end{align}
\end{subequations}
where $\alpha_{\bm kb}$ is the amplitude to find an electron in a band labeled
by $b$ with momentum $\bm k$, $c_{\bm k,b, i}^\dagger$ creates a fermion of
momentum $\bm k$ and spin projection $i,j= 1 {\rm\ or\ }2 = \ \uparrow$ or
$\downarrow$ in band $b$, $\bm Q$ is an AF ordering vector, $\Omega$ is the
effective lattice degeneracy, which is the maximum allowed number of doped
electrons that can form coherent SU(4) pairs (explained further below), and
$g(\bm k)$ is a pairing formfactor. We may attach a simple physical
interpretation to the operators in \eqs{operatorset}:
\begin{itemize}
 \item 
The vector $\vec S$ is the electron spin operator. 
\item
The vector $\vec \AF $ is the staggered magnetization characterizing the
antiferromagnetism. 
\item
$\vec \pi^\dagger$ ($\vec \pi$) is a vector  of creation
(annihilation) operators for bondwise spin-triplet pairs. 
\item
$\singletPair^\dagger$ ($\singletPair$) is a creation (annihilation) operator
for bondwise singlet pairs. 
\item
$\hat{n}$  is the electron number
operator.
\item
$Q_+$ is a commensurate charge density wave operator.
\end{itemize}
It will  sometimes prove useful to replace the number operator $\hat n$ in
\eq{operatorset_n} with 
\begin{equation}
    M=\tfrac12(S_{11}+S_{22})=\tfrac12 (\hat n-\Omega),
\label{chargeOp}
\end{equation}
where we may interpret $M$ physically as the charge operator.

The preceding operators may receive contributions from more than one band.  If
we introduce effective one-band creation and annihilation operators through
\begin{equation}
 a^\dagger_{ki} = \sum _b \alpha_{\bm k b} c^\dagger_{\bm k b i}
\qquad
a_{\bm k i} = (a^\dagger_{ki})^\dagger
\qquad
\sum_b |\alpha_{\bm k b}|^2 = 1,
\label{oneBandEff}
\end{equation}
then \eqs{E1} may we written as
\begin{subequations}
\label{E3}
\begin{align}
\singletPair^\dagger&=\sum_{\bm k} g(\bm k) c_{\bm k\uparrow}^\dagger
c_{-\bm k\downarrow}^\dagger
\qquad \singletPair=(\singletPair^\dagger)^\dagger
\label{E3.1}
\\
q_{ij}^\dagger &= \sum_{\bm k} g(\bm k) c_{\bm k+\bm Q,i}^\dagger
c_{-\bm
k,j}^\dagger
\qquad q = (q^\dagger)^\dagger
\label{E3.2}
\\
Q_{ij} &= \sum_{\bm k} c_{\bm k+\bm Q,i}^\dagger c_{\bm k,j} \qquad
S_{ij} =
\sum_{\bm k}
c_{\bm k,i}^\dagger c_{\bm k,j} - \tfrac12 \Omega \delta_{ij} ,
\label{E3.3}
\end{align}
\end{subequations}
which is the form of the  operators for the original single-band SU(4) model
introduced in Ref.~\cite{guid01}. Although \eqs{E3} are an adequate
starting point for discussion of cuprate superconductivity, the multiband
expressions in \eqs{E1} are more appropriate for applications such as
the iron-based superconductors where 
multiband pairing is important.

\subsection{\label{su4Algebra} The SU(4) Algebra and Subalgebras}

Inserting the AF ordering vector $\bm Q = (Q_x, Q_y) = (0, \pi)$ appropriate for
the observed FeAs magnetic structure, or $\bm Q = (Q_x, Q_y) = (\pi, \pi)$
appropriate for the cuprates, and calculating all commutators for the 16
operators in \eqs{operatorset}, the set is found to be closed under commutation
if three conditions are satisfied by the pairing formfactor:
\begin{equation}
g(\bm k) = g(-\bm k)\qquad g(\bm k + \bm Q) = \pm g(\bm k)
\qquad |g(\bm k)| = 1.
\label{conditions}
\end{equation}
(We shall elaborate on the physical meaning of these
constraints in \S\ref{closureConditions}.)
If these conditions are met, the operators defined in
\eq{operatorset} close a $\ufour\supset\uone \times \sufour $ Lie algebra,
where the U(1) factor is generated by the commensurate charge density wave
operator $Q_+$, which commutes with all other generators \cite{guid01,guid04}.
Because of the direct-product structure, one can without loss of generality view
the theory as an SU(4) theory describing superconductivity and
antiferromagnetism, and global charge and spin conservation, with the U(1)
charge-density wave sector treated independently. This SU(4) group has three
independent subgroup chains 
%
\newcommand{\supsetRL}{\mathbin{\rotatebox[origin=c]{45}{$\supset$}}}
\newcommand{\supsetRR}{\mathbin{\rotatebox[origin=c]{-45}{$\supset$}}}
\begin{subequations}
\begin{align}
&\supsetRL \ \sofour \times \uone \supset \sutwo\tsub{s} 
\times \uone
\label{so4chain}
 \\ 
\sufour \ &\ \supset \sofive
\supset \sutwo\tsub{s} \times \uone
\label{so5chain}
\\ &\supsetRR \ \sutwo\tsub{p}
\times \sutwo\tsub{s} \supset \sutwo\tsub{s} \times \uone
\label{su2chain}
\end{align}
\label{eq3}%
\end{subequations}
ending in the subgroup $\sutwo\tsub{s} \times \uone$ representing spin
and charge conservation, with the U(1) subgroup being generated by the charge
operator $M$ and the $\sutwo\tsub{s}$ subgroup being generated by the three spin
operators $S_1$, $S_2$, and $S_3$. Some important properties of this group and
subgroup structure are summarized in Tables \ref{table:su4Properties} and 
\ref{table:symmetryLims} of Appendix \ref{appendix}. 


\setcounter{table}{2}

As we shall discuss further below, the subgroup chains defined in 
\eqs{eq3} imply three {\em fermion dynamical symmetries} \cite{FDSM} that will
permit  exact many-body solutions to be obtained for particular ratios of
antiferromagnetic and pairing coupling strengths. The methodology is that of
{\em symmetry-dictated truncation,} as outlined
in \S\ref{dynamicalSymmetryMethod}.
The subgroup chains of \eqs{eq3} and the corresponding many-body solutions imply
that charge and spin are conserved.  In \S\ref{generalizedCoherentStates} we
shall find it useful to introduce {\em approximate solutions} through coherent
state methods that are generalizations of the BCS solution and lead to
spontaneous symmetry breaking and to intrinsic states violating particle number
conservation, but the starting point \eqnoeq{eq3} conserves both charge and
spin.

\subsection{\label{collSub} Collective Subspace and Associated Hamiltonian}

The group SU(4) is rank-3 and the irreducible representations (irreps)  may be
labeled by three weight-space quantum numbers, $(\sigma_1,\sigma_2,\sigma_3)$
\cite{su4so6}. We assume  a collective subspace illustrated  in
\fig{truncatedSpace}%
\singlefig
{truncatedSpace}       
{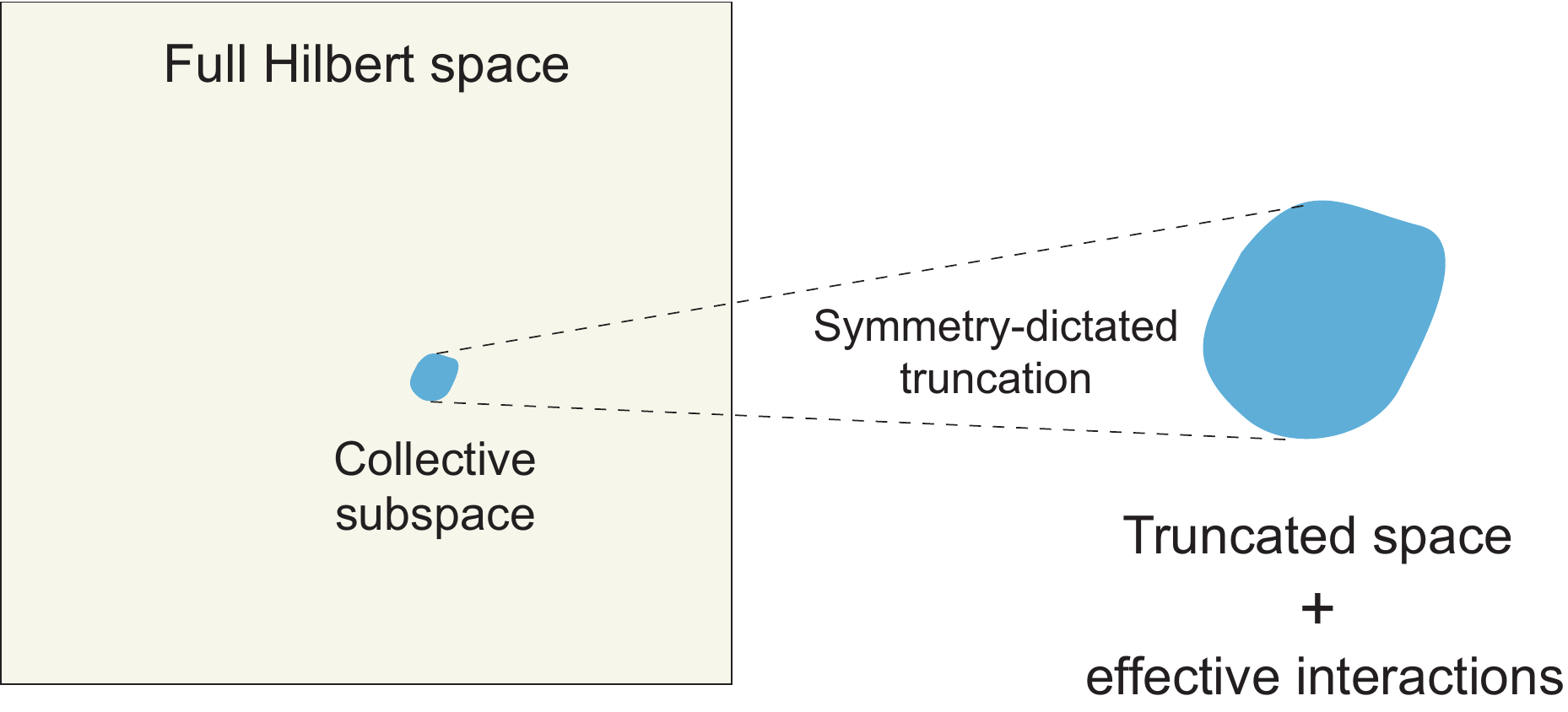}
{0pt}         
{0pt}         
{0.45}         
{Symmetry-dictated truncation of the full Hilbert space to a small collective
subspace. 
The actual collective subspace for the present
discussion is miniscule compared with the full space.
}
that is spanned by the vectors
\begin{equation}
\ket{S} = \ket{n_x n_y n_z n_s} = 
(\pi_x^\dagger)^{n_x}
(\pi_y^\dagger)^{n_y}
(\pi_z^\dagger)^{n_z}
(\singletPair^\dagger)^{n_s}
\ket{0} .
\label{collsubspace}
\end{equation}
If there are no unpaired particles, the collective subspace is associated with
``fully-stretched'', and therefore maximally collective, irreducible 
representations (irreps) of the form
\begin{equation}
(\sigma_1,\sigma_2,\sigma_3) = \left(\frac \Omega2,0,0 \right) .
\label{fullyStretched}
\end{equation}
(More general representations having broken pairs of particles are discussed in
Ref.~\cite{sun05}.) A subspace associated with such maximally-collective
configurations is an obvious candidate for describing the lowest-energy states
of the system. This wavefunction represents a coherent superposition of pairs
and implies a rather rich structure, as illustrated for the real space in
\fig{linearCoordinateSpace}.%
\singlefig
{linearCoordinateSpace}       
{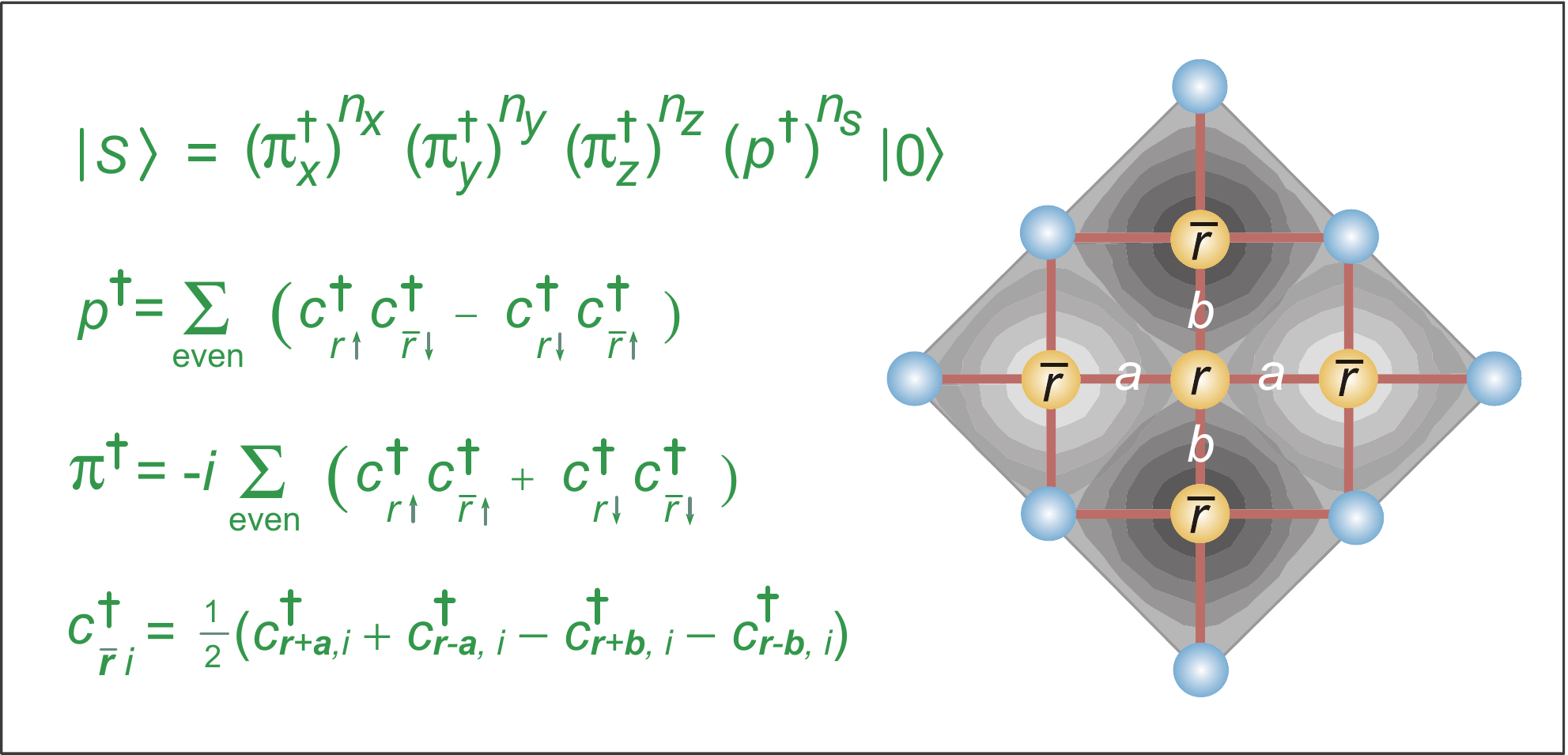}    
{0pt}         
{0pt}         
{0.43}         
{Linear superpositions in the coordinate space implied by the wavefunction
\eqnoeq{collsubspace}. The inset figure is discussed more extensively in
connection with \fig{schematicHolePair}.}

The commensurate charge density wave operator $Q_+$ defined in
\eq{operatorsetQ+} is the generator of the $\uone$ factor in $\ufour \supset
\uone \times \sufour$.  It commutes with all generators so it annihilates the
state $\ket S$ and
\begin{equation}
\mel S{Q_+}S = 0 ,
\end{equation}
which means that charge-density wave excitations are excluded in the symmetry
limits for the low-lying collective subspace of the effective theory. 

Thus we
shall ignore the U(1) charge-density wave for much of the subsequent discussion
and concentrate on the antiferromagnetic--superconductivity competition embodied
in the SU(4) symmetry. However, we shall address the issue of possible charge
degrees of freedom induced by perturbations of the SU(4) subspace later in
\S\ref{criticalInhomo}.

The group SU(4)  has a quadratic Casimir operator (see Table 
\ref{table:su4Properties} of 
Appendix \ref{appendix}),
\begin{equation}
C\tsub{su(4)}=\pidagpi + p^\dagger p +
\SdotS + \QdotQ + M(M-4) ,
\label{csu4}
\end{equation}
and the corresponding expectation value of the  SU(4) quadratic Casimir 
operator evaluated in the
irreducible representations \eqnoeq{fullyStretched} is a constant,
\begin{equation}
\ev{C\tsub{su(4)}}=\frac\Omega2\left(\frac\Omega2 + 4\right).
\label{constantSU4Cas}
\end{equation}
The most general 2-body Hamiltonian within the collective pair subspace 
consists of
a linear combination of (lowest-order) Casimir operators $C\tsub g$ for all
subgroups g (Table \ref{table:su4Properties} of Appendix \ref{appendix}) 
\cite{threebodynote}.  These are
\begin{subequations}
\begin{align}
C\tsub{so(5)} &= \vec \pi^\dagger \cdot \vec \pi + \vec S \cdot \vec
S +M(M-3)
\label{so5C}  
\\
C\tsub{so(4)} &= \QdotQ + \vec S
\cdot\vec S  
\label{so4C}
\\
C_{\rm\scriptstyle su(2)\tsub{p}} &= \singletPair^\dagger  \singletPair
+M(M-1)
\label{su2pC} 
\\
C_{\rm\scriptstyle su(2)\tsub s} &= \vec S \cdot \vec S
\label{su2sC}
\\
C\tsub{u(1)} &= M \mbox{ and }M^2   .
\label{u1C}
\end{align}
\label{casimirOps}%
\end{subequations}
We may utilize that the SU(4) Casimir expectation value \eqnoeq{constantSU4Cas}
is constant to eliminate terms in $\pidagpi$, and the most general SU(4) 2-body
Hamiltonian can be expressed as \cite{guid01,guid11}
\begin{equation}
H =H_0 -\tilde{G}_0\, [ (1-\pratio) \singletPair^{\dagger }\singletPair +
\pratio \QdotQ ]
+ g' \vec S\cdot \vec S,
\label{eq1}
\end{equation}
where $H_0$, $\tilde G_0$ and $g'$ are effective interaction parameters, 
$\singletPair^\dagger$
creates singlet pairs, $\vec\AF$ is the staggered magnetization, $\vec S$ is
spin, $ \tilde G_0 = \chi(x)+G_0(x), $ and the parameter $\sigma$, given by
\begin{equation}
\pratio = \sigma(x) = \frac{\chi(x)}{\chi(x)+G_0(x)}, \label{pratio}
\end{equation}
with $G_0(x)$ and $\chi(x)$ the effective SC and AF coupling strengths,
respectively, governs
the relative strength of antiferromagnetic and pairing interactions in the 
subspace \cite{sun06,sun07}. In these expressions, doping is characterized by a
parameter 
\begin{equation}
 x = 1-\frac{n}{\Omega}
\label{dopingParam}
\end{equation}
for an $n$-electron system, with $\Omega$ the maximum number of doped holes (or
doped electrons for electron-doped compounds) that can form coherent pairs,
assuming the normal state at half filling ($n=\Omega$, implying $M=0$) to be the
vacuum. Since $\Omega-n$ is the hole number when $n<\Omega$, positive $x$
represents the case of hole-doping, with $x=0$ corresponding to half filling (no
doping) and $x=1$ to maximal hole-doping. Negative $x$ ($n>\Omega$) is then the
relative doping fraction for electron-doping. 

The parameter $x$ is the effective doping concentration, corresponding to the
ratio of the number of doped pairs $\tfrac12 (\Omega-n)$  to the pair
degeneracy  $\tfrac12 \Omega$ of the collective subspace. The physical doping 
fraction,
defined as $P = (\Omega-n)/\Omega_e$, where $\Omega_e$ is the number of physical
lattice sites, is related to $x$ through 
\begin{equation}
P=x\,\frac\Omega{\Omega_e}=x\,P_{\mbox{\scriptsize f}}, \qquad 
P_{\mbox{\scriptsize
f}}\equiv\frac\Omega{\Omega_e}, \label{P-x} 
\end{equation} 
with $P>0$ for hole
doping and $P <0$ for electron doping. $P_{\mbox{\scriptsize f}}$ can be
regarded as the maximum value of the physical doping fraction $P$ that is
consistent with \sufour\ symmetry, as we shall explain further below.

\section{\label{dynamicalSymmetryLimits} The Dynamical Symmetry Limits}

Each of the three dynamical symmetry limits in \eqs{eq3} defines a basis
in which the effective Hamiltonian \eqnoeq{eq1} is diagonal. Thus, they have
exact solutions \cite{guid99,guid01,lawu03} that may be constructed using the
methods developed in Ref.\ \cite{FDSM}. These solutions result from special
choices of the parameter $\sigma$ in the Hamiltonian \eqnoeq{eq1} and are
summarized in \tableref{symlim}.%
\begin{table}[t]
  \centering
  \caption{Dynamical Symmetries of the Hamiltonian \eqnoeq{eq1}}
  \label{table:symlim}
  \begin{normalsize}
    \begin{centering}
      \setlength{\tabcolsep}{4 pt}
      \begin{tabular}{cll}
        \hline
            $\sigma$ &
            Symmetry &
            Physical interpretation

        \\        \hline
            0 &
            SO(4) &
            Antiferromagnetic Mott insulator

        \\[0pt]        
            $\frac12$ &
            SO(5) &
            Critical dynamical symmetry

        \\ [0pt]      
            1 &
            SU(2) &
            $d$-wave singlet superconductor

        \\        \hline
      \end{tabular}
    \end{centering}
  \end{normalsize}
\end{table}

\begin{enumerate}
 \item 
The SO(4) limit corresponds to choosing $\pratio=1$ in \eq{eq1}, which
eliminates the pairing terms in the Hamiltonian. It represents a collective 
Mott-insulator, AF
state defined by the subgroup chain of \eq{so4chain}. 
\item
The $\sutwo\tsub p$ limit [which we will term the SU(2) limit for
brevity] corresponds to choosing $\pratio=0$ in \eq{eq1}, which eliminates the
antiferromagnetic terms in the Hamiltonian. It represents a collective 
$d$-wave singlet SC state
defined by the subgroup chain of \eq{su2chain}.
\item
The SO(5) limit corresponds to choosing $\pratio= \tfrac12$ in \eq{eq1}, which
leads to a Hamiltonian in which pairing and antiferromagnetism enter with equal
strengths. The SO(5) limit represents a {\em critical dynamical symmetry} that
interpolates dynamically between the SC and AF limits.  It is defined by the
subgroup chain of \eq{so5chain}.
\end{enumerate}
The detailed mathematical properties of these symmetry limits are summarized in
Table \ref{table:symmetryLims} of Appendix \ref{appendix} and their schematic 
relationships are displayed in
\fig{dynamicalSymmetryTriangle}.%
\singlefig
{dynamicalSymmetryTriangle}       
{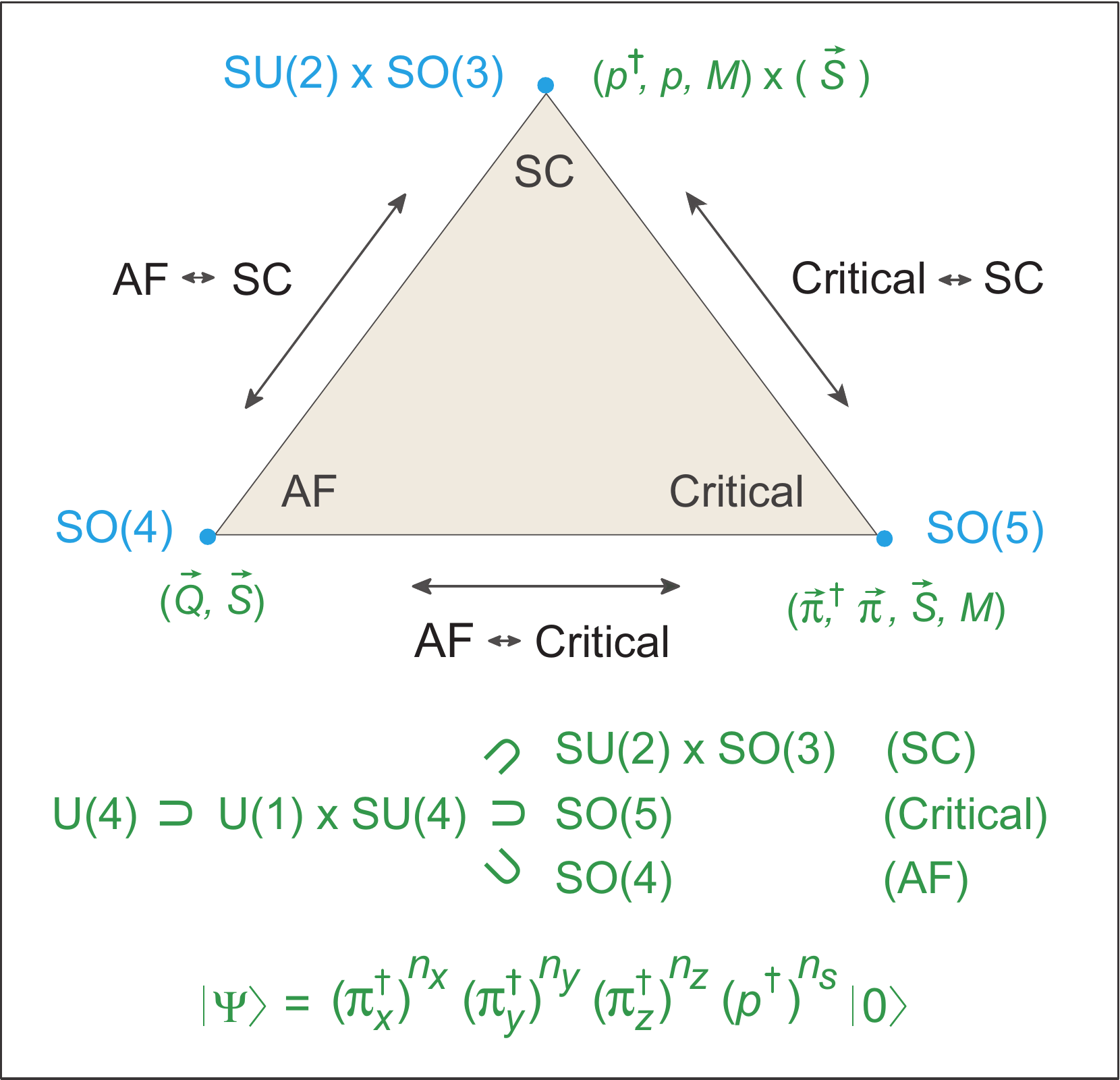}    
{0pt}         
{0pt}         
{0.48}         
{Relationships among the SU(4) dynamical symmetries.}
A generic doping--temperature phase diagram corresponding to these dynamical 
symmetries is
displayed in \fig{genericSU4Phase}.
\singlefig
{genericSU4Phase}       
{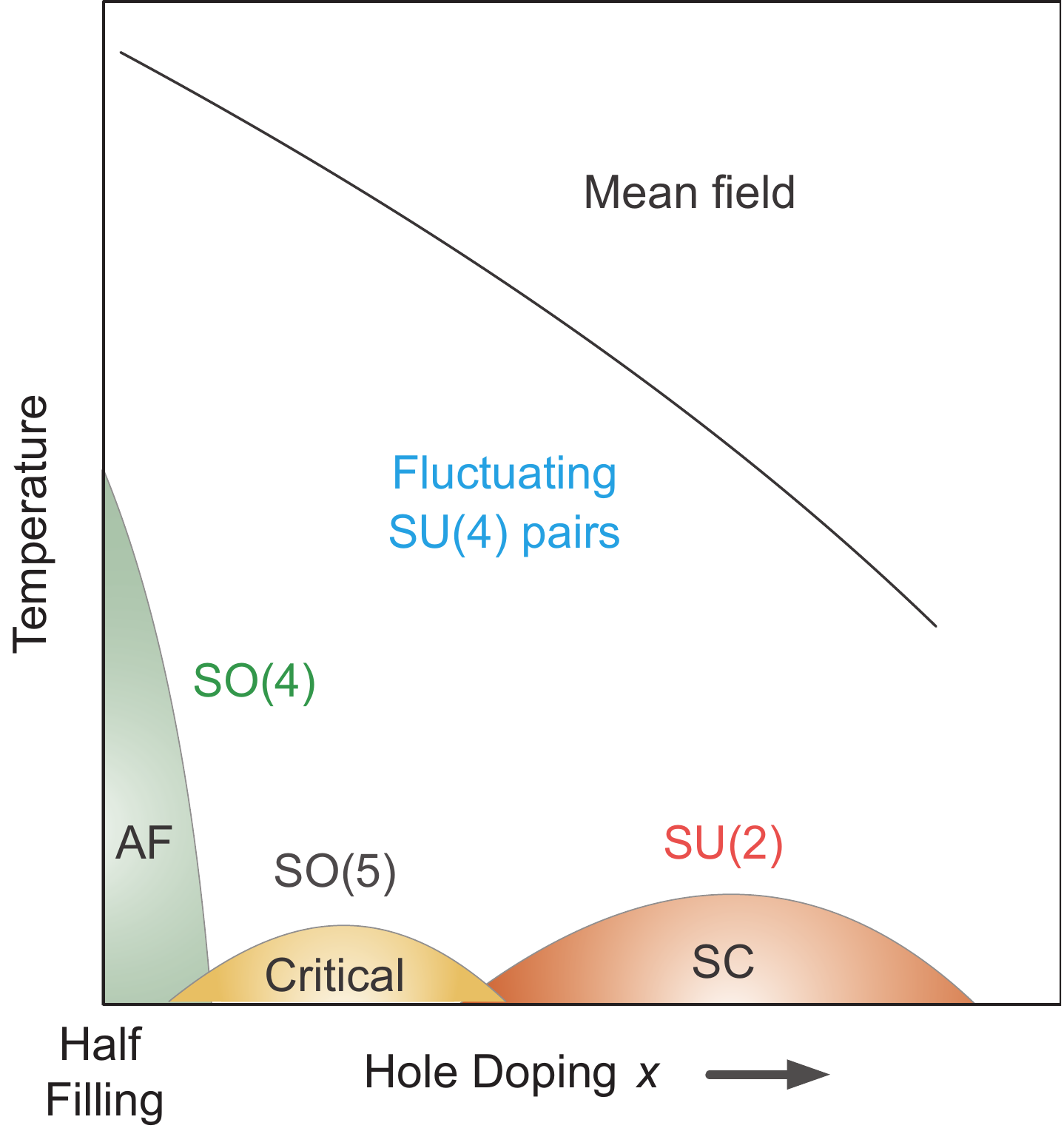}    
{0pt}         
{0pt}         
{0.50}         
{A schematic SU(4) phase diagram. The SO(4) symmetry corresponds to an
antiferromagnetic Mott insulator, the SU(2) symmetry corresponds to a $d$-wave
singlet superconductor, and the SO(5) symmetry corresponds to a critical
dynamical symmetry that interpolates between the AF and SC phases.  A more
quantitative discussion of the expected
phase structure will be given in \S\ref{su4phase}.}
We now discuss the justification for the above identifications and the important
physical properties of the exact solutions in these symmetry limits.

\subsection{\label{so4Dynamical} The SO(4) Dynamical-Symmetry Limit}

The  SU(4) subgroup chain
$$
 \sufour \supset \sofour \times \uone \supset \sutwo\tsub{s} \times \uone, 
$$
which we term the {\em SO(4) dynamical symmetry} for brevity, is the symmetry
limit of \eq{eq1} when $\pratio=1$.  As we now argue, it corresponds to
a collective state having long-range antiferromagnetic order and  Mott insulator
character. The \sofour\ subgroup is locally isomorphic to the product group
$\sutwo_F\times \sutwo_G$ that is generated by the linear combinations
\begin{equation}
\vec F= \tfrac12 (\vec\AF + \vec S)
\qquad
\vec G = \tfrac12 (\vec\AF-\vec S)
\label{su2xsu2}
\end{equation}
of the original \sofour\ generators $\vec\AF$ and $\vec S$. The new generators
$\vec F$ and $\vec G$ may be interpreted physically by noting that if  $Q_{ij}$
and $S_{ij}$ defined in momentum space in \eq{E1} are transformed to the 
physical
coordinate lattice we obtain, 
\begin{subequations}
\begin{align}
            Q_{ij} &= \sum_r (-)^r  c^{\dag}_{ri} c_{rj}
                     =\sum_{r={\rm \scriptsize even}} c^{\dag}_{ri} c_{rj}
                      -\sum_{r={\rm \scriptsize odd}} c^{\dag}_{ri} c_{rj}
\label{Qcoord}
\\
              S_{ij} &= \sum_r c^{\dag}_{ri} c_{rj}
                     =\sum_{r={\rm \scriptsize even}} c^{\dag}_{ri} c_{rj}
                      +\sum_{r={\rm \scriptsize odd}} c^{\dag}_{ri} c_{rj},
\label{Scoord}
\end{align}
\end{subequations}
implying that $\vec F$ is the generator of total spin on even sites and $\vec G$
is the generator of total spin on odd sites, as illustrated schematically in 
\fig{oddEvenSites}.%
\singlefig
{oddEvenSites}       
{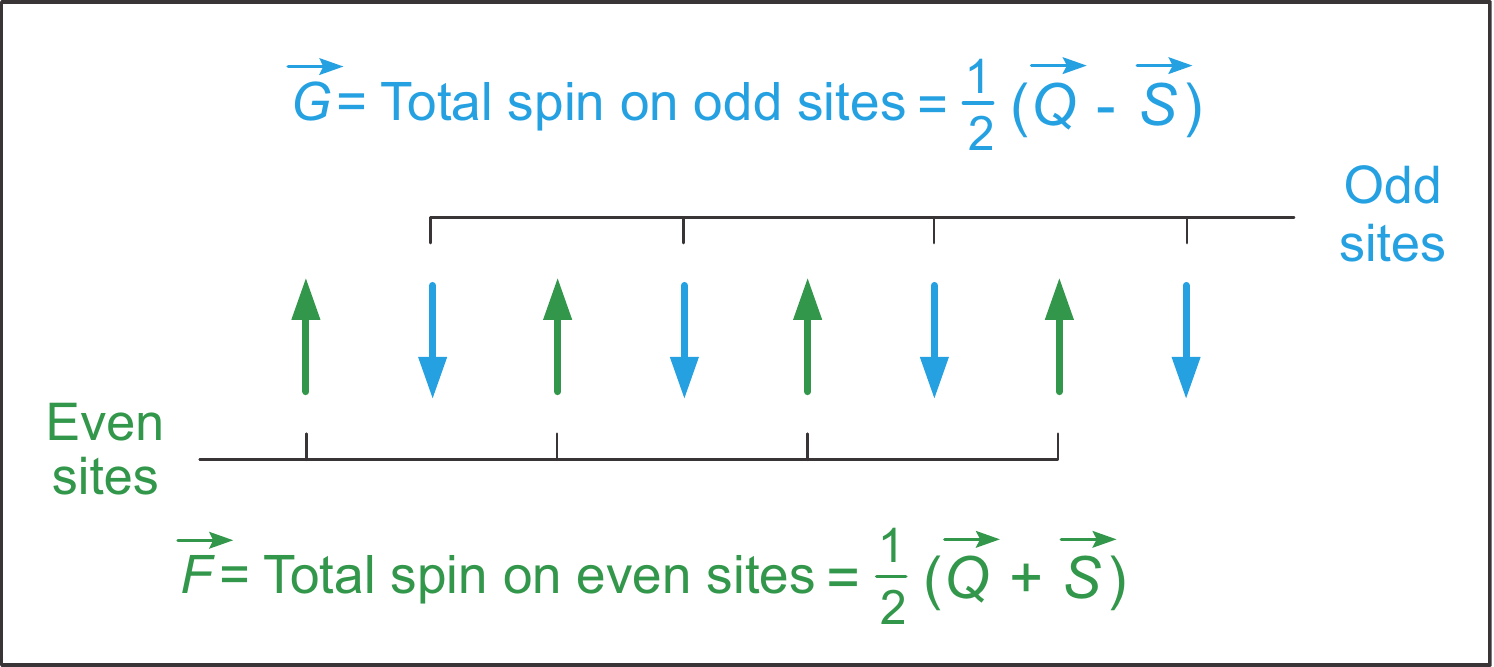}    
{0pt}         
{0pt}         
{0.56}         
{Geometrical interpretation of the SO(4) dynamical symmetry as an
antiferromagnetic state.}
Thus,  \sofour\ is generated by
two independent spin operators:  one that is the total spin on all sites and one
that is the difference in spins on even and odd sites of the spatial lattice.
This clearly is an algebraic version of the physical picture associated with
antiferromagnetic long-range order.

The \sofour\ Casimir operator may be expressed as 
\begin{equation}
C_{\scriptstyle\sofour} =2(\vec{F}\, ^2 +\vec{G}\, ^2).
\end{equation}
The \sofour\ representations can be labeled by the spin-like quantum numbers $(
F=\tfrac12 w, G=\tfrac12 w)$, where
$w=N-\mu$ with $\mu=0,2,\ldots,N$. 
Eigenstates are labeled by $w$ and the spin $S$,
$$
\psi(\sofour)=|N,w,S,m_s\rangle,
$$
and are of dimension $(w+1)^2$. 

The ground state corresponds to $\omega=N$ and $S=0$, and has $\tfrac12 n$
spin-up electrons on the even sites ($F=N/2$) and $\tfrac12 n$ spin-down
electrons on odd sites ($G=N/2$), or vice versa. Thus it has maximal staggered
magnetization,
\begin{equation}
Q=\tfrac12 \Omega(1-x)= \tfrac12 n ,
\end{equation}
and a large energy gap associated with the antiferromagnetic correlation
$\QdotQ$, 
\begin{equation}
\Delta E = 2\chi\tsub{eff}(1-x)\Omega.
\end{equation} 
In addition, the pairing gap 
\begin{equation}
\Delta =  
\tfrac12 G^{0}\tsub{eff} \Omega \sqrt{x(1-x)}
\end{equation}
 is small near half filling ($x=0$), and we shall demonstrate below that the
\sufour\ symmetry requires that the lattice have no double occupancy at half
filling and thus be a Mott insulator. We conclude that near half filling these
\sofour\ states are identified naturally with a collective, antiferromagnetic,
Mott insulating state. This identification will be strengthened below by
examination of the ground-state energy surface evaluated in this limit 
(illustrated in \fig{eSurfaces1D} and discussed in \S\ref{so4Surface}).

\subsection{\label{su2Dynamical}The SU(2) Dynamical-Symmetry Limit}

The \sufour\ subgroup chain
$$
 \sufour \supset \sutwo\tsub{p} \times \sutwo\tsub{s} \supset 
\sutwo\tsub{s} \times \uone,
$$
which we shall term the {\em SU(2) dynamical symmetry} for brevity, corresponds
to the $\pratio=0$ symmetry limit of \eq{eq1}. The eigenstates are labeled by
$v$ and spin $S$, 
$$
\psi(\sutwo)=|N,v,S,m_s\rangle,
$$ 
and are of dimension
$\tfrac12 (v+1)(v+2)$. The seniority-like quantum number $v$ is the number of
particles that do not form singlet pairs (see Table \ref{table:symmetryLims} of 
 Appendix \ref{appendix}). The
ground state has $v=0$, implying that all electrons are singlet-paired. In
addition, there exists a large pairing gap 
\begin{equation}
\Delta E = G^{(0)}\tsub{eff} \Omega 
\end{equation}
(Table \ref{table:symmetryLims} of 
 Appendix \ref{appendix}), the pairing correlation is the largest among the
three symmetry limits, and the staggered magnetization vanishes in the ground
state:
\begin{equation}
\Delta = \tfrac12 G^{0}\tsub{eff} \Omega \sqrt {1-x^2}
\qquad
Q=0 .
\label{DeltaQ}
\end{equation}
Thus we interpret this dynamical symmetry as a $d$-wave pair condensate 
associated with a
collective spin-singlet superconducting state. This identification will be
strengthened below by examination of the ground-state energy surface evaluated
in this limit (illustrated in \fig{eSurfaces1D} and discussed in
\S\ref{su2Surface}).

\subsection{\label{so5Dynamical} The SO(5) Dynamical-Symmetry Limit}

The \sufour\ subgroup chain
$$
 \sufour \supset \sofive \supset \sutwo \tsub{s} \times \uone,
$$ 
which we shall term the {\em SO(5) dynamical symmetry,} has the nature of a
transitional or {\em critical dynamical symmetry}. This
symmetry limit results when $\pratio=\tfrac12$ in \eq{eq1}. The \sofive\
irreps are labeled by a quantum number $\tau$ and the eigenstates may be labeled
by $\tau$ and the spin $S$, 
$$
\psi(\sofive)=|N,\tau,S,m_s\rangle,
$$
with $N=\Omega/2-\tau +\lambda$, where $\lambda$ is the number of $\pi$ pairs. 
For given $N$ the irreducible representations are of dimensionality  $\tfrac12 
(\lambda +1)( \lambda +2)$ and  the ground state has $\lambda=0$ and $S=0$. 

The \sofive\ dynamical symmetry has very unusual character. Although the
expectation values of $\Delta$ and $Q$ for the ground state in this symmetry
limit are the same as those of \eq{DeltaQ} for the \sutwo\ case, there exist
many states with different values of $\lambda$ (the number of $\pi$ pairs) that
can mix easily with the ground state when $x$ is small because the excitation
energy in this symmetry limit is  
\begin{equation}
\Delta E=\lambda G^{(0)}\tsub{eff}\Omega_{} x
\end{equation}
(see Table \ref{table:symmetryLims} of Appendix \ref{appendix}). In particular, 
at half filling ($x=0$) the
ground state is highly degenerate with respect to $\lambda$ and mixing different
numbers of $\pi$ pairs in the ground state costs no energy.  The $\pi$ pairs
must be responsible for the antiferromagnetism in this phase, since within the
model space only $\pi$ pairs carry spin if there are no unbroken pairs. Thus the
ground state in this symmetry limit has large-amplitude fluctuation in the AF
(and SC) order.  As such, it will be seen to play a role as a doorway between
antiferromagnetic and superconducting order, as illustrated in \fig{so5Doorway}.
\singlefig
{so5Doorway}       
{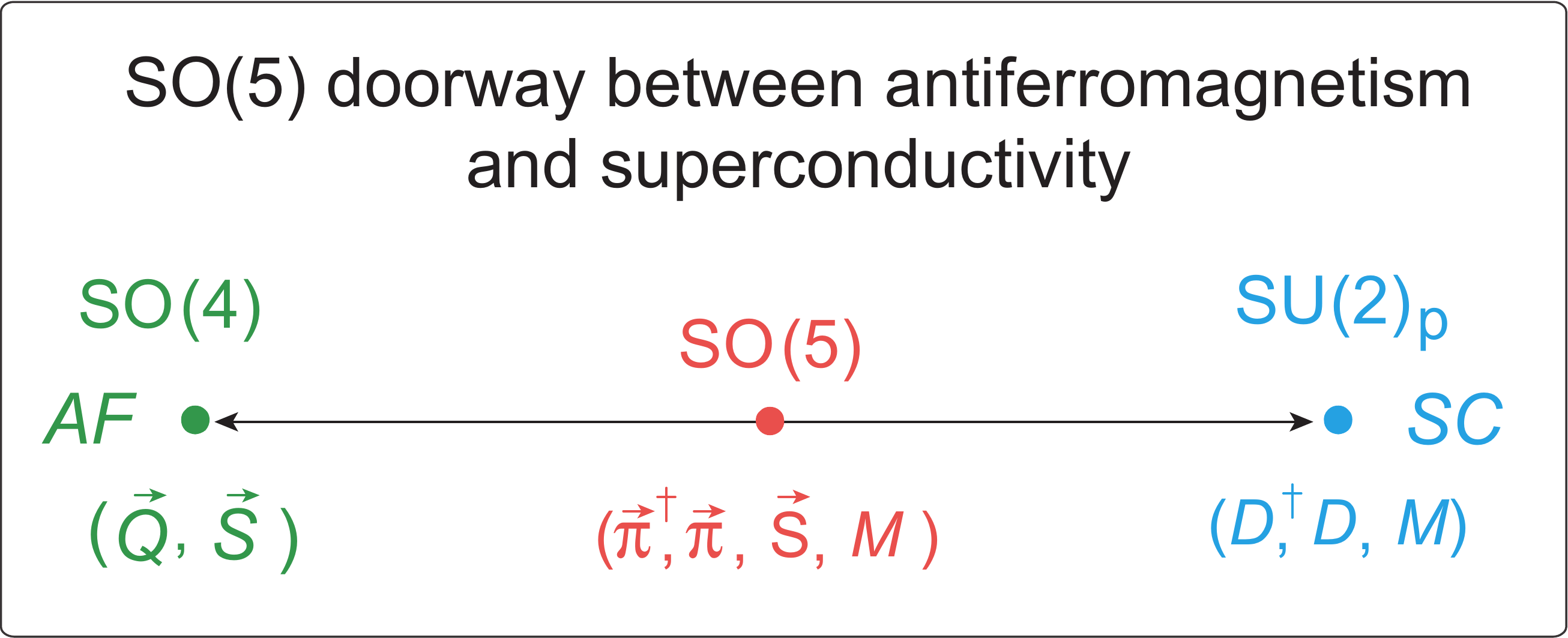}    
{0pt}         
{0pt}         
{0.28}         
{The SO(5) critical dynamical symmetry as a doorway connecting
antiferromagnetism and superconductivity.}
This identification will be strengthened below by examination of the
ground-state energy surface evaluated in this limit (illustrated in
\fig{eSurfaces1D} and discussed in \S\ref{so5Surface}).

\section{\label{generalizedCoherentStates} Generalized SU(4) Coherent States}

For values of the coupling-strength ratio $\pratio$ not equal to the special
choices $\{0, \tfrac12, 1\}$, the SU(4) symmetry model does not have an exact
solution. However, an approximate solution can be obtained using the {\em
generalized coherent-state method} \cite{zhan90,wmzha88,wmzha87,wmzha88a}, which
relates a many-body algebraic theory with unbroken symmetry to an approximation
of that theory exhibiting spontaneously-broken symmetry. The methodology is
outlined in \fig{generalizedCoherentStates} and explained more extensively
below.%
\singlefig
{generalizedCoherentStates}       
{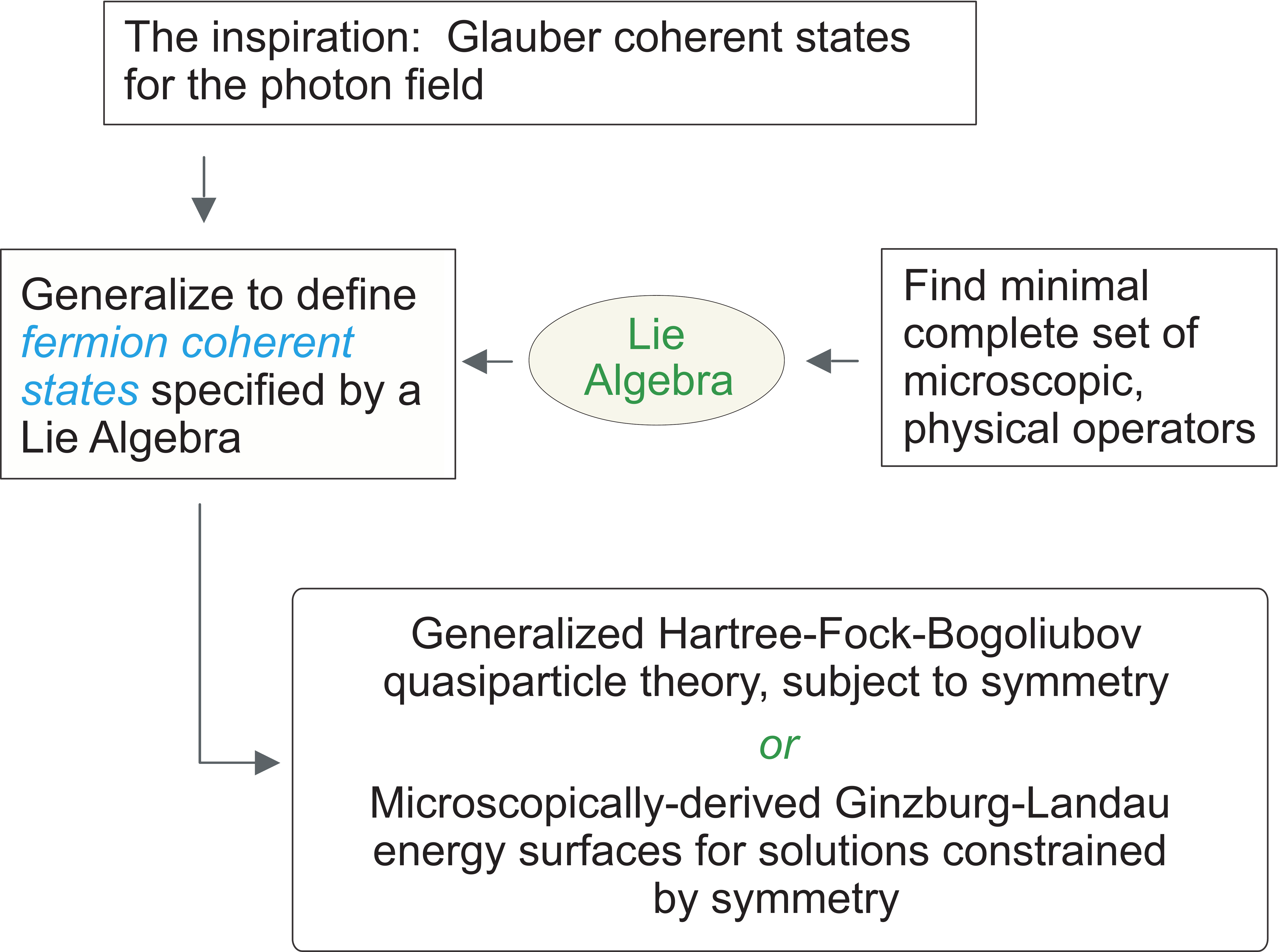}
{0pt}         
{0pt}         
{0.155}         
{The method of generalized coherent states.}

\subsection{Associating Coherent States with Lie Algebras}

The work of Gilmore \cite{gil72,gil74} and Perelomov \cite{per72} (see also
Klauder \cite{kla63}) showed that Glauber coherent states \cite{gla63} of the
electromagnetic field could be generalized to coherent states specified by an
arbitrary Lie algebra.  Specifically, it was found that the original Glauber
theory for coherent photon states could be expressed in terms of an SU(2) Lie
algebra by examining the commutation properties of the second-quantized
operators of the theory, and then this formalism could be generalized  to
encompass a set of such operators closed under any Lie algebra. The resulting
theory can apply to either fermion or boson fields, provided that the relevant
creation and annihilation operators close a Lie algebra. This extension of the
Glauber theory to arbitrary Lie algebras is termed the {\em generalized coherent
state method.}  A comprehensive review exists \cite{zhan90}, so we proceed
directly to application of the generalized coherent state method to fermions
described by the SU(4) algebra of \S\ref{su4Algebra}.

\subsection{SU(4) Coherent States}

The SU(4) coherent state $|\psi\rangle$ can be expressed as
\begin{equation}
|\psi\rangle={\cal T}\mid 0^* \rangle,
\label{eq10}
\end{equation}
where the operator ${\cal T}$ is given by
\begin{equation}
{\cal T} = \exp (\eta_{00} p_{12}^{\dagger }+\eta_{10}
q_{12}^{\dagger }-{\rm h.\,c.}).
\label{eq10b}
\end{equation}
In \eq{eq10}, $|0^{*}\rangle$ is the physical vacuum (ground state) of
the system, the real parameters $\eta_{00}$ and $\eta_{10}$ are
symmetry-constrained variational parameters, and h.\,c.\ stands for hermitian
conjugation. The variational parameters weight the elementary excitation
operators $p^\dagger_{12}$ and $q^\dagger_{12}$ in \eq{eq10b}, so they
represent collective state parameters for a pair subspace truncated under the
SU(4) symmetry \cite{coherent-parameters}. We then express a
symmetry-constrained variational Hamiltonian as
\begin{equation}
H' = H-\lambda\hat{n},
\label{1.2}
\end{equation}
where $H$ is the Hamiltonian given in \eq{eq1} and $\lambda$ is the chemical
potential, determined by requiring that the average particle number be 
conserved. The parameters $\eta_{00}$ and $\eta_{10}$ in \eq{eq10b} are
determined by the variational condition $\delta\langle H'\rangle=0$, where 
\begin{equation}
\langle H'\rangle\equiv\langle 0^*| H'|0^*\rangle
\label{1.2b}
\end{equation}
is the expectation value of $H'$ with respect to the ground state $|0^*\rangle$.

\subsection{Generalized Quasiparticle Transformation}

As shown in Appendix A of Ref.\ \cite{sun05}, it is convenient to evaluate the
variation $\delta\langle H'\rangle=0$ using a 4-dimensional matrix
representation that was employed in Refs.\ \cite{lawu03,wmzha88}. In this
representation the unitary operator ${\cal T}$ implements a transformation from
the original particle basis to a quasiparticle basis and the variational
parameters $\eta_{00}$ and $\eta_{10}$ are replaced, respectively, by $u_\pm$
and $v_\pm$, subject to a unitarity condition 
\begin{equation}
 u^2_\pm+v^2_\pm=1.
\label{unitarityuv}
\end{equation}
Under this transformation the physical vacuum state $\ket{0^*}$ is transformed
to a quasiparticle vacuum state $\ket\psi$ and the basic fermion operators
$c_{\bm r i}$ are converted to new quasiparticle operators $a_{\bm r i}$ through
the transformation
\begin{equation} {\cal T}
\left (
\begin{array}{c} c_{{{\bf r}}\uparrow}
\\ c_{{{\bf r}}\downarrow}
\\ c^\dagger_{\bar{\bf r}\uparrow}
\\ c^\dagger_{\bar{\bf r}\downarrow}
\end{array}
\right) \left |0^*\right\rangle=
\left(
\begin{array}{c} a_{{{\bf r}}\uparrow}
\\ a_{{{\bf r}}\downarrow}
\\ a^\dagger_{\bar{\bf r}\uparrow}
\\ a^\dagger_{\bar{\bf r}\downarrow}
\end{array}
\right)\left |\psi\right\rangle.
\label{transfA}
\end{equation}
(see Appendix A of Ref.\ \cite{sun05}), with
\begin{align}
\left( u_+ c_{{\bf r}\uparrow} + v_+ c^\dagger_{{\bar{\bf
r}}\downarrow} \right) |0^*\rangle &=  a_{{\bf r}\uparrow}
|\psi\rangle
\nonumber\\
\left( u_- c_{{\bf r}\downarrow} - v_- c^\dagger_{{\bar{\bf
r}}\uparrow} \right) |0^*\rangle &=  a_{{\bf r}\downarrow}
|\psi\rangle
\nonumber\\
\left( u_- c^\dagger_{{\bar{\bf r}}\uparrow} + v_- c_{{\bf
r}\downarrow} \right) |0^*\rangle &=  a^\dagger_{{\bar{\bf
r}}\uparrow} |\psi\rangle
\nonumber\\
\left( u_+ c^\dagger_{{\bar{\bf r}}\downarrow} - v_+ c_{{\bf
r}\uparrow} \right) |0^*\rangle &=  a^\dagger_{{\bar{\bf
r}}\downarrow} |\psi\rangle .
\nonumber
\end{align}
Thus this is a  transformation of the Bogoliubov form: each quasiparticle state
is a mixture of a particle and a hole, and the coherent state $|\psi\rangle$ is
a quasiparticle vacuum constrained to respect SU(4) symmetry. It follows that
the generalized coherent state method described here is equivalent to the most
general Hartree--Fock--Bogoliubov (HFB) variational method, but subject to an
SU(4) symmetry constraint \cite{guid99,guid01,lawu03}. By using the
4-dimensional matrix representation, the expectation value for any operator
$\hat O$ in the coherent state representation can be calculated through the
transformation
\begin{equation}
\langle 0^*|\hat
O|0^*\rangle = \langle\psi |{\cal T}\hat O{\cal
T}^{-1}|\psi\rangle, 
\label{matrixElements}
\end{equation}
as detailed in Appendix A of Ref.\ \cite{sun05}.

\subsection{Temperature Dependence}

At finite temperature $|\psi\rangle$ will generally no longer be a quasiparticle
vacuum state and the quasiparticle annihilation operators acting on
$|\psi\rangle$ do not necessarily give zero. In Appendix B of Ref.\ \cite{sun05}
a formalism is derived to deal with the finite-temperature case, by replacing
the state $|\psi\rangle$ with a state $|\psi(T)\rangle$ in which the
quasiparticles at temperature $T$ may be thermally excited. 

For a temperature $T$ we assume that
the single-particle levels $\varepsilon_{r\pm}$ [which are defined in \eq{qse}
below] are degenerate and contain $\tilde{n}_{r+} +\tilde{n}_{r-}$
quasiparticles.  The quasiparticle number densities  are assumed given by the
Fermi--Dirac distribution
\begin{equation}
\tilde{n}_{\pm} (T) =\frac
2{\Omega}\sum_{r={\rm even}}\tilde{n}_{r\pm}(T) =\frac{2}{1+\exp (R
e_{\pm}/k\tsub B T)} ,
\label{thermal}
\end{equation}
where $e_\pm$ is the quasiparticle energy defined in \eq{qpe} below and
$R$ is an empirical scaling factor of order one that corrects on average for
approximations that we shall make for the single-particle spectrum. 

We may then evaluate the expectation value of one-body operators at finite
temperature, with the result (see Appendix B of Ref.\ \cite{sun05})
\begin{align}
\langle \singletPair^{\dagger}\rangle &= \langle
\singletPair\rangle=-\frac{\Omega}{2}
\left[P_+(T)u_+v_+ + P_-(T)u_-v_-\right]
\nonumber
\\
\langle \pi^{\dagger}_z\rangle &= \langle \pi_z\rangle=-\frac
{\Omega}{2} \left[P_+(T)u_+v_+ -P_-(T)u_-v_-\right]
\nonumber
\\
\langle \AF_z\rangle &=  \frac{\Omega}{2} \left[ P_+(T)v_+^2
- P_-(T)v_-^2\right]
\label{eqone2}
\\
\langle\hat n\rangle &=  \frac{\Omega}{2} \left[P_+(T)(2v_+^2-1)
+P_-(T) (2v_-^2-1)+2\right] \nonumber
\\
\langle \pi_x\rangle &=  \langle \pi_y\rangle=\langle
\vec{S}\rangle = \langle \AF_x\rangle=\langle \AF_y\rangle=0 ,
\nonumber
\end{align}
where in these expressions we have employed the definitions
\begin{equation}
P_{\pm}(T)\equiv 1-\tilde{n}_{\pm}(T) =\tanh \left(\frac{R
e_{\pm}}{2k\tsub B T} \right).
\label{PT}
\end{equation}
If a large-$\Omega$ approximation is invoked by ignoring terms of order
$1/\Omega$ relative to the leading terms, the scalar products of these one-body
operators reduce to products of the corresponding one-body ones,
\begin{equation}
\begin{array}{c}
\langle \singletPair^{\dagger }\singletPair\rangle =  \langle
\singletPair\rangle^2
\qquad
\langle \vec{\pi}^{\dagger
}\cdot\vec{\pi}\rangle =  \langle \pi_z\rangle^2
\\[5pt]
\langle \QdotQ \rangle =  \langle \AF_z\rangle^2
\qquad
\langle \vec{S} \cdot \vec{S}\rangle = \langle S \rangle^2 = 0 .
\end{array}
\label{eqtwo2}
\end{equation}
If $T\rightarrow 0$, then $P_\pm(T)\rightarrow 1$ in \eq{PT} and 
\eqs{eqone2} and \eqnoeq{eqtwo2} reduce to the simpler forms
\begin{subequations}
\begin{align}
\langle \singletPair^{\dagger }\rangle &=  \langle
\singletPair\rangle
= -\tfrac12 \Omega (u_+v_+ + u_-v_-)
\label{zeroTMel1}
\\
\langle {\pi}_z^\dagger\rangle &=  \langle \pi_z\rangle
= -\tfrac12 \Omega (u_+v_+ - u_-v_-)
\label{zeroTMel2}
\\
\langle \AF_z \rangle &=
 \tfrac12 \Omega (v_+^2 - v_-^2)
\label{zeroTMel3}
\\
\langle \hat n \rangle &= \Omega (v_+^2 + v_-^2)
\label{zeroTMel4}
\\
\langle \pi_x\rangle &=  \langle \pi_y\rangle=\langle
\vec{S}\rangle = \langle \AF_x\rangle=\langle \AF_y\rangle=0 
\label{zeroTMel5}
\\
\langle \singletPair^{\dagger }\singletPair\rangle &=  \langle
\singletPair\rangle^2
= \tfrac14 \Omega^2 (u_+v_+ + u_-v_-)^2
\label{zeroTMel6}
\\
\langle \vec{\pi}^{\dagger
}\cdot\vec{\pi}\rangle &=  \langle \pi_z\rangle^2
= \tfrac14 \Omega^2 (u_+v_+ - u_-v_-)^2
\label{zeroTMel7}
\\
\langle \QdotQ \rangle &=  \langle \AF_z\rangle^2
= \tfrac14 \Omega^2 (v_+^2 - v_-^2)^2
\label{zeroTMel8}
\\
\langle \vec{S} \cdot \vec{S}\rangle &= \langle S \rangle^2 = 0 
\label{zeroTMel9}
\\
\langle M^2\rangle &= \tfrac14 (n-\Omega)^2+\tfrac12 \Omega
[(\mbox{u}_+\mbox{v}_+)^2 + (\mbox{u}_-\mbox{v}_-)^2\,].
\label{zeroTMel10}
\end{align}
\label{zeroTMel}%
\end{subequations}
which are valid only for $T=0$ (and large $\Omega$ for the 2-body terms).

\subsection{Energy Gaps and Gap Equations}

The preceding results may be used to find the expectation value of
the variational Hamiltonian $\langle H'\rangle$ in the coherent state
representation. In terms of the energy gaps defined by
\begin{subequations}
\begin{align}
\Deltad &= G_0\sqrt{\left\langle \singletPair^\dag
\singletPair\right\rangle}
\label{1.11a}
\\
\Deltapi &= G_1\sqrt{\left\langle\vec{\pi}^\dag\cdot
            \vec{\pi}\right\rangle}
\label{1.11b}
\\
\Deltaq &= \chi\sqrt{\left\langle \QdotQ \right\rangle} ,
\label{1.11c}
\end{align}
\label{1.11}
\end{subequations}
one obtains from \eqs{1.2}, \eqnoeq{1.2b}, and \eqnoeq{eq1},
\begin{equation}
\langle  H' \rangle = (\varepsilon-\lambda) n -\left (\
\frac{\Deltadsq}{G_0} +\frac{\Deltapisq}{ G_1}+\frac{\Deltaqsq}{
\chi}\ \right),
\nonumber
\end{equation}
where $\epsilon$ is the single-particle energy. Variation of $\langle H'\rangle$
with respect to $u_\pm$ or $v_\pm$ yields
\begin{equation}
2u_{\pm}v_{\pm} (\varepsilon_{ \pm}-\lambda)-\Deltapm(u^2_{\pm}
-v^2_{\pm})=0,
\nonumber
\end{equation}
which is satisfied by
\begin{equation}
\begin{array}{c}
\displaystyle u^2_{\pm} = \frac{1}{2}\left
(1+\frac{\varepsilon_{\pm}-\lambda}{e_{\pm}}\right)
\quad
\displaystyle v^2_{\pm} = \frac{1}{2}\left
(1-\frac{\varepsilon_{\pm}-\lambda}{e_{\pm}} \right),
\end{array}
\label{uv}
\end{equation}
where
\begin{equation}
e_{\pm} = \sqrt{(\varepsilon_{\pm}-\lambda)^2+{\Deltapm}^2}
\label{qpe}
\end{equation}
and
\begin{equation}
\Deltapm = \Deltad\pm\Deltapi \qquad
\varepsilon_{\pm}=\varepsilon\mp\Deltaq .
\label{qse}
\end{equation}
Inserting \eq{uv} into \eqs{eqone2}--\eqnoeq{eqtwo2} and employing the gap
definitions \eqnoeq{1.11}, one obtains the temperature-dependent gap equations
\begin{subequations}
\label{gap:whole}
\begin{align}
\Deltad &= \frac{G_0\Omega}{4}\left ( w_+ \DeltaPlus  + w_-
\DeltaMinus\right )
\label{subgapeq:d}\\
\Deltapi &= \frac{G_1\Omega}{4} \left (  w_+ \DeltaPlus - w_-\
\DeltaMinus \right )
\label{subgapeq:pi}\\
\frac{4\Deltaq}{\chi\Omega} &= w_+ ( \Deltaq+\lambdaPrime ) + w_-
( \Deltaq-\lambdaPrime )
\label{subgapeq:q}\\
-2x &= w_+ ( \Deltaq+\lambdaPrime ) - w_- ( \Deltaq-\lambdaPrime ),
\label{subgapeq:la}
\end{align}
\end{subequations}
where we define
\begin{equation}
w_\pm \equiv \frac {P_\pm(T)}{e_\pm} 
\qquad
\lambdaPrime \equiv \lambda-\varepsilon\ .
\label{wpm}
\end{equation}
Solution of the above algebraic equations yields all the gaps and the chemical
potential $\lambdaPrime$ (we shall discuss methods of solution in \S
\ref{solutionGapEquations} and \S \ref{finiteTsolutions} below). Then the total
energy is
\begin{align}
E = \left \langle H'\right\rangle+\lambda n 
=\varepsilon n -\left(\frac{\Deltadsq}{G_0}
        +\frac{\Deltapisq}{G_1}+\frac{\Deltaqsq}{\chi} \right).
\nonumber
\end{align}
To simplify the discussion, we shall ignore the single-particle energy
in the above equation by setting $\varepsilon=0$, since this term has been
approximated as a state-independent constant and thus plays no role
in the phase competition. The energy density $E/\Omega$ is then
\begin{equation}
 \frac{E}{\Omega}=
-\left ( \frac{{\Deltad}^2}{G_0\Omega
}+\frac{{\Deltapi}^2}{G_1\Omega } +\frac{{\Deltaq}^2}{\chi\Omega }
\right ) .
\label{eqED}
\end{equation}

The three gaps $\Deltad$, $\Deltapi$ and $\Deltaq$ in the above equations are
defined in \eqnoeq{1.11} and represent the characteristic energy scales of
spin-singlet pairing correlations, spin-triplet pairing correlations, and
antiferromagnetic correlations, respectively. Hence, these correlations
determine the ground state energy. Once the gaps and the chemical potential
$\lambdaPrime$ are known from solution of the gap equations \eqnoeq{gap:whole},
the quasiparticle energies $e_\pm$ and the amplitudes $u_\pm$ and $v_\pm$ can
all be determined through \eqs{uv}--\eqnoeq{qse}, permitting other ground state
properties to be calculated.

In the following sections we shall give analytical solutions for
the gap equations \eqnoeq{gap:whole}, first for zero temperature
and then for finite temperatures. As we shall see, a rich phase
structure emerges naturally in these solutions as a consequence of
competition between the various energy scales.

\subsection{\label{BCSrelationship} Relationship to Ordinary BCS and
N\'eel Theory}

The preceding results are formally analogous to those of the BCS theory with
$v_{\pm}^2$ the probability of single particle levels $\varepsilon_{\pm}$ being
occupied, $\Deltapm$ the energy gaps, and $e_{\pm}$ the quasiparticle energies. 
The essential difference from normal BCS theory is that conventional pairing 
theories
deal with one energy gap and one kind of quasiparticle; here we have two kinds
of quasiparticles and several energy gaps, implying a large variety of new
physics. As should be clear from the derivations leading to \eq{eqED}, this new
physics arises as a natural consequence of a BCS-like theory in which the
pairing may be unconventional, Coulomb repulsion plays a significant role, and 
antiferromagnetic correlations are allowed to
enter on a footing equal with that of the pairing correlations.

In the formalism describing this more sophisticated pairing the quantities
$e_{\pm}$ are energies for two kinds of quasiparticle excitation, corresponding
to two sets of non-degenerate single particle energy spectra
$\{\varepsilon_{\pm}\}$ separated by an energy $2\Deltaq$, as illustrated in
\fig{AFsplitting}.
\singlefig
{AFsplitting}       
{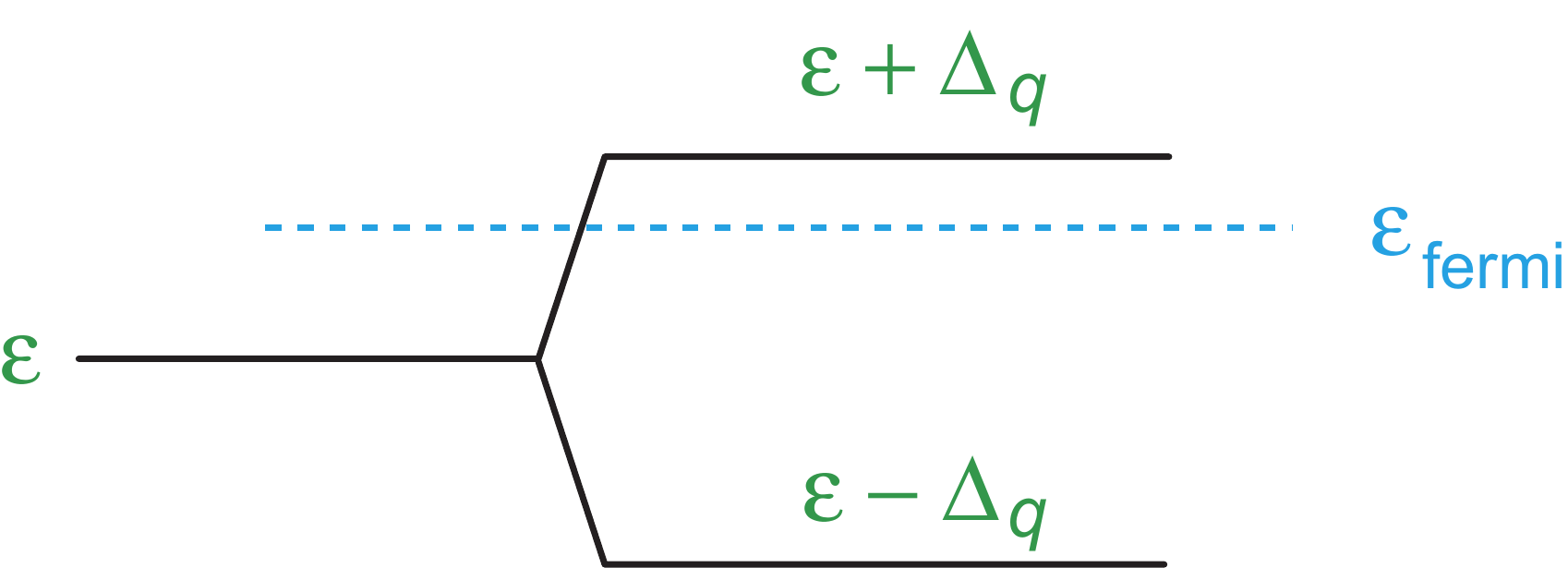}    
{0pt}         
{0pt}         
{0.38}         
{Modification of the quasiparticle energies by interaction with the
antiferromagnetism in the SU(4) coherent-state gap equations.}
Each level can be occupied by only one electron of either up or down spin. The
corresponding pairing gaps are $\Deltapm$, which are linear combinations of the
two gaps $\Deltad$ and $\Deltapi$. The probabilities for single-particle levels
to be occupied or unoccupied are $v^2_{\pm}$ and $u^2_{\pm}$, respectively.

The resulting theory has the limits that we would expect for a symmetry that
unites superconductivity and antiferromagnetism at a fundamental level.  The
gap equations reduce exactly to the gap equations of the ordinary BCS theory in
the limit that the antiferromagnetic interactions can be neglected. 
Conversely, in the limit that the pairing interactions can be neglected we
obtain the equations expected for a N\'eel antiferromagnet.

\subsection{\label{solutionGapEquations} Solution of the Gap Equations at Zero
Temperature}

The three parameters, $\chi$, $G_0$, and $G_1$ in the  algebraic
equations \eqnoeq{gap:whole} correspond to the three basic effective 
interactions in
the SU(4) model: 
\begin{enumerate}
 \item 
The antiferromagnetic correlation, with strength $\chi$. 
\item
The spin-singlet pairing correlation, with strength $G_0$.
\item
The spin-triplet pairing correlation, with strength $G_1$.
\end{enumerate} 
Experimental evidence suggests that these three interactions in cuprates are all
attractive, and that their strengths are ordered
\begin{equation}
\chi > G_0 > G_1 > 0 .
\label{condition}
\end{equation}
Solutions for the gap equations assuming this condition to be satisfied can be
obtained as follows.

\subsubsection{\label{criticalDopingPoint} The Critical Doping Point}

The gap solutions for \eqs{gap:whole} at $T=0$ can be written explicitly for two
doping regimes separated by a special doping value given by 
\begin{equation}
x_q=\sqrt{\frac{\chi-G_0}{\chi-G_1}}. 
\label{xq} 
\end{equation} 
This value of
$x_q$ is a {\em critical doping point} marking a quantum phase transition,
because we shall find that at zero temperature the wavefunctions and physical
properties of the two doping regions lying on either size of this point  differ
qualitatively. Specifically, one finds the following solutions
(general derivations are given in Appendix C of Ref.\
\cite{sun05}).

\subsubsection{The All Gaps Finite Solution for $T=0$} 

Equations~\eqnoeq{gap:whole} have a solution for all gaps
nonzero that corresponds to:
\begin{subequations}
\label{gapT0:whole}
\begin{align}
\Deltaq &= \frac{\chi\Omega}{2}\sqrt{(x^{-1}_q- x)(x_q-x)}
\label{udopDq}
\\
\Deltad &= \frac{G_0\Omega}{2}\sqrt{x (x_q^{-1}- x)}
\label{udopDd}
\\
\Delta_\pi &= \frac{G_1\Omega}{2}\sqrt{x (x_q- x)}
\label{udoppi}
\\
\lambdaPrime  &= -\frac{\chi\Omega}{2} x_q (1-x_q
x)-\frac{G_1\Omega}2x.
\label{udopL}
\end{align}
\end{subequations}
This solution illustrates the central role of the critical doping point $x_q$
because {\em it exists only for the doping range $x\le x_q$.} We shall find
below that it corresponds to the ground state of the system at zero temperature
if the doping lies in the range $0< x\le x_q$.

\subsubsection{The Pure Singlet-Pairing Solution for $T=0$}  

Equations~\eqnoeq{gap:whole} also have a solution corresponding to
$\Delta_q=\Delta_\pi=0$ that is given by
\begin{subequations}
\label{gapDd:whole}
\begin{align}
\Deltaq &= \Delta_\pi= 0
\label{odopDq}\\
\DeltaZero &\equiv \Deltad =\frac{G_0\Omega}{2}\sqrt{1- x^2}
\label{odopDd}\\
\lambdaPrime\hspace{3pt}  &= -\frac{G_0\Omega}{2}x.
\label{odopL}
\end{align}
\end{subequations}
This solution is valid for the {\em entire physical doping range $0\le x\leq
1$}, but  we shall 
see later that it corresponds to the ground state  at zero
temperature only for  $ x_q < x <1$.

\subsubsection{The Pure Triplet-Pairing Solution for $T=0$}

A solution to \eqs{gap:whole} exists if all gaps except the triplet
pairing vanish:
\begin{subequations}
\label{gapDpi:whole}
\begin{align}
\Deltaq &= \Deltad= 0
\label{eqDqd}\\
\Deltapi  &= \frac{G_1\Omega}{2}\sqrt{1- x^2}
\label{eqDpi}\\
\lambdaPrime  &= -\frac{G_1\Omega}{2}x.
\label{eqLpi}
\end{align}
\end{subequations}
This solution is valid for the entire range of physical doping.

\subsubsection{The Pure AF Solution for $T=0$}

A solution to \eqs{gap:whole} that we shall term the {\em pure
antiferromagnetic solution} exists if all gaps except the
antiferromagnetic correlation vanish:
\begin{subequations}
\label{gapDaf:whole}
\begin{align}
\Deltad &= \Deltapi= 0
\label{eqDdpi}\\\
\Deltaq  &= \frac{\chi\Omega}{2}(1- x)
\label{eqdaf}\\
\lambdaPrime  &= -\frac{\chi\Omega}{2}(1-x).
\label{eqLaf}\
\end{align}
\end{subequations}
This solution is valid for the entire range of physical doping.

\subsubsection{The Uncorrelated Solution for $T=0$ }

If all gaps vanish, \eqs{gap:whole} have a trivial solution that we
shall term the {\em uncorrelated solution} in which all antiferromagnetic and
pairing correlations are zero:
\begin{subequations}
\label{gapMetal:whole}
\begin{align}
\Deltaq &= \Deltad=\Deltapi=0,
\label{gap0metal} \\[5pt]
\lambdaPrime &= -2kT\ \atanh(x) = 0.
\label{metal}
\end{align}
\end{subequations}
This solution is valid for the entire range of physical doping.

\subsubsection{The Zero-Temperature Ground State}

Among the five sets of gap solutions for $T=0$ presented above, we can
determine which corresponds to the ground state at a given doping $x$ by
inserting the gap solutions directly into \eq{eqED} to ascertain
which has the lowest energy. We shall discuss the resulting gap diagram in
more detail below, but---assuming the ordering of coupling strengths given by
\eq{condition}---the general result for $T = 0$ is that 

\begin{itemize}
 \item 
The all gaps finite solution is the ground state for the doping range $0<x \le
x_q$.  
\item
The pure singlet-pairing solution is the ground state for $x_q < x < 1$.  
\item
The pure triplet pairing solution is never the ground state under any
conditions.

\item
The uncorrelated solution is never the ground state if there are
finite
AF or pairing correlations.
\item
The AF correlation solution is never the ground state, except possibly for
$x\sim 0$.
\end{itemize}
Thus, our primary interest at zero temperature will be in the all gaps finite
and the pure singlet-pairing solutions for general values of $x$, and in the
pure antiferromagnetic solution near $x=0$.

\subsection{\label{finiteTsolutions} Solution of the Gap Equations for Finite
Temperature}

The gap equations for $T>0$ differ from those at $T=0$ in that the terms $w_\pm$
depend on temperature.  From \eqs{wpm} and \eqnoeq{PT},
\begin{equation}
w_\pm =\frac{P_\pm(T)}{e_\pm}=\frac{\tanh(R e_{\pm} /2k\tsub
BT)}{e_{\pm}}.
\label{wpmT}
\end{equation}
For finite temperature the gap equations could have a variety of solutions, even
for a fixed doping $x$.  Some general solutions have been derived in Appendix C
of Ref.\ \cite{sun05} using the physically-motivated assumption
\eqnoeq{condition} for the relative correlation strengths. By inserting these
solutions in \eq{eqED}, one can then find the ones having the lowest
energy and thus determine the physical ground-state solution for a given
temperature $T$ and doping $x$. Several representative cases are examined below 
below.

\subsubsection{The $\Deltad+\Deltaq+\Deltapi$ Finite-Temperature Case}

This case generalizes to finite $T$ the all-gap solution found at $T=0$ for the
doping range $0\leq x\leq x_q$.  The temperature and doping dependent gap
solutions are derived in Appendix C of Ref.\ \cite{sun05}, with the results
\begin{subequations}
\label{Tgap:whole}
\begin{align} \Deltaq &= \frac
{\chi\Omega}{2} \sqrt{(x_q^{-1}-y)(x_q-y)}\ \frac xy
\label{eqDqyT}
\\
\Deltad &=  \frac{G_0\Omega}{2} \sqrt {x(x_q^{-1}-y)}\, g(y)
\label{eqDdyT}
\\
\Deltapi &=  \frac{G_1\Omega}2 \sqrt {x(x_q-y)} \, g(y)
\label{eqDpiyT}
\\
\lambdaPrime  &= -\frac{(\chi-G_1)\Omega}{2}
x_q\left(\frac{x}{y}-x_qx\right)-\frac{G_1\Omega}2x \label{eqLyT}
\end{align}
\end{subequations}
where $y$ and $g(y)$ are defined through
\begin{subequations}
\begin{align}
y &= \frac {x}{\sqrt{I_+(T)+I_-(T) \Gamma(y)}}\
\label{eqyT}
\\
g(y) &= \sqrt{\frac xy+\frac{I_+(T)-(x/y)^2}{2x(\bar{x}_q-y)}}
\label{eqgT}
\end{align}
\end{subequations}
with
\begin{subequations}
\begin{align}
\Gamma(y) &= \frac{\bar{x}_q-y}{\sqrt{(x_q-y)(x_q^{-1}-y)}}
\label{Gamma_yT}
\\
I_\pm(T) &= \frac {P_-^2(T)\pm P_+^2(T)}2
\label{eqIT}
\end{align}
\end{subequations}
and $\bar{x}_q\equiv (x_q^{-1}+x_q)/2$.
For a given value of doping $x$ and temperature $T$, the gaps and chemical
potential may be found as follows. From \eqs{wpmT} and
\eqnoeq{qpe},
\begin{equation}
T=\frac{R \sqrt{ (\Deltaq\pm\lambdaPrime)^2+{\Delta_\pm}^2}}
{2k\tsub B\,\atanh(w_\pm e_\pm)}\ ,
\label{3.6}
\end{equation}
which implies that
\begin{equation}
\frac{\sqrt{(\Deltaq+\lambdaPrime)^2+{\Delta_+}^2}}{\atanh(w_+e_+)}
=\frac{\sqrt{(\Deltaq-\lambdaPrime)^2+{\Delta_-}^2}}{\atanh(w_-e_-)}.
\label{3.5}
\end{equation}
By solving \eqs{subgapeq:q}--\eqnoeq{subgapeq:la} directly, we obtain
\begin{equation}
w_\pm= \frac{(2\Deltaq/\chi\Omega) \mp x}
{\Deltaq\pm\lambdaPrime}.
\nonumber
\end{equation}
Now from Eq.~(C2) in Appendix C of Ref.\ \cite{sun05},  $\Deltapi$
and $\Deltad$ are related by
\begin{equation}
 \Deltapi =\left (\frac{w_- -2/G_0\Omega}{w_- -2/G_1\Omega}\right )\Deltad,
\label{DpitoDd}
\end{equation}
and thus, utilizing the first of \eqs{qse}, $\Delta_+$ and $\Delta_-$
can be related to $\Deltad$ through
\begin{equation}
 \Delta_\pm =\left [1 \pm\left (\frac{w_- - 2/G_0\Omega}{w_- -2/G_1\Omega
}\right )\right ]\Deltad.
\label{DpmtoDd}
\end{equation}
Hence, for a given doping $x$, we can obtain a solution by the following
algorithm:

\begin{enumerate}
 \item 
Choose different values of $\Deltaq$. For each choice of $\Deltaq$ one can solve
for $y$ from \eq{eqDqyT}, and then for $\lambdaPrime$ from \eq{eqLyT}.  
\item
The pairing gap $\Deltad$ can then be obtained by substituting 
\eq{DpmtoDd} into \eq{3.5} and solving the resulting equation for
$\Deltad$. 
\item
With $\Deltad$ determined, $\Deltapi$ can be found from \eq{DpitoDd} and
the corresponding temperature $T$ follows from \eqs{3.6} and
\eqnoeq{DpmtoDd}. 
\end{enumerate} 
By this procedure, we can find complete solutions for each choice of $x$ and
$\Delta_q$.

\subsubsection{The $\Deltad$ Finite-Temperature Case}

This solution generalizes the pure singlet-pairing solution at zero temperature.
Following a similar procedure as in the $T=0$ case, one obtains
the temperature and doping dependent gap solutions
\begin{subequations}
\label{TgapDd:whole}
\begin{align} \Deltaq  &=  \Deltapi = 0
\label{eqDpiq}
\\
\Deltad  &=  \frac{G_0\Omega}{2} \sqrt{I_+(T)-x^2}
\label{eqDdG}
\\
\lambdaPrime  &= - \frac{G_0\Omega}{2}x ,
\label{eqLT}
\end{align}
\end{subequations}
where in the present case,
$$
I_+(T)=P^2_+(T)=P^2_-(T).
$$
By using \eqs{qpe}), \eqnoeq{wpmT}, and \eqnoeq{eqLT}, \eq{eqDdG} may be
expressed as
\begin{equation}
T=\frac{R \sqrt{\left(\dfrac{G_0\Omega}2x\right)^2+{\Deltad}^2} }
{2k\tsub B\ \atanh\left[
\sqrt{\left(\dfrac{2\Deltad}{G_0\Omega}\right)^2+x^2}\,\right]}\ .
\label{3.8}
\end{equation}
Therefore, for given $x$ and $T$ the gap $\Delta_d$ can be obtained from 
\eq{3.8}, and $\lambdaPrime$ follows from \eq{eqLT}.

\subsubsection{ The $\Deltaq$ Finite-Temperature Case}

In this case both $\Deltad$ and $\Deltapi$ are zero, corresponding
to a solution that applies only for temperatures $T>T\tsub c$:
\begin{subequations}
\label{TgapDq:whole}
\begin{align}
\Deltad &= \Deltapi = 0 \\
\Deltaq  &= \frac{\chi\Omega}2(P_-(T)-x).
\label{eqDqPx}
\end{align}
\end{subequations}
These results can be derived from \eqs{subgapeq:q} and
\eqnoeq{subgapeq:la}, which in the present case reduce to
\begin{align}
\frac{4\Deltaq}{\chi\Omega} &= P_-(T)+P_+(T), \label{eqDqP}
\\
2x\  &= P_-(T)-P_+(T).
\label{eqxP}
\end{align}
By solving \eqs{eqDqP} and \eqnoeq{eqxP} one obtains
\begin{align}
T &= \frac{R \Deltaq }{k\tsub B A_+}
\label{3.9}
\\
\lambdaPrime &= \frac{k\tsub B T A_-}{R},
\label{3.10}
\end{align}
with
$$
A_\pm \equiv \left[\atanh\left(\dfrac{2\Deltaq}{\chi\Omega}-x\right)
\pm \atanh\left(\dfrac{2\Deltaq}{\chi\Omega}+x\right)\right ] .
$$
Thus, for a temperature $T$, $\Deltaq$ can be obtained from 
\eq{3.9} and $\lambdaPrime$ can be obtained from \eq{3.10}.

\subsection{\label{k-dependentSU4} Momentum-Dependent SU(4) Solutions}

The SU(4) generators discussed to this point are summed over the momentum  $\bm 
k$.  They are
appropriate for data that are not momentum-selected.  However, some data, such
as ARPES Fermi-arc measurements exhibit explicit dependence on $\bm k$.

\subsubsection{Momentum-Dependent Generators of the Algebra}

The SU(4) formalism has been extended \cite{sun07} to deal with this case by
viewing the individual $\bm k$ components of the operators defined in 
\eq{operatorset} as the symmetry generators ${\cal G}(\bm k)$:
\begin{equation}
\{{\cal G}(\bm k)\} \equiv \{\singletPair^\dagger(\bm k),\singletPair(\bm
k),\vec{\pi}^\dagger(\bm
k),\vec{\pi}(\bm k),\vec{Q}(\bm k),\vec{S}(\bm k),n_{\bm k}  \}
\label{kgenerators1.1}
\end{equation}
where, for example, the singlet pair generator $\singletPair^\dagger$ in
\eq{operatorset} is related to the $\bm k$-dependent generators
$\singletPair^\dagger(\bm k)$ by 
$$
 \singletPair^\dagger=\sum_{\bm
k>0}\singletPair^\dagger(\bm k), 
$$
and $\bm k>0$ means either $k_x>0$ or $k_y > 0$.

Instead of the global SU(4) symmetry generated by 
\eqs{operatorset}, the symmetry now is a direct product of
$k$-dependent SU(4) groups, $\prod_{\bm k>0}\otimes$SU(4)$_{\bm k}$.
The corresponding Hamiltonian is
\begin{align}
    H &= \sum_{\bm k>0}\epsilon_{\bm k}\,n_{\bm k}
    -\sum_{\bm k,\bm k'>0}\{\chi_{\bm k \bm k'}\vec{Q}(\bm k)\cdot\vec{Q}(\bm
k')
\nonumber
\\
&\quad +G^{(0)}_{\bm k \bm k'}\singletPair^\dagger(\bm k)\,\singletPair(\bm k')
+G^{(1)}_{\bm k \bm k'}\vec{\pi}^\dagger(\bm k)\cdot\vec{\pi}(\bm k')
\} ,
\label{Hk1.1}
\end{align}
where $\epsilon_{\bm k}$ and $n_{\bm k}$ are single-particle energies and
occupation numbers, respectively, and the interaction strengths are
\begin{equation}
    \chi_{\bm k \bm k'}=\chi^0 |g(\bm k)g(\bm k')|
\qquad
G^{(i)}_{\bm k \bm k'}=G^{(i)}|g(\bm k)g(\bm k')| ,
\label{kdependentStrength}
\end{equation}
where the index $i$ takes the values 0 and 1.

\subsubsection{Zero-Temperature, Momentum-Dependent Gap Equations}

The formalism may now be developed in a manner parallel to that described in
Refs.~\cite{guid01,sun07}. For example, one obtains a set of $k$-dependent gap
equations that generalize the BCS gap equations. To keep the discussion simple,
we shall consider only the $T=0$ solutions here.   Employing the critical doping
point $x_q$ defined in \eq{xq} with the revised definitions 
$$
G_i \equiv G_i^0 \bar g^2 \qquad \chi \equiv \chi^0 \bar g^2,
$$
with $\bar g$ an averaged $g_k$, solutions for the gap equations at temperature
$T=0$ and momentum $\bm k$ for doping  $x \le x_q$ are found to be \cite{guid09}
\begin{subequations}
\label{gapsUnderdoped}
\begin{align}
   \singletGap (\bm k) &= \displaystyle\frac\Omega2 G_0 \frac{g(\bm
k)}{\bar g}
                    \sqrt{x(x_q^{-1} -x)}
\label{gapsUnderdoped1}
\\[2pt]
   \tripletGap (\bm k) &= \displaystyle\frac\Omega2 G_1 \frac{g(\bm
k)}{\bar g}
                    \sqrt{x(x_q -x)}
\label{gapsUnderdoped2}
\\[2pt]
   \Delta_q(\bm k) &= \displaystyle\frac\Omega2 \chi \frac{g(\bm
k)}{\bar g}
                    \sqrt{x(x_q^{-1} -x)(x_q-x)}
\label{gapsUnderdoped3}
\\[2pt]
   \lambda_k^\prime  &= -\displaystyle\frac\Omega2 \frac{g(\bm k)}{\bar
g}
     \left[
        (\chi-G_1)x_q(1-x_qx)+G_1x
     \right]
\label{gapsUnderdoped4}
\end{align}
\end{subequations}
and the corresponding solutions for $x > x_q$ are
\begin{subequations}
\label{gapsOverdoped}
\begin{align}
   \Delta_q(\bm k) &=
   \tripletGap (\bm k) = 0
\label{gapsOverdoped1}
\\
   \singletGap (\bm k) &= \displaystyle\frac\Omega2 G_0 \frac{g(\bm
k)}{\bar g}
                    \sqrt{1-x^2}
\label{gapsOverdoped2}
\\
   \lambda_k^\prime  &= -\displaystyle\frac\Omega2 \frac{g(\bm k)}{\bar
g}
     \, G_0 x.
\label{gapsOverdoped3}
\end{align}
\end{subequations}
As before, $\singletGap$ and $\tripletGap$ correspond to correlation energies
for singlet and triplet pairing, respectively, $\Delta_q$ corresponds to
correlation energy in the pseudogap state that is fluctuating AF in nature, and
$\lambda'$ denotes the chemical potential. But now each of these quantities
depends on the momentum.

These solutions may then be used to determine other physically important
quantities using the methods described in
Refs.~\cite{guid01,lawu03,guid04,sun05,sun06,sun07,guid08,sun09,guid09,guid09b,
guid10,guid11}. For example, the superconducting transition temperature $T\tsub
c$ is
\begin{equation}
   T\tsub c (\bm k) = G_0 \frac{g(\bm k)}{\bar g} \Omega
        \frac{Rx}{4 k\tsub B\, \atanh(x)},
\label{Tc}
\end{equation}
where the parameter $R$ is of order one and defined in Ref.~\cite{sun07}, and
the pseudogap temperature is
\begin{equation}
  T^* (\bm k) = \chi \frac{g(\bm k)}{\bar g} \Omega
        \frac{R(1-x^2)}{4 k\tsub B},
\label{Tstar}
\end{equation}
and is seen to depend on $k$, which we shall show below leads to a quantitative
description of Fermi arcs.

Equations \eqnoeq{gapsUnderdoped}--\eqnoeq{Tstar} define a $k$-dependent SU(4)
model that can accommodate multiband physics. They are appropriate for
comparison with experimental data that can resolve $\bm k$.  If one averages
these expressions over all momenta $\bm k$ near the Fermi surface, then the
averaged factors $\langle g(\bm k)/\bar g \rangle \rightarrow 1$ and
\eqs{gapsUnderdoped}--\eqnoeq{Tstar} reduce to the equations of the original
SU(4) model.  These are appropriate for comparison with experimental quantities
that do not resolve $\bm k$.

\section{Global Implications of SU(4) Symmetry}

We shall discuss below in some detail that the SU(4) model  permits various
quantitative comparisons with data.  However, because
of the constraints implied by the non-abelian algebra and subalgebras, there are
some important physical consequences of SU(4) symmetry that follow directly from
the symmetry structure itself, without the need of significant calculation. We
shall summarize some of those results in this section.

\subsection{\label{closureConditions} Physical Conditions for Closure of the
SU(4) Algebra}

The preceding formalism is predicated on the operators of \eq{operatorset}
closing an \sufour\ Lie algebra, which requires that the pairing formfactor
satisfy the conditions of \eq{conditions}. For purposes of illustration, let us
discuss the closure condition for the specific case of cuprate superconductors,
where there is essentially uniform agreement on the geometry of the pairing
formfactor. (We shall address the corresponding situation for the iron
superconductors in \S\ref{ironSC}.)

\subsubsection{Closure of the SU(4) Algebra in Momentum Space}

For the cuprates, the pairing formfactor takes the $d$-wave form
\begin{equation} 
g(k)=(\cos k_x -\cos k_y) . 
\label{dWaveFormfactor} 
\end{equation} 
However, $g(k)$ defined in this way does not satisfy the third 
closure criterion of \eq{conditions}. We may obtain closure under commutation if
we
approximate the momentum-space formfactor by a step function,
\begin{equation} 
g(k)=(\cos k_x -\cos k_y) \approx {\rm sgn} (\cos k_x -\cos k_y), 
\label{dWaveApprox} 
\end{equation}
where sgn is the sign operator. The physics  implied by this approximation
becomes more transparent if we transform to coordinate space.

\subsubsection{Closure of SU(4) on the Real-Space Lattice}

Using the {\em exact form} of \eq{dWaveFormfactor}, \eqs{E3} transformed to
coordinate space take the form \cite{guid04}
\begin{equation}
\begin{alignedat}{2} 
p_{12}^\dagger &= \sum_r  c_{{\bf r}\uparrow}^\dagger 
c^\dagger_{\bar{\bf r}\downarrow}\   
\qquad &p_{12} &= \sum_r  
c_{\bar{\bf r}\downarrow}c_{{\bf r}\uparrow} 
\\  
q_{ij}^\dagger &= \sum_r(-)^r c_{{\bf r},i}^\dagger 
c^\dagger_{\bar{\bf r},j}  
&\qquad q_{ij} &= \sum_r (-)^r 
c_{\bar{\bf r},j}c_{{\bf r},i}  
\label{rspace} 
\\  
{Q}_{ij} &= \sum_r(-)^r  c_{{\bf r},i}^\dagger 
c_{{\bf r},j}   
&\qquad S_{ij} &= \sum_r c_{{\bf r},i}^\dagger 
c_{{\bf r},j} - \frac12\Omega\delta_{ij}
\end{alignedat}
\end{equation}
%
where (see \fig{schematicHolePair})
\begin{itemize}
 \item 
$c^\dagger_{{\bf r},i}$ ($c_{{\bf r},i}$) creates (annihilates) an electron of
spin $i$ located at ${\bf r}$, and 
\item
$c^\dagger_{\bar{{\bf r}},i}$ ($c_{\bar{{\bf r}},i}$) creates (annihilates) an
electron of spin $i$ at the four neighboring sites, ${\bf r}\pm{\bf a}$ and
${\bf r}\pm{\bf b}$,
\end{itemize}
with equal probabilities (${\bf a}$ and ${\bf b}$ are lattice constants in the
${\bf x}$ and ${\bf y}$ directions, respectively), 
\begin{equation} 
c^\dagger_{\bar{\bf r},i}=\frac 12 
\left(c^\dagger_{{\bf r} 
+{\bf a},i}+c^\dagger_{{\bf r}-{\bf a},i}  
-c^\dagger_{{\bf r}+{\bf b},i} 
-c^\dagger_{{\bf r}-{\bf b},i}\right) . 
\label{cbar} 
\end{equation}  
The factor $(-)^r$ in \eq{rspace} is $(-)^{n_x+n_y}$  and ($n_x,n_y$) are the
coordinates of a lattice  site on the copper oxide plane, ${\bf r}=n_x{\bf
a}+n_y {\bf b}$,  which is 
\begin{itemize}
 \item 
positive for even sites ($n_x+n_y=$ even) 
\item
negative for the odd sites ($n_x+n_y=$ odd). 
\end{itemize}
This factor originates from the assumption $e^{i{\bf Q}\cdot{\bf r}}\approx
(-)^r$ and implies Mott insulator properties:  the electrons are localized at
lattice sites by strong Coulomb repulsion, with small overlap between orbitals 
of electrons on neighboring
lattice sites.

From the coordinate representation \eqnoeq{rspace}  we see that  spin-singlet
and spin-triplet pairs are formed by holes on adjacent sites, as illustrated in
\fig{schematicHolePair}.%
\singlefig
{schematicHolePair}       
{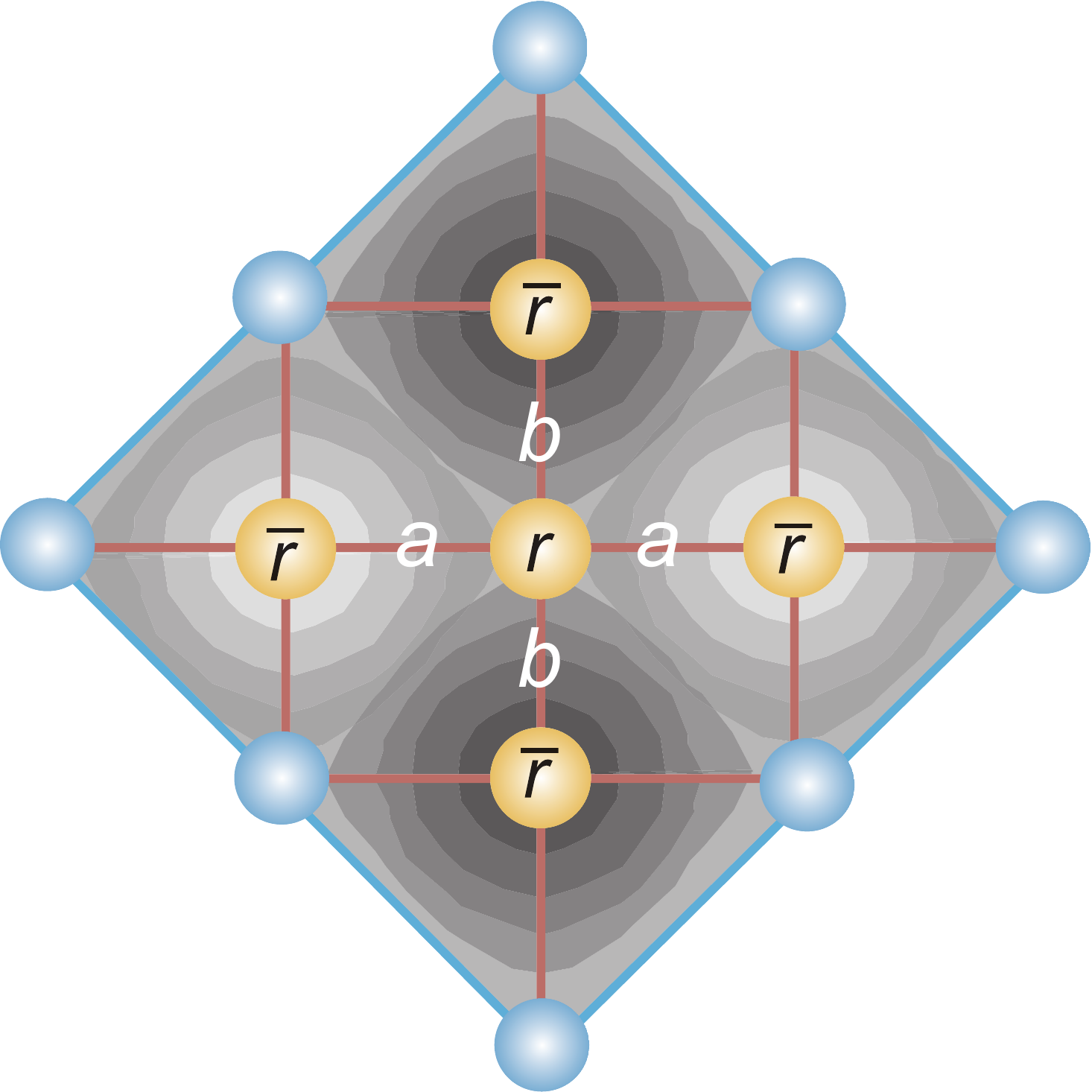}    
{0pt}         
{0pt}         
{0.34}         
{A schematic hole pair.  The yellow-shaded balls are  sites where electron holes
form a pair:   one hole at {\bf r}, the other with equal probability
($\tfrac14$) at the four neighboring sites  (${\bf \bar{r}=r\pm a}$ and ${\bf
r\pm b}$). Balls  on the outer boundary of the
figure (connected by diagonal blue lines) are empty sites where the presence of 
a hole would imply double occupancy (see \fig{maxPairOccupancy}).}
If one hole is located at ${\bf r}$, the other hole occupies the four adjacent
sites (${\bf r}\pm{\bf a}$ and ${\bf r}\pm{\bf b}$), each with equal
probability. These pairs are highly coherent by virtue of the summation over
${\bf r}$ in the pair creation (annihilation) operators. It also can be seen
that 
\begin{align*}  
\hat{n} &= \hat{n}^{\rm (e)}+\hat{n}^{\rm(o)}   
\qquad Q_+=\hat{n}^{\rm (e)}-\hat{n}^{\rm (o)} 
\\  
S_{ij} &= S_{ij}^{\rm (e)}+S_{ij}^{\rm (o)}   
\qquad {Q}_{ij}=S_{ij}^{\rm (e)}-S_{ij}^{\rm (o)}  ,
\end{align*}
where $\hat{n}^{\rm (e)}$ and  $S_{ij}^{\rm (e)}$ are the
total electron number and spin operators at even sites,  and
$\hat{n}^{\rm (o)}$ and 
$S_{ij}^{\rm (o)}$ are the corresponding quantities at odd sites: 
$$
\begin{alignedat}{2} 
\hat{n}^{\rm (e)} &= \sum_{i,r= {\rm even}}  c_{{\bf r},i}^\dagger
c_{{\bf
r},i} 
&\qquad \hat{n}^{\rm (o)}&=\sum_{i,r= {\rm odd}} c_{{\bf r},i}^\dagger c_{{\bf 
r},i}\  
\\ 
S_{ij}^{\rm (e)} &= \sum_{r= {\rm even}}  c_{{\bf r},i}^\dagger c_{{\bf
r},j} 
&\qquad S_{ij}^{\rm (o)}&=\sum_{r= {\rm odd}} c_{{\bf r},i}^\dagger c_{{\bf
r},j} . 
\end{alignedat} 
$$  
Thus the collective operator $Q_+$ may be interpreted as the difference in total
charge, and the collective operator $\vec{Q}$ may be interpreted as the 
difference in
total spin, between even and odd sites on the lattice. From this we may 
conclude that in the U(4)$\supset$SU(4) algebra $Q_+$ will be associated 
with charge density waves and $\vec Q$ will be associated with 
antiferromagnetism.

\subsubsection{SU(4) Symmetry and Double Occupancy of Sites}

The real-space operator set that is defined in \eq{rspace} will be closed under
the operation of commutation for all operators in the set only
if
\begin{equation} 
\{ c_{\bar{\bf r}',i}\ , c^\dagger_{\bar{\bf r},j} \}= 
\delta_{\bf r'r}\delta_{ij} \qquad 
 \{ c_{\bar{\bf r}',i}\ , c_{\bar{\bf r},j} \}=0  
\label{cbarcom1} 
\end{equation} 
(that is, $c_{\bar{\bf r},i}^\dagger$ ($c_{\bar{\bf r},i}$) constitutes  a basis 
for
particles occupying sites adjacent to {\bf r}).  This divides sites of the
lattice into categories A and B, with 
\begin{enumerate}
 \item 
$r$ = even corresponding to the A sites, with operators $c_{{\bf r},i}^\dagger$ 
and $c_{{\bf r},i}$, 
\item
$r$ = odd corresponding to the  B sites, with operators 
$c_{\bar{\bf r},i}^\dagger$ and $c_{\bar{\bf r},i}$
\end{enumerate}
(or vice versa).  Then \eq{cbarcom1} permits \eq{rspace} to be written as 
\begin{align}  
p_{12}^\dagger &= \sum_{{\bf r}\in A} \left( c_{{\bf 
r}\uparrow}^\dagger c^\dagger_{\bar{\bf r}\downarrow}\ -\ c_{{\bf 
r}\downarrow}^\dagger c^\dagger_{\bar{\bf r}\uparrow}\right) 
\nonumber 
\\  
q_{ij}^\dagger &= \pm\sum_{{\bf r}\in A} \left( c_{{\bf 
r},i}^\dagger c^\dagger_{\bar{\bf r},j}+c_{{\bf r},j}^\dagger 
c^\dagger_{\bar{\bf r},i}\right) 
\nonumber 
\\ 
S_{ij} &= \sum_{{\bf r}\in A} \left( c_{{\bf
r},i}^\dagger 
c_{{\bf r},j}-c_{\bar{\bf r},j} c_{\bar{\bf r},i}^\dagger\right) 
\label{rspace2}
\\   
\tilde{Q}_{ij} &= \pm\sum_{{\bf r}\in A} \left( c_{{\bf r},i}^\dagger
c_{{\bf 
r},j}+c_{\bar{\bf r},j}c_{\bar{\bf r},i}^\dagger\right)
\nonumber
\\  
p_{12} &= (p_{12}^\dagger)^\dagger \qquad  
q_{ij}\hspace{3pt}=(q_{ij}^\dagger)^\dagger  
\nonumber 
\end{align} 
with $\tilde{Q}_{ij}\equiv Q_{ij}+\tfrac12 \Omega \delta_{ij}$, where the 
sign is $+$ ($-$) if A is chosen to be even (odd) sites.   (Whether A sites are
taken to be even or odd is a labeling choice and  does not influence the
physics.) Then by explicit commutation the operators \eqnoeq{rspace2} close an
SU(4) algebra if \eq{cbarcom1} is satisfied. But by \eq{cbar}, 
\begin{subequations}
\begin{align}
  \{c_{\bar{\bf r}',i}\ , c^\dagger_{\bar{\bf r},j} \} &= 
\delta_{\bf r'r}\delta_{ij}+ \tfrac14 \delta_{ij}
\sum_t g(\bf t)\ \delta_{\bf r',r+t}\label{cbarcom2}
\\
g({\bf t}) &=
\left\{\begin{array}{l} 
+1\mbox{ for }{\bf t=\pm 2a, \pm 2b}
\\ [3pt]
-1\mbox{ for }{\bf t=+a \pm b,-a\pm b} 
\label{eq15b} 
\end{array}\right.
\end{align}
\end{subequations}
and  \eq{cbarcom1} is generally {\em not} satisfied unless the second term on
the right side of \eq{cbarcom2} can be neglected. This term vanishes if we
require that whenever there is a hole pair $c^\dagger_{{\bf
r}i}c^\dagger_{\bar{\bf r}j}$ at ${\bf r}$ (see \fig{schematicHolePair}), no
pair is permitted at ${\bf r'}={\bf r+t}$, leaving nothing  to be annihilated by
$c_{\bar{\bf r}'i}$.  This is equivalent to a {\em no-double-occupancy
constraint,} because there is a finite amplitude for double occupation of
lattice sites unless it is imposed. For example, if a pair is located at ${\bf
r'}={\bf r}+2{\bf a}$ and a second pair is located at ${\bf r}$,  there is a
probability of 1/16 for two holes to be located at the site ${\bf r}+{\bf a}$
(see \fig{schematicHolePair}). We conclude that {\em closure of the SU(4)
algebra is a direct consequence of no double occupancy} in the copper oxide
conducting plane. 

The validity of \eq{cbarcom1} actually follows from the more general condition
that no pairs overlap spatially in the components of allowed configurations, 
which is a consistency requirement ensuring that the pair space and the pairing
correlations be well defined.  This no-pair-overlap constraint implies
naturally
that a pair centered at ${\bf r}$ precludes a pair being located at ${\bf
r'}={\bf r+t}$ with ${\bf t}$ given in \eq{eq15b}, and hence ensures that
\eq{cbarcom1} holds.

\subsubsection{Pair Formfactors and Closure of the Algebra}

In light of the preceding discussion, let us now revisit the momentum-space
closure conditions for the SU(4) algebra given in \eq{conditions}. The first
requirement $g(\bm k) = g(-\bm k)$ is generally satisfied by
physically-reasonable formfactors. As discussed  in  the preceding section and
in Ref.~\cite{guid04}, the condition $|g(\bm k)| = 1$ necessary to close the
algebra in momentum space may be interpreted as an occupation constraint on the
{\em full formfactor without this condition} in the real space. Specifically,
for cuprates the $d$-wave formfactor $g(\bm k) = \cos k_x - \cos k_y$ must be
approximated by ${\rm sgn\ } (\cos k_x -\sin k_x)$ to close the algebra in
momentum space. However, if the operators are Fourier transformed to the real
space retaining the  {\em full formfactor} $\cos k_x - \cos k_y$, the algebra
closes, but only if the lattice is restricted to no double pair occupancy. 

This suggests that  $|g(\bm k)| = 1$ is not an approximation but rather is a
physically-necessary corollary in momentum space to no double occupancy (by
pairs) for the collective wavefunction in the real space. Therefore,  an
alternative statement of the closure conditions \eqnoeq{conditions} is the
requirement of no double occupancy of the real-space lattice and
\begin{equation}
 g(\bm k + \bm Q) = \pm g(\bm k) 
\qquad
g(\bm k) = g(-\bm k)
\label{conditions2}
\end{equation}
applied to the full momentum-space formfactor, {\em without} the condition
$|g(\bm k)|=1$ of \eq{conditions}.

\subsection{\label{reduction_so8-su4} Reduction from SO(8) to SU(4) Symmetry} 

If all possible bilinear particle--hole and particle--particle (pair) operators
are taken as generators, the minimal closed algebra for an $N$-dimensional basis
is SO(2$N$) \cite{unitaryAlgebra}. The simplest basis for cuprate
superconductors may be regarded as 4-dimensional, since electrons can exist only
in four basic states,  
\begin{enumerate}
 \item 
On A-sites with spin up.
\item
On A-sites with spin down.
\item
On B-sites with spin up.
\item
On B-sites with spin down
\end{enumerate}
Thus, in the absence of further constraints,  the minimal Lie algebra for a
set of generators that can describe high-$T_c$ superconductivity and
antiferromagnetism simultaneously in a cuprate system is ${\rm SO}(2N) = {\rm
SO}(8)$ and not its subgroup SU(4).  The 28 generators of SO(8) are the 16
operators of \eq {rspace2}, plus the 12 additional operators 
\begin{align}  
\bar{p}_{12}^\dagger &= \sum_{{\bf r}\in A} \left( c_{{\bf 
r}\uparrow}^\dagger c^\dagger_{{\bf r}\downarrow}\ -\ c^\dagger_{{\bf 
\bar{r}}\downarrow}c_{ \bf 
\bar{r}\uparrow}^\dagger\right) 
\nonumber 
\\  
\bar{q}_{12}^\dagger &= \pm\sum_{{\bf r}\in A} \left( c_{{\bf 
r}\uparrow}^\dagger c^\dagger_{{\bf r}\downarrow}\ + \ c^\dagger_{{\bf 
\bar{r}}\downarrow}c_{ \bf 
\bar{r}\uparrow}^\dagger\right) 
\nonumber 
\\    
\bar{S}_{ij} &= \sum_{{\bf r}\in A}  \left( c_{{\bf 
r},i}^\dagger c_{{\bf \bar{r}},j}-c_{{\bf r},j}c_{{\bf 
\bar{r}},i}^\dagger\right) 
\label{so8}
\\   
\bar{Q}_{ij} &= \pm\sum_{{\bf r}\in A} \left( c_{{\bf r},i}^\dagger
c_{{\bf 
\bar{r}},j}+c_{{\bf r},j}c_{{\bf \bar{r}},i}^\dagger\right)
\nonumber
\\ 
\bar{p}_{12} &= (\bar{p}_{12}^\dagger)^\dagger \qquad 
\bar{q}_{12}=(\bar{q}_{12}^\dagger)^\dagger  ,
\nonumber 
\end{align} 
where the $\pm$ signs depend on the even--odd choice for  A-sites; see
\eq{rspace2}. Equation (\ref{so8}) contains two new kinds of spin-singlet pairs
created by $\bar{p}_{12}^\dagger$ and $\bar{q}_{12}^\dagger$, which we shall
term $S$ and $S^*$ pairs, respectively.  Unlike the bondwise pairs associated
with the SU(4) subalgebra within the SO(8) algebra, these new pairs are {\em
onsite}, where the two electrons (or two holes) occupy the same site, with equal
probability to appear anywhere in the lattice coherently. Figure
\ref{fig:su4-so8compare} illustrates.%
\singlefig
{su4-so8compare}       
{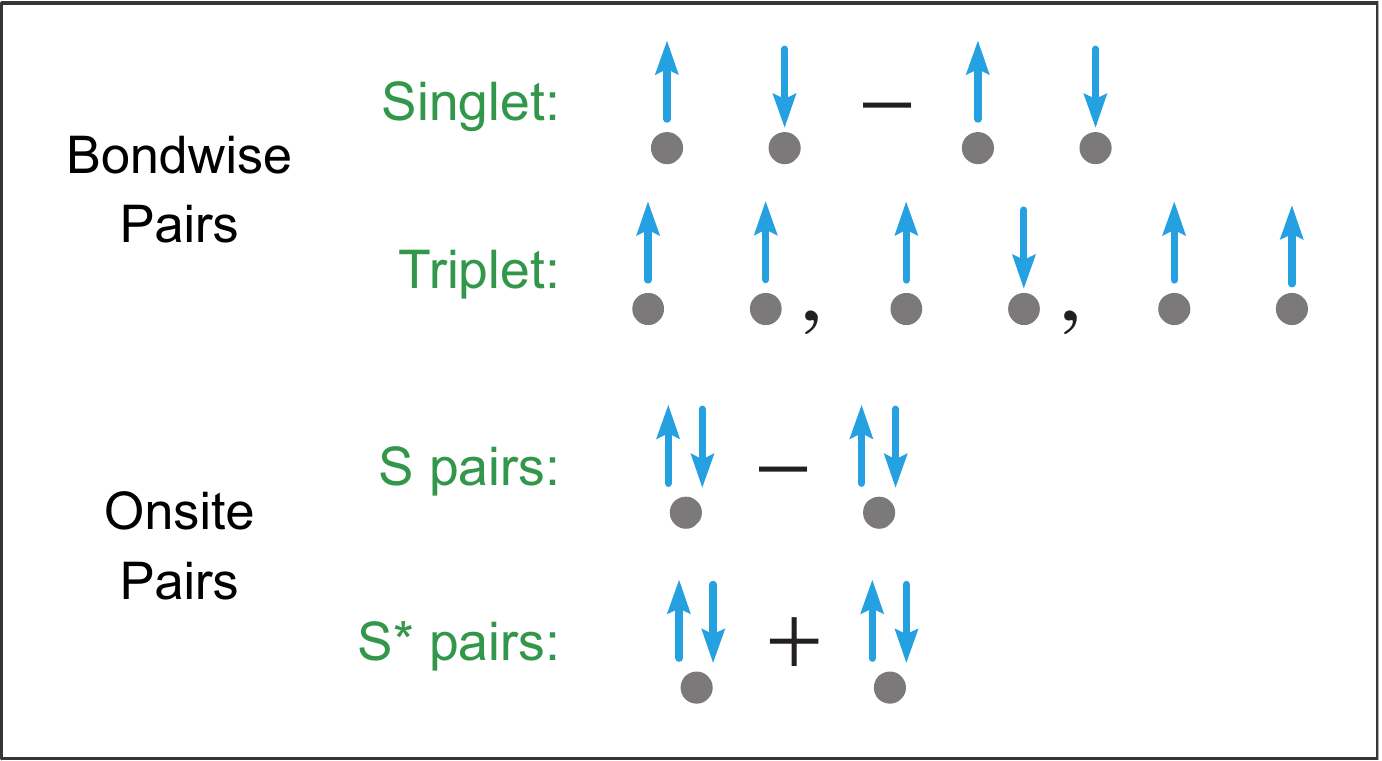}    
{0pt}         
{0pt}         
{0.57}         
{Schematic difference between bondwise and onsite  airs in the SO(8) symmetry.}
The $S^*$ and $S$ pairs differ from each other only in phases. The operators
$\bar{S}_{ij}$ are the hopping operators with and without spin flip, and
$\bar{Q}_{ij}$ is the staggering of the hopping. These operators interchange 
singlet
and triplet pairs with $S$ and $S^*$ pairs. 
 
The SO(8) algebra reduces to the subalgebra SU(4) if the $S$ and $S^*$ pairs may
be neglected. This will be a reasonable approximation if we assume onsite
Coulomb repulsion pushing the $S$ and $S^*$ pairs to sufficiently high energy,
as illustrated in \fig{su4-so8separate}.%
\singlefig
{su4-so8separate}       
{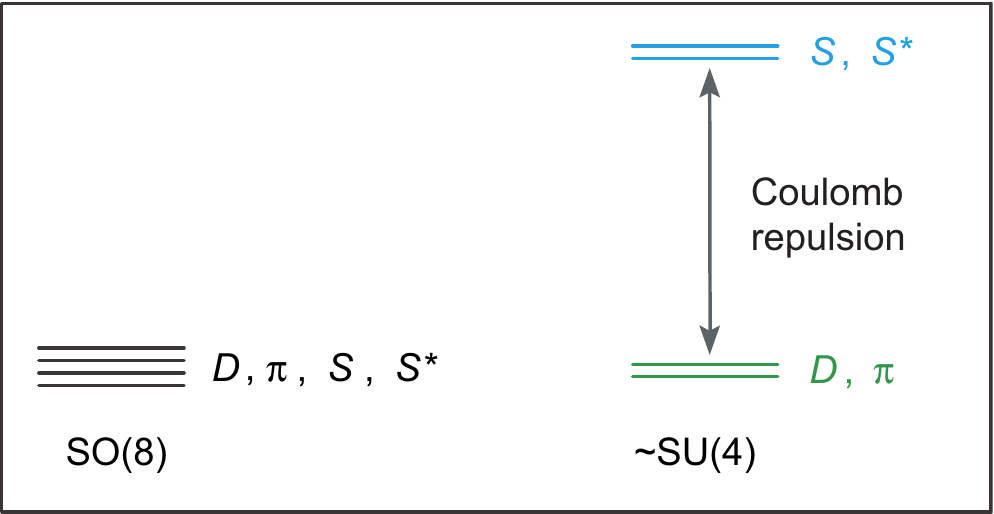}    
{0pt}         
{0pt}         
{0.79}         
{Separation of the onsite pair ($D, \pi$) and bondwise pair ($S, S^*$) 
energy scales by onsite Coulomb
repulsion, which has the effect of reducing SO(8) symmetry to an effective 
SU(4) low-energy symmetry.}
Thus, restriction to no double occupancy effectively allows the operators in
\eq{so8} to be ignored and reduces SO(8) to its subalgebra SU(4). We conclude
that the minimal Lie algebra  that can describe antiferromagnetism and $d$-wave
superconductivity in a  cuprate system is in general SO(8), but under the
constraint of no double occupancy the symmetry effectively reduces to SU(4).
The assumption of an SU(4) symmetry in a cuprate system automatically implies
the imposition of a {\em no-double-occupancy constraint} on the more general
SO(8) symmetry in the copper--oxygen planes. 

The general relationship between SO(8) and SU(4) symmetry, and the
corresponding relationship of SO(8) to the symmetry of a conventional BCS
superconductor, are illustrated in \fig{su4-so8_relationship}.%
\singlefig
{su4-so8_relationship}       
{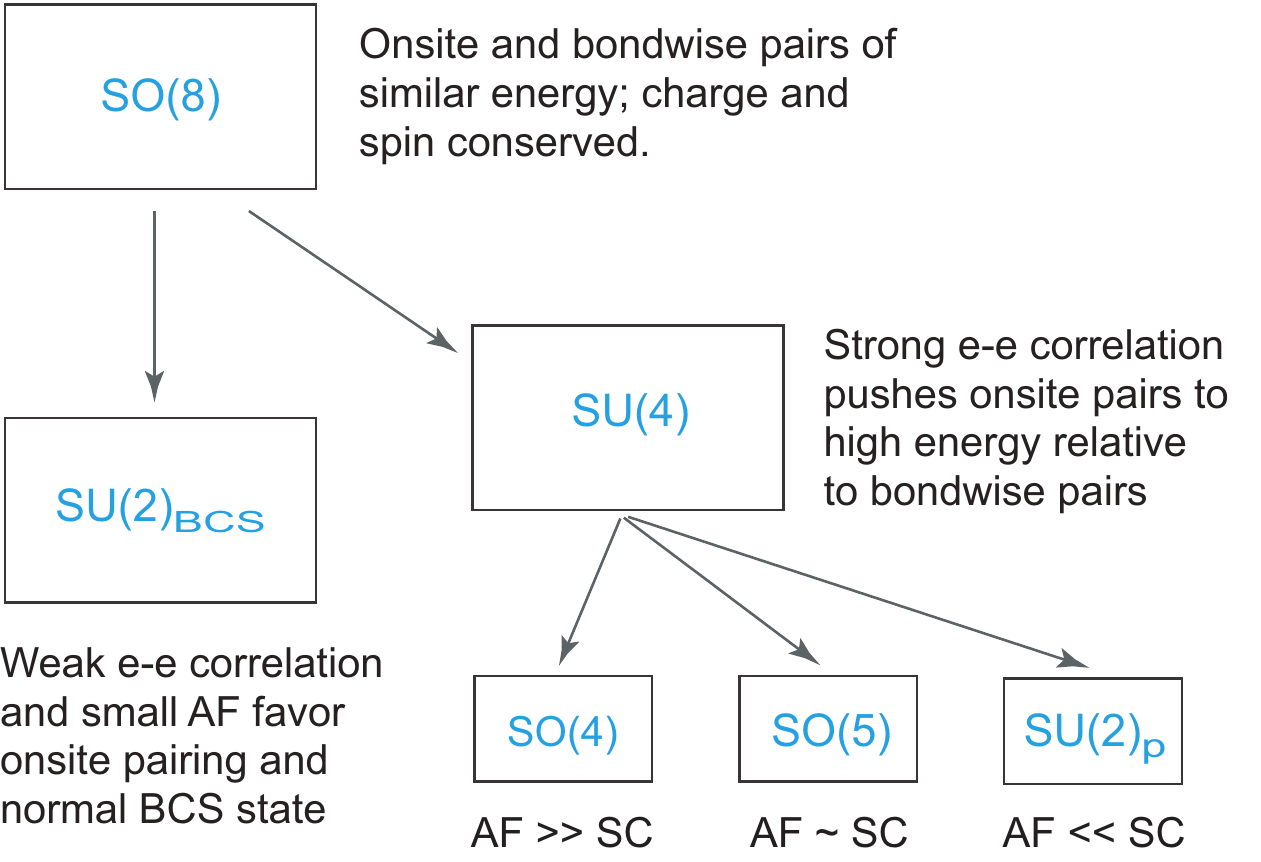}    
{0pt}         
{0pt}         
{0.68}         
{The general relationship among SU(4),  SO(8), and the
pseudospin SU(2) symmetry of a conventional BCS superconductor. Both SU(2)
subgroups have pseudospin pair generators, but differ in the pairs being
onsite for SU(2)$\tsub{BCS}$ and bondwise for SU(2)$\tsub p$, implying 
different orbital formfactors for pairing in the two cases.  The symmetry 
SU(2)$\tsub{BCS}$ with conventional formfactor is favored only if onsite 
Coulomb repulsion and any collective modes like AF competing with pairing can 
be neglected.}
If physically we assume weak enough electron--electron correlations and neglect
antiferromagnetism, the SO(8) states favor a dynamical symmetry 
\begin{equation}
\soeight \supset \ldots \supset \sutwo\tsub{BCS}
\label{su4so8Relationships} 
\end{equation} 
that indicates a spontaneously-broken symmetry selecting the conventional
superconductor direction in the SO(8) space (the ellipses in the subgroup chain
denote possible intervening subgroups).  The final SU(2) subgroup in this chain
is that of normal pseudospin for onsite pairs (which are compatible with binding
by the lattice phonon interaction characteristic of conventional BCS
superconductivity).

\subsection{\label{upperDopingLimit} SU(4) Symmetry and an Upper Doping Limit
for the Superconducting State}

The implicit SU(4) occupancy constraint discussed in the preceding section
dictates an upper limit for the doping fraction in SU(4)-conserving states
\cite{guid04}. This is illustrated in \fig{maxPairOccupancy},%
\singlefig
{maxPairOccupancy}      
{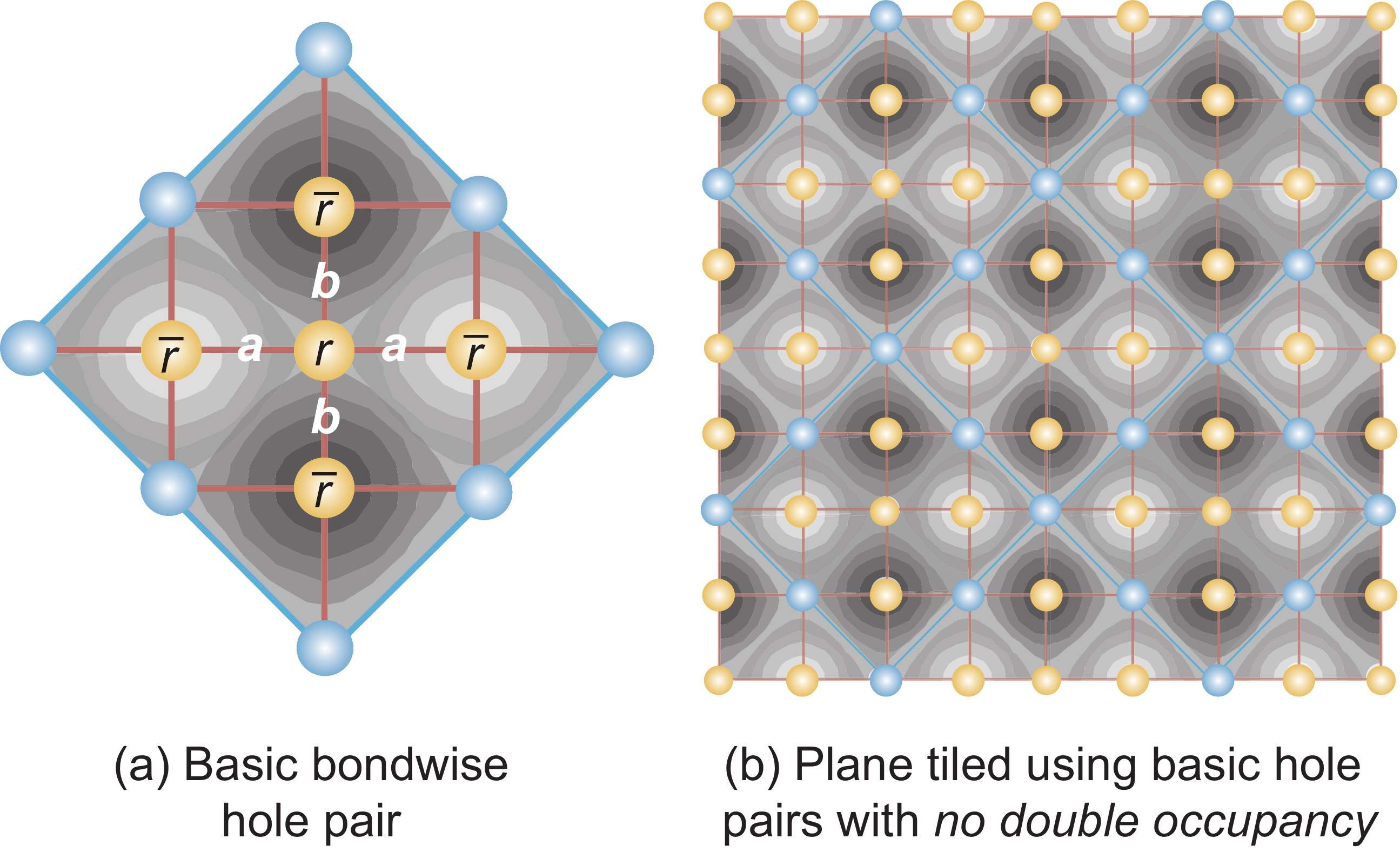} 
{0pt}         
{0pt}         
{0.328}         
{(a)~The schematic hole pair of \fig{schematicHolePair}.
(b)~Tiling of the plane by the hole 
pairs of (a).  Each diamond outlined by dashed boundaries corresponds to one 
unit pair from (a).  Lighter-colored balls connected by solid horizontal and 
vertical lines are  sites where the electron holes  form pairs. Darker balls 
connected by diagonal dashed lines indicate sites where the presence of a hole 
would imply average double occupancy of some sites on the lattice, which would 
break SU(4) symmetry. By counting, the lattice can ensure no double occupancy 
[and thus preserves SU(4) symmetry] only if it is not more than $\tfrac14$ 
occupied by electron holes.}
%
%
which shows the spatial distribution of a representative configuration 
when the hole-pair number is maximal.  By counting,  the maximum 
number of
holes consistent with SU(4) symmetry is $\Omega=\tfrac14 \Omega_e$, where 
$\Omega_e$ is the total number of
lattice sites. Thus the largest doping fraction  preserving SU(4) symmetry is
$P\tsub f=\Omega/\Omega_e= \tfrac14$.  The maximum hole-doping fraction
($0.23\sim 0.27$) that is seen experimentally for cuprate superconductivity may 
then be
interpreted as a direct consequence of physical constraints on the 
realization of SU(4) symmetry.

\subsection{\label{AF-SCcompetition} The Antiferromagnetic--Superconducting
Transition}

The \sofive\ subgroup corresponds to a critical dynamical symmetry interpolating
between AF  and SC order  for a  range of intermediate doping parameters
\cite{lawu03}.  We now show that the emergent SU(4) symmetry implies a differing
dependence on doping for SC and AF order, and that the dynamical symmetry
structure itself controls the transition between the superconducting \sutwo\
symmetry and the antiferromagnetically ordered \sofour\ symmetry.  Thus, we
shall  show that the \sufour\ symmetry has a natural propensity to favor
antiferromagnetic Mott order at half-filling and singlet-pair superconductivity 
 as
the system is doped away from half-filling, for a broad range of Hamiltonian
parameters.

\subsubsection{Competing Antiferromagnetism and Superconductivity}

We drop the common dependence of both phases on the spin and charge generators
and consider the competition between  \sofour\ stabilization energy arising from
$\QdotQ$ and \sutwo\ stabilization energy associated with the term $\DdagD$ in
the Hamiltonian \eqnoeq{eq1}.  These differ in their dependence on particle
number and thus on doping.  If we evaluate the AF correlation energy in the AF 
limit and the
pairing correlation energy in the SC limit, we obtain as a function of doping
$x$
\begin{subequations}
 \label{compare}
\begin{align}
\ev{\chi \QdotQ} &= \tfrac14 \chi\Omega^2 (1-x)^2
\label{compare1}
\\[3pt]
\ev{G_0 \DdagD} &= \tfrac14 G_0\Omega^2 (1-x^2) ,
\label{compare2}
\end{align}
\end{subequations}
so that their ratio is
\begin{equation}
 \frac{\ev{G_0 \DdagD}}{\ev{\chi \QdotQ}} = \frac{G_0}{\chi}
\frac{(1+x)}{(1-x)}.
\end{equation}
The competition between AF and pairing correlation energy is illustrated in
\fig{SC-AFcompetition}.%
\singlefig
{SC-AFcompetition}       
{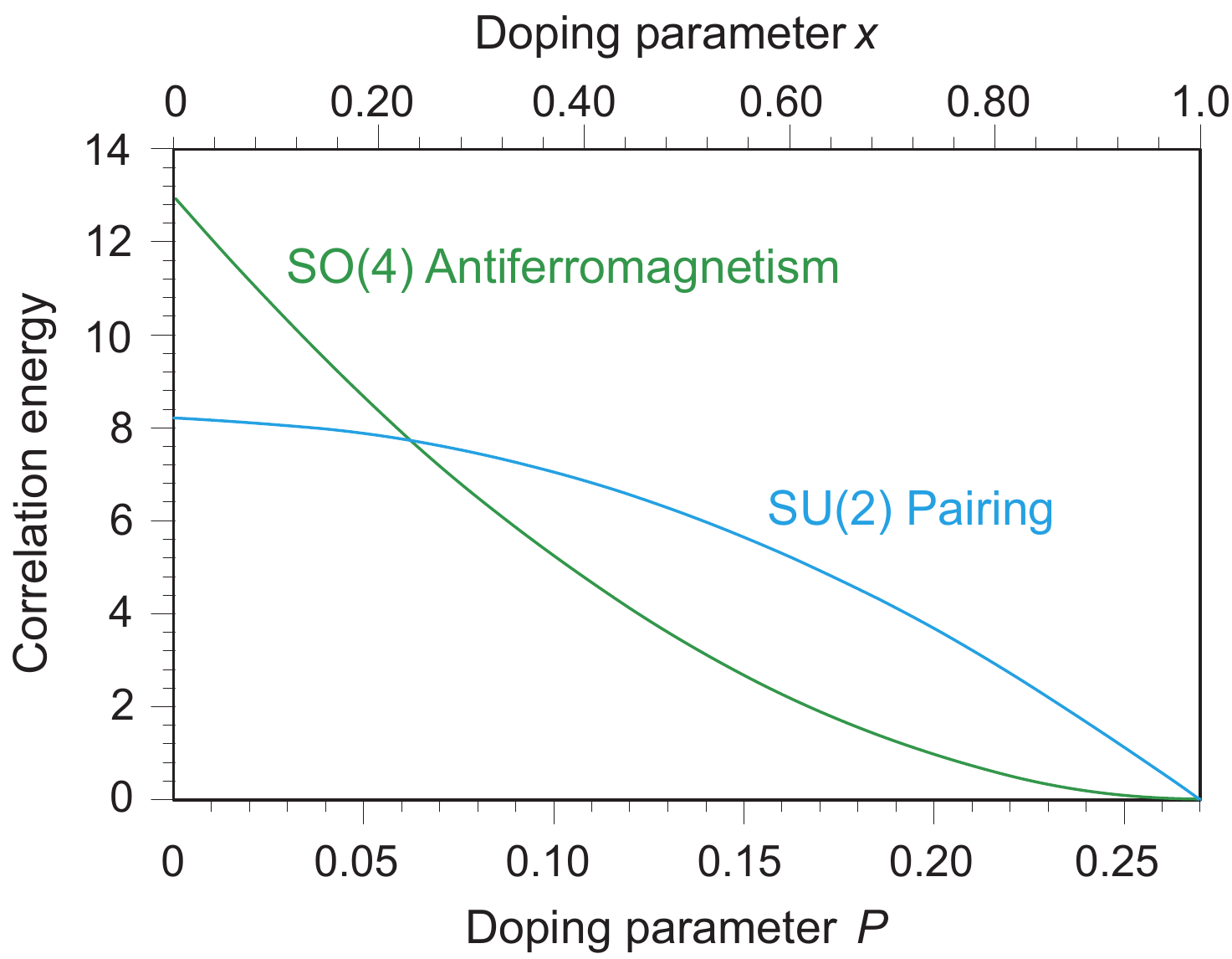}    
{0pt}         
{0pt}         
{0.55}         
{Competition between antiferromagnetic and pairing correlation energy as
a function of doping assuming the coupling strength parameters used in
\fig{Energy}.}
At half filling ($x=0$), the Hamiltonian exhibits effective \sofour\ symmetry if
$\chi >G_0$ because the \sofour\ correlation energy \eqnoeq{compare1} is
dominant. With increasing hole-doping $x$, the \sofour\ correlation energy
decreases more rapidly than the singlet pairing correlation energy
\eqnoeq{compare2}. Thus, the pairing correlation eventually dominates and the
Hamiltonian exhibits effective \sutwo\ symmetry. This argument concerning the
SC--AF competition is only qualitative, particularly in the vicinity of the
crossing curves in \fig{SC-AFcompetition}, since the expressions
\eqnoeq{compare} are evaluated in the respective symmetry limits, not in the
actual physical ground state. We shall deal more 
quantitatively with the antiferromagnetic to superconducting transition in 
later sections but these 
qualitative arguments are useful to illustrate
the general trend of the AF--SC competition with doping. 

These features imply immediately that if $\chi /G_0>1$, antiferromagnetism
tends to dominate at half-filling but pairing tends to dominate as holes are
doped into the system. Thus, the AF ground state at half filling and the
tendency to superconductivity as the system is doped away from half-filling
follow directly from the dynamical symmetry structure of the Hamiltonian,
independent of detailed parameter choices, and independent of detailed
underlying microscopic physics [as long as it is consistent with emergent SU(4)
symmetry and the condition $\chi /G_0>1$]. That is a significant
conclusion but the implications of the SU(4) symmetry are even more dramatic
than that. As we shall show in \S\ref{pairingInstability}, because of the
\sufour\ dynamical symmetry structure the AF insulating state 
at half
filling hides within it a superconductor that can emerge spontaneously with
infinitesimal doping if there is a finite singlet pairing interaction.

\subsubsection{Analogies in Nuclear Structure Physics}

The competition between antiferromagnetism and superconductivity has many 
parallels with the competition between spherical and deformed structure for 
nuclei that is a central paradigm of nuclear structure physics. The transition 
from spherical nuclei,  which dominate the beginnings and endings of neutron 
and 
proton shells, to deformed nuclei, which tend to dominate the middle of shells, 
is governed by a microscopic competition between long-range 
quadrupole--quadrupole interactions favoring deformation and short-range 
monopole pairing interactions that favor spherical vibrational structure.  This 
competition may be expressed algebraically as a competition 
between a dynamical symmetry that favors  pairing (particle--particle) 
interactions and a 
dynamical symmetry that favors multipole (particle--hole) interactions 
\cite{FDSM}.  

The essential physics of the spherical--deformed transition in nuclear 
structure 
is determined by the dependence of the dynamical symmetries on particle 
number:  nuclear pairing energy increases linearly with particle number from 
closed shells but the quadrupole deformation energy is approximately quadratic 
in particle number.  Thus, the dynamical symmetry structure implies that  
spherical vibrational nuclei (which are favored by pairing energy) dominate the 
beginning and ends of shells and deformed nuclei (which are favored by the 
deformation energy) 
dominate the middle of shells \cite{footnote2}. 
 
This behavior  is a close analog of the competition
between AF dominating the half-filled lattice and SC
dominating the hole-doped lattice that was discussed above, 
suggesting that these problems from  different fields of physics
may have a common dynamical symmetry basis. Although they involve different
constituent particles, different basic interactions, and fundamentally different
energy and length scales, they share the common general features of being
strongly-correlated quantum systems of Fermi--Dirac particles with essential
interactions in both particle--hole and particle--particle channels.  The work
presented here and in Refs.~\cite{FDSM,guid13,guid2017b} suggests  a unified 
picture of strongly correlated
condensed matter and nuclear structure physics in terms of dynamical symmetries 
of
their respective Hamiltonians, despite the obvious differences between the
ingredients making up those Hamiltonians and the fundamentally different energy 
and
length scales.

\subsubsection{Analogies with Graphene Quantum-Hall Physics}

Although we will not discuss it in any detail, the present methodology 
has been applied successfully to describing collective states 
for monolayer graphene in a strong magnetic field 
\cite{lawu2016,lawu2017,guid2017}. Remarkably, these graphene states also 
exhibit a dynamical symmetry with a Lie algebra similar to that 
described here, and the total energy surfaces for those graphene states were 
found to be almost identical to those found for high-temperature 
superconductors and nuclear structure physics, even though the underlying 
microscopic physics could hardly be more different among these cases.  This 
suggests a universality of emergent behavior through Lie algebras that even 
goes beyond the universality of superconductivity and superfluidity proposed 
here
\cite{guid13,guid2017b}.

\section{\label{esurfaces}Ground-State Energy Surfaces}

Let us now turn to more quantitative applications, beginning with ground-state
energy surfaces.
The total energy of the \sufour\ state may be found in coherent state
approximation by evaluating the expectation value of the Hamiltonian
\eqnoeq{eq1}.  There are only two independent variational parameters in the
coherent state variational equations because of the unitarity condition
\eqnoeq{unitarityuv}. They may be chosen as either $\mbox{v}_+$ and
$\mbox{v}_-$,  or as $\alpha$ and $\beta$, using the definitions
\begin{equation}
\mbox{v}_+ \equiv \alpha+\beta\hspace{24pt}\mbox{v}_- \equiv
\alpha-\beta
\label{abuv}
\end{equation}
However, from  \eq{zeroTMel4} the squares of $\mbox{v}_\pm$ (or of
$\alpha$ and $\beta$) are constrained by 
\begin{equation}
n = \langle \hat n\rangle 
= \Omega (\mbox{v}_+^2 + \mbox{v}_-^2)
=2\Omega (\alpha ^2+\beta ^2).
\label{eqnv}
\end{equation}
Thus, for a fixed particle number $n$ we may evaluate matrix elements with only
a single variational parameter, say $\beta$, which may in turn be related to
standard order parameters by comparing matrix elements. For example, the $z$
component of the staggered magnetization is  related to $\beta$ and $v_\pm$ by
\begin{align}
Q \equiv \langle \AF_z \rangle 
&= \tfrac12 \Omega (\mbox{v}_+^2 - \mbox{v}_-^2)
\nonumber
\\
&= 2\Omega \beta (n / (2\Omega) - \beta^2)^{1/2}, 
\label{eqq}
\end{align}
and these measures of AF order are in turn related to the
superconducting order parameter $\alpha$ through \eq{eqnv}. From 
\eqs{eqq} and \eqnoeq{eqnv}, the ranges of $\beta$ and $\alpha$ are 
\begin{equation}
0\le\beta\le\sqrt{n/4\Omega} 
\qquad
\sqrt{n/4\Omega}\le\alpha\le \sqrt{n/2\Omega}.
\label{alphabetaranges}
\end{equation}
From \eqs{eqnv} and \eqnoeq{eqq}, one can show that
\begin{equation}
\mbox{v}^2_\pm=\frac{n}{2\Omega}\pm\frac{Q}{\Omega} .
\end{equation}
Equations \eqnoeq{zeroTMel1} and \eqnoeq{zeroTMel2} can be written as
\begin{subequations}
\begin{align}
\Delta &\equiv \langle \singletPair^{\dagger}\rangle =\langle
\singletPair\rangle=\sqrt{ \DdagD}=
\Delta_+ +\Delta_-  
\label{pairOrder1}
\\
\Pi &\equiv \langle \pi^{\dagger}_z\rangle =\langle \pi_z\rangle =\sqrt{
\pidagpi}=\Delta_+ -\Delta_- ,
\label{pairOrder2}
\end{align} 
\label{pairOrder}
\end{subequations} 
where we define
\begin{equation}
\Delta_\pm \equiv \frac\Omega2
\sqrt{\frac14-\left(\frac{Q}{\Omega}\mp\frac x2\right)^2},
\end{equation}  
and $x$ is the effective hole concentration that was introduced in 
\eq{dopingParam}. The quantities $\Delta$ and $\Pi$ present the spin-singlet
and  spin-triplet pairing correlations, respectively.  The former is
proportional to the singlet pairing gap and thus is directly related to the
superconducting order. The latter is a measure of collectivity for triplet
pairing, but it also measures the \sofive\ correlation since from Table 
\ref{table:su4Properties} of
Appendix \ref{appendix},
\begin{equation} 
C_{\sofive}=\pidagpi+\vec{S}\cdot\vec{S}+M(M-3) 
\label{so5correlationE}
\end{equation}
is the \sofive\ Casimir operator, which from \eq{pairOrder2} is proportional to
$\Pi^2$ for fixed charge and spin. 

Using \eqs{zeroTMel} and \eqnoeq{abuv}--\eqnoeq{eqq}, one can then
evaluate the energy surface as a function of the order parameters. For example,
if we assume that $\ev{\vec S \cdot \vec S} = 0$, 
\eqs{zeroTMel6}--\eqnoeq{zeroTMel8} may be used to obtain a general expression
for the \sufour\ energy surface that takes the form \cite{lawu03}
\begin{equation}
E(Q)=
\langle H\rangle-H_0=-\tilde{G}_0\, [ (1-\pratio)\Delta^2 + \pratio Q^2] ,
\label{esurfaceQ}
\end{equation} 
in the limit $\Omega \rightarrow \infty$, where $\pratio$ is a parameter varying
between 0 and 1. Equation \eqnoeq{eqq} may be used to
convert this to an energy surface  as a function of the alternative AF order
parameter $\beta$ at fixed particle number $n$,
\begin{align}
E(\beta)&=\langle H\rangle-H_0=-\frac{\tilde{G}_0\Omega^2}4 
\nonumber
\\
&\hspace*{-20pt}\times\left\{
(24\pratio-8)\beta^2\left(\frac{n}{2\Omega}-\beta^2\right)+2(1-\pratio)
\left[\
\frac{n}{2\Omega}\left(1-\frac{n}{2\Omega}\right)\nonumber
\vphantom{\sqrt{ \left( \left( \frac{n}{2\Omega} \right) ^ 2 \right) } }
\right.\right. 
\\
&\hspace*{-20pt}+\left.\left.\left(
\frac{n}{2\Omega}-2\beta^2
\right)
\sqrt{
\left(1-\frac{n}{2\Omega}\right)^2-4\beta^2\left(\frac{n}{2\Omega}
-\beta^2\right)}\
\right]\right\}  
\label{eqeb}
\end{align}
and this also may be expressed in terms of the superconducting order
parameter $\alpha$ using \eq{eqnv}. 

The SU(4) coherent state solutions are valid for arbitrary ratios of the pairing
and antiferromagnetic coupling strengths, so the corresponding ground-state
total energy surfaces can be computed for arbitrary coupling strength ratio.
However, it is instructive to first examine the energy surfaces in the dynamical
symmetry limits of \S\ref{dynamicalSymmetryLimits}. Figures
\ref{fig:eSurfaces1D} and \ref{fig:eSurfaces} illustrate.
\singlefig
{eSurfaces1D}       
{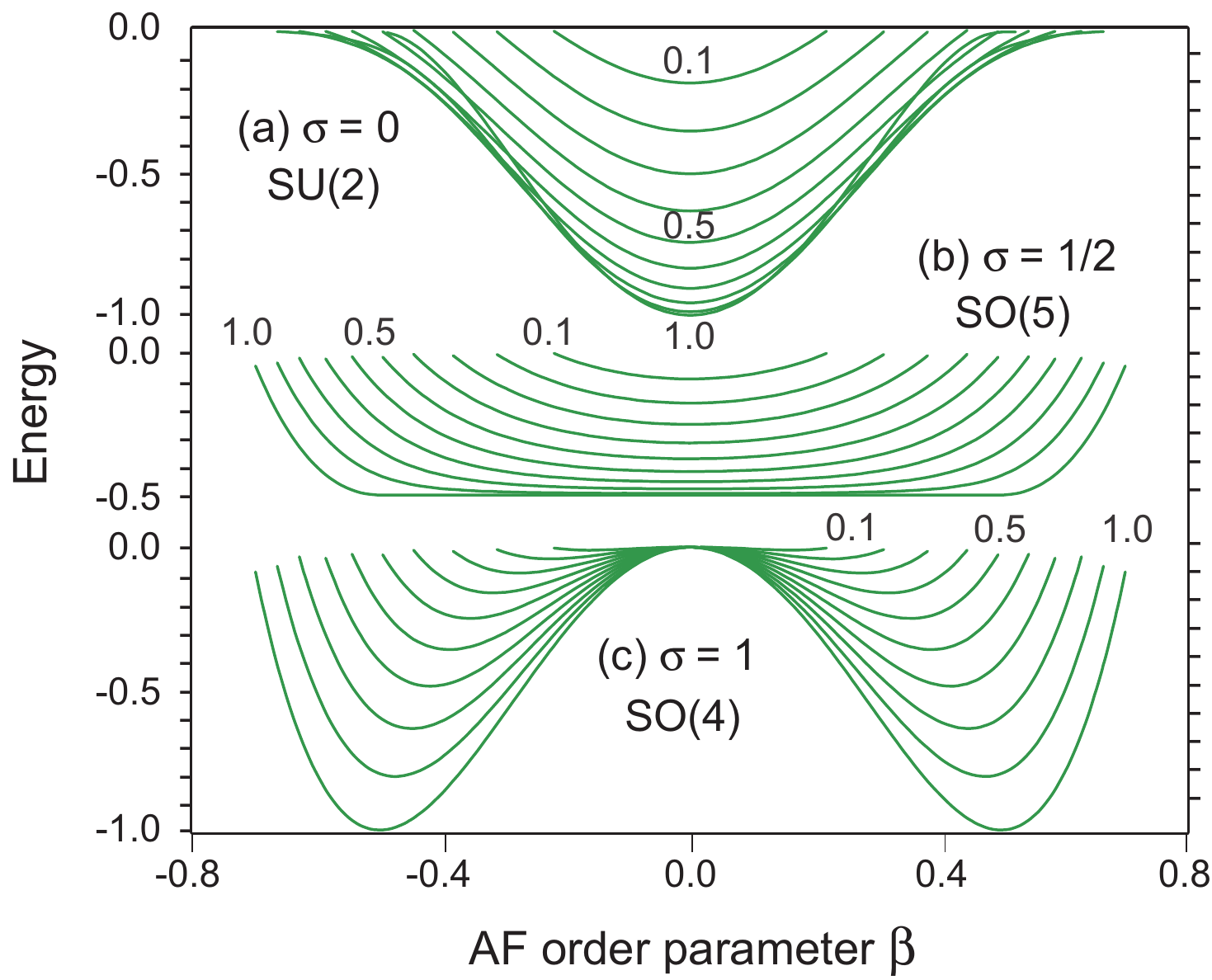}    
{0pt}         
{0pt}         
{0.57}         
{SU(4) Coherent-state energy surfaces for the three symmetry limits of the SU(4)
Hamiltonian. The energy unit is $\tfrac14 \tilde{G}_0\Omega^2$ (see \eq{eqeb}).
$H_0$ is taken as the energy zero point.  Numbers on curves are  the lattice
occupation fractions $n/\Omega = 1-x$, with $n / \Omega = 1$ corresponding to
half filling and $0 < n / \Omega < 1$ to finite hole doping. The
antiferromagnetic \sofour\ symmetry corresponds to $\sigma=1$, the critical
\sofive\ symmetry corresponds to $\sigma=\tfrac12$, and the superconducting
\sutwo\ symmetry corresponds to $\sigma=0$ for the Hamiltonian \eqnoeq{eq1}. The
allowed range of $\beta$ is
$[-\frac12 (n/\Omega)^{1/2},\frac12(n/\Omega)^{1/2}]$, 
which depends on $n$. The order parameter $\beta$ is related to the order
parameter $Q \equiv \langle \AF_z \rangle$ (staggered magnetization) and the
electron number $n$ through \eq{eqq}.}
\singlefig
{eSurfaces}       
{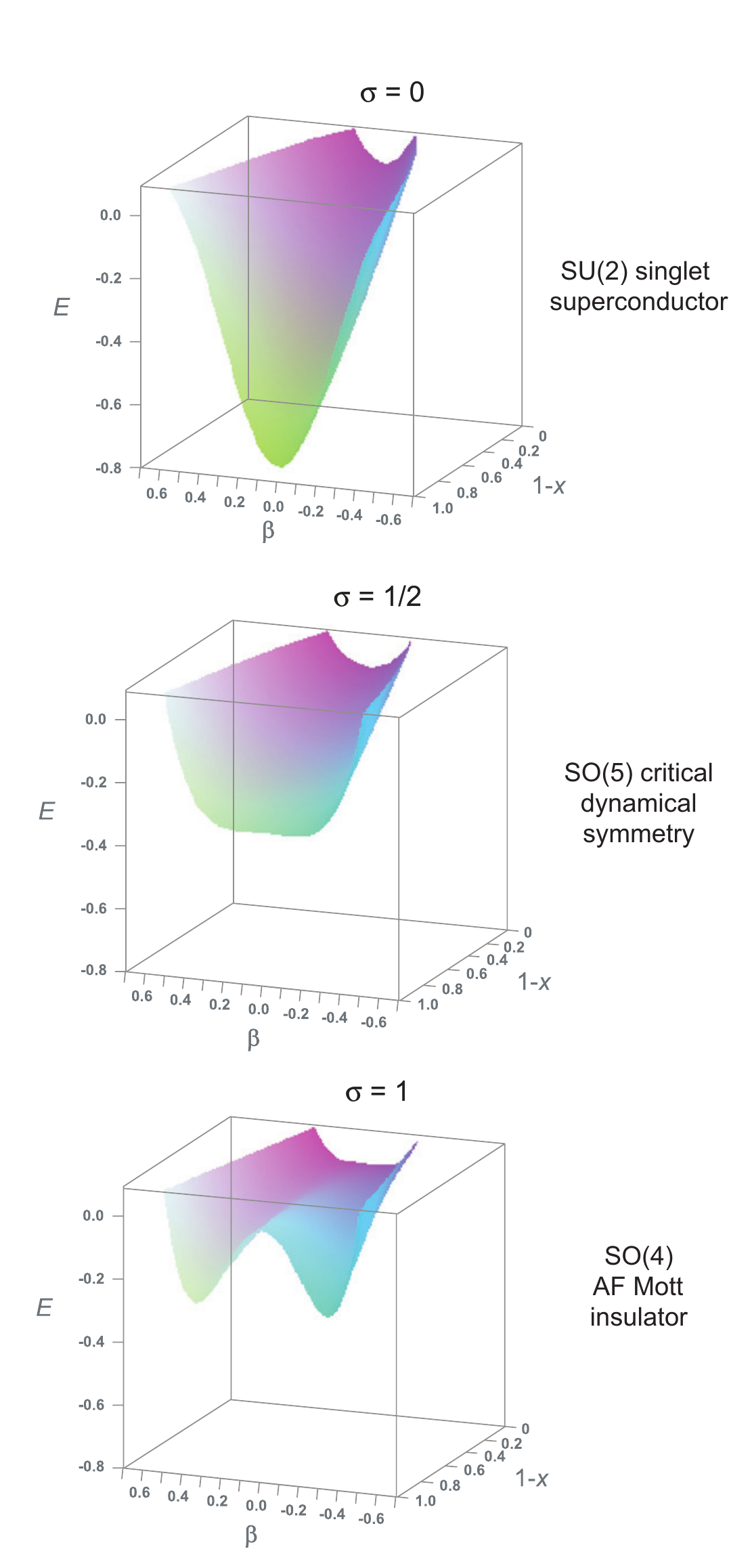}    
{0pt}         
{0pt}         
{0.78}         
{SU(4) energy surfaces as a function of the AF order parameter $\beta$ and the
doping $x$.}

\subsection{\label{so4Surface} Energy Surfaces in the SO(4) Limit}

The identification of the SO(4) limit in \S\ref{so4Dynamical} as an
AF state is strengthened by examining the 
energy surfaces illustrated in \fig{eSurfaces1D}. For $\sigma = 1$ [\sofour\
limit; see \fig{eSurfaces1D}c], $\beta = 0$ is an unstable point and an
infinitesimal fluctuation will drive the system to the energy minima at finite 
$
\beta = \pm \tfrac12 (n/\Omega)^{1/2}. 
$
Thus, $|Q|$ reaches its maximum value of $n/2$, indicating a
spontaneously-broken symmetry and a state having AF order.

\subsection{\label{su2Surface} Energy Surfaces in the SU(2) Limit}

Further insight into the SU(2)  limit discussed in \S\ref{su2Dynamical} follows
by examining the ground-state energy surfaces illustrated in \fig{eSurfaces1D}.
For $\sigma = 0$ [\sutwo\ limit; see \fig{eSurfaces1D}a], the minimum energy
occurs at $\beta = 0$ (equivalently, $Q=0$) for all values of $n$. Thus,
$\Delta$  reaches its maximum value of
\begin{equation}
\Delta_{\rm max} = \tfrac12 \Omega \sqrt{1-x^2}, 
\end{equation}
indicating a state having superconducting order but vanishing antiferromagnetic
order.

\subsection{\label{so5Surface} Energy Surfaces in the SO(5) Limit}

The unusual nature of the \sofive\ critical dynamical symmetry  discussed
in \S\ref{so5Dynamical} is brought into sharp focus by examination of the 
ground-state
energy surfaces that are illustrated in \fig{eSurfaces1D}. From
\fig{eSurfaces1D}b, the \sofive\ dynamical symmetry is seen to have extremely
interesting behavior: the minimum energy occurs at $\beta = 0$ for all values of
$n$, as in the \sutwo\ case, but there are large-amplitude fluctuations in
antiferromagnetic and superconducting order. In particular, when $n$ is near
$\Omega$ (near half filling), the system has an energy surface almost flat for
broad ranges of $\beta$ (or $Q$ or $\alpha$). This suggests a phase very soft
against fluctuations in the AF and SC order parameters. 
However, as $n/\Omega$
decreases, fluctuations become smaller and the energy surface tends
to the \sutwo\ (superconducting) limit.

\subsection{\label{criticalD} Critical Dynamical Symmetries}

Dynamical symmetries that interpolate between other dynamical symmetries are 
termed 
{\em critical dynamical symmetries} \cite{wmzha87}.  The $\sufour
\supset \sofive$ symmetry exhibits such transitional properties. At
half filling the energy surface is completely flat under variations of the
antiferromagnetic order parameter $\beta$ (see the $n=1.0$ curve of
\fig{eSurfaces1D}b), implying large fluctuations in the order parameters. But as
hole doping is increased the \sofive\ energy surface changes smoothly into one
localized around $\beta = 0$ (see the $n=0.1$ curve of \fig{eSurfaces1D}b).

Under an exact $\sufour \supset \sofive$ symmetry the antiferromagnetic and
superconducting states are degenerate at half filling, there is no barrier
between AF and SC states, and one can fluctuate into the other at zero cost in
energy (see the $n/\Omega =1$ curve of \fig{eSurfaces1D}b). To see in more
detail how in the \sufour\ model an \sofive\ symmetry can interpolate
between AF and SC states as particle number varies,  let us examine a
case that is perturbed slightly away from the \sofive\ limit of $\pratio =
\tfrac12$ in \eq{eq1}.

\subsection{\label{weaklyBrokenSO5} Weakly-Broken SO(5) Symmetry}

In \fig{so5fig2}a,
\singlefig
{so5fig2}       
{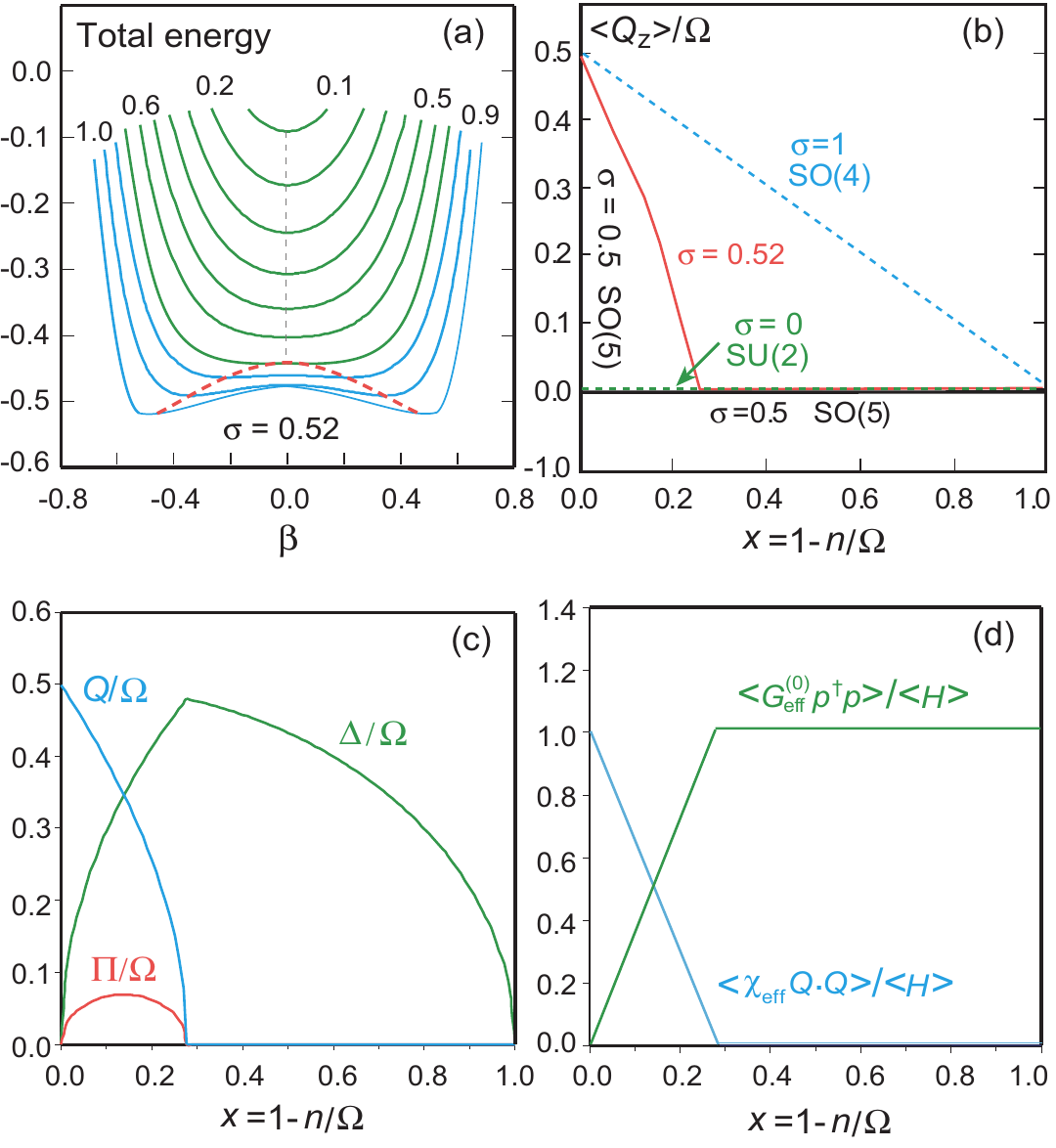}    
{0pt}         
{0pt}         
{0.80}         
{(a)~As for \fig{eSurfaces1D}, but for slightly perturbed \sofive\ corresponding
to $\sigma = 0.52$ \cite{lawu03}. The red dotted line indicates the location of 
the
ground state
in $\beta$ as $n$ varies. (b)~Variation of the AF order parameter with effective
occupation number for different values of $\sigma$. (c)~Variation of the AF, SC
and \sofive\ correlations $Q$, $\Delta$, and $\Pi$ as functions of the
effective hole concentration $x$. (d)~Variation of the ratio of pairing and
$\QdotQ$ interactions to the total energy of the system  as functions of the
effective hole concentration $x$.} 
\sufour\ coherent state energy surfaces as a function of the AF correlation
$\beta$ are shown for the case $\pratio = 0.52$, which corresponds to SO(5)
symmetry very weakly perturbed in the AF direction \cite{lawu03}. Numbers on
curves are  the lattice occupation fractions, with $n / \Omega = 1$
corresponding to half filling and $0 < n / \Omega < 1$ to finite hole doping.
The corresponding variation of the AF correlation parameter 
$Q=\langle\AF_z\rangle$
with hole doping $x$, and its comparison with the variation in various symmetry
limits, are summarized in \fig{so5fig2}b. The variations of the AF, SC
and SO(5) correlations ($Q$, $\Delta$  and $\Pi$, respectively) with the hole
doping $x$ are shown in \fig{so5fig2}c, while the variations of the
contributions of each term in the Hamiltonian to the total energy are shown in
\fig{so5fig2}d. From these results we see quite clearly the rapid evolution
of the weakly-broken \sofive\ symmetry from an energy surface that looks AF-like
to one that looks SC-like, as the hole doping is increased from zero.

Thus, \sofive\ is a critical dynamical symmetry that interpolates
continuously between the \sofour\ antiferromagnetic dynamical symmetry and the 
\sutwo\
superconducting dynamical symmetry. Such symmetries are well known in nuclear 
structure
physics \cite{wmzha87,wmzha88a,zhang89,zhan90} and the \sofive\ 
critical dynamical
symmetry discussed here in a condensed matter context has many formal
similarities with critical dynamical symmetries of the (nuclear) Fermion
Dynamical Symmetry Model \cite{FDSM}. We remark in passing that 
application of dynamical symmetry methods to quantum Hall states for 
monolayer graphene in a magnetic field 
exhibits also such critical dynamical symmetries 
\cite{lawu2016,lawu2017,guid2017}. These examples suggest that critical 
dynamical
symmetries may be a fundamental organizing principle for strongly 
correlated fermionic systems in
which an emergent order is found adjacent to another form of emergent
order in the phase diagram.

\section{\label{su4gaps} SU(4) Energy Gaps}

Let us now examine the correlation energies associated with SU(4) symmetry in
more detail. We begin by analyzing the
expected energy-gap structure as a function of doping, using parameters
characteristic of the cuprate  superconductors. The solutions of the \sufour\
energy gap equations discussed in \S\ref{solutionGapEquations} imply a rich
physics as a function of doping. Among the five sets of $T=0$ gap solutions
[\eqs{gapT0:whole}--\eqnoeq{gapMetal:whole}], the one with the lowest energy at
each doping corresponds to the physical ground state. We can calculate these
energies by inserting the gap solutions directly into \eq{eqED}, and then
investigate how these different sets of solutions compete with each other at
$T=0$. Some results  are illustrated in \fig{Energy}.

\subsection{Energy-Ordering of Gaps}
 
In \fig{Energy},
\singlefig
{Energy}       
{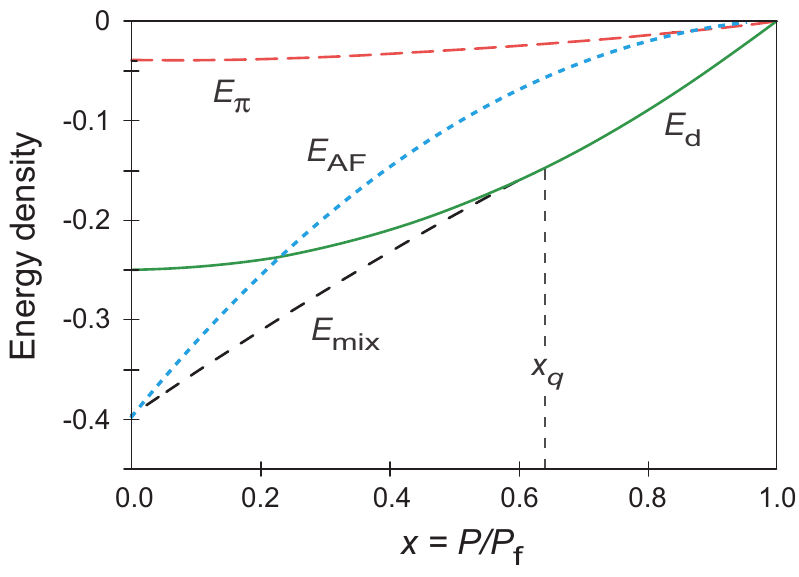}    
{0pt}         
{0pt}         
{0.95}         
{Total energy associated with  gap solutions at $T=0$. $E\tsub{mix}$ is the
energy calculated with the all-gap solution \eqnoeq{gapT0:whole}, while
$E\tsub{d}$ [calculated from the solution in \eqs{gapDd:whole}], $E_\pi$
(\eqs{gapDpi:whole}), and $E\tsub{AF}$ (\eqs{gapDaf:whole}) represent,
respectively, the energy density of the spin-singlet pairing, the spin-triplet
pairing, and the AF solutions. The energy of the uncorrelated solution is set to
zero and taken as the energy reference. Interaction strengths are assumed
constant, with $\chi = 13$, $G_0 = 8.2$, and $G_1 = 1.3$ (in an arbitrary energy
unit), but reasonable values that satisfy the condition \eqnoeq{condition} will
give similar results.} 
we see that the all-gap solution $E\tsub{mix}$ is always lowest in energy and
thus is the physical ground state for the doping range $x\leq x_q$. For $x>x_q$,
the pure singlet pairing state becomes the ground state because $E\tsub{d}$ is
the lowest energy for this doping range. All the other possible solutions lie
higher in energy. They may be regarded as collective excited states but they
cannot become the physical ground state at $T=0$.

We shall find that for $T>0$ the AF or the uncorrelated state could become the
ground state in certain temperature and doping ranges. However, this can never
happen for the pure spin-triplet pairing state, as long as $G_1$ is the weakest
of the three coupling parameters in \eq{condition}. Although spin-triplet
pairing plays an important role as a component of the wavefunction in the SU(4)
theory, the pure spin-triplet state is never found to be the ground state if the
coupling-strength hierarchy of \eq{condition} is satisfied.

\subsection{Generic Features of SU(4) Gaps}

A generic gap diagram at $T=0$ describing features of the energy gaps as
functions of doping $x$ is shown in \fig{Gaps}.
\singlefig
{Gaps}       
{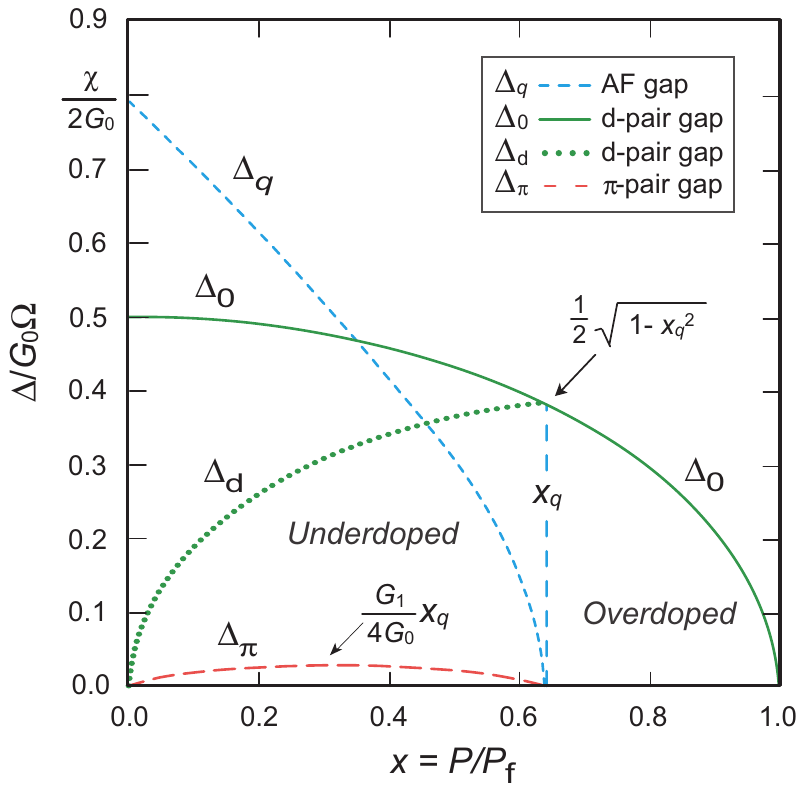}    
{0pt}         
{0pt}         
{1.0}         
{Generic diagram for energy gaps vs.\ doping, as predicted by the \sufour\
model at $T=0$. Energy gaps are scaled by $G_0\Omega$ and the doping
parameter is scaled by the maximum doping $P\tsub f$ (consistent with data, we 
assume $P\tsub f=1/4$
\protect\cite{guid04}). Interaction strengths are assumed to be independent of
doping, with $\chi = 13$,
$G_0 = 8.2$, and $G_1 = 1.3$ (arbitrary energy units), which, according to
\eq{xq}, requires the critical doping point to be $x_q=0.64$,
corresponding to a critical physical doping parameter $P_q = x_qP\tsub f \sim
0.16$.}
The diagram is constructed using \eqs{gapT0:whole} and \eqnoeq{gapDd:whole}. It
is {\it generic} because the  basic forms of the gaps are dictated entirely by
the algebraic structure which, in turn, is determined by the
physically-motivated choice of generators given in \eq{operatorset}. The precise
values of the coupling strengths affect only details.  Four doping-dependent
energy scales are predicted:

(1)~The gap $\Deltaq$ is defined in \eq{udopDq} and measures AF correlations. 
It 
is maximal at $x=0$, decreases rapidly  to the region of the pairing gaps with 
increased doping, crosses the pairing gaps, and vanishes at the critical doping 
$x_q$.

(2)~The spin-singlet pairing gap $\Deltad$ defined in \eq{udopDd} is the
superconducting gap for $x<x_q$.

(3)~The spin-singlet pairing gap $\DeltaZero= \Deltad$ defined in \eq{odopDd} is
the superconducting gap for $x>x_q$, but is not the ground-state order parameter
when $x\leq x_q$.

(4)~The spin-triplet pairing gap $\Deltapi$, is defined in \eq{udoppi}. Like
$\Deltaq$, it exists only in the doping range $x \le x_q$. It reaches its
maximum value at $x_q/2$ and vanishes at $x=0$ and $x=x_q$.

\subsection{The Critical Doping Point}

The spin-singlet pairing gap exhibits qualitatively different behavior for
doping less than or greater than $x_q$. For $x > x_q$, it corresponds to a
monotonic curve labeled $\DeltaZero$, but below $x_q$ the spin-singlet gap
splits into two curves (labeled $\Deltad$ and $\DeltaZero$) having very
different doping dependence, with the splitting increasing for decreasing
doping.

The critical doping point and the splitting of the SC pairing gap result from
competing SC pairing and AF correlation for  $x<x_q$.  At small doping the AF
correlation dominates the SC pairing and a state with large AF correlations and
suppressed pairing can become the ground state. Therefore, the superconducting
gap $\Deltad$ for the ground state is smaller and the larger pairing gap
$\DeltaZero$ is associated with an excited state in this doping range. However,
as doping increases the pairing correlation grows quickly and the AF correlation
decreases. The point $x_q$ marks the doping fraction at which the AF correlation
is fully suppressed, leaving complete dominance of the superconducting
correlations for $x_q < x < 1$.

The critical doping point $x_q$ defines a natural boundary between regions where
the wavefunctions are qualitatively different, as illustrated further in
\fig{separationChargeSpinSU4}.%
\singlefig
{separationChargeSpinSU4}
{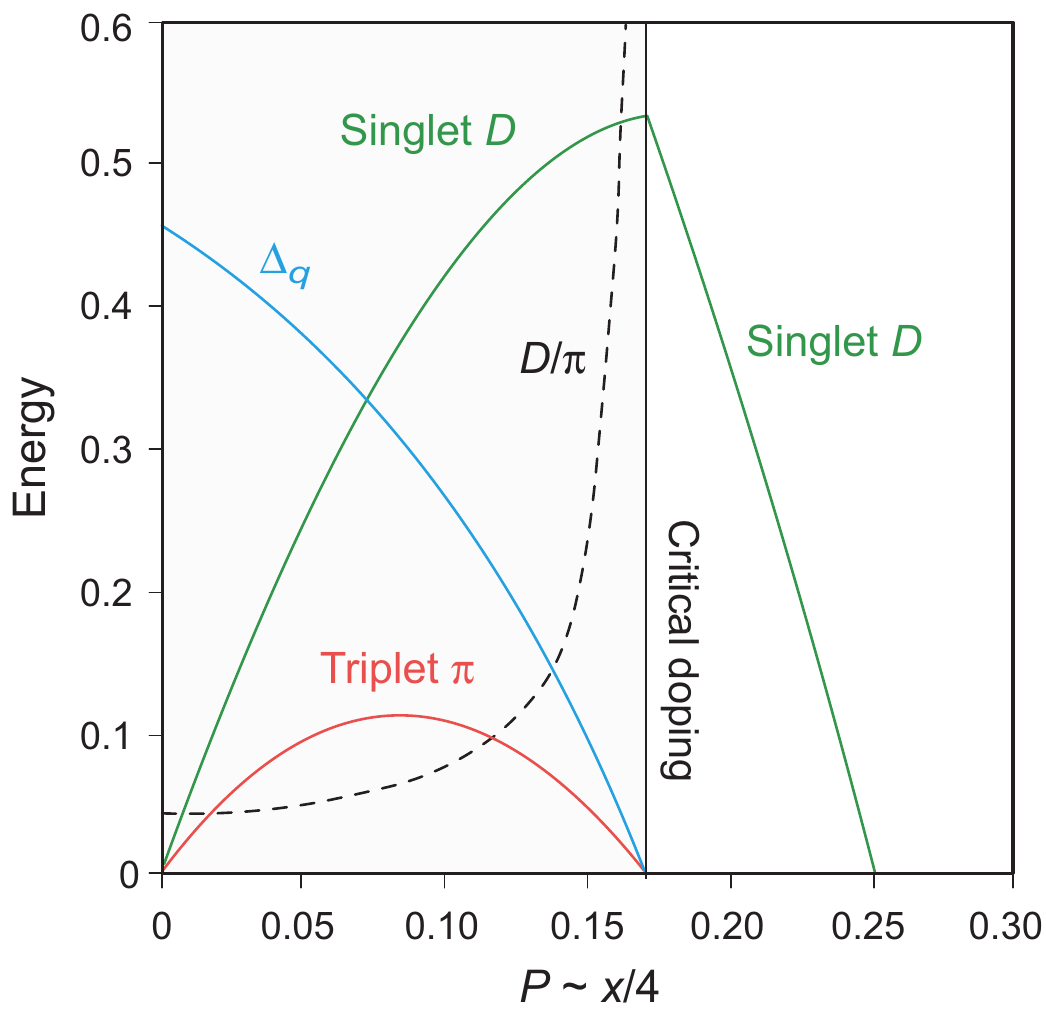}
{0pt}
{0pt}
{0.73}
{Behavior of various correlation energies with doping in the SU(4) model. The 
ratio of singlet to triplet pairing energy $D/\pi$ diverges as critical doping 
is approached.}
It separates a doping regime characterized by weak superconductivity and reduced
pair condensation energy from a doping regime characterized by strong
superconductivity and maximal pair condensation. The optimal doping point
(maximum of the SC pairing gap) is often used empirically to demarcate 
underdoped and
overdoped superconductors. It is the doping point where competition between the
AF and SC correlations leads to the maximal $T\tsub c$. Figure
\ref{fig:separationChargeSpinSU4} suggests that optimal doping and the 
quantum phase transition at critical doping are closely associated.

\subsection{Comparison with Gap Data}

Figure~\ref{fig:gapDataCompare}
\singlefig
{gapDataCompare}       
{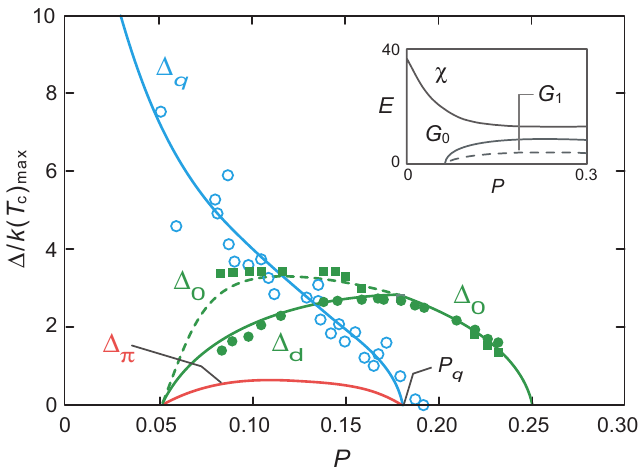}    
{0pt}         
{0pt}         
{1.34}         
{Comparison with data for energy gap diagrams at $T=0$ as a function of doping
in hole-doped cuprates. The doping rate is defined as $P=(\Omega - n)/\Omega_e$
with $\Omega_e$ and $\Omega$ being the number of lattice sites and the maximum
allowed number of holes, respectively. Doping-dependent strengths indicated by
the inset to the figure are used and $P_q = 0.18$ was assumed. Data were taken
from Refs.\ \cite{tall01} (open blue circles) and \cite{tall03} (green
squares and red filled circles).} 
is reproduced from Ref.\ \cite{sun06} and compares computed \sufour\ gaps with
measured ones for the hole-doped cuprates. The gap data, which are taken from
Refs.\ \cite{tall01,tall03}, appear to support the complex gap structure
suggested by the \sufour\ quasiparticle solutions.  In particular, the predicted
splitting of the singlet pairing gap and termination of the $\Delta_q$ gap at a
critical doping $P_q \sim x_q/4$ are consistent with the data points plotted. A
simple variation of the coupling-strength parameters (inset to
\fig{gapDataCompare}) has been allowed to facilitate a more precise fit to 
data, but it was demonstrated in Ref.~\cite{sun06} (and will be discussed 
further below) that the basic features
of the gap diagrams are reproduced even with values of the coupling strengths in
the highly-renormalized subspace that are assumed (probably unphysically) to be
independent of doping. This suggests  that the gap structure observed in the
cuprate superconductors is a generic feature of the \sufour\ symmetry,
independent of microscopic details.

\subsection{Competing Order and Preformed Pairs}

In the {\em competing-order picture} \cite{tall01}, the pseudogap (PG) is an
energy scale for an order that competes with superconductivity and vanishes at a
critical doping point. From  \fig{gapDataCompare}, $\Deltaq$ has precisely these
properties. But the AF operators entering into $\Deltaq$ are generators of
SU(4), so $\Deltaq$ {\em also} is the stabilization energy for a mixture of
``preformed'' singlet and triplet SU(4) pairs that condense into a strong
superconductor only after AF and triplet-pairing fluctuations are suppressed by
doping.  This is a non-abelian generalization of the phase-fluctuation
model \cite{EK95} for {\em preformed pairs}. Preformed pair and competing order
models  have generally been viewed as mutually antagonistic explanations of the
pseudogap.  However, we see that the SU(4) PG state results from
competing AF and SC order expressed in a basis of singlet and triplet fermion
pairs, which may be viewed as a unification of the competing order and preformed
pair pictures of the cuprate pseudogap state.

We conclude that the nature of the pseudogap state requires {\em both} competing
order and preformed pairs, but the requisite preformed pairs are more
sophisticated than those of a simple phase fluctuation model because of the
strong and complex correlations that are present in the realistic system.
BCS-like pairs correspond to an \sutwo\ subgroup of the full SU(4) algebra.
Their phase fluctuations represent phase rotations about a single axis for an
abelian U(1) subgroup. In contrast, the full SU(4) algebra has 15 generators and
various non-abelian subgroups. Thus, phase fluctuations of SU(4) pairs involve
(non-commuting) rotations around axes in a multidimensional space, implying a
rich structure for the preformed pairs.  

For example, we have seen that in the \sofive\ limit at low doping there is a
significant pairing correlation energy but no long-range order because there is
no barrier to phase fluctuations between the SC and AF directions in the SU(4)
space. This is a particularly clear example of states that have a significant
pairing correlation energy (preformed pairs) but no long-range order of either
SC or AF form, because of quantum fluctuations in the SU(4) solutions (competing
order modified by quantum fluctuations).

\subsection{The Role of Triplet Pairs}

Triplet pairs are an essential component of the SU(4) many-body wavefunction.
The SU(4) algebra doesn't even close if the corresponding operators are omitted,
implying that the operator set and corresponding Hilbert space are
quantum-mechanically incomplete in the absence of triplet pairs (see
\fig{completeSet} and the general discussion in \S\ref{su4Operators}). However,
\fig{gapDataCompare} indicates that the triplet-pair correlation energy is small
for underdoped cuprates and vanishes for doping larger than the critical doping
$x_q$.  

The primary role of SU(4) triplet pairs in the hole-doped cuprates lies in
fluctuations mediating the AF--SC competition at lower doping. This
interpretation is supported by the observation that the triplet pair operators
are fundamental generators of the \sofive\ critical dynamical symmetry [see
Table \ref{table:su4Properties} of Appendix \ref{appendix} and 
\eq{so5correlationE}], which acts physically as a
doorway between AF and SC order (\fig{so5Doorway}). Thus, although the
superconductivity in the cuprates is implemented in terms of a condensate of
singlet pairs and triplet pairs play no direct role in the charge transport of
the superconducting state, the triplet pairs are central to the nature of
cuprate superconductivity because they are key to mediating fluctuations and the
rapid transition from AF to SC order with hole doping.

Hence, below the critical doping point $x_q$ and below the superconducting
critical temperature $T\tsub c$, the superconducting state is dominantly
spin-singlet but it necessarily has a small spin-triplet admixture
because the AF correlations are non-vanishing for $x<x_q$ and they scatter
singlet pairs into triplet pairs and vice versa. Only for $x>x_q$ does the
SU(4) symmetry force the AF correlations to vanish identically, leaving a
condensate of pure-singlet Cooper pairs for  $T < T\tsub c$.

\section{\label{su4phase} SU(4) Phase Diagrams}

We may use the formalism developed in \S\ref{finiteTsolutions} to construct
doping--temperature phase diagrams corresponding to SU(4) coherent-state
solutions.  In Figs.~\ref{fig:Phase1} and \ref{fig:Phase2}
\singlefig
{Phase1}       
{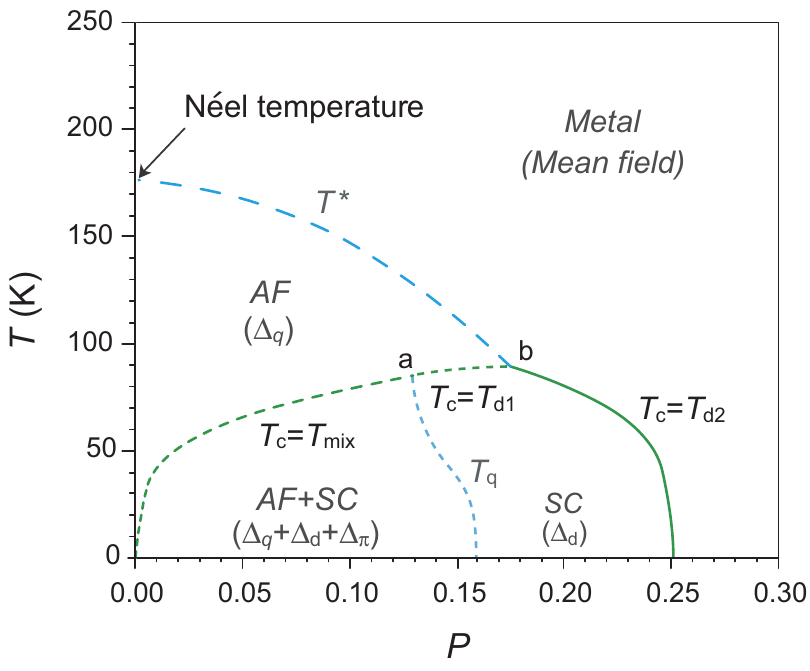}    
{0pt}         
{0pt}         
{1.02}         
{Phase diagram predicted by the SU(4) model with $R=0.6$. Interaction strengths
are the same as those used in \fig{Gaps}, but in units of $k_B(T\tsub
c)\tsub{max}$, where $(T\tsub c)\tsub{max}$ is taken to be 90 K. The critical
doping point is  $P_q=0.16$ (corresponding to $x_q=0.64$).}
\singlefig
{Phase2}       
{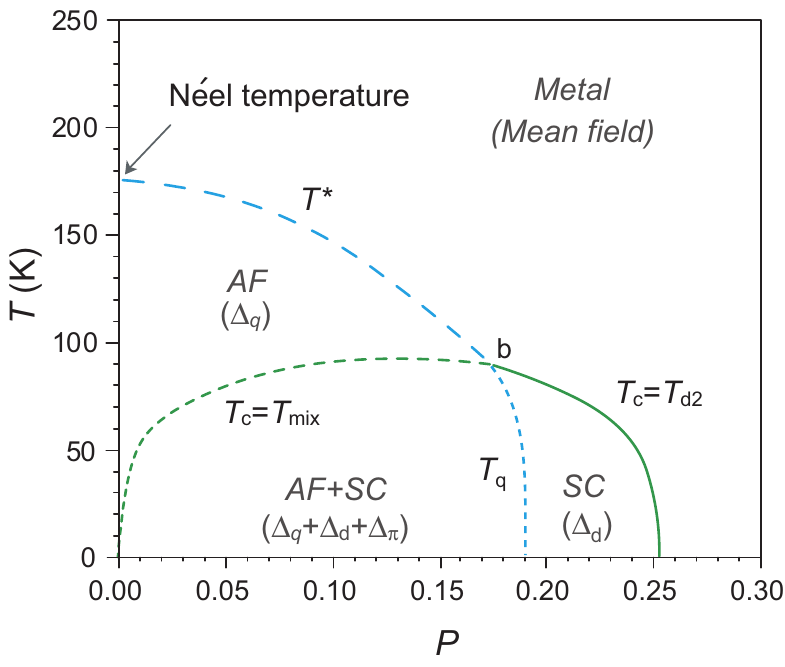}    
{0pt}         
{0pt}         
{1.02}         
{Phase diagram predicted by the SU(4) model for $P_q>P_b$. $P_q=0.19$
(corresponding to $x_q=0.76$) is chosen in this figure while $P_b=0.175$
(corresponding to $x_q=0.7$) is the same as that in \fig{Phase1}. All parameters
remain unchanged from \fig{Phase1}, except that $G_1$ was increased from 1.3 to 
4.7
(in units of $k_B({T\tsub c})\tsub{max}$), giving a larger value of $P_q$.} 
we show typical phase diagrams resulting from solution of the SU(4)
finite-temperature gap equation. Both figures represent possible physical
solutions. They differ primarily in the assumed strength of triplet pairing,
which is not a well-determined parameter. In the calculations the interaction
strengths are kept constant, with the same values as those in the
zero-temperature case discussed in Figs.~\ref{fig:Energy} and \ref{fig:Gaps}.
The only adjustable parameter is $R$ (appearing in \eq{thermal}).

\subsection{The Predicted Phases}

Four distinct phases may be identified in Figs.~\ref{fig:Phase1}
or \ref{fig:Phase2}: 

\begin{enumerate}
 \item
An antiferromagnetic phase labeled AF. 
\item
A superconducting phase labeled SC. 
\item
A transitional phase with all three correlations present, which we label a
mixed phase (marked as AF+SC).
\item
A phase without net antiferromagnetic or superconducting order that is labeled 
metal (mean field).
\end{enumerate}
The correlations (energy gaps) associated with each phase are indicated in
parentheses. The doping-dependent transition temperatures $T\tsub c$, $T^*$, and
$T_q$ define the boundaries for these phases, and  the points $a$ and $b$ mark
the intersections of multiple phases.

\subsection{Comparison with Data}

Figure~\ref{fig:gapsResolve_k}
\singlefig
{gapsResolve_k} 
{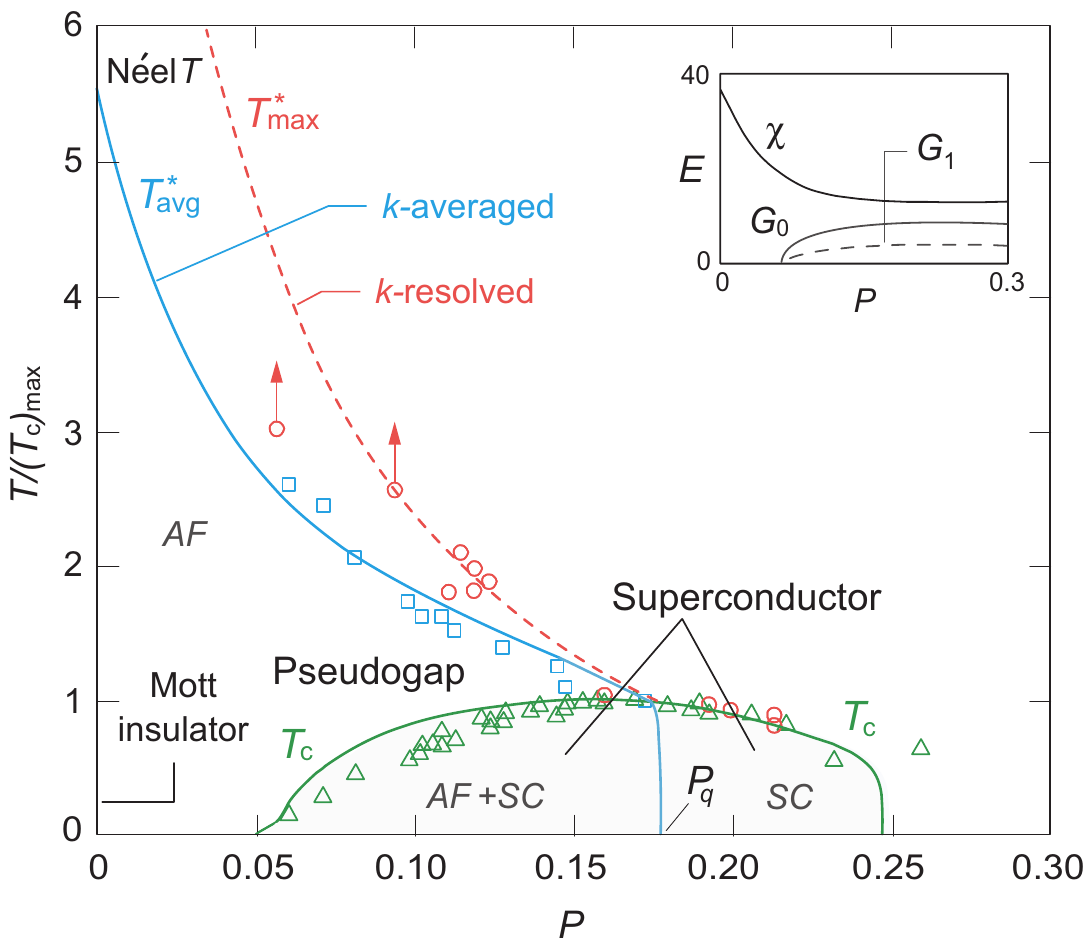}      
{0pt}                          
{0pt}                          
{0.79}
{SU(4) cuprate phase diagram compared with data. Strengths of the AF and 
singlet 
pairing correlations were determined in Ref.\ \cite{sun06} by global fits to 
cuprate data. The PG temperature is $T^*$ and the SC transition temperature is 
$T\tsub c$. The AF correlations vanish, leaving a pure singlet $d$-wave 
condensate, above the critical doping  $P_q$. Dominant correlations in each 
region are indicted by italic labels. Data in green (open triangles) and blue 
(open squares) are taken from Ref.\ \cite{dai99}, and those in red (open 
circles) from Ref.\ \cite{camp99} (arrows indicate that the point is a lower 
limit).  The two different curves for the theoretical pseudogap temperature are 
for experiments in which the momentum $k$ is either resolved 
($T^*_{\rm\scriptstyle max}$) or not ($T^*_{\rm\scriptstyle avg}$). Data in 
blue 
do not resolve $k$; data in red resolve $k$.}
compares with cuprate data and indicates that the coherent state solution can
give a quantitative description of both the superconducting transition
temperature $T\tsub c$ and the pseudogap transition temperature $T^*$. Two
theoretical pseudogap temperatures are shown because the predicted value of
$T^*$ depends on whether the experiment resolves the momentum $k$.  The \sufour\
operators that we have defined in \eqs{operatorset}--\eqnoeq{E3} average over
the momentum $k$. Thus, they are appropriate for comparison with experiments
that do not resolve $k$.  We denote the PG temperature computed in this
approximation $T^*_{\scriptstyle\rm avg}$. However, as discussed in depth in
Ref.\ \cite{sun07}, for experiments that resolve $k$ the appropriate form of the
\sufour\ generators carries a momentum index and the pseudogap temperature
computed in the \sufour\ formalism for specified $k$, which we denote as
$T^*_{\scriptstyle\rm max}$, is in general higher than the PG temperature
computed in the $k$-averaged \sufour: $T^*_{\scriptstyle\rm max} >
T^*_{\scriptstyle\rm avg}$.


The neutron-scattering data shown as blue squares in \fig{gapsResolve_k} do
not resolve $k$. The corresponding \sufour\ pseudogap temperature
$T^*_{\scriptstyle\rm avg}$ was determined using the standard $k$-averaged
version of the \sufour\ model. The ARPES PG data shown in red circles in
\fig{gapsResolve_k} do resolve $k$. The corresponding \sufour\ pseudogap
temperature $T^*_{\scriptstyle\rm max}$ was computed using the $k$-dependent
\sufour\ model described in \S\ref{k-dependentSU4} and in Refs.\
\cite{sun07,guid09}, assuming a single band contributing to the pairing.
Particularly in the underdoped region, this comparison of theory with data
suggests that measurements on the {\em same system} but with different
experimental methods having different $k$ resolutions may find different
pseudogap temperatures $T^*$.

Figures \ref{fig:Phase1}--\ref{fig:gapsResolve_k} may explain the vortex-like
Nernst signal \cite{xu00} observed above $T\tsub c$. In the present SU(4)
picture the singlet pair gap vanishes there but the many-body SU(4) wavefunction
has finite pair content in the region between the curves $T^*$ and $T\tsub c$
that decreases with increasing $T$ and decreasing doping.  Thus, contours for
pair fluctuations above $T\tsub c$ may be expected to be similar to observed
contours for Nernst signal strength, but a Meissner effect is expected only
below $T\tsub c$. Ong et al \cite{ong04} have concluded that a consistent
explanation of pseudogap and Nernst data requires PG and SC pairing states that
are distinct but related by symmetry, as proposed here.

Perhaps the most remarkable result obtained from phase calculations is 
illustrated in \fig{phaseDiagFixed-Adjust},
%
\doublefig
{phaseDiagFixed-Adjust}   
{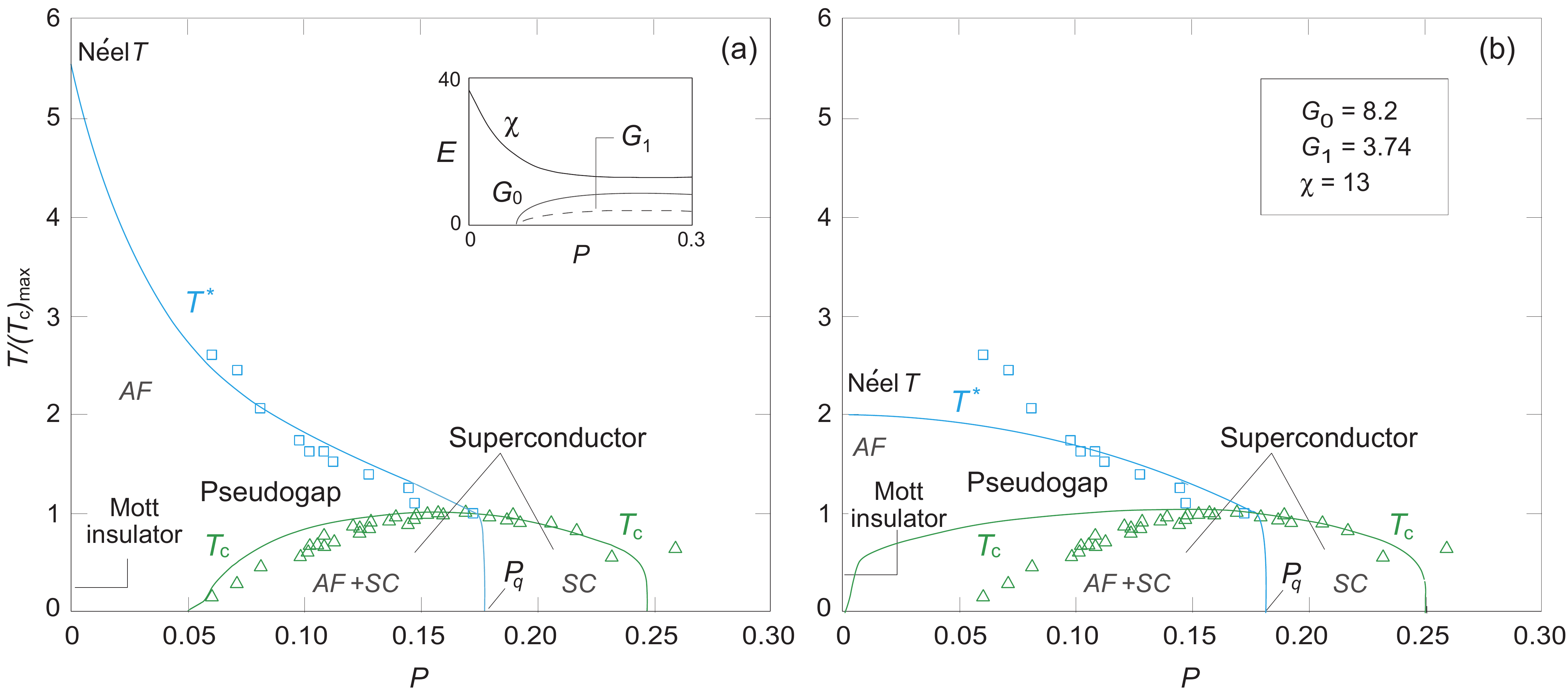}
{0pt}         
{0pt}         
{0.365}         
{
Camparison of cuprate phase diagram calculated (a)~using an effective 
interaction varying smoothly 
with doping and (b)~calculated using constant effective interaction 
strengths, independent of doping. Although an effective interaction that 
does not depend on doping is almost certainly not quantitatively realistic, the 
two diagrams are seen to be qualitatively similar. This suggests that the basic 
features 
of the cuprate phase diagram are a consequence of SU(4) dynamical symmetry.
}
where we compare  a phase diagram calculated using an effective 
interaction varying 
smoothly 
with doping (as in \fig{gapsResolve_k}) and one using a constant 
effective interaction. An effective interaction that 
does not depend on doping is almost certainly not realistic since the Hilbert 
space excluded by  truncation is affected by the doping.  Nevertheless,  the 
two diagrams are {\em qualitatively similar.}  This implies that the basic 
structure of the cuprate phase diagram is {\em determined entirely by the 
dynamical symmetry,} with parameters reflecting the underlying microscopic 
structure affecting only quantitative details in a smooth way. Specifically, we 
see that a quantititatively better description of data is obtained if the ratio 
of antiferromagnetic to singlet pairing coupling is larger at low doping, as 
might be expected on physical grounds, but that the qualitative properties of 
the phase diagram don't depend on such parameter adjustment.

\section{Fundamental Instabilities}

The SU(4) symmetry implies two fundamental instabilities that  may
play a key role in understanding the properties of high-temperature
superconductors. The first provides a natural explanation for the propensity of
cuprate Mott insulator states to become superconductors with only modest hole
doping.  The second provides an explanation of how high-temperature
superconductors can exhibit a rather universal phase diagram and at the same
time display substantial local inhomogeneity, particularly at lower doping.

\subsection{\label{pairingInstability} Pairing Instability with Doping} 

The pairing instability of cuprate superconductors has generally been viewed as
difficult to understand. As illustrated in \fig{cooperInstability},
\singlefig
{cooperInstability}
{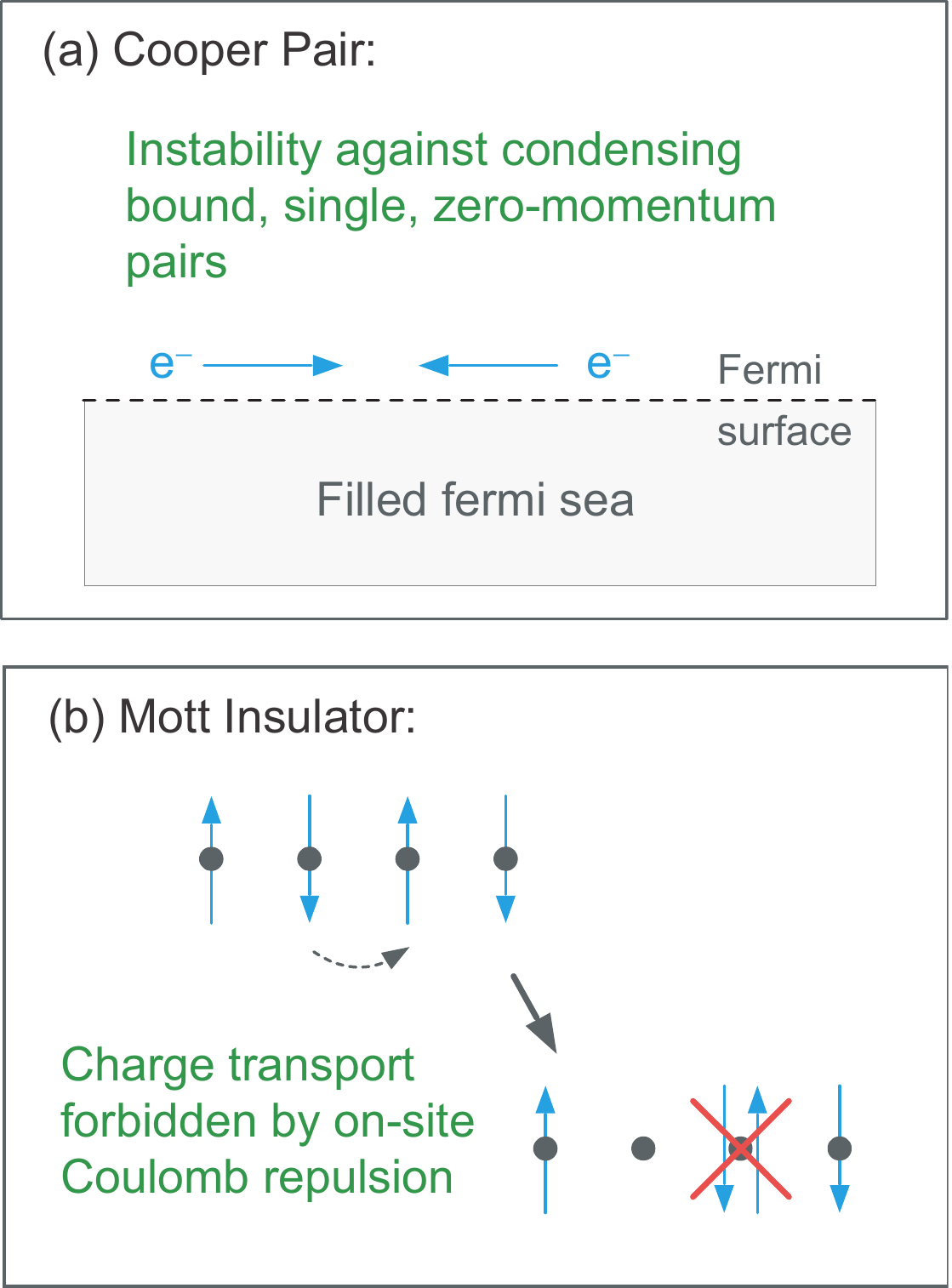}
{0pt}
{0pt}
{0.57} 
{The Cooper instability and Mott insulators. (a)~Basic properties of normal
superconductors arise from an instability of the Fermi liquid normal state and
are well described by the BCS formalism. (b)~High-temperature superconductors
call this picture into question because superconductivity in the cuprates
derives from a Mott insulator normal state, not a Fermi liquid.}
normal superconductors develop from a Fermi liquid normal state, but cuprate
superconductors appear to arise from a (non-Fermi-liquid) antiferromagnetic Mott
insulator state.  Furthermore, this Mott insulator state seems to harbor a
secret propensity to superconductivity, since these compounds can be turned from
AF Mott insulators into high-temperature superconductors by  hole doping at a
modest 3-5\% level. Why do the Mott insulator normal states for the cuprates
become superconducting with only a small amount of hole doping?  The emergent
\sufour\ symmetry provides a natural explanation.   From the $T=0$ solution for 
$\Delta$ given in \eq{udopDd},
\begin{equation}
    \left. \frac{\partial\Delta}{\partial x} \right|_{x=0} = 
        \left. \frac14 \frac{x_q^{-1} -2x}{(x(x_q^{-1} -x))^{1/2}}
\right|_{x=0}
        = \infty ,
\label{analyticalDelderiv}
\end{equation}
displaying explicitly a fundamental pairing instability at $x=0$. Thus, the
\sufour\ symmetry implies that the ground state at half filling is an
antiferromagnetic Mott insulator, but that this state is fundamentally unstable
against condensing singlet hole pairs under infinitesimal hole doping if there
is a finite attractive pairing interaction.  We have termed this {\em precocious
pairing} \cite{guid10}.

This  instability of the \sufour\ symmetry against condensing pairs as the
system is doped away from half filling may also be illustrated graphically using
the \sufour\ coherent-state energy surfaces. Figure~\ref{fig:esurfaces}
\singlefig
{esurfaces}
{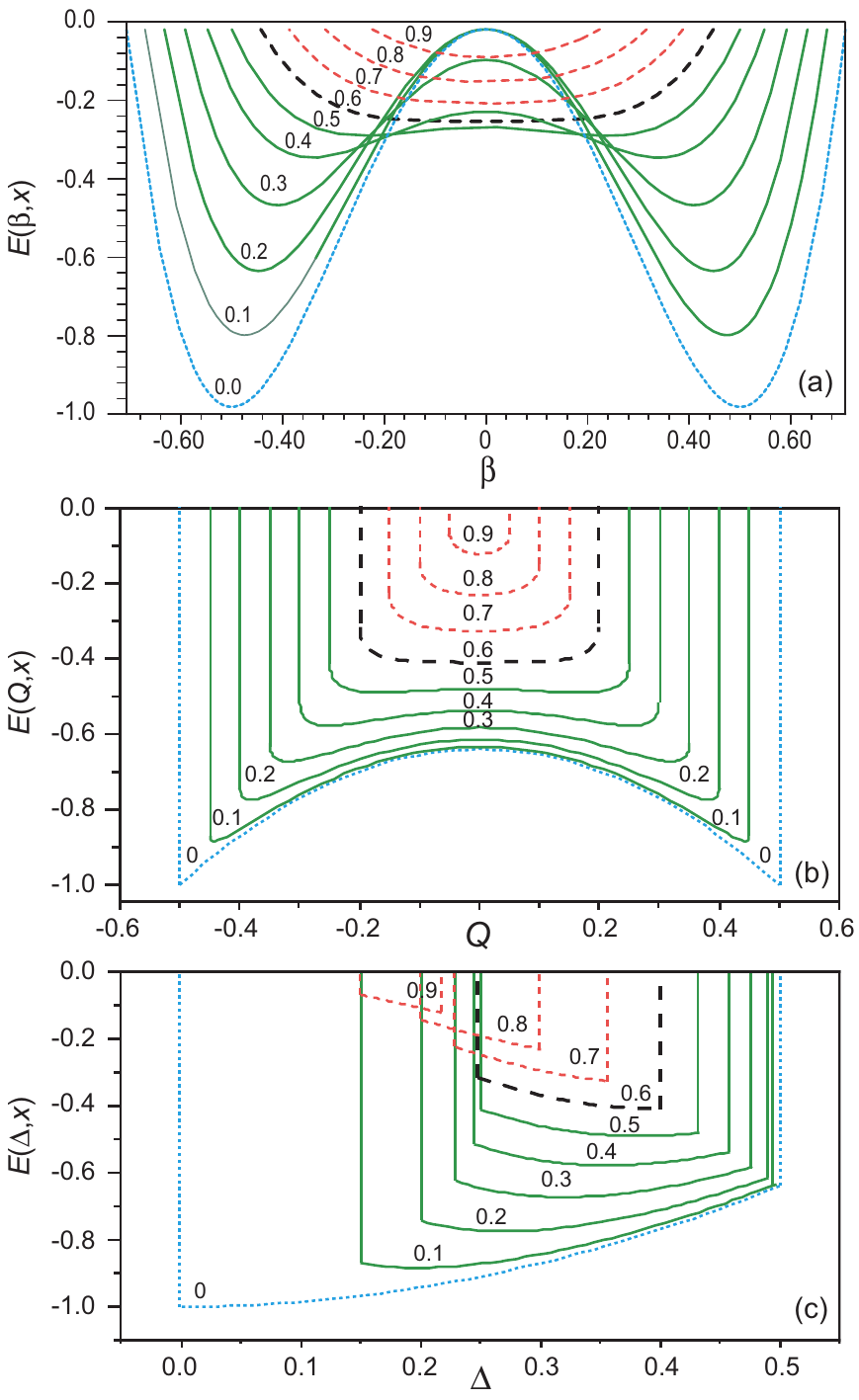}
{0pt}
{0pt}
{1.0} 
{Total energy vs.\ AF order parameters $\beta$ (top) and $Q$ (middle), and SC 
order $\Delta$ (bottom); curves labeled by hole-doping $x\simeq 4P$, where $P$ 
is the number of holes per copper site. Energy is in units of $\chi\Omega^2/4$, 
with $\chi$ the AF coupling strength. The heavy black dashed line indicates the 
critical doping $x=x_q \sim 0.6$ (see Ref.\ \cite{sun05}); color coding 
indicates energy surfaces favoring SC (red, short-dash, from $x=0.7-1.0$), AF 
(blue, dotted, $x\sim 0$), and AF + SC (green, solid, $x \sim 0.1-0.5$). The 
finite, doping-dependent ranges of the energy contours in the order parameters 
reflect the finite valence space (single band) of the microscopic model.}
shows the SU(4) total energy surface in coherent state approximation as a
function of AF and SC order parameters
$$
Q = \frac1\Omega \langle \QdotQ \rangle^{1/2}
\qquad
\Delta = \frac1\Omega \langle \DdagD\rangle^{1/2}
$$ 
as doping $x\simeq 4P$ (for $P$ holes per copper lattice site) is varied. From
\eq{esurfaceQ}, the explicit expression for the energy surface is given by
\begin{subequations}
\begin{align}
E&=-\chi \Omega^2[(1-x_q^2)\Delta^2 + Q^2]
\\
\Delta &\equiv \frac12 \left[\frac14-\left(Q-\frac x2\right)^2 
\right]^{1/2} 
+ \frac12 \left[\frac14-\left(Q+\frac x2\right)^2\right]^{1/2}
\end{align}
\end{subequations}
where $x_q$ is the critical doping (see \fig{esurfaces} and
\S\ref{criticalDopingPoint}). The vertical lines bounding the curves for
different doping in \fig{esurfaces} represent the constraints 
$$
|Q| \le \frac12(1-x) \qquad  \left(x(1-x)\right)^{1/2} 
\le 2\Delta \le  (1-x^2)^{1/2}
$$
that result from SU(4) symmetry realized within a finite valence space.
Specifically, $|Q|$ must lie between 0 and $\tfrac12 n$ (where $n$ is electron
number) because of the finite number of spins available per lattice site,  and
SU(4) symmetry then relates this constraint on $Q$ to the one on $\Delta$. 

The expectation values of the order parameters $Q_0$ for antiferromagnetism and
$\Delta_0$ for singlet pairing from the energy surfaces in \fig{esurfaces} are
illustrated in \fig{cooperInstability2}.%
\singlefig
{cooperInstability2}
{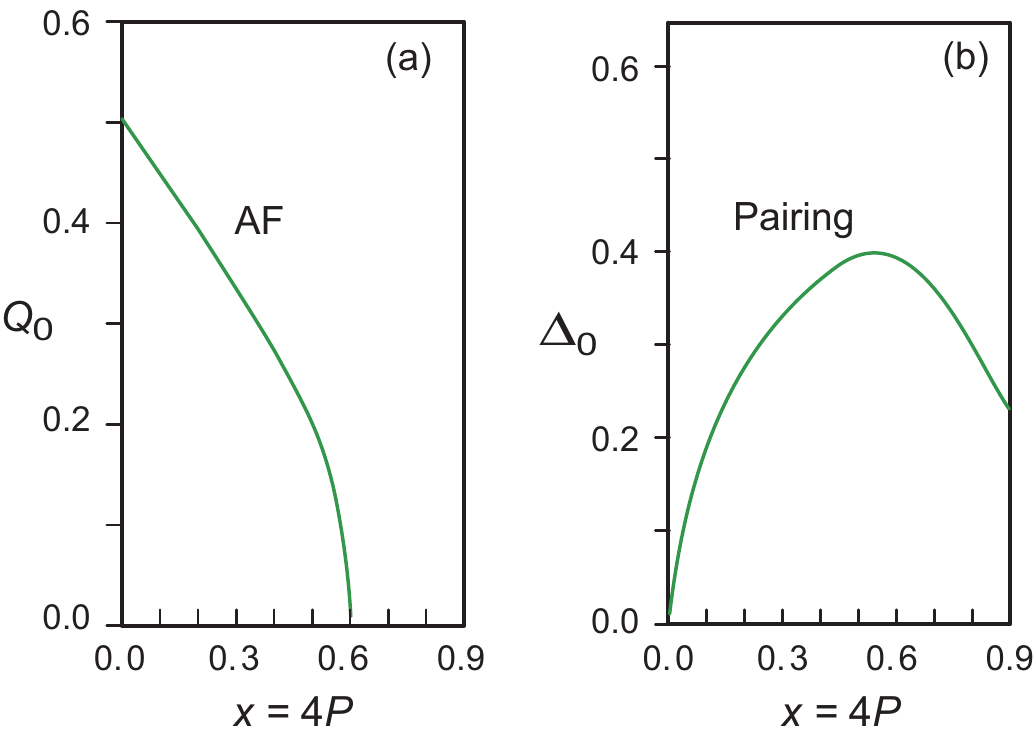}
{0pt}
{0pt}
{0.80} 
{Expectation values of the order parameters (a)~$Q_0$ for antiferromagnetism and
(b)~$\Delta_0$ for singlet pairing, as a function of hole doping.  Evaluated 
from the energy surfaces in
\fig{esurfaces}.}
From \fig{esurfaces} and \fig{cooperInstability2}, the energy surface at half
filling ($x=P=0$) implies a $T=0$ ground state with  AF order but no pairing
order ($Q_0 \ne 0$ and $\Delta_0 = 0$, where the subscript zero denotes the
value  at the minimum of the energy surface).  The ground state for $x=0$ may be
interpreted as an antiferromagnetic Mott insulator \cite{guid01,lawu03}.  From
\fig{esurfaces}(a), the energy surface retains strong AF character for small $x$
with $Q_0 \ne 0$, but from \fig{esurfaces}(b) the ground state differs
qualitatively from that at half filling even for infinitesimal hole-doping. 
Specifically, for any non-zero attractive pairing strength a finite singlet
$d$-wave pairing gap develops spontaneously for {\em any non-zero $x$,} and 
$\Delta_0$ has increased to half its value at optimal doping by $P \simeq 0.03$
($x \simeq 0.12$).

Thus, at half filling the SU(4)-symmetric lattice is a Mott insulator with
long-range AF order and no pairing order, but upon infinitesimal hole-doping  a
finite singlet pairing gap and a ground state representing strong competition
between AF and SC order appears. This spontaneous development of a finite
singlet pairing gap for infinitesimal hole-doping has been obtained in the
coherent-state approximation subject to SU(4) symmetry.  Doping dependence of
the effective interactions, AF and SC fluctuations, or weak  breaking of SU(4)
symmetry could delay the onset of the instability from $P \sim 0$ to a small
doping fraction, as observed. However, we propose that the pair-condensation
instability of the SU(4) symmetry limit represents the essential physics
governing the rapid emergence of superconductors from doped Mott insulators in
the cuprates. 

The remarkable conclusion that a weakly-doped antiferromagnetic Mott insulator 
can
rearrange itself spontaneously into a superconductor follows directly from SU(4)
invariance,
which requires that for any SU(4)-symmetric solution, 
\begin{equation}
Q^2 + \Delta^2 + \Pi^2 = \frac14 (1-x^2),
\label{AFpairingGaps}
\end{equation} 
where $\Pi = \langle \vec\pi^\dagger \vec\pi \rangle^{1/2}/\Omega$ is the 
triplet pair correlation \cite{sun05}.  But for the pure antiferromagnetic 
SU(4) solution, 
$
Q^2=\tfrac14 (1-x)^2.
$  
Thus, comparing with 
\eqnoeq{AFpairingGaps}, we see that even the antiferromagnetic limit has finite 
pairing gaps $\Delta^2 
+ \Pi^2$ unless $x$ vanishes identically.

\subsubsection{Implications for Cuprates at Low Hole Doping}

Cuprate data for low doping suggest that normal compounds at half filling are 
AF 
Mott insulators, that a finite pairing gap develops   when doping reaches 
$P\simeq 0.05$, and that a pseudogap develops in the underdoped region having 
doping dependence for the PG temperature $T^*$ opposite that of $T\tsub c$ for 
the singlet pairing gap.  These results are consistent with a Mott insulator 
state at half filling that evolves rapidly into a state with a finite singlet 
pairing gap at very low hole-doping.  However, since at low doping both the 
singlet pairing and antiferromagnetic correlation energies are substantial, the 
AF fluctuations 
prevent development of full-strength superconductivity until the critical 
doping point $x_q$, 
where the zero-temperature AF correlations are completely suppressed at a 
quantum phase transition.  

Below $T\tsub c$  for doping less than the critical doping, this leads to a 
$d$-wave SC state weakened by antiferromagnetic fluctuations.  For a range of 
temperatures 
above $T\tsub c$ but for doping less than critical, the pairing gap vanishes 
but 
strong AF correlations in a basis of fermion pairs leads to a pseudogap that 
may 
be interpreted in terms of preformed pairs having a structure strongly 
influenced by competing AF and SC order. Finally, the AF competition weakens 
with hole doping until the pure superconductor emerges and the pseudogap 
disappears near the critical doping point (which is typically near optimal 
doping).

\subsubsection{The Generalized Cooper Instability}
 
Our results show that an inherent instability against condensing Cooper
pairs as doping is increased occurs naturally in a minimal model of singlet 
bond-wise
pairing interacting with AF correlations on a lattice with no double occupancy. 
Thus, the rapid onset of superconductivity with hole-doping in the cuprates
results from a Cooper-like instability against condensing pairs for
non-zero attractive pairing interaction, but for $d$-wave pairs in an AF Mott
insulator.  This solution reduces formally to ordinary $d$-wave BCS theory if
the AF interaction vanishes, and to an antiferromagnetic Mott
insulator  if the pairing vanishes (see \S\ref{su4BCS}, \S\ref{su4neel}, and
Ref.~\cite{sun05}); thus it represents a minimal self-consistent generalization
of the Cooper instability to doped Mott insulators.  To summarize:  there is 
nothing mysterious  about the
rapid onset of superconductivity upon doping a  Mott insulator with 
holes.  This is just the Cooper instability, but for a Fermi sea 
polarized by onsite Coulomb
repulsion and antiferromagnetic correlations.

\subsubsection{\label{RVBprecocious} Implications for Resonating Valence Bond
Models}

The resonating valence bond (RVB) model \cite{RVB} assumes  that
quantum antiferromagnets should exhibit superconductivity, at odds with the
observation that cuprate ground states at half-filling appear to have 
long-range AF order and no
superconductivity \cite{vakn87}. Thus, resonating valence bond models assume
implicitly that the half-filled state is in some sense practically a spin
liquid, though it looks like an AF state.  Our results give independent support
for a picture rather similar to this, but without explicit RVB assumptions
\cite{guid08}:  the SU(4) ground state at half-filling is an antiferromagnetic
Mott insulator, but its wavefunction can reorganize spontaneously into a
superconductor when perturbed by a vanishingly-small hole doping in the presence
of a non-zero pairing interaction. We shall have more to say about the
relationship between the RVB and SU(4) models in \S\ref{su4RVB}.

\subsection{\label{criticalInhomo} Critical Dynamical Symmetry and
Inhomogeneity}

We now demonstrate that SU(4) symmetry implies a second fundamental instability 
near the
critical doping point $x_q$, and that this instability may also play a key role
in the observed properties of cuprate superconductors. From the $T=0$
solution for $Q$ given by \eqs{udopDq} and \eqnoeq{1.11c}, we find
\begin{equation}
    \left. \frac{\partial Q}{\partial x} \right|_{x=x_q} = 
        \left. -\frac14 \frac{x_q + x_q^{-1}-2x}
        {[(x_q-x)(x_q^{-1} -x)]^{1/2}} \right|_{x=x_q}
        = - \infty ,
\label{analyticalDelderivb}
\end{equation}
and a small change in doping will cause a large change in antiferromagnetic 
correlations near $x=x_q$.  This is a consequence of SU(4) symmetry, which
requires that $Q$ vanish for $x\ge x_q$, and be finite for $0<x<x_q$. The
physical implications of this instability are most easily demonstrated using
the coherent-state energy surfaces discussed in \S\ref{esurfaces}.

The energy surfaces at constant doping in Figs.~\ref{fig:eSurfaces1D} and
\ref{fig:eSurfaces} fall into three general classes: AF+SC (e.g., $x=0.1$), SC
(e.g., $x=0.9$), and critical (e.g., $x=0.6$, which marks a quantum phase
transition), as illustrated in \fig{esurfaceClasses}.%
\singlefig
{esurfaceClasses}
{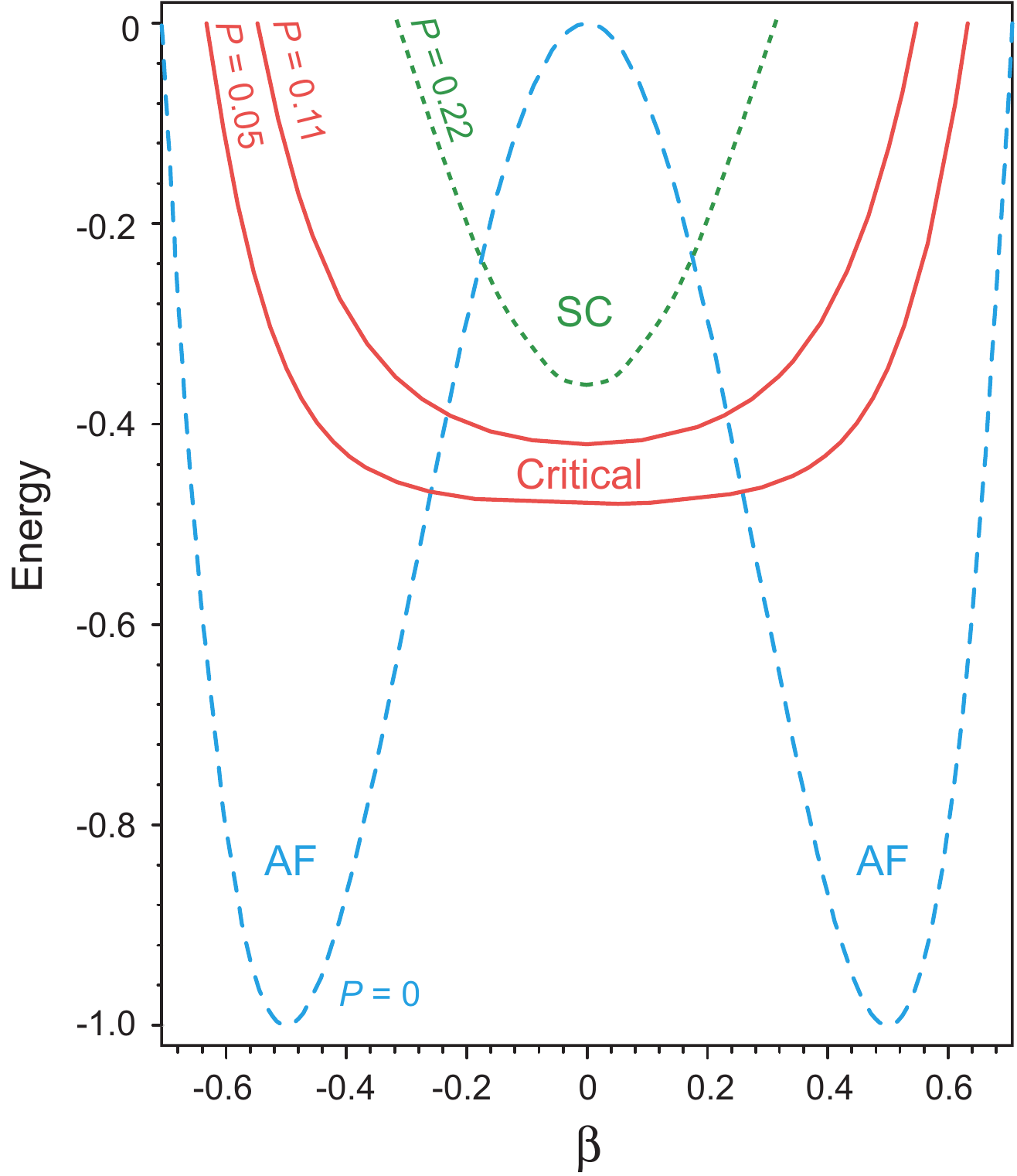}
{0pt}
{0pt}
{0.58}
{Three classes of SU(4) energy surfaces as a function of the AF order 
parameter $\beta$.  Cases  labeled by the hole-doping parameter
$P$: superconducting (SC, dotted green), antiferrromagnetic (AF, dashed blue), 
and critical (solid red).}
Curves in the AF+SC class have minima at finite and large $\beta_0$, and small
but finite $\Delta_0$, where the subscript zero denotes the value of the order
parameter at the minimum of the energy surface.  Curves in the class SC are
characterized by $\beta_0=0$ and finite $\Delta_0$.  Of most interest in the
present context are the surfaces that are near critical in
\fig{esurfaceClasses}, which correspond to broken SU(4) $\supset$ SO(5)
dynamical symmetry \cite{guid01,lawu03} and are rather {\em flat} over large
regions of parameter space.  This implies that there are many states lying near
the ground state with very different values for $\beta$ and $\Delta$.  Thus the
surface is critically balanced between AF and SC order, and small perturbations
can drive it strongly from one to the other.  This defines a critical dynamical
symmetry of the SU(4) algebra, as discussed in \S \ref{criticalD},
\S\ref{weaklyBrokenSO5}, and Refs.~\cite{guid01,lawu03}; we shall term this
situation {\em dynamical criticality}.  Figure \ref{fig:esurfaces} suggests that
underdoped cuprates have near-critical energy surfaces. Thus, critical dynamical
symmetry may be central to the discussion of inhomogeneity and to the general
issue of understanding pseudogap states in underdoped cuprates.

\subsubsection{Dynamical Criticality and Sensitivity to Perturbations}

The extreme sensitivity of critical surfaces to perturbations is
illustrated in \fig{perturbations}.  
\singlefig
{perturbations}
{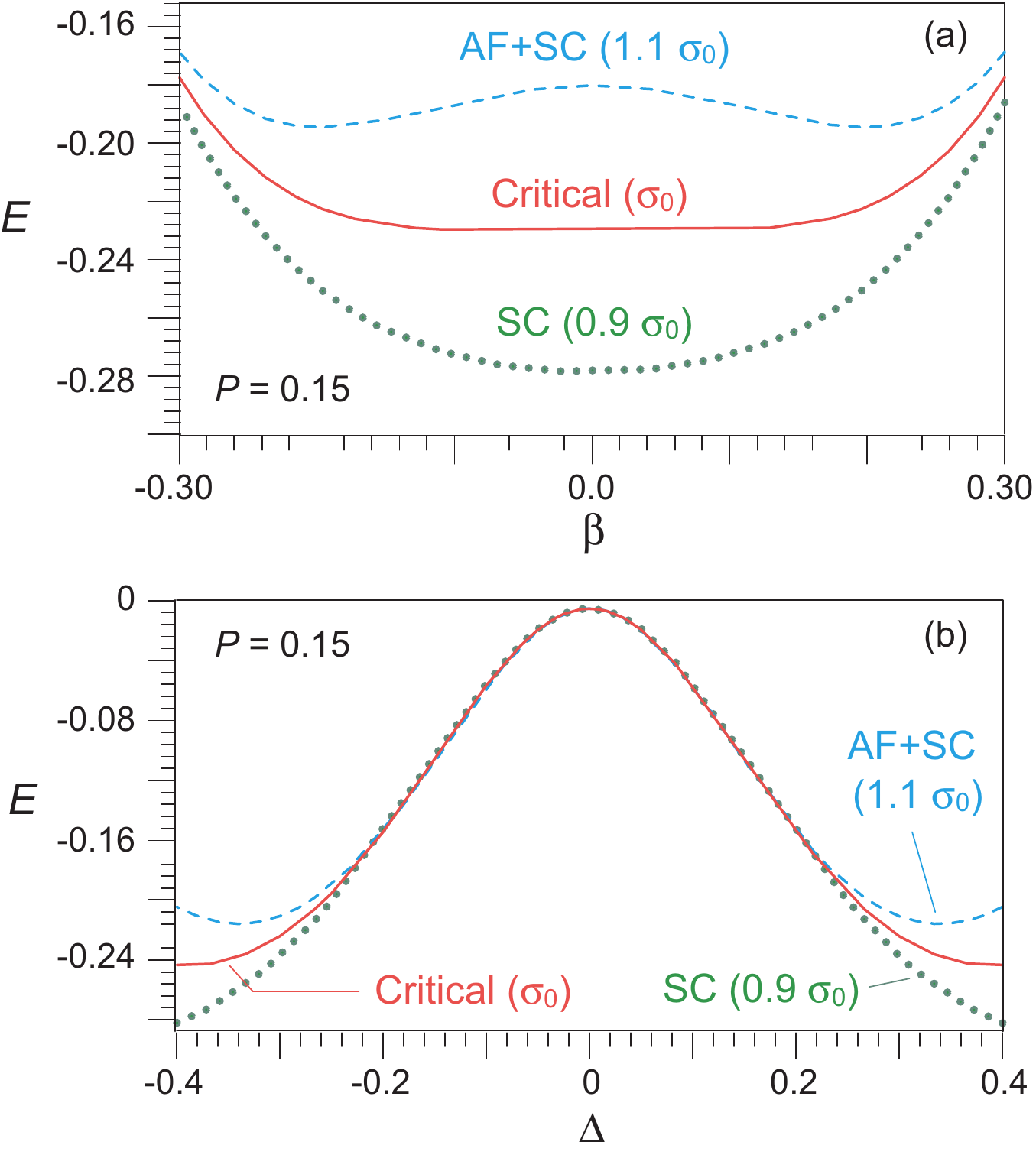}
{0pt}
{0pt}
{0.62}
{Energy surfaces as a function of (a)~the AF order parameter $\beta$ and (b)~the
singlet pairing order parameter $\Delta$, for fixed doping $P = \tfrac 14x =
0.15$.  The solid red curve corresponds to $\pratio=\pratio_0 = 0.6$, the upper
dashed blue curve to a 10\% increase in $\pratio$, and the lower dotted 
green curve to a 10\% reduction
in $\pratio$. }
Each set of curves is associated with a fixed doping $x=0.6$
($P=0.15)$, with the solid line corresponding to $\pratio=0.6$,
the dashed line to a 10\% increase in $\pratio$ (AF perturbation), and the
dotted line to a 10\% reduction in $\pratio$ (SC perturbation).  In
\fig{perturbations}(a) we see that the perturbation changes the AF character
of the ground state. The effect on
 $\Delta$  in \fig{perturbations}(b) is less dramatic:
$\Delta_0$ is shifted, but remains finite in all three cases.  We see that this
small fluctuation in $\pratio$ can alter the energy surface between AF+SC
(finite $\beta_0$ and $\Delta_0$) and SC ($\beta_0=0$ and finite $\Delta_0$).
This sensitivity is specific to the critical [broken SO(5)] dynamical symmetry.
The antiferromagnetic region near $x=0$  and the superconducting region at
larger hole doping (see \fig{esurfaces}) are very stable against such
perturbations, as illustrated in \fig{compareEsurfacePerturbations}.%
\singlefig
{compareEsurfacePerturbations}
{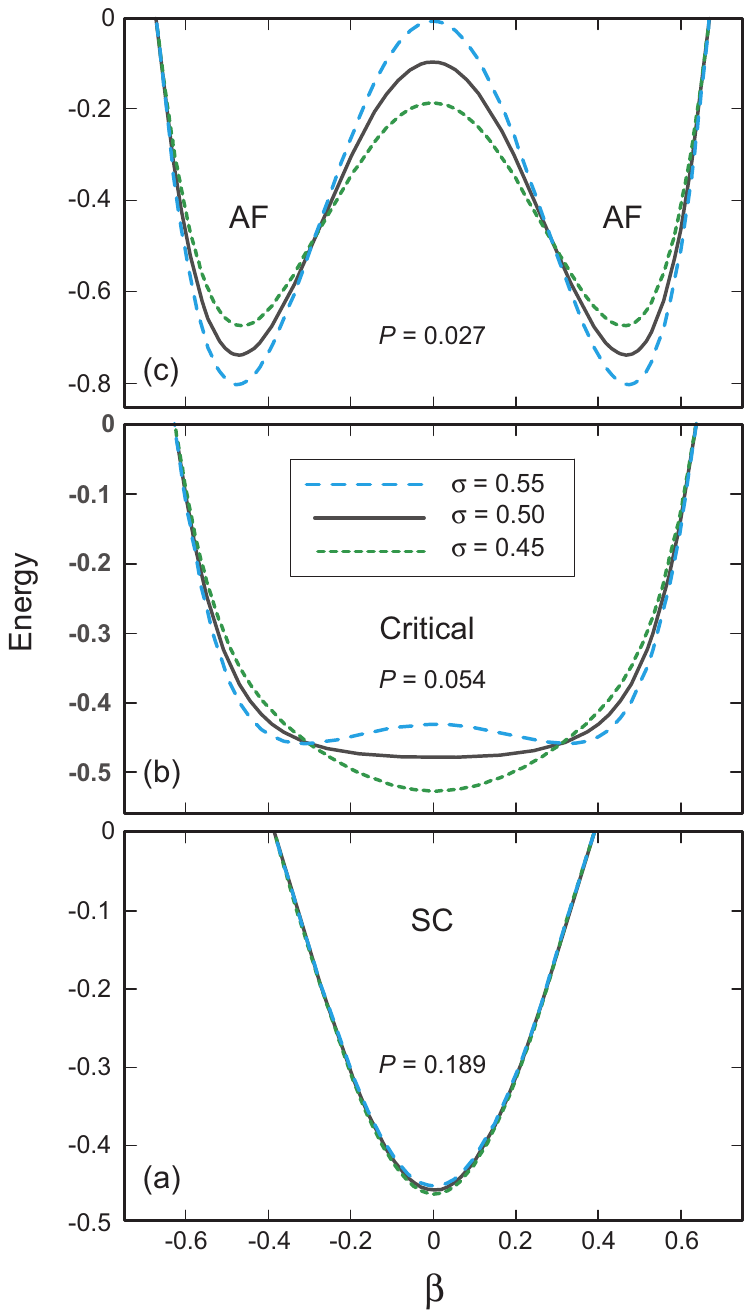}
{0pt}
{0pt}
{1.02}
{Effect of altering the ratio $\sigma$ of antiferromagnetic to superconducting 
coupling strengths for three 
values of
hole doping in the cuprates. In (b) the perturbation changes the nature of the
ground state, but in (a) and (c) it hardly alters the location of the
deep energy minima, implying that the effects of dynamical criticality are 
specific to underdoped
compounds. }

Various mechanisms  could alter the ratio of 
AF to SC coupling locally.  For example, it was found in
Ref.\ \cite{mcel05} that nanoscale disorder is tied to 
influence of dopant impurities, and a number of authors have discussed possible 
reasons for this in terms of mechanisms like lattice-distortion 
leading to modification 
of electron--phonon coupling or superexchange
\cite{nunn05,fang05,he2006,mask2009,peti2009,foye2009,john2009,okam2010,
khal2010}.

\subsubsection{Spatial Inhomogeneity Induced by Background
Perturbations}

Figure~\ref{fig:stripeOrigin} is constructed  from the expectation value of
\eq{eq1} in coherent state approximation, assuming a periodic one-dimensional
spatial perturbation of form $\Delta\sigma \propto \sin(2\pi L)$ around
$\sigma=0.6$. This figure illustrates schematically how a small (10\%) periodic
fluctuation in the antiferromagnetic and superconducting coupling for a
critically-symmetric underdoped compound  can lead to inhomogeneity.  In this
example, one-dimensional spatial variations of the coupling ratio $\pratio$ give
fluctuations in order parameters leading to stripes in which AF+SC
($\pratio>0.6$) and SC ($\pratio<0.6$) are favored alternatively. Also shown are
the responses of  AF fluctuations $d\beta/dL$ to this variation in $\pratio$.
(We do not intend this as a realistic model of a stripe phase, but rather as a 
cartoon
indicating how such emergent structures can arise in the underdoped region 
because of
background perturbations on energy surfaces that have been rendered critical by
the SU(4) symmetry constraints.)
\singlefig
{stripeOrigin}
{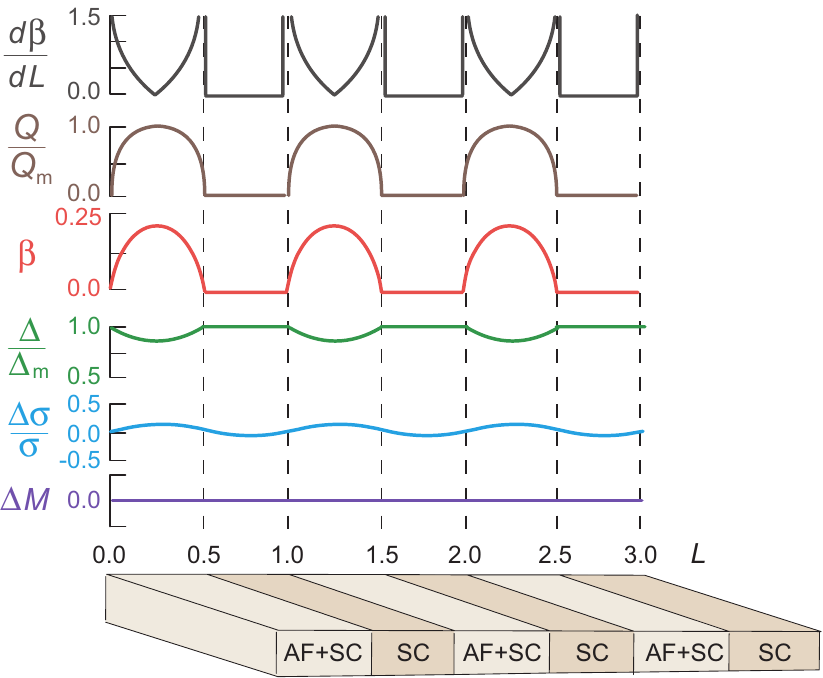}
{0pt}
{0pt}
{1.05}
{A small spatial variation $\Delta\sigma/\sigma$ in the AF and SC coupling can 
produce
inhomogeneity without necessarily implying significant charge variation. The 
spatial scale is $L$, the charge is $M$, $\Delta$ is the SC order parameter, 
$\beta$ and $Q$ are AF order parameters, and a
subscript m denotes maximum values. }

Figure~\ref{fig:stripeOrigin} indicates that a small spatial variation in the
coupling ratio $\pratio$ can produce regions having large AF and weaker SC
order, interspersed with regions having significant SC correlations  but no AF
order.   The nature of the underdoped energy surface dictates that the primary 
fluctuation
between regions is in the AF order, which can jump between zero and the maximum
allowed by the SU(4) model in adjacent stripes; the pairing order changes
are smaller and pairing is finite in both the SC and SC+AF regions.  Finally,
the variation of $d\beta/d L$ indicates that appreciable softness in the
AF and SC may occur on the boundaries between
regions, implying that such induced structure could be static or dynamic.

The minimal patch size that can support SU(4) coherent states is of 
interest in the present context.  We have shown  (for example, Ref.\ 
\cite{FDSM})  that
dynamical symmetry can be realized in strongly-correlated fermion valence spaces
having as few as several particles. This would be consistent with
inhomogeneity on scales comparable to atomic dimensions, as required by data
\cite{mcel05}. The preceding example employed a periodic modulation of $\sigma$ 
as 
a 
simple demonstration but it
is clear that a similar mechanism can operate for non-periodic perturbations.

\subsubsection{The Role of Charge}

The expectation value of the charge $M$ is not a function of the coupling ratio
$\pratio$ in the SU(4) coherent state, so the critical energy surface
fluctuations responsible for alternating AF+SC and SC stripes in
\fig{stripeOrigin} cause no charge variation ($\Delta M=0$). Data indicate that
the relative charge variation for Bi-2212 surface nanoscale patches is less than
10\%, implying heterogeneity that is not strongly coupled to charge 
\cite{mcel05}. This
view is supported by analyses of heat capacity and NMR data on Bi-2212 and YBCO
that find a universal phase behavior for cuprates, with little static charge
modulation in evidence \cite{lora04, bobr02}.

Of course, the variation in $\pratio$ could be {\em caused} by a charge
modulation. From the preceding discussion, we may expect that if a charge
modulation occurs in either the antiferromagnetic region near half filling or 
the superconducting region at
larger doping, its effect will be small because the energy surfaces are not
critical there.  However, if a charge modulation occurs in the underdoped region
where energy surfaces are near critical, its effect may be amplified by
dynamical criticality even if $\pratio$ is not altered significantly, since this
is equivalent to a doping modulation.  

Thus, we suggest a mechanism operating only in underdoped materials that could 
produce strong inhomogeneity without
necessarily invoking charge fluctuations, but that could in particular cases be 
associated indirectly with a charge modulation. Such a
mechanism could explain strong local inhomogeneity in the face of a universal 
overall phase
diagram, and resolve competing experimental claims regarding the role of charge
variation in producing inhomogeneity.  In our view the central 
issue is spatial modulation of AF and SC coupling; spatial modulation of charge 
is a possible but not a necessary corollary.

\subsubsection{Self-Organization Versus Dopant Impurities}

In the literature, inhomogeneity caused by electronic self-organization is often
contrasted with that caused by dopant impurities.  Our discussion implicates
both as sources of nanoscale structure.  The immediate cause may be impurities
that perturb the SU(4) energy surface but the criticality of that surface, which
can greatly amplify the influence of the impurities, results from the
self-organizing, doped Mott insulator  encoded in the SU(4) algebra. 
Note that Ref.~\cite{fang05} suggests, from a different perspective, that
intrinsic amplification of impurity effects is required to explain nanoscale
structure in Bi-2212.

\subsubsection{Amplification of Proximity Effects}

Giant proximity effects are observed in the cuprates where non-superconducting
copper oxide material sandwiched between superconducting material can carry a
supercurrent, even for a thickness much larger than the coherence length
\cite{bozo04}. Phenomenology indicates that pre-existing nanoscale
SC patches can precipitate such effects \cite{gonz05}.  We find
similar possibilities here, but also suggest that the
inhomogeneity need not pre-exist. Dynamical criticality renders even a
homogeneous pseudogap phase unstable against fluctuations in the
AF and SC order. Thus, proximity of superconducting
material to pseudogap material, coupled with perturbations from background
impurity fields, can trigger nucleation of nanoscale structure and giant
proximity effects dynamically, even if no static inhomogeneity exists
beforehand.

Because electronic disorder in cuprate superconductors can exist on a scale
that is comparable to the coherence length ($\sim 15$ \AA), a proximity effect 
may be
expected once static nanoscale disorder forms whereby SC character
leaks between patches.  This effect can be greatly enhanced if patches have
energy surfaces that are critical, since a small perturbation that flips the
energy surface of a patch has a non-perturbative influence far out of proportion
to its size (a small tail wagging a large dog).

In experiments with one unit cell thickness La$_2$CuO$_4$ antiferromagnetic 
barrier layers between superconducting La$_{1.85}$Sr$_{0.15}$CuO$_4$ samples, 
it 
was found that the two phases did not mix, with the barrier layer completely 
blocking a supercurrent \cite{bozo03}.  These results were interpreted  to rule 
out models of high-temperature superconductors like the Zhang SO(5) model 
\cite{deml98} in which SC and AF phases are nearly degenerate. The absence of a 
proximity effect between the antiferromagnetic and superconducting phases  (but 
the presence of a strong proximity effect between pseudogap and SC material) is 
plausibly consistent with the SU(4) model.   The SU(4) antiferromagnetic phase 
is not rotated directly into the SU(4) superconducting phase phase but instead 
evolves with increased doping into a competing SC and AF phase, which then is 
transformed into a pure superconducting state by a quantum phase transition at 
a 
critical doping point near optimal doping, where the antiferromagnetic 
correlations vanish identically \cite{sun05}.

\subsubsection{Dynamical Criticality, Emergence, and Complexity}

As we have noted, the spontaneous appearance of properties in a many-body 
system 
that do not pre-exist in its elementary components is termed emergence;  
systems 
with emergent behavior are said to exhibit complexity \cite{dag05}.  Complexity 
results when there are multiple potential ground states and the choice between 
them is sensitive to weak external perturbations. The amplification 
effect implied by SU(4) dynamical criticality can facilitate emergent behavior 
and associated complexity, as exemplified by giant proximity effects and the 
perturbatively-induced structure of \fig{stripeOrigin}. More generally, 
critical 
dynamical symmetry may be of fundamental importance for 
complexity in various strongly-correlated Fermi systems, not just those of 
condensed matter. 
For example, critical dynamical 
symmetries have long been known in nuclear structure physics 
\cite{wmzha87,wmzha88a}.

\subsubsection{Varied Inhomogeneity but a Universal Phase Diagram}

The SU(4) coherent state method that yields the variational energy surfaces
discussed here admits quasiparticle solutions that generalize the BCS equations,
giving a rich, highly universal phase diagram in agreement with much available
data \cite{sun05}. Therefore, the propensity of the pseudogap state to a broad
variety of inhomogeneity, and a quantitative model of the cuprate phase
diagram (including pseudogaps) that exhibits highly universal character, both
follow directly from a model that implements antiferromagnetic and $d$-wave 
superconductor competition in a
doped Mott insulator.  This natural coexistence of a universal phase diagram
with a rich susceptibility to disorder in a limited region of its 
control-parameter
space could reconcile many seemingly contradictory observations in the cuprate
superconductors.  For example, since the SU(4) pseudogap state has both pairing
fluctuations and critical dynamical symmetry, Nernst vortex states could appear
as perturbations on a homogeneous phase that is unstable against developing
nanoscale disorder.

\subsubsection{Critical Dynamical Symmetries Near Magnetic Vortices and Magnetic
Impurities}

Critical dynamics also may produce inhomogeneities in the vicinity of magnetic
vortices and magnetic impurities.  Because magnetic fields should suppress
superconductivity, distance from a vortex $d$ may
be expected to alter the average value of $\pratio$ in a similar way as changing
the doping $P$ would. Therefore, regions near magnetic vortices or magnetic
impurities may exist where the symmetry is critical and exhibits sensitivity to
perturbations similar to that shown in Fig.\ \ref{fig:perturbations}, with
distance $d$ from the vortex or impurity modulating $\pratio$ rather than $P$.
This possibility is illustrated schematically in
\fig{dynamicalSymmetryNearVortex},%
\singlefig
{dynamicalSymmetryNearVortex}
{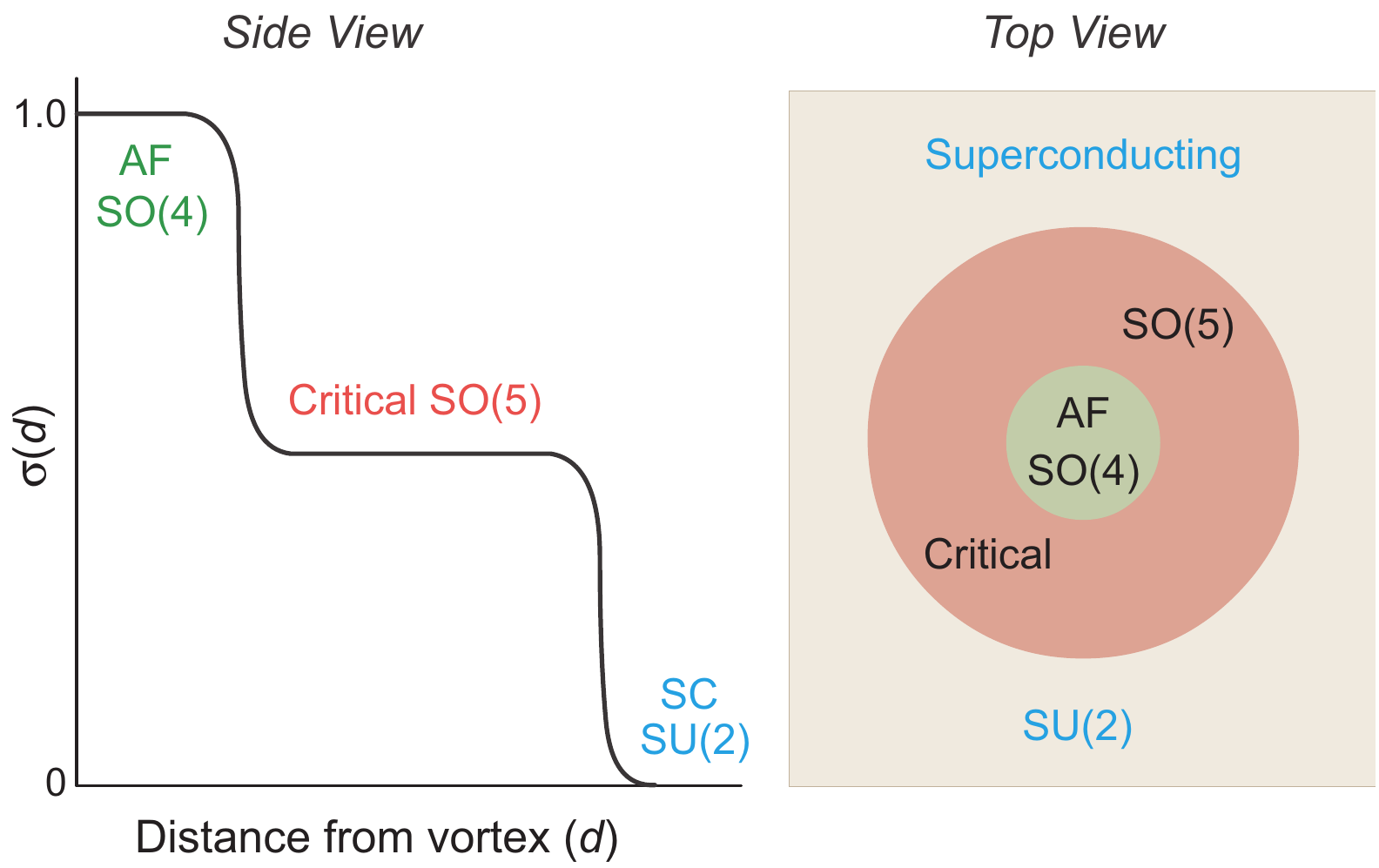}
{0pt}
{0pt}
{0.54}
{Illustration of critical dynamical symmetry in the vicinity of a
magnetic vortex. }
which suggests that vortices and magnetic impurities may be surrounded by
regions strongly susceptible to AF and SC fluctuations.

\subsubsection{Summary}

The generality of the SU(4) solution discussed here implies that any realistic 
theory deriving
superconductivity from a doped Mott insulator should contain features similar to
those discussed here.  Then the existence of complex inhomogeneities for
compounds having universal phase diagrams suggests that (1)~properties of
superconductors in the underdoped region, and near magnetic vortices and
impurities, are largely determined by critical dynamical symmetry, and
(2)~inhomogeneity is a strong diagnostic for the mechanism of high-temperature
superconductivity, but has little direct relationship with its cause.

\section{The Pseudogap and Mean Fields}

Let us now address whether the pseudogap is explicable within a mean-field
context.  In particular,  can the pseudogap be described in terms of competing
order when little conclusive experimental evidence supports a distinct phase
associated systematically with the pseudogap? The coherent-state approximation
is mean-field, in a similar sense as BCS.  However, the corresponding mean field
is quite rich because it is defined in terms of singlet and triplet fermion (not
boson) pairs that are strongly influenced by antiferromagnetic and pairing
correlations, and that are constrained to obey SU(4) symmetry.  This leads to
unusual properties in the resulting solutions. In
addition, although the general coherent state solution is mean field, the SU(4)
model has exact solutions in three (physically relevant) dynamical symmetry
limits that do not correspond to mean-field approximations.  These observations
have the following implications.

(1)~The SU(4) theory is in its essentials a model of competing order, since its
algebra closes only if there are operators corresponding to AF and pairing modes
that enter the theory on an equivalent footing. However, because of the basis
\eqnoeq{collsubspace} of singlet and triplet fermion pairs in the collective
subspace, the PG state may be interpreted also in terms of preformed
pairs \cite{sun05}:  The Hamiltonian implements a competition between
antiferromagnetism and pairing interactions, but the eigenstates in the
PG region correspond to rich superpositions of fermion pairs 
with a structure dictated by the AF--SC competition.  

(2)~Because the SU(4) coherent state solution is mean field in nature, it does
not include the effect of quantum fluctuations on the ground-state wavefunction.
However, comparison of the coherent state solution with the three exact SU(4)
solutions indicates that in all cases the coherent state energy of the ground
state coincides with that of the exact solution, but the wavefunction of the
coherent state solution provides only the expectation value of the energy, not
the quantum fluctuations.

We may illustrate with two specific examples:  (a)~The solutions of the coherent
state gap equations for a N\'eel state and for the PG state above $T\tsub c$ are
formally equivalent, but (as noted above) they lead to fundamentally different
dependence of the corresponding energy surfaces on antiferromagnetism and
pairing, implying that the full wavefunction of the two states (including
quantum fluctuations) are very different, but that the exact solution still
corresponds to a superposition of states described by mean fields having
approximate SU(4) symmetry.  (b)~The solutions for the coherent-state gap
equations for the SU(2) superconducting limit and the SO(5) limit are formally
equivalent, but again their energy surfaces are qualitatively different,
implying very different roles for quantum fluctuations. This explains why the
SU(4) coherent state solution is successful in describing quantities such as
energy gaps and the critical temperatures $T\tsub c$ and $T^*$ that are
determined primarily by the ground-state energy.  

(3)~The SU(4) energy surfaces provide a guide to the expected quantum 
fluctuations.
The coherent state solutions imply that the ground-state in much of the 
pseudogap
region is inherently unstable against both antiferromagnetic and pairing
fluctuations, because many solutions having different combinations of
antiferromagnetism and superconductivity become nearly degenerate.  Thus, 
although
the exact SU(4) coherent state contains no fluctuations, the form of the 
solution
itself implies large fluctuations in both pairing and antiferromagnetism for 
states
slightly perturbed away from the exact pseudogap solution.  This  instability
(coinciding with broad, flat regions of the energy surface) occurs predominantly 
in
the underdoped pseudogap region, implying unique features for the pseudogap 
state. For our
N\'eel state solution at half filling and the superconducting solution in 
optimally
doped and overdoped regions the energy surfaces have sharp minima in both the AF 
and
pairing directions, suggesting that these define stable phases with small 
influence
from quantum fluctuations (see \fig{compareEsurfacePerturbations}). For the
underdoped region below $T\tsub c$ our solutions indicate an energy surface that 
is
soft in the AF direction but localized around a finite value of superconducting
order, giving a solution that is superconducting, but with the superconductivity
modified significantly by AF fluctuations. 

(4)~Thus, although the SU(4) coherent state is a mean field solution, the energy
surfaces evaluated in this approximation (and compared with exact many-body
solutions in the three symmetry limits) already give a strong guide to the
expected role of quantum fluctuations around the mean-field solution. An
extended theory including fluctuations associated with the underdoped
instability described above may be expected to produce a PG state that is more
complex than in the mean-field limit. 

This mean field plus quantum fluctuations PG state would be a
superposition of SU(4) states having expectation values near zero for SC and AF
order parameters, but with large fluctuations in both, and strong susceptibility
to external perturbations.  The transition to this pseudogap
state would be expected to exhibit a rapid evolution in properties instead of a
clear phase transition. The resulting theory should retain the correct
predictions of \fig{gapsResolve_k}. However, quantum fluctuations of the
PG state, coupled with the predicted extreme sensitivity to
perturbations, would set the stage for a complex set of phenomena (Nernst vortex
fluctuations, spatial inhomogeneity, giant proximity effects, and other emergent
properties), all described within a framework unifying the competing order and
preformed pairs pictures of the pseudogap.

\section{\label{fermiArcs} Anisotropy of the Pseudogap}

For normal superconductors the Fermi surface  is key to the understanding 
superconductivity.  In optimally doped and overdoped high-$T\tsub c$ compounds 
a ``normal'' Fermi surface exists and BCS theory utilizing $d$-wave singlet 
hole pairs seems applicable \cite{bonn06}. But in underdoped cuprates PG states 
(lying between $T^*$ and $T\tsub c$) have anomalous  Fermi surfaces, with 
indications that the large Fermi surface found in the overdoped region has been 
reduced to arc-like vestiges, or to small pockets in $k$-space.

\subsection{Fermi Arcs and Magnetic Quantum Oscillations}

Angle-resolved photoemission spectroscopy (ARPES) \cite{dama03,norm03} probes
electronic properties \cite{norm03,dag94} of states in high-temperature
superconductors. These data suggest decreased state density near the Fermi
energy for temperature $T<T^*$ that is anisotropic in momentum and has a strong
temperature dependence.  A full Fermi surface is observed for $T > T^*$; as the
temperature decreases below $ T^*$, only arcs centered on the $d$-wave nodal
lines survive \cite{norm98,zhou04,kani06}, scaling as $T/ T^*$ to zero length
as $T$ tends to zero \cite{kani06}. Thus gapping  is anisotropic in momentum
space, with ungapped Fermi surface surviving only as temperature-dependent
{\em Fermi arcs}.

The validity of the Fermi arcs interpretation of ARPES measurements has been
challenged by  measurements of Shubnikov--de Haas  and de Haas--van Alphen
effects that probe the cuprate Fermi surface directly
\cite{doi07,leb07,yel08,ban08,jau08}.  These magnetic quantum oscillation
experiments  indicate that the Fermi surface in the underdoped region consists,
not of disconnected arcs, but of small closed regions (``pockets''). These
measurements determine the area of the Fermi surface rather precisely and
indicate that in the underdoped cuprates the Fermi surface is small, though they
cannot determine the location of the Fermi surface in momentum space. This
differs substantially from the large Fermi surface found for ordinary
superconductors and for overdoped cuprate samples. It is generally expected that
the state seen by quantum oscillation experiments under high magnetic field
conditions corresponds to the normal state from which the superconductivity
develops.

In this section we shall show that SU(4) symmetry for the cuprate
superconductors implies that 

\begin{enumerate}
 \item 
The correlation energy associated with opening the pseudogap necessarily has a
strong angular dependence in the $k$-space.
\item
This momentum-space anisotropy leads to strong temperature-dependent and
doping-dependent restrictions on regions of the Brillouin zone where ungapped
Fermi surface can exist. 
\end{enumerate}
We shall demonstrate explicitly that this prescription yields a description of
the length of Fermi arcs as a function of temperature that is in quantitative
agreement with the ARPES results of Ref.\ \cite{kani06}, if one assumes
as a starting point a full hole Fermi surface.  Conversely, if one assumes the
Fermi surface to correspond to the closed pockets suggested by quantum
oscillation experiments, we shall argue that our results place strong
restrictions on possible location and size of those pockets.

\subsection{Momentum-Dependent SU(4) and the Pseudogap}

The appropriate version of the SU(4) theory to address the questions of this
section is the \sufourk\ model discussed in \S\ref{k-dependentSU4} and
Refs.~\cite{sun07,guid09b}, which uses momentum-dependent generators of the
algebra. All results of the  $k$-independent SU(4) model
\cite{guid99,guid04,sun05,sun06}, including the $k$-averaged results in
\fig{gapsResolve_k}, are recovered as a special case of the \sufourk\ model for
observables dominated by contributions from near the Fermi surface ($\tilde k=
k\tsub f$) and averaged over all $\bm k$ directions. However, general solutions
of the \sufourk\ model give new $k$-anisotropic properties, one of which is that
the temperature for the pseudogap closure $T^*(\bm k)$ becomes anisotropic in
$\bm k$:
\begin{equation}
    T^*(\bm k)\equiv T^*(k\tsub f, \theta) = \left| \frac{g(k\tsub
f,\theta)}{g_0(k\tsub f)} \right| \, T^* ,
\label{TPG1.1}
\end{equation}
where $T^*$ is the gap closure temperature measured by ARPES along the antinodal
($\theta = 0,\pi/2$) direction. See  Ref.\ \cite{sun07} for details.  The
physical reason for this temperature and doping dependence of the
pseudogap closure temperature is that the  pairs interacting by
AF interactions in the SU(4) PG state each carry a $g(\bm k)$
formfactor. This introduces a $\bm k$ dependence in the effective AF coupling,
and thus in $T^*$. The behavior of $g(\bm k)$ as a function of ($k_x,k_y$) is
illustrated in \fig{gkPotentialComposite}(a).
\singlefig
{gkPotentialComposite}
{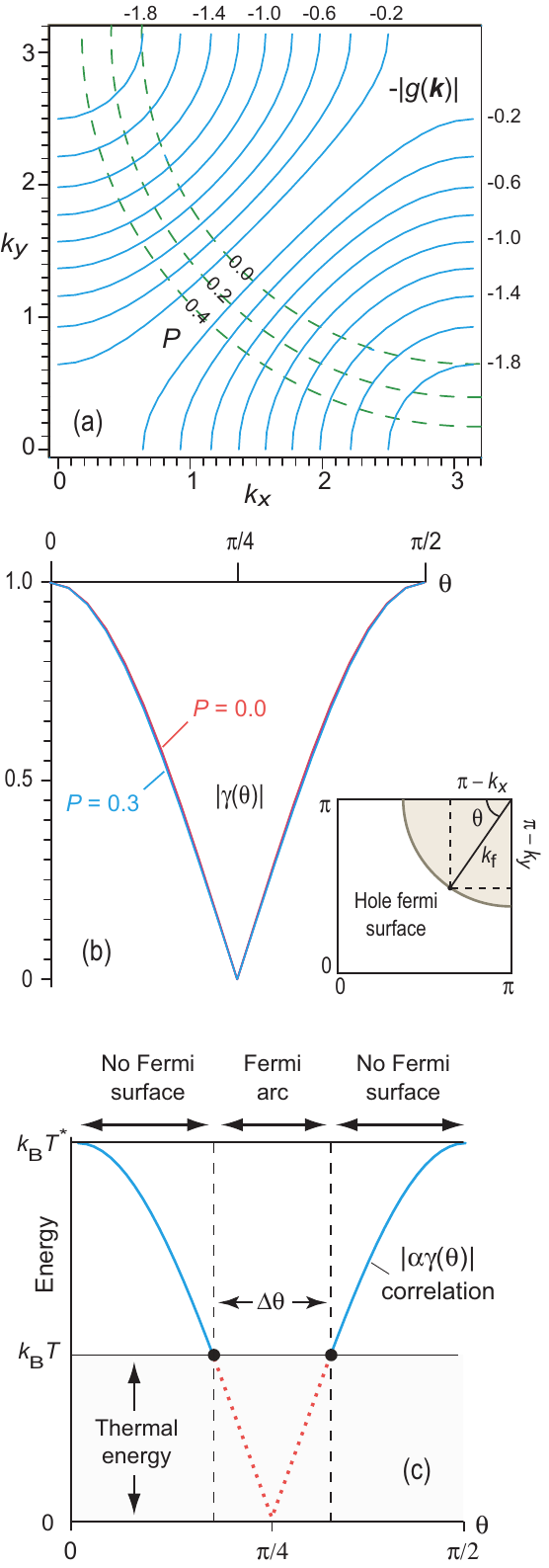}
{0pt}
{0pt}
{1.12}
{The $\bm k$-anisotropic factor, pseudogap correlation energy, and graphical
Fermi-arc solution, assuming a hole-like Fermi surface centered on $(k_x,
k_y)=(\pi,\pi)$. (a)~The $\bm k$-anisotropic factor $|g(\bm k)|$ (solid blue 
lines)
and contours of equal hole doping $P$ (dashed green lines), in the first
Brillouin zone. (b) Correlation energy $\alpha \gamma(\theta)$ for unit
$\alpha=kT^*$ evaluated along the Fermi surface [curves of constant $P$ in part
(a)] as a function of the angle $\theta$ (defined in inset).  Curves
corresponding to doping $P= 0-0.3$ are almost coincident, indicating that
$\gamma(\theta)$ is insensitive to doping. (c)~Graphical solution for Fermi
arcs.  The curve defines the PG correlation energy and horizontal lines
correspond to constant thermal energy scales $k\tsub B T$.  Their intersections
(black dots) represent points in the momentum space where the pseudogap is just
closed by thermal fluctuations; these bracket arcs $\Delta\theta$ of surviving
Fermi surface. }

The strong  dependence of the pseudogap temperature $T^*$ on $\bm k$ that is
implied by Eq.\ (\ref{TPG1.1}) and Fig.\ \ref{fig:gkPotentialComposite}(a)
represents a theoretical prediction with potential observational implications
that is independent of the experimental controversy over whether the Fermi
surface of underdoped cuprates consists of small close pockets or disconnected
arcs.  Nevertheless, we now examine the possible implications of this finding
for the competing Fermi arc and magnetic oscillation pictures of the cuprate
Fermi surface at low hole doping.

\subsubsection{Implications for Fermi Arcs}

In $g(\bm k)$ the components ($k_x,k_y$) or $(k\tsub f,\theta)$ are constrained
by the Fermi surface ($\tilde k^2 = k^2_{\scriptstyle\rm f}$). A more realistic 
shape for the Fermi surface will be considered below but if we assume for
illustration an isotropic hole surface centered around $(\pi, \pi)$ [dashed 
lines in Fig.\ \ref{fig:gkPotentialComposite}(a) and inset to Fig.\
\ref{fig:gkPotentialComposite}(b)], then 
$$
    (\pi - k_x)^2 + (\pi-k_y)^2 = k_{\scriptstyle\rm f}^2 = 2\pi(1+P) .
$$
For a given doping $P$ (thus $k\tsub f$) and temperature $T$, with
$T^*(\bm k)=T$ by virtue of  Eq.\ (\ref{TPG1.1}) under the  above
constraint, we obtain the angles  $\theta_1$ and $\theta_2$ at which
the PG closes [the angle $\theta$ is defined in the inset to Fig.\
\ref{fig:gkPotentialComposite}(b)], and the length of the surviving
Fermi arc is $ k\tsub f |\theta_2-\theta_1| \equiv  k\tsub
f\Delta\theta$.

This result may be interpreted graphically.  In Fig.\
\ref{fig:gkPotentialComposite}(b) we show $\gamma(\theta) \equiv
|g(k\tsub f, \theta)/g_0(k\tsub f)|$ versus angle $\theta$ along
different Fermi surfaces [dashed lines in Fig.\
\ref{fig:gkPotentialComposite}(a)].  We see that $\gamma(\theta)$
is almost independent of doping $P$.  Thus, solving Eq.\
(\ref{TPG1.1}) with the Fermi surface constraint  is equivalent to
solving
\begin{equation}
   k\tsub BT^*(\bm k)=\alpha\gamma(\theta) \qquad \alpha=k\tsub BT^* ,
\label{TPG1.2}
\end{equation}
with $k\tsub B$ the Boltzmann constant. The PG correlation energy
$\alpha \gamma(\theta)$ depends on the $\bm k$ direction $\theta$
[Fig.\ \ref{fig:gkPotentialComposite}(c)].  The pseudogap closes
when the thermal excitation energy $k\tsub B T$ is comparable to the
pseudogap correlation energy.  The intersections of horizontal lines of
fixed $k\tsub B T$ with the correlation energy curve [heavy black dots in
Fig.\ \ref{fig:gkPotentialComposite}(c)] define Fermi-arc solutions
$\theta_1$ and $\theta_2$.  Outside those points (solid  part of the curve),
the correlation energy is larger than the energy of thermal
fluctuations (shaded region), the pseudogap opens, and the Fermi surface is
gapped.  Inside these points (the dotted  part of the curve),  thermal  
energy
exceeds the correlation energy, the PG closes, and the Fermi surface
exists in an arc $\Delta\theta$ between the dots. When $T>T^*$, the
PG is closed in all directions and there is a full Fermi surface
since $k\tsub B T >\alpha$, and $\alpha$ is the maximum pseudogap
correlation energy.

For a given temperature $T$, the arc solution that is exemplified graphically in
Figs.\ \ref{fig:gkPotentialComposite}(b)--(c), or algebraically in Eq.\
(\ref{TPG1.1}), implies an anisotropy of  $\gamma(\theta)$ that partitions the
$k$-space uniquely into regions that can have a Fermi surface [$T>T^*(k\tsub
f,\theta)$] and regions that cannot [$T<T^*(k\tsub f,\theta)$]. As Fig.\
\ref{fig:gkPotentialComposite}(c) suggests,  arc lengths decrease with $T$ at
fixed doping, with a full Fermi surface at $T=T^*$ but only the nodal points at
$T=0$. Doping dependence enters primarily through the maximum pseudogap
temperature $T^*$ and the Fermi momentum $k\tsub f$; the temperature dependence
enters through $T/T^*$.  For fixed $T$, arcs shrink toward the nodal points with
decreased doping because $T^*$ in $T/T^*$ increases at smaller doping
(\fig{gapsResolve_k}). However, if  the length of Fermi arcs is measured by the
fraction $\Delta\theta/(\pi/2)$ and  the temperature is scaled by $T^*$, the
weak doping dependence of $\gamma (\theta)$ ensures that $\Delta\theta/(\pi/2)$
versus $T/T^*$ is rather universal, 
almost independent of both doping and compound.

\subsubsection{Temperature Dependence of Fermi Arcs}
 
In \fig{fermiArcLength}%
\singlefig
{fermiArcLength}
{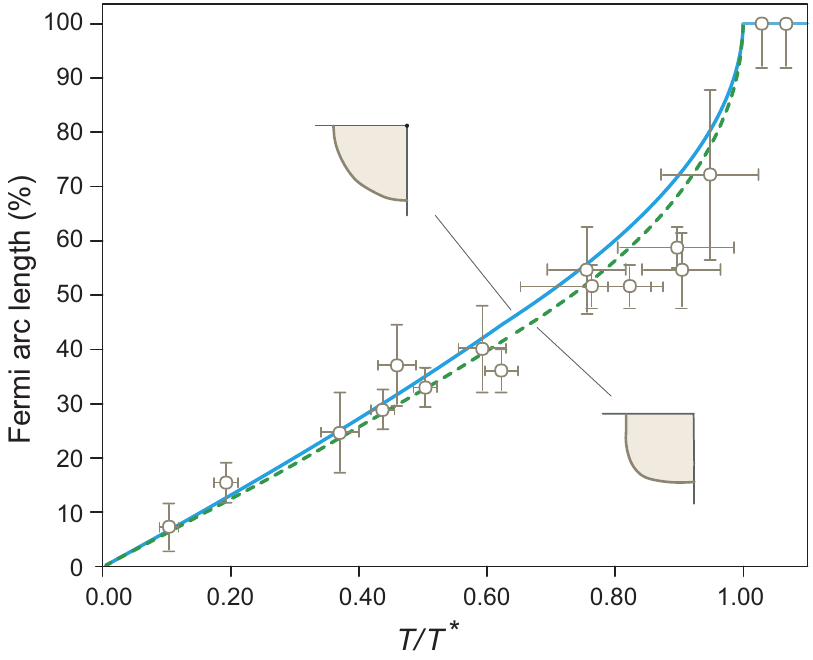}
{0pt}
{0pt}
{1.05}
{Fermi arc length vs.\ temperature. Experimental arc length is displayed as a
percentage of full Fermi surface length vs.\ $T/T^*$ for underdoped Bi2212
\cite{kani06};  the solid curve is our prediction for an isotropic Fermi surface
(inset upper left); the dashed curve assumes a Fermi surface with flatter
antinodal segments (inset lower right). No parameters (beyond those already fit
to gaps in \fig{gapsResolve_k}) were adjusted, and the
curves are almost independent of doping.}
we show experimentally-determined fractional arc lengths versus $T/T^*$ for 
Bi2212 \cite{kani06}. The  theoretical solution  for the fractional arc lengths 
[obtained from \fig{gkPotentialComposite}(c) or \eq{TPG1.1}] is the solid curve 
in \fig{fermiArcLength}. The agreement between theory and data is rather 
good, given that the theoretical curve has no adjustable parameters (it is 
determined completely by the parameters fixed previously in 
\fig{gapsResolve_k}) 
and that the theory {\em predicts} the scale $T^*$ implicit in the data 
(\fig{gapsResolve_k}). Notice that the rapid drop in arc length for decreasing 
$T\sim T^*$, transitioning to a linear decrease to zero arc-length at $T=0$, is 
explained entirely by geometry  in \fig{gkPotentialComposite}(c).

The dashed curve in \fig{fermiArcLength} repeats the analysis using the flatter 
Fermi surface  shown inset lower right. The similarity of dashed and solid 
curves indicates that solutions depend only weakly on  differing curvature in 
nodal and antinodal regions. Absolute arc lengths depend on doping and 
temperature, but the scaled arc lengths of \fig{fermiArcLength} {\em are 
near-universal functions only of the ratio} $T/T^*$, largely independent of 
compound, doping, and Fermi surface details, as suggested by data.

\subsubsection{Discriminating among Theories for Fermi Arcs}

The preceding discussion shows that a  viable theory of Fermi arcs  must make
two correct predictions: (1)~the scale $T^*$ and its doping dependence, and
(2)~that $T^*(\bm k) \propto \gamma(\theta)$. {\em Any theory} having a $T^*(\bm
k)$ consistent with Eq.\ (\ref{TPG1.2}) can describe the  scaled data of
\fig{fermiArcLength}, if $T^*$ is taken from data. Scaled ARPES data
(\fig{fermiArcLength}) test whether a pseudogap has a nodal structure similar to
that of the superconducting state. But discriminating among different theories
meeting this condition requires a quantitative, self-consistent description of
Fermi-surface disappearance and doping dependence of the PG temperature scale 
$T^*$
(\fig{fermiArcLength}). Thus, only a highly-restricted set of
models can be consistent with the aggregate properties of Fermi arcs.

\subsubsection{Implications for Fermi Surface Pockets}

Finally, let us consider the case where the Fermi surface for underdoped
cuprates is interpreted in terms of small pockets instead of Fermi arcs, as
is favored by magnetic quantum oscillation experiments.  This discussion must
necessarily be more qualitative, since quantum oscillation experiments cannot
localize the inferred small pockets definitively in momentum-space.

We hypothesize that cuprate normal states deviate from Fermi liquid behavior
primarily because of correlations producing the pseudogap, which in the present
model represents fluctuating AF correlations in a singlet and triplet pair
basis.  Thus suppression of AF correlation in cuprates should lead to a normal
Fermi surface and BCS-like superconductivity. There are no AF correlations
beyond the critical
doping $P\tsub c \simeq 0.18$, consistent with experiments suggesting that SC is
($d$-wave) BCS-like with a Fermi liquid normal state.  Below the critical doping
the AF correlations compete with pairing in general, but the
present analysis suggests the additional feature that this competition has
strong angular localization in the $\bm k$-space.

The actual Fermi surface of the physical system depends on how
interactions rearrange the occupied and unoccupied orbitals of the
non-interacting system. However, the preceding arguments indicate
that, independent of details, at low doping the momentum space
available to form a normal Fermi surface becomes small and
restricted to limited regions of $\bm k$-space by the angular
formfactors displayed graphically in
\fig{gkPotentialComposite}.

Thus, we expect  large conventional Fermi surfaces near critical
doping and beyond, but at low doping the strong  anisotropy of the
pseudogap correlations severely restricts the volume of $\bm
k$-space available to produce Fermi pockets, or Fermi surfaces of any
form. This argument also suggests that any ungapped Fermi surface is
likely to lie near the nodal region in either the ARPES or quantum
oscillation interpretations. This is consistent with the conjecture
of Ref.~\cite{jul07} for location of the Fermi pockets. Hence, the
most important implication of both Fermi arc and quantum oscillation
data may be that they provide comprehensive evidence for localized,
anisotropic correlations responsible for producing the pseudogap,
independent of detailed interpretation.

\subsection{Summary: Anisotropy, Arcs, and Pockets}

We have demonstrated that an SU(4) solution for cuprate superconductivity
implies a strong $\bm k$-space anisotropy and corresponding localization for
correlations associated with the pseudogap, leading to a pronounced $\bm
k$-dependence in the pseudogap temperature $T^*$.  We believe this to reflect
general properties expected for any realistic model of the cuprates, and  to
have significant consequences for gapping of the cuprate Fermi surface,
particularly in the underdoped region.

If the Fermi surface in the underdoped region is interpreted in terms of Fermi
arcs, the \sufourk\ theory reproduces quantitatively  the observed variation of
the length of the arc segments with temperature, with no parameter adjustment.
This suggests that ARPES experiments are measuring quantities related directly
to the anisotropic structure of the pseudogap correlations. We can make less
definitive statements about the Fermi surface pockets favored by quantum
oscillation experiments because  the proposed small pockets have not been
localized in $\bm k$-space. However, our results suggest that an ungapped Fermi
surface of {\em any type} is unlikely to survive the pseudogap correlations as
the temperature is lowered below $T^*$, except in increasingly localized nodal
regions of $\bm k$-space. 

Thus, our analysis suggests that ARPES and quantum oscillation experiments are
seeing (perhaps different aspects of) physics associated with strong $\bm
k$-space anisotropy of the pseudogap correlations in underdoped cuprates.
The essential physics of Fermi arcs and quantum oscillation results for
underdoped cuprates may lie, not in the underlying Fermi surface itself, but
rather in how that Fermi surface is modified by the anisotropic pseudogap
correlations.

\section{\label{ironSC} The Iron-Based Superconductors}

\begin{minipage}{3.0in}
\begin{small}
``SU(4) symmetry, not $d$-wave pairing \ldots is the
ultimate cause of cuprate behavior, implying that systems could exist having
non-$d$ pairing but cuprate-like dynamics \ldots [This] prediction may be tested
by searching experimentally for compounds having pairing structure other than
$d_{x^2-y^2}$ that still satisfy the SU(4) algebra.''

\vspace{4pt}
\rightline{{\em M. W. Guidry, Y. Sun, and C.-L. Wu (2004)} \cite{guid04}}
\end{small}
\end{minipage}

\vspace{10pt}

In Ref.~\cite{guid04} we proposed that cuprate superconductivity was
characterized by more complex behavior than normal BCS superconductivity because
the symmetry structure associated with the superconductivity was non-abelian.
Therefore, we termed this new form of superconductivity {\em non-abelian
superconductivity,}  and we predicted that new forms of superconductivity having
many of the characteristics of the cuprates but not necessarily with $d$-wave
formfactors were possible.  In this section we shall provide evidence that the 
new class of
high-temperature superconductors based on the iron arsenides and selenides that
was discovered beginning in 2008
\cite{FeAsDiscovery,che08,wen08,che08b,che08c,ren08b,che08d,
FeSeDiscovery} represents the first examples of the new non-abelian
high-temperature superconductors predicted in Ref.~\cite{guid04}.

\subsection{Non-Abelian Superconductors}

Non-abelian superconductivity differs from conventional superconductivity in the
richness of the pair structure for condensed states and in the appearance of
competing sources of long-range order \cite{guid04}. The key issues for SU(4)
non-abelian superconductivity are that coherent pairs are formed by electrons or
holes on adjacent or nearby sites (rather than the same site), so that both
singlet and triplet pair states can in principle contribute, and that
alternative long-range order  enters with the same
standing as the superconductivity.  
 
In contrast to BCS superconductivity, which is described by a single dynamical
symmetry chain having only abelian subgroups  [$\sutwo \supset \uone$], the
minimal symmetry consistent with cuprate data is SU(4), which has a much richer 
structure (three dynamical symmetries having non-abelian subgroups and differing
fundamentally in their properties). We propose that the differences in
observational characteristics for these two types of superconductivity originate
in this difference in dynamical symmetry structure. 

The important role of SU(4) symmetry in non-abelian cuprate superconductivity
that we have documented in this review suggests  that any pairing structure
leading to the SU(4) algebra entails dynamics similar to that of cuprates.
Therefore, $d$-wave symmetry of the pairs need not be critical to non-abelian
superconductivity in general and SU(4) superconductivity in particular.   Pairs
with any internal symmetry  could exhibit SU(4) superconductivity (or a similar
form of superconductivity based on other non-abelian groups) if the
no-double-occupancy constraint is valid and correlations can form  bondwise (not
on the same site) pairs.  

For example, the operator $c^\dagger_{\bar{\bf r},i}$  defined in
\eq{cbar} specifically for cuprate $d$-wave pairs may be generalized to  
\begin{equation} 
c^\dagger_{\bar{\bf r},i}=\sum_{\bf t} g({\bf t}) 
c^\dagger_{{\bf r}+{\bf t},i}\ \qquad 
\sum_{\bf t} \left | g({\bf t})\right |^2=1 
\label{gt} 
\end{equation}  
where {\bf t} is a few finite lattice displacements of ${\bf r}$ and $g({\bf
t})$ is the form factor.  The structure (\ref{cbar}) of the $d$-wave pairs is
only a  special case of (\ref{gt}) with  
$$
{\bf t}=\pm a , \pm b
\qquad
 g(\pm{\bf
a})=\tfrac12
\qquad
 g(\pm{\bf b})=-\tfrac12.
$$ 
Different internal symmetries of the pairs lead to different forms of $g({\bf
t})$, but they all can satisfy the condition (\ref{cbarcom1}) under  no double
occupancy and thus preserve the SU(4) algebra and the  Hamiltonians
implied by its dynamical symmetry chains.

\subsection{Extending SU(4) to Iron-Based Superconductors}

SU(4) symmetry provides a comprehensive understanding of the cuprate
superconductors. We shall now show that a quantitative extension of this
approach to Fe-based SC requires only that (1)~the relevant collective degrees
of freedom are superconductivity and antiferromagnetism, and (2)~the
superconductivity involves bondwise (not onsite) pairing. Let us begin by
listing some of the similarities and differences between these two classes of
high-temperature superconductors.

\subsubsection{Cuprate and Fe-Based Phenomenology}

There are many similarities between cuprate and Fe-based superconductors.
(1)~The superconductivity in both seems to be unconventional, involving strong
electron correlations.
(2)~The superconductivity in both seems to be in close proximity to
antiferromagnetism and there is strong evidence for competing order.
(3)~The superconducting states in both appear to involve singlet pairing.

On the other hand, there are some substantial differences between cuprate and Fe
superconductors.  For example, (1)~The parent state in the cuprates is an AF
Mott insulator; the parent state in the Fe-based compounds is a (poor) metal,
though it may be near a Mott transition. (2)~The antiferromagnetism in the
Fe-based compounds differs from that of the cuprates. (3)~There are multiple
bands near the Fermi surface in Fe-based compounds, implying the possibility of
more complex multiband physics for the gaps. (4)~Because As atoms are out of
plane in Fe-based compounds, next-nearest neighbor interactions may be as
important as nearest-neighbor ones. (5)~The superconductivity in Fe-based
compounds is more often influenced by pressure, and there is a larger variety of
SC with both hole and particle doping in the Fe compounds than for the cuprates.
(6)~Cuprate SC is highly 2-D; there is more evidence in Fe-based SC for 3D
($c$-axis) effects.

Prior to the discovery of Fe-based SC, many had thought that the Mott insulator 
parentage of cuprate superconductors was critical to the mechanism of 
high-temperature superconductivity.  Difference (1) above calls this into 
question since parent compounds for the Fe-based high-temperature 
superconductors are typically metals (albeit poor ones), not Mott insulators. 
We 
now  argue that the essential point is not Mott behavior 
but rather SU(4) symmetry, which is highly compatible with a Mott insulator 
parent state for the cuprates, but can also produce a superconductor from a 
parent state that need not be a Mott insulator, as for  Fe-based SC.

\subsubsection{Mott Insulator versus Poor-Metal Parents}

Large onsite Coulomb repulsion leads to a Mott insulator normal state for
cuprates. In Ref.~\cite{guid04} we demonstrated that suppression of double site
occupancy for pairs in the real space is a sufficient condition to guarantee
that the minimal closed algebra is SU(4). The situation in the Fe-based
compounds is less clear. The onsite repulsion $U$ must lie in
an intermediate range between no correlations and the strong onsite repulsion
found for the cuprates, for if $U$ were too large the parent states would
develop a charge gap and be good insulators and if it were too small the parent
states would be good metals. Most are  in fact poor metals, suggesting an
intermediate range of $U$.

The presence of strongly-competing antiferromagnetism and superconductivity
ensures that Fe-based superconductors correspond to a non-abelian symmetry
\cite{guid04}, but the non-abelian algebra need not be the SU(4) algebra 
found for the cuprates. The minimal algebra for the iron arsenide 
superconductors depends on whether the
pairing leading to the superconductivity involves onsite pairs. If we simplify
by restricting attention to a single kind of nearest neighbor or next nearest
neighbor bondwise pair, the arguments of Ref.~\cite{guid04} indicate that the
minimal closed algebra is SU(4) if there are only bondwise (no onsite) pairs and
they don't overlap spatially. (That is, the wavefunction is a superposition of
bondwise pairs where no lattice site is occupied significantly by particles from
two different pairs.)

For the cuprate superconductors, strong onsite repulsion  opens a 
large energy gap between
the bondwise and onsite pairs, implying that bondwise pairs dominate the ground
state at low temperature (see \fig{su4-so8separate}). The iron arsenides have
onsite repulsion of intermediate strength relative to the cuprates and the
situation is less clear.  However, the issue is not whether the onsite repulsion
suppresses double occupancy in general (and thus produces a Mott insulator), but
only whether the correlations are sufficient to push onsite collective pairs to
substantially higher energy than bondwise collective pairs (which could be
compatible with a poor-metal).  

The key distinction is between the coherent 
pairs of the SU(4)
symmetry-truncated basis (which are responsible for charge transport in the
superconducting state) and additional valence particles that are not part of the
coherent pairs and and may contribute to charge transport in the normal state at
zero temperature. The SU(4) Lie algebra respects the Mott insulator
characteristics of the cuprate normal states in that it represents optimal
configurations for competing AF and SC in the presence of strong onsite
repulsion.  However, the SU(4) solution in general {\em need not correspond to
an insulator,} since the (normal-state) charge transport properties may be
strongly influenced by the properties of particles not in the collective pairs.

For example, the onsite repulsion could be sufficiently strong to make it 
energetically unfavorable either  to form onsite collective pairs or to have 
double site occupancy by unpaired particles, in which case the symmetry is 
SU(4) 
and in addition the material would be expected to be insulating in the normal 
state. This is representative of the situation in the cuprates.  But it could 
also be the case that somewhat weaker same-site repulsion strongly disfavors 
onsite pairs over bondwise pairs, but does not forbid some charge transport by 
unpaired particles in the normal state.  The resulting superconducting 
material would again be described by SU(4), but now is expected to be a metal 
or 
poor metal in the normal state. This situation may be representative of the 
FeAs compounds.

If that is the case, a minimal low-energy theory may be constructed using only 
bondwise pairs and the algebra of that theory will be SU(4) if these pairs do 
not overlap on the spatial lattice \cite{guid04}. Since numerical calculations 
\cite{dag08,mor09} indicate that the dominant pairing channels in the FeAs 
compounds involve nearest neighbor or next nearest neighbor {\em bondwise 
pairing}, we conclude that a minimal description of their superconductivity 
corresponds to the same SU(4) symmetry as was found for the cuprates, though 
values of the effective interaction parameters would likely differ from those 
of the cuprates.

\subsubsection{An SU(4) Model for Iron-Based Superconductivity}

The iron superconductors exhibit a different form of antiferromagnetism than
that found in the cuprates, as illustrated in \fig{spinPlaneCompact}.%
 \singlefig
     {spinPlaneCompact}
     {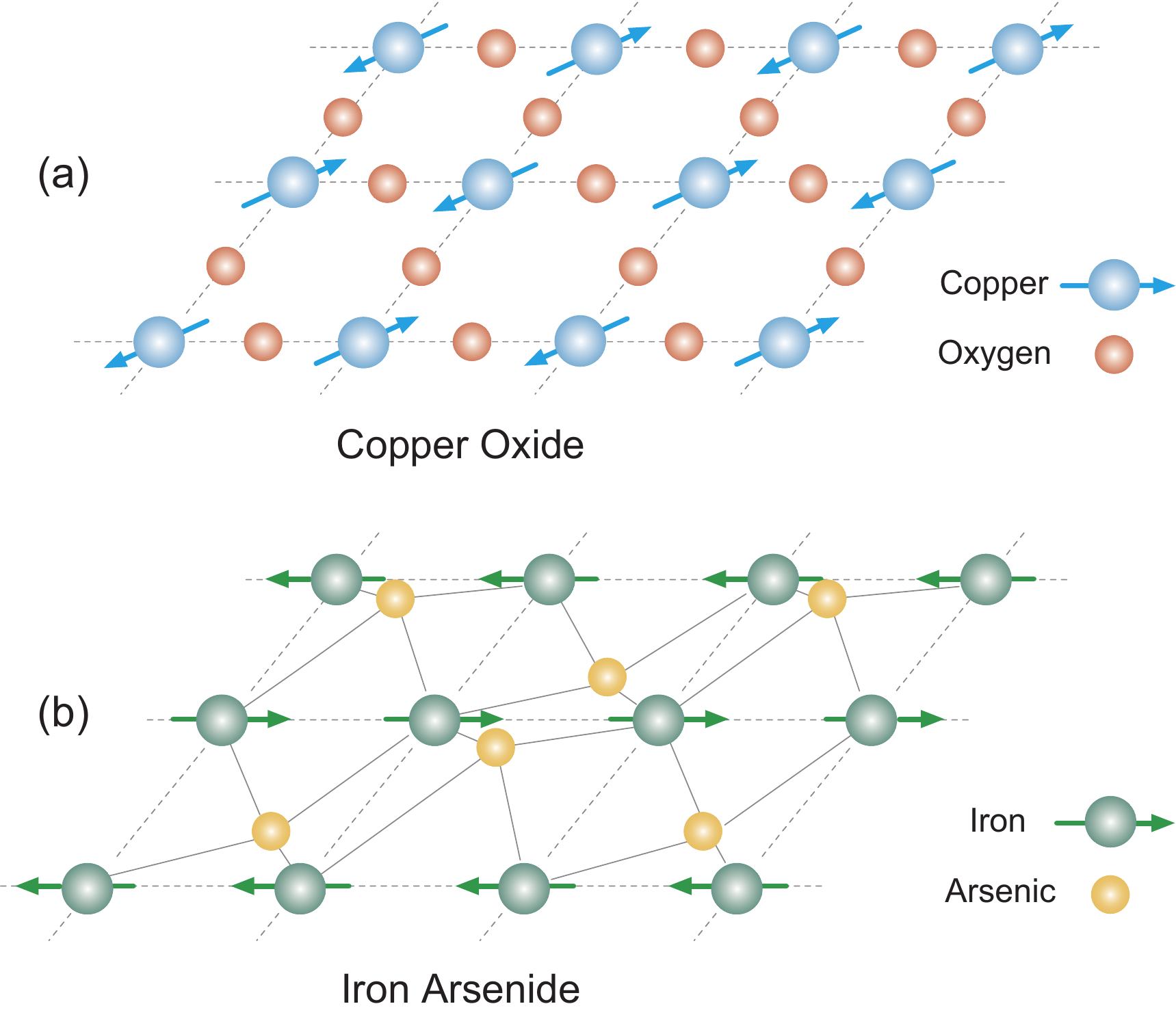}
     {0pt}
     {0pt}
     {0.50}
     {Schematic spin structure for (a)~cuprate and (b)~FeAs compounds. The
undoped iron arsenides are characterized by a ``stripe antiferromagnetism''
magnetic structure that differs from the AF observed in the cuprates.  The 
oxygen atoms are 
generally in the same plane as the copper atoms in the cuprates, but the 
arsenic 
atoms are located
above and below the plane of the iron atoms in the iron arsenides.}
Following the hint of a large amount of data, we shall assume that in the iron
superconductors the superconductivity and antiferromagnetism are related, and
that the former develops out of the latter through doping or pressure. Because
in the FeAs superconducting compounds the arsenic atoms are out of the Fe
plane, general arguments suggest that next nearest neighbor (NNN) lattice
interactions may compete favorably with nearest neighbor (NN) interactions. 
Construction of NN singlet pairs by addition to the magnetic background
corresponding to the FeAs compounds exhibited in \fig{spinPlaneCompact} is
illustrated schematically in \fig{pairs2OrbitNN},%
 \singlefig
     {pairs2OrbitNN}
     {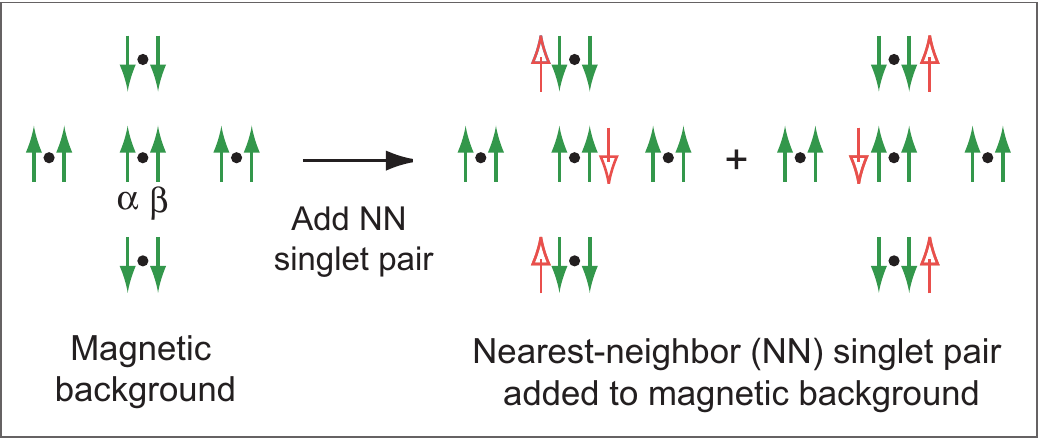}
     {0pt}
     {0pt}
     {0.82}
     {Possible SU(4) nearest neighbor (NN) spin-singlet pair structure in a
2-orbital model for a pair centered on a particular site. Arrows indicate
spin-up and spin-down particles, with an arrow to the left of a lattice point
signifying a particle in orbital $\alpha$ and an arrow to the right of a lattice
point signifying a particle in  orbital $\beta$. Solid green arrows represent
the undoped magnetic background state. The open red arrows represent the added
pair. The collective SU(4) NN pair would then correspond to a coherent sum over
the lattice of such pairs centered on individual lattice sites.}
and the corresponding construction of an NNN singlet pair is illustrate in
\fig{pairs2OrbitNNN}.
 \singlefig
     {pairs2OrbitNNN}
     {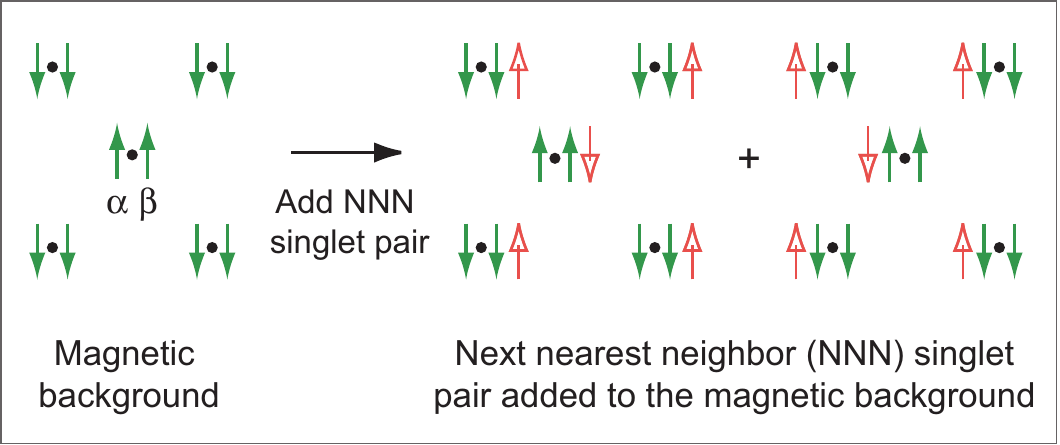}
     {0pt}
     {0pt}
     {0.82}
     {As for \fig{pairs2OrbitNN} but for a possible SU(4) next nearest neighbor
(NNN) spin-singlet pair structure in a 2-orbital model for a pair centered on a
particular site. }

By our previous arguments, if we assume that the superconductivity in the
Fe-based superconductors involves singlet Cooper pairs interacting with
antiferromagnetism, charge, and spin, the corresponding emergent symmetry will
be SU(4), just as in the cuprates, provided that  (1)~the pairs are bondwise and
not onsite pairs, (2)~electron correlations are strong enough to suppress
overlap of pairs (suppression of double site occupancy in the collective pairs),
and (3)~the condition \eqnoeq{conditions2} is satisfied for the pairing
formfactor.  The requirement that $g(\bm k)$ satisfy \eqnoeq{conditions2}
places immediate constraints on its possible form.  In
\tableref{FeGapTable} %
%
%
\begin{table}[t]
  \centering
  \caption{Some pairing gap orbital symmetries, whether they satisfy
\eq{conditions2} and thus close the SU(4) algebra for Fe-based and cuprate 
compounds, and
maximum doping fraction $P_f$ for allowed FeAs symmetries.}
  \label{table:FeGapTable}
\vspace{1pt}
  \begin{normalsize}
    \begin{centering}
      \setlength{\tabcolsep}{5 pt}
      \begin{tabular}{cccc}
        \hline
            $g(\bm k)$ &
            Fe-based &
            Cuprate & $P_f$

        \\[1 pt]        \hline
            $s_{x^2+y^2}=\cos k_x+\cos k_y$ &
            No &
            Yes & --

        \\[1 pt]
            $d_{x^2-y^2} = \cos k_x-\cos k_y$ &
            No &
            Yes & --

        \\[1 pt]
            $s_{x^2y^2}=\cos k_x\cos k_y$ &
            Yes &
            Yes & $1/3$

        \\[1 pt]
            $d_{xy} = \sin k_x\sin k_y $ &
            Yes &
            Yes & $1/3$
    \\[1 pt]
            $s_{x^2+y^2} \pm d_{x^2-y^2} $ &
            Yes &
            Yes & $2/3$

    \\[1 pt]
            $s_{x^2+y^2} \pm id_{x^2-y^2} $ &
            No &
            Yes & --

        \\[1 pt]        \hline
      \end{tabular}
    \end{centering}
  \end{normalsize}
\end{table}
%
%
we apply these constraints to some gap symmetries that have been proposed for
the FeAs superconductors, indicating whether the SU(4) algebra closes for
each case. For reference, we carry out the same procedure for the cuprates,
though in that case we already know experimentally that the gap symmetry is
$d_{x^2-y^2}$. Some of the formfactors that are tested in \tableref{FeGapTable}
are also illustrated in \fig{contour4filled}.%
\singlefig 
{contour4filled} 
{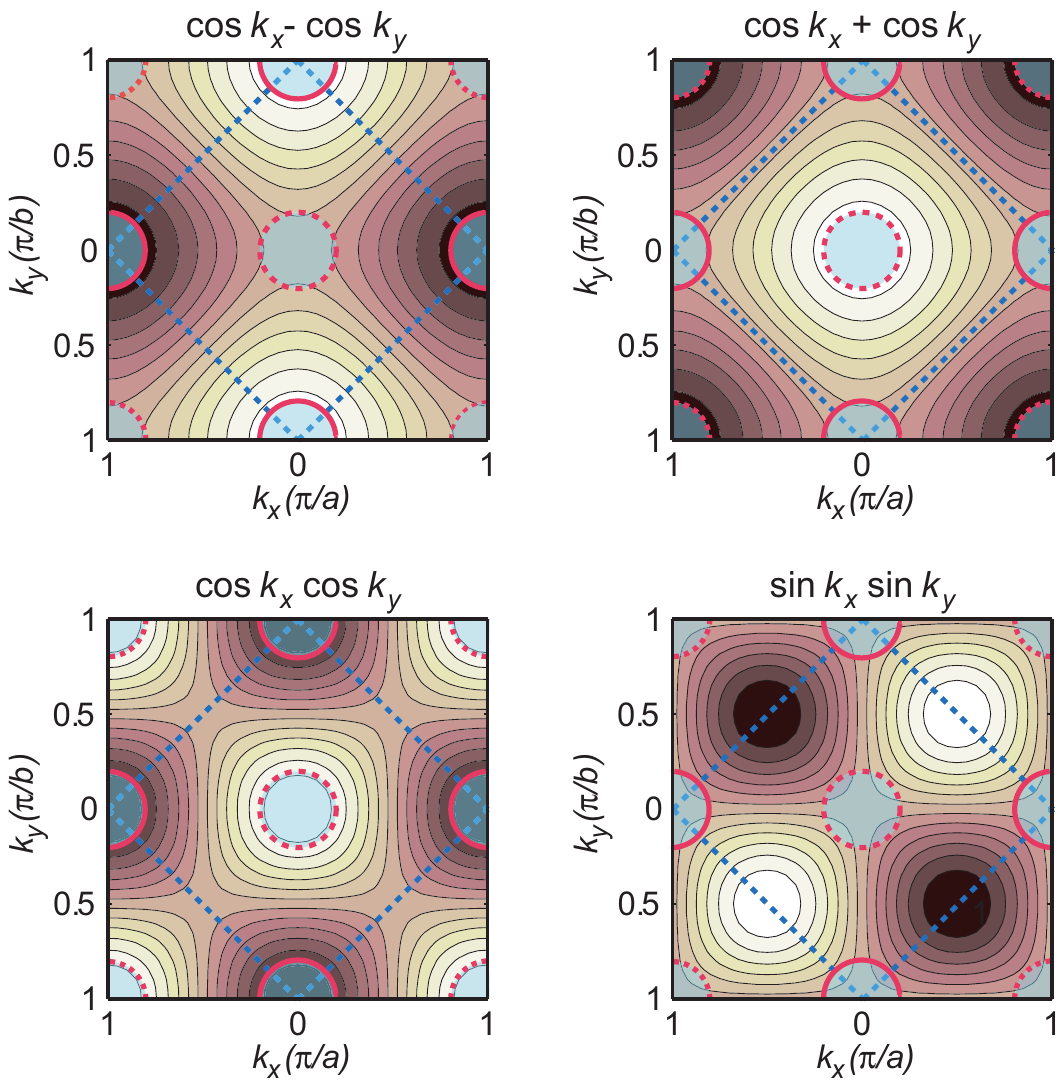} {0pt} {0pt}
{0.80} 
{Momentum-space formfactors for some  cases listed in \tableref{FeGapTable}.
The outer black square for each diagram is the large Brillouin zone  associated
with the Fe-only real-space lattice. The dashed blue diamond inset in each box
is the small Brillouin zone associated with the true real-space lattice. Very
schematic locations for the small Fermi surface pockets obtained from typical
calculations are sketched as heavy red curves: solid for regions with electron
pockets and dashed for regions with hole pockets. }

From \tableref{FeGapTable} we find that symmetries such as $s_{x^2y^2}$ and
$d_{xy}$ can close the SU(4) algebra, but symmetries such as $s_{x^2+y^2}$ and
$d_{x^2-y^2}$ do not. Thus, in the simplest symmetry-limit 
picture, neither $s_{x^2+y^2}$ nor $d_{x^2-y^2}$
are valid orbital symmetries for the Fe-based superconductors because their
failure to close the SU(4) algebra indicates that pairing with that geometry is 
 incompatible with the observed magnetic structure for the iron
superconductors.

We conclude that a unified SU(4) model of cuprate and Fe-based superconductivity
suggests that the corresponding orbital symmetry of the pair gap for Fe-based
compounds is likely to differ from the cuprate $d_{x^2-y^2}$ symmetry (in the 
symmetry
limits of the theory). It is also seen from \tableref{FeGapTable} that the more
symmetric antiferromagnetism of the cuprates (see \fig{spinPlaneCompact}) is
compatible with many possible pairing formfactors (though most seem not be
realized physically), but the asymmetric AF of the iron arsenides provides
stronger constraints on a compatible pairing structure.

The SU(4) symmetry allows a further constraint to be placed on the orbital
formfactor.  The coherent pair states for those formfactors allowed
for Fe-based SC in \tableref{FeGapTable} have a general structure
\newcommand{\op}[1]{c^{\dagger}_{#1}}
$$
p^\dagger = \sum_{r=\{x,y\}} \op{r\uparrow}
   \op{\bar r \downarrow}
$$
in the real space, where $\op{r i}$ is the electron creation operator $a_{\bm k
i}^\dagger$ in the coordinate representation. The $\op{\bar r i}$ for
$$
s_{x^2+y^2} + d_{x^2-y^2} = 2\cos k_x \qquad
s_{x^2+y^2} - d_{x^2-y^2} = 2\cos k_y
$$
 pairs are formed from nearest-neighbors,
\begin{align*}
 \op{\bar ri} &= 2^{-1/2} (
\op{(x+a,y)i} + \op{(x-a,y)i} ) \qquad g(\bm k) = 2\cos k_x
\\
 \op{\bar ri} &= 2^{-1/2} (
\op{(x,y+a)i} + \op{(x,y-a)i} ) \qquad g(\bm k) = 2\cos k_y ,
\end{align*}
and for $\cos k_x \cos k_y$ or $\sin k_x \sin k_y$ pairs are formed
from next-nearest neighbors,
\begin{align}
 \op{\bar r\downarrow} &= {\tfrac12} (
\op{(x+a,y+b)\downarrow} + \op{(x-a,y-b)\downarrow}
\nonumber\\
&\quad\pm \op{(x+a,y-b)\downarrow} \pm \op{(x-a,y+b)\downarrow} ),
\label{pairnnn}
\end{align}
with $(+)$ corresponding to $\cos k_x \cos k_y$ and $(-)$ to $\sin
k_x \sin k_y$. These are illustrated in Fig.\
\ref{fig:primitivePair}.

\singlefig
{primitivePair}       
{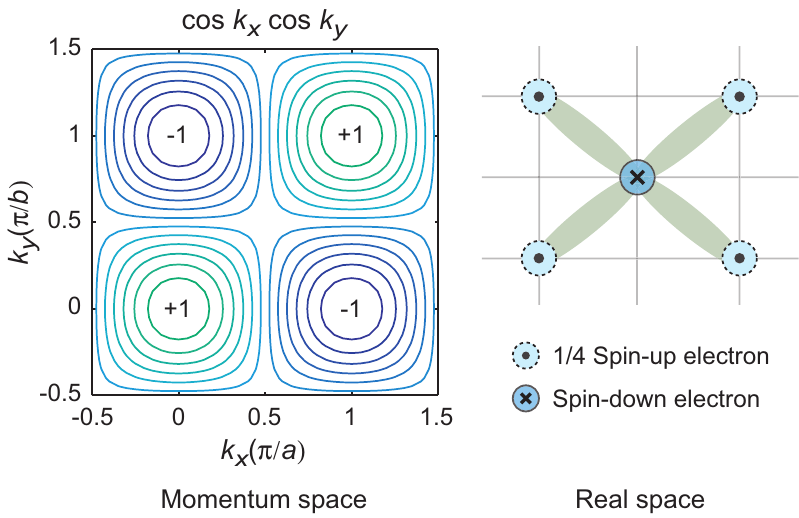}    
{0pt}         
{0pt}         
{1.07}         
{Pairing gap corresponding to a $\cos k_x \cos k_y$ formfactor in
momentum space and the corresponding schematic real-space pair
structure for a singlet electron pair. The shorthand notation ``$
1/4$ electron'' indicates  that the spin-up electron is distributed
with equal probability on four next-nearest neighbor sites in the
pair wavefunction of Eq.~(\ref{pairnnn}). The spatial pair structure
for $\sin k_x \sin k_y$ is similar to that for $\cos k_x \cos k_y$,
differing only in phases.}

As discussed for the cuprates in \S\ref{upperDopingLimit} and Ref.\
\cite{guid04}, the SU(4) requirement of no double occupation by pairs implies a
lattice occupancy restriction for the superconducting state. Counting the
maximum number of pairs that can be placed on the lattice without overlap, as
illustrated in Fig.~\ref{fig:pairCount},%
\singlefig
{pairCount}
{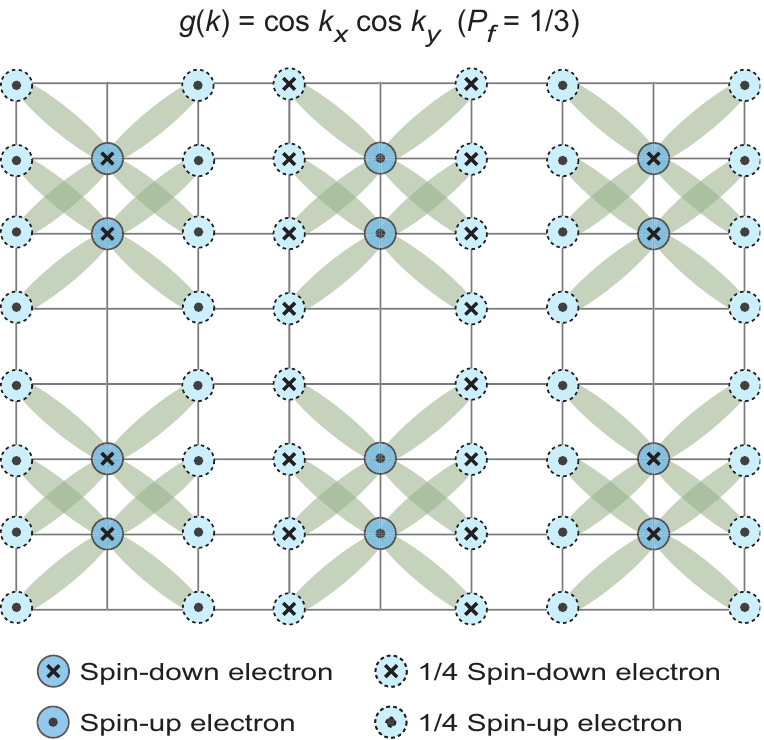}
{0pt}
{0pt}
{1.08} 
{Schematic count of maximum pair density consistent with SU(4) symmetry assuming
electron-doped material with a singlet $\cos k_x \cos k_y$ pair gap formfactor.
For this segment of the lattice, no additional pairs of this structure can be
added without causing a finite amplitude for double site occupancy by pairs,
which would break SU(4) symmetry. By counting of occupied and unoccupied sites,
the maximum fraction of lattice sites that can be occupied by $\cos k_x \cos
k_y$ pairs without double occupancy is $\tfrac13$.  The realistic wavefunction
will be a superposition of such configurations, each with a maximum pair
occupancy of $\tfrac13$. The spatial pair structure and maximum doping for $\sin
k_x \sin k_y$ is the same as for $\cos k_x \cos k_y$, since they differ only in
phases (see Eq.~(\ref{pairnnn})).}
indicates that the largest doping fraction consistent with SU(4) symmetry is 
$P_f = \tfrac23$ for $\cos k_x$ and $P_f = \tfrac13$ for $\cos k_x \cos k_y$ or
$\sin k_x \sin k_y$. These are summarized in the last column of
\tableref{FeGapTable}. Since current data suggest that the superconductivity
does not extend much beyond $P_f = \tfrac13$ for most FeAs compounds, this
favors $\cos k_x \cos k_y$ or $\sin k_x \sin k_y$, among the allowed
orbital symmetries for iron arsenides in \tableref{FeGapTable}.

\subsubsection{Multiple Pairing Gaps in the Iron Superconductors}

The iron superconductors involve pairs that receive contributions from multiple
Fe orbitals.  Their Fermi surfaces often correspond to disconnected sheets in
the Brillouin zone and this implies that their pairing properties are generally
$k$-dependent.  A generalization of the SU(4) coherent-state formalism to handle
this situation has been described in \S\ref{k-dependentSU4}, with the relevant
gap equations given in \eqs{gapsUnderdoped} and \eqnoeq{gapsOverdoped}, and the
predicted superconducting transition temperature as a function of doping in
\eq{Tc}. We now apply this formalism to an analysis of iron superconductors
\cite{guid09}.

In the preceding we argued that requiring (1)~closure of the SU(4) algebra and
(2)~consistency of pairing with  the observed antiferromagnetism favors a
pairing formfactor $g(\bm k) = \cos k_x \cos k_y$ or $ \sin k_x \sin k_y$. First
consider $g(\bm k) = \cos k_x \cos k_y$. Assuming the doping $x$ to be less than
the critical value $x_q$, Eq.~(\ref{gapsUnderdoped1}) gives for the singlet
pairing gap
\begin{equation}
  \singletGap (\bm k) = \Delta_0 \cos k_x \cos k_y
   \quad
  \Delta_0 \equiv \frac{G_0\Omega}{2\bar g}
   \sqrt{x(x_q^{-1} -x)}.
\label{gapcoskxcosky}
\end{equation}
ARPES measurements on ${\rm Ba}_{0.6} {\rm K}_{0.4} {\rm Fe}_2 {\rm As}_2$ 
find four sheets of Fermi surface within the Brillouin zone \cite{nak08}.  Let
us introduce a simple model that assumes the four pockets of Fermi surface
(labeled $\alpha$, $\beta$, $\gamma$, and $\delta$) to be spheres centered at
the appropriate momentum, with the radii $k_\alpha$, $k_\beta$, $k_\gamma$, and
$k_\delta$ determined by fits to  ARPES data.
Figure~\ref{fig:FeAsSphericalFermiSurfaces} illustrates.%
\singlefig
{FeAsSphericalFermiSurfaces}       
{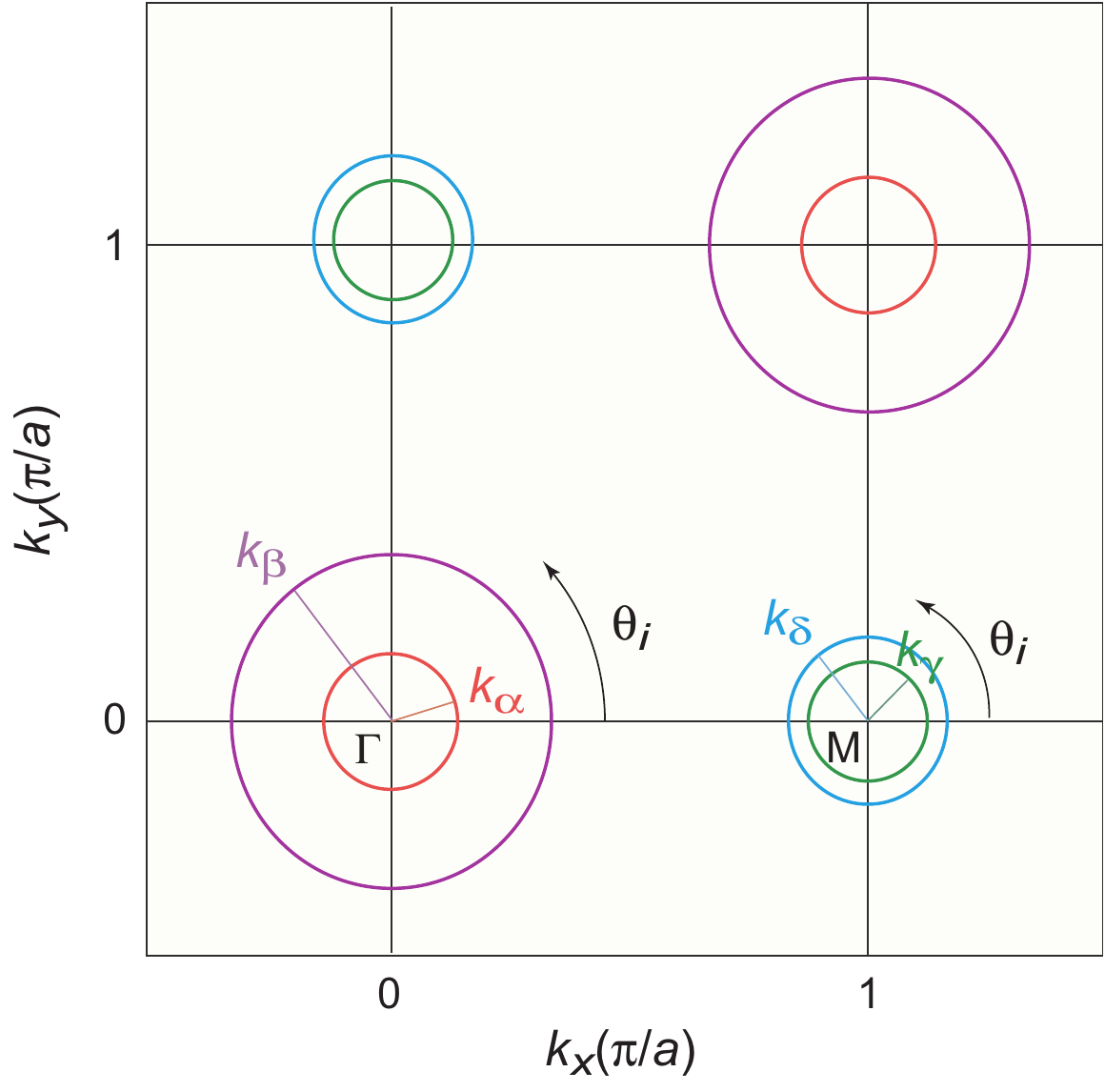}    
{0pt}         
{0pt}         
{0.67}         
{
Approximate Fermi surfaces. The four Fermi surface pockets,
two around the $\Gamma$ point and two around the X point, are approximated by
circles.}
Then from Eq.~(\ref{gapcoskxcosky}) the pairing gaps on the four sheets of Fermi
surface are given by
\begin{equation}
  \Delta_i \equiv \singletGap (k_i, \theta_i) = \Delta_0
     \cos(k_i \cos \theta_i) \cos(k_i \sin \theta_i),
\label{gaps4sheets}
\end{equation}
where $i= \alpha, \beta, \gamma, \delta$ labels the sheets and the polar angles
$\theta_i$ are centered at the $\Gamma$ and $M$ points
(see Fig.~\ref{fig:FeAsSphericalFermiSurfaces}). Writing this out explicitly for
the four cases yields Table \ref{FeGapTable2}.%
%
%
%
{\renewcommand\arraystretch{1.20} 
\begin{table}[t]
  \centering
  \caption{Pairing gaps on four sheets of the idealized Fermi surface.}
  \label{FeGapTable2}
\vspace{1pt}
  \begin{normalsize}
    \begin{centering}
      \setlength{\tabcolsep}{5 pt}
      \begin{tabular}{ccc}
        \hline
            Label &
            Fermi surface &
            Pairing gap

        \\        \hline
            $\Delta_\alpha$ &
            $k_x^2 + k_y^2 = k_\alpha^2$ &
            $\Delta_0 \cos(k_\alpha \cos \theta_\alpha)
           \cos(k_\alpha \sin \theta_\alpha)$

        \\[1 pt]
            $\Delta_\beta$ &
             $k_x^2 + k_y^2 = k_\beta^2$&
            $ \Delta_0 \cos(k_\beta \cos \theta_\beta)
           \cos(k_\beta \sin \theta_\beta)$

        \\[1 pt]
            $\Delta_\gamma$ &
             $k_x^2 + k_y^2 = k_\gamma^2$&
             $\Delta_0 \cos(k_\gamma \cos \theta_\gamma)
           \cos(k_\gamma \sin \theta_\gamma)$

        \\[1 pt]
            $\Delta_\delta$ &
            $k_x^2 + k_y^2 = k_\delta^2$ &
            $\Delta_0 \cos(k_\delta \cos \theta_\delta)
           \cos(k_\delta \sin \theta_\delta)$
        \\[1 pt]        \hline
      \end{tabular}
    \end{centering}
  \end{normalsize}
\end{table}
}
%
%

A fit of the parameters $\Delta_0$ and the $k_i$ to the data of
Ref.~\cite{nak08} gives the description of the pairing gaps illustrated in
Figs.~\ref{fig:4gaps}--\ref{fig:gapVscoskxcosky},%
\singlefig
{4gaps}       
{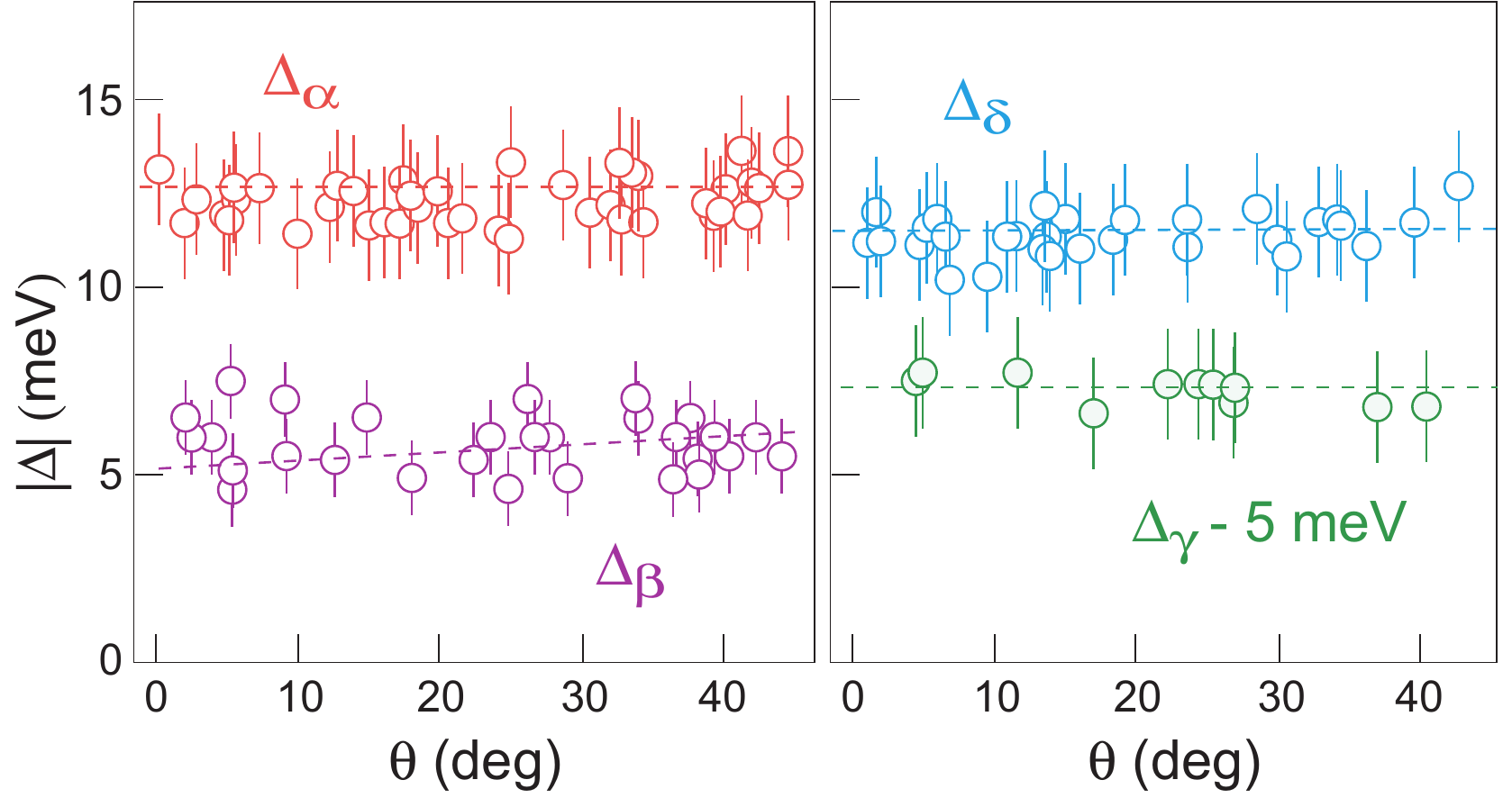}    
{0pt}         
{0pt}         
{0.51}         
{Pairing gaps on four sheets of the Fermi surface for ${\rm Ba}_{0.6} {\rm
K}_{0.4} {\rm Fe}_2 {\rm As}_2$. Note that $\Delta_\gamma$ is displaced by 5 meV
for plotting purposes. Circles are data from Ref.~\cite{nak08} and dashed lines
are theoretical using Eq.~(\ref{gapcoskxcosky}). Parameters $\Delta_0 = 13.5$
meV, $k_\alpha/\pi = 0.135$, $k_\beta/\pi = 0.370$, $k_\gamma/\pi = 0.141$, and
$k_\delta/\pi = 0.181$ were determined by fitting to the data.}
\singlefig
{FSoverlay}       
{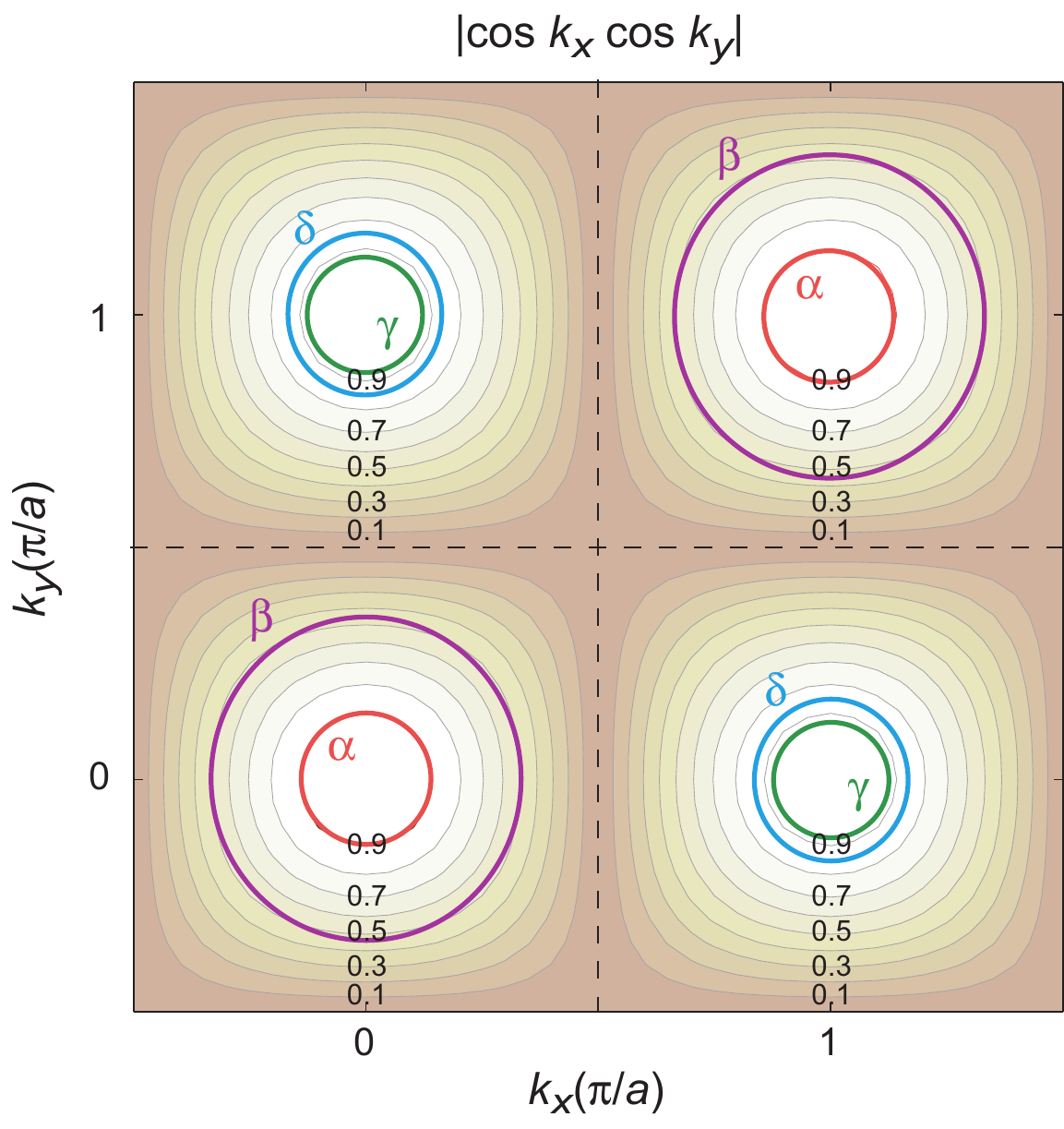}    
{0pt}         
{0pt}         
{0.68}         
{Fermi surfaces $\alpha$, $\beta$, $\gamma$, and $\delta$ (thick 
curves) superposed
on contours of the pairing formfactor $|\cos k_x \cos k_y|$.  Gap
nodes are indicated by dashed lines.}
\singlefig
{gapVscoskxcosky}       
{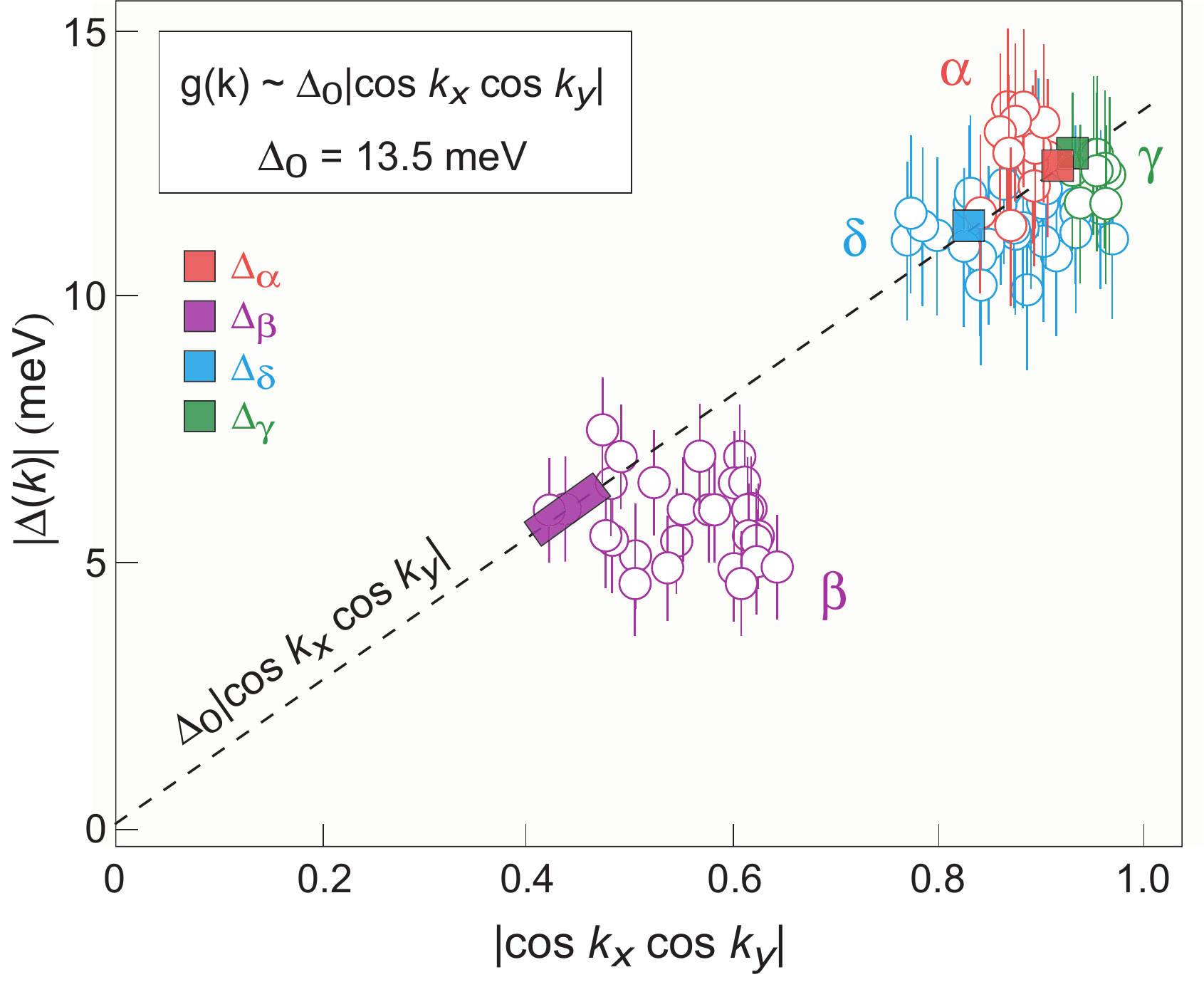}    
{0pt}         
{0pt}         
{0.505}         
{Pairing gaps on four sheets of the Fermi surface versus $|\cos k_x \cos k_y|$
for ${\rm Ba}_{0.6} {\rm K}_{0.4} {\rm Fe}_2 {\rm As}_2$.  Data from
Ref.~\cite{nak08} and the squares and rectangles indicate theoretical values for
gaps on the four sheets deduced from Fig.~\ref{fig:FSoverlay}. The same
parameters as for Fig.~\ref{fig:4gaps} were used.}%
where in Fig.~\ref{fig:4gaps} circles indicate data and the dashed
lines  represent the gaps calculated from Eq.~(\ref{gaps4sheets}).
We see that the ARPES measurements of Ref.~\cite{nak08} are at
least approximately consistent with the $\cos k_x \cos k_y$ pairing
gap formfactor that is deduced in the present paper by requiring
self-consistency of antiferromagnetism and superconductivity within
an SU(4) symmetry.

The $\sin k_x \sin k_y$ formfactor  compatible with the antiferromagnetism
according to \tableref{FeGapTable} would not be compatible with the data
displayed in Fig.~\ref{fig:4gaps}, since it would imply nodes on the Fermi
surfaces (compare Figs.~\ref{fig:contour4filled} and \ref{fig:FSoverlay}) that
are not observed in the data.  Consistency of the observed antiferromagnetism
with the superconductivity in the FeAs compounds is possible with either $\cos
k_x \cos k_y$ or $\sin k_x \sin k_y$ pairing formfactors, but we find
that requiring in addition consistency with the ARPES data of Ref.~\cite{nak08}
would restrict to the $\cos k_x \cos k_y$ choice.

\subsection{Unified Cuprate and Fe-Based Superconductivity}

We have presented evidence that the Fe-based high-temperature superconductors
represent the second example (after the cuprates) of the non-abelian
superconductors proposed in Ref.~\cite{guid04}.   The identification of a common
non-abelian superconductivity in these two classes of compounds permits a
unified model of cuprate and Fe-based superconductors based on an SU(4) group
and dynamical-symmetry subgroup chains corresponding to emergent degrees of
freedom in the strongly-correlated electron system.

Requiring that the SU(4) algebra simultaneously close under commutation and be
consistent with the magnetic structure inferred from neutron scattering
experiments constrains the orbital symmetries for the pairing gap in FeAs
compounds.  We find that in the symmetry limits neither $s_{x^2+y^2}$ nor 
$d_{x^2-y^2}$ symmetries are
compatible with the neutron scattering data but $s_{x^2y^2}$ or $d_{xy}$ could
be, and comparing the predicted gaps with ARPES data restricts the choice
uniquely to $d_{x^2y^2}$ (that is, $\cos k_x \cos k_y$). Thus, we find  that a
unified SU(4) model of iron-based and cuprate high temperature superconductivity
is possible, but consistency with neutron scattering and ARPES data suggests
that the pairing in the two cases corresponds to {\em different} orbital
formfactors at the microscopic level. 

There is widespread evidence from ARPES measurements  for a relatively universal
effective pairing formfactor of approximate $\cos k_x \cos k_y$ form in both 122
and 111 families of FeAs (iron-pnictide) superconductors
\cite{nak08,ding08,zhao08,umez2012,liu2011}. This is puzzling from a microscopic
point of view (see the discussion in Ref.~\cite{hu2012}), since there is good
reason to believe that all five $d$-orbitals of Fe will contribute in this
region and there is little reason to believe that this contribution would be
uniform across compounds.  As we now discuss, the discovery of the FeSe
(iron-chalcogenide) superconductors \cite{FeSeDiscovery} makes this situation
even more puzzling.

Given the widespread belief that superconductivity in the iron-based
superconductors is a consequence of pair binding by electron--electron
correlations, an additional minus sign is required in the pairing matrix
elements to turn the repulsive electron--electron interaction into an effective
attractive one.  A nodal gap function permitting pair scattering between nodes
of different signs is one possibility, as is believed to be the case for the
cuprates.  In the FeAs compounds, instead data often suggest that the gap has no
nodes, but that there are electron and hole pockets of fermi surface in the
Brillouin zone separated approximately by lattice vectors.  Thus, in that case
it has been proposed that the attraction could come from scattering between
these particle and hole pockets (which have gaps of opposite sign). This is the 
motivation for the extended $s$-wave gap symmetry assumed by many authors for
the FeAs compounds \cite{hirs2011}.

However, data on the chalcogenides show high-temperature superconductivity for
electron-doped compounds in which there appear to be only electron and not hole
pockets of Fermi surface. This calls into question any mechanism based on
scattering between electron and hole pockets as a general explanation of
superconductivity in the iron-based compounds.  We conclude that it is difficult
to justify a unified picture of even the iron-based superconductors (much less
unifying the iron superconductors with the cuprate superconductors) based on the
standard microscopic approaches. 

Unless we are content to assume that pnictide, chalcogenide, and cuprate
superconductivity are all due to separate mechanisms, there must be broader
symmetries at work than those manifest from the usual microscopic pictures of
these compounds.  As we have discussed here (and proposed originally in
Refs.~\cite{guid04,guid09}), any microscopic conditions that lead to the
realization of emergent SU(4) symmetry will provide a unified picture of
superconductivity across all of these compounds, independent of microscopic
details (which change only the values of parameters for the emergent collective 
modes).

The preceding discussion is another example of a recurring theme of this review:
understanding the origin of superconductivity, whether conventional or
unconventional, and whether in various occurrences in condensed matter or in
manifestations in a variety of other disciplines, lies in understanding the
general conditions that lead to the Cooper instability in such a diverse set of
systems, not in the detailed microscopic properties of each system. We believe
that the iron-based superconductors have been extremely important in this regard
because they---perhaps more so than for any other class of
superconductors---have shown that unconventional superconductivity seems to be
compatible with a quite broad range of microscopic details, as long as those
details are consistent with emergence of superconductivity within a system
exhibiting other forms of collective behavior, such as antiferromagnetism.  That
this should be the case is, of course, central to the point of view presented in
this review.

\section{Relationship with other Models}

The SU(4) model discussed in this review uses the mathematical tools of Lie
algebras and Lie groups, which are in some respects rather different from the
mathematical tools most commonly applied to the strongly-correlated electron
problem. However, as we have emphasized in many contexts, the resulting physical
picture has considerable resonance with many ideas found from application of
other more standard condensed-matter methodologies. Formally it is always
possible to make such identifications because either the present Lie algebra and
Lie group based approach, or more conventional approaches for condensed matter,
have as their common final results the calculated matrix elements corresponding
to physical observables. 

These matrix elements provide a common denominator for
comparison, irrespective of the differences in mathematical techniques used to
obtain them, as illustrated in \fig{matrixElements}.%
\singlefig
{matrixElements}       
{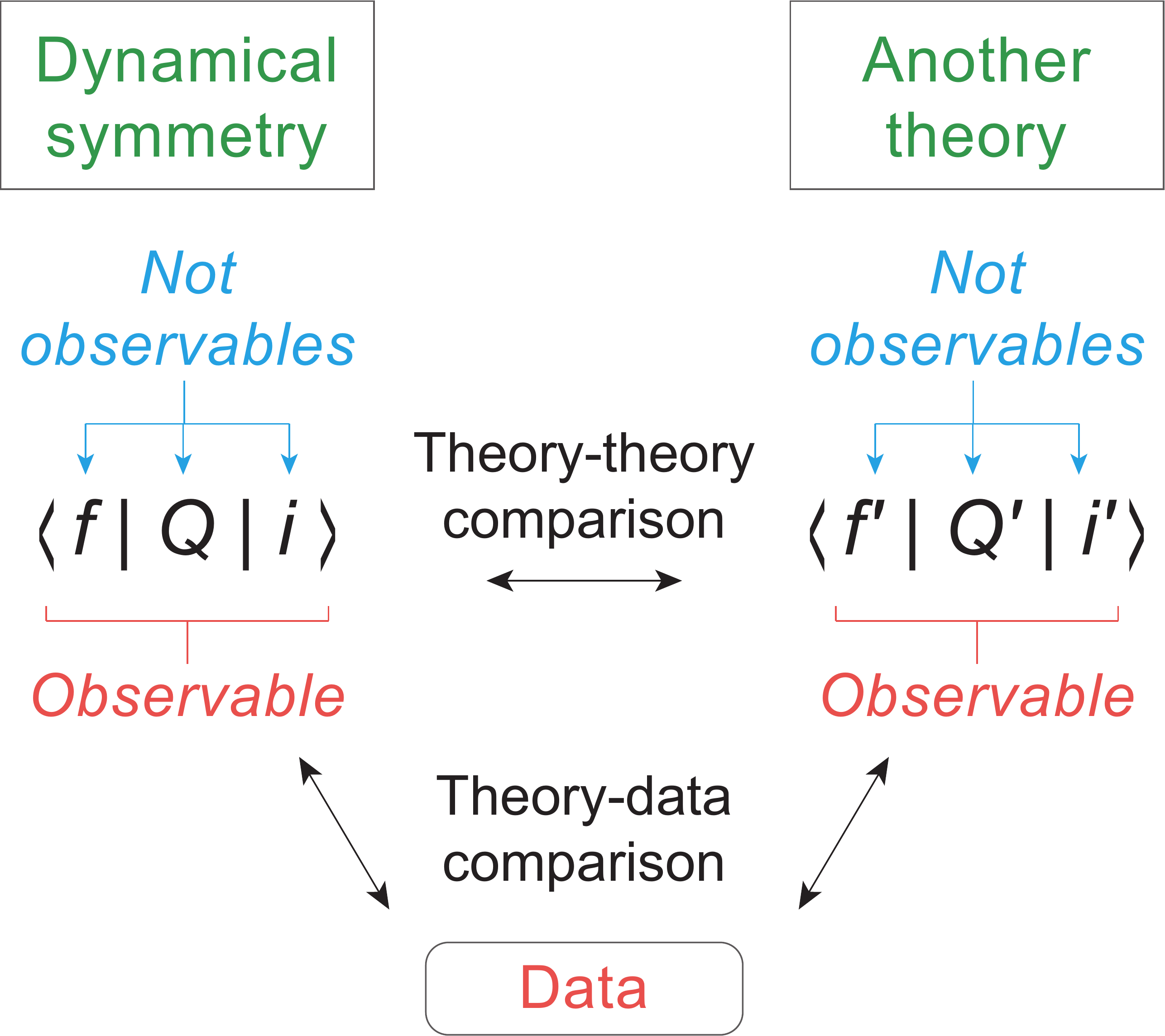}    
{0pt}         
{0pt}         
{0.24}         
{Comparison of matrix elements among different theories and data. Wavefunctions 
and operators are not observables.  Only matrix elements are directly related 
to 
experimental data.  Even though a dynamical symmetry theory and some other 
theory may use very different methodologies applied to a given problem, they 
both must produce matrix elements of observables as their physical output.  
Thus 
the valid comparisons between theories, and with data, are through matrix 
elements; wavefunctions and operators separately are relevant only in that they 
may be helpful padagogically, and that they are tools to produce matrix 
elements.
}
In this section we use 
such considerations to summarize the
relationship of the SU(4) model of nonabelian superconductivity to some other
methodologies that have been widely discussed for understanding
superconductors. 

\subsection{\label{su4BCS}SU(4) and BCS Models}

As we have explained in some depth, the standard BCS description of conventional
superconductors may be viewed as the limit of the methodology developed here
when (1)~other non-pairing order such as antiferromagnetism may be neglected,
(2)~the electron--electron correlations are weak, and (3)~as a result of the
preceding assumptions bondwise singlet pairs have no energetic advantage over
onsite singlet pairs. In that case the SU(4) symmetry is enlarged to SO(8) with
the addition of onsite pairs and their interactions, and the physical conditions
then favor an SO(8) dynamical symmetry chain ending in the SU(2) pseudospin
symmetry for onsite pairs, which is well known to describe a conventional
($s$-wave) BCS superconductor. Under these physical conditions the dynamical
symmetry does not specify uniquely the underlying microscopic interaction, but
is compatible with a weak phonon-based binding of the Cooper pairs.

However, for a different set of physical conditions the SU(4) formalism leads 
to 
another picture that is BCS in the form of the wavefunction, but differs from 
the preceding example in that the resulting superconductivity may be 
unconventional. In the limit of strong electron--electron correlations and 
significant antiferromagnetism the favored highest symmetry is SU(4) because 
the 
correlations favor bondwise pairs, and if the AF is sufficiently weakened by 
doping the favored dynamical symmetry becomes $\sufour \supset \sutwo\tsub p$. 
By comparison of physical matrix elements, we have shown that this dynamical 
symmetry has the properties of a singlet-pairing condensate having a BCS-like 
wavefunction, but a possibly unconventional order parameter since the 
formfactor 
$g(\bm k)$ need not be conventional $s$-wave. Again, the dynamical symmetry 
does not specify uniquely the nature of the pair binding, but the context in 
which it arises suggests strongly correlated-electron rather than phonon 
binding 
for the Cooper pairs.

This second form of BCS superconductivity is also evident in the
SU(4) coherent-state solutions, since we showed in \S\ref{BCSrelationship} that
the SU(4) gap equations reduce exactly to the BCS gap equations if the
antiferromagnetic correlations can be neglected, but with a formfactor that is
possibly unconventional.  It is widely believed that the superconducting state
in the overdoped cuprates has this BCS but with unconventional formfactor
property, and it is clear that both data and the microscopic doping dependence
of the SU(4) model favor the suppression of AF correlations and thus the
$\sufour \supset \sutwo\tsub p$ dynamical symmetry in this region.

\subsection{\label{su4neel}SU(4) and N\'eel Antiferromagnetism}

The $\sufour \supset \sofour$ dynamical symmetry has matrix elements
corresponding to the usual picture of a N\'eel antiferromagnet (see
\S\ref{so4Dynamical}), and for cuprates this state is  favored at half
filling by the intrinsic doping properties of the SU(4) solutions.  By virtue 
of the properties for the bondwise pairs of the parent SU(4)
symmetry, this antiferromagnetic state is also a Mott insulator at half filling 
for cuprates,
since SU(4) enforces no double occupancy of lattice sites by electrons or holes
making up the SU(4) pairs. As we have discussed in \S\ref{pairingInstability},
this AF Mott insulator state becomes strongly disfavored when doped away from
half filling because it is unstable against condensing Cooper pairs if there is
a non-zero effective pairing interaction in the collective subspace.

\subsection{\label{su4Mott}SU(4) and Mott Insulators}

It is known that the parent state of a cuprate superconductor corresponds to an 
antiferromagnetic state with no double occupancy of the lattice, implying 
Mott insulator character, and there is  evidence that the parent 
states of some iron superconductors lie near a Mott transition. As noted above, 
the 
SU(4) symmetry itself implies that fundamentally the lattice is not 
doubly-occupied because closure of the SU(4) Lie algebra requires that sites 
contained in the pairs not have double occupancy. Physically, this is because 
the SU(4) symmetry 
results from the more general SO(8) symmetry under the assumption that 
onsite pairs are suppressed by Coulomb repulsion.  Thus, Mott character for the 
parent states of high-temperature superconductors is a natural outcome of the 
SU(4) dynamical symmetry itself, and requires no additional constraints such as 
Gutzwiller projection.

Furthermore, one may argue that the Mott character of the cuprate and 
iron-based parent states is fundamental, in that the subgroup chains 
(collective states) corresponding to the three possible dynamical symmetries 
inherent the no double occupancy constraint from the parent SU(4) group.  This 
is in contrast to a Gutzwiller projection of the unperturbed basis, which 
ensures no double occupancy of the basis, but does not ensure that emergent 
collective states (antiferromagnetic or superconducting) are described by a 
basis with no double occupancy.  Hence, the SU(4) symmetry represents the 
consistent implementation of the role of Coulomb repulsion in suppressing 
double occupancy of lattice sites in highly-collective states.  The parent 
state for a cuprate superconductor is then both an antiferromagnet and a Mott 
insulator because of {\em two separate properties} of the overall SU(4) 
symmetry:  (1)~the SU(4) symmetry is broken by double occupancy of lattice 
sites, and (2)~the SU(4) symmetry has an SO(4) subgroup with the matrix 
elements of a N\'eel antiferromagnetic state and a doping dependence that 
favors 
it as the ground state at zero doping.

\subsection{\label{su4RVB} SU(4) and Resonating Valence Bond States}

As we have already mentioned in \S\ref{RVBprecocious}, the properties of the 
SU(4) coherent state at low hole-doping presumably share many features with 
resonating valence bond (RVB) states.    Spin-triplet pairs are essential for a 
complete set of operators in the minimal SU(4) model [for example, no double 
occupancy is enforced by the SU(4) Lie algebra, which fails to close without 
triplet pairing operators], and a mixture of singlet and triplet pairs is 
essential to describe the AF states at half filling in the highly-truncated 
SU(4) fermion basis. But the significance of triplet pairs relative to singlet 
pairs is small and decreases rapidly at higher doping, as illustrated in 
\fig{separationChargeSpinSU4}. Therefore, SU(4) ground states could have 
significant overlap with a singlet spin liquid.

The SU(4) coherent state justifies many features of RVB models, but it is 
richer 
than typical RVB applications because it accounts even-handedly for both AF and 
SC on a lattice with no double occupancy.  As a consequence, the SU(4) 
variational wavefunction is more complex than that of a singlet spin liquid. 
Conversely, the SU(4) coherent-state model is simpler in many respects than RVB 
models because superconductivity and antiferromagnetism are accounted for 
quantitatively in a minimal theory having only (dressed) electron degrees of 
freedom:  the theory requires no explicit introduction of pair bosons,  gauge 
fields, or spinons and holons (which have formal justification in one 
dimension, 
but are less obviously justified in higher dimensions, and for which there is 
little direct evidence in cuprate superconductors). 

The SU(4) coherent state represents a minimal extension of the BCS formalism to 
incorporate $d$-wave pairing in the presence of strong AF correlations and 
large 
effective onsite electron repulsion.  It requires no Gutzwiller projection 
because the symmetry enforces no double occupancy on the lattice.  It exhibits 
a 
type of spin--charge separation (see the discussion in Ref.\ \cite{ande97}), 
but 
not through topological spinons and holons:  in the fermion basis, charge is 
carried both by singlet fermion hole pairs having a spin of 0 and charge $-2$, 
and triplet fermion hole pairs having a spin of 1 and charge $-2$, but spin is 
carried solely by the triplet hole pairs. From this point of view, 
``spin--charge separation'' in emergent degrees of freedom is a fairly mundane 
consequence of the complete set of operators argument in \S\ref{su4Operators} 
and \fig{completeSet}. Antiferromagnetic operators scatter spin-singlet pairs 
into spin-triplet pairs, and spin-triplet pairs into spin-singlet pairs.  Thus, 
both types of excitations (singlet pairs carrying charge but no spin and 
triplet 
pairs carrying charge and spin) must be present as fundamental excitations in 
the realistic collective subspace if the system exhibits antiferromagnetic 
correlations.

\subsection{SU(4) and the Zhang SO(5) Model}

Ideas having some similarity to those discussed in this review  have been
proposed by S. C. Zhang and collaborators \cite{zha97}. To distinguish from the
SO(5) dynamical symmetry of the SU(4) model discussed in this review, we shall
term this the Zhang SO(5) model.  In the Zhang SO(5) model the AF and SC order
parameters are assembled into a 5-dimensional vector order parameter that is
rotated between AF and SC order by SO(5) generators. The methodology described
here is different, starting instead from identification of a closed algebra
associated with a general set of fermion pairing and particle--hole operators
defined on a periodic lattice, with order parameters arising as matrix elements
of various bilinear forms for these operators rather than being introduced as
fundamental entities, and a corresponding truncation of the full Hilbert space 
to a symmetry-dictated collective subspace. Nevertheless, we find that we 
recover Zhang's SO(5)
symmetry as a subgroup of a more general SU(4) symmetry if various
approximations are made in the full SU(4) theory. 

Our SO(5) subgroup is embedded in a larger SU(4) group defined microscopically
in the fermion degrees of freedom, which implies constraints on the SO(5)
subgroup. Our SU(4) model and Zhang's SO(5) model start from the same building
blocks (the operator set (\ref{operatorset}), but see note \cite{u1note}).
However, we implement the full quantum dynamics (the commutator algebra) of
these operators exactly, while in Ref.\ \cite{zha97} a subset of 10 of the
operators acts as a rotation on the remaining 5 operators
$\{\singletPair^\dagger,\singletPair,\Qvector\}$, which are treated
phenomenologically as 5 independent components of an order-parameter vector.
Thus only 10 of the 15 generators of our SU(4) group are treated dynamically in
the Zhang SO(5).

The embedding of SO(5) as a subgroup in our larger SU(4) group has various 
physical
consequences that do not appear if an SO(5) group is considered in
isolation.

1. A transition from antiferromagnetism to superconductivity at zero temperature
that is controlled by the doping emerges naturally and microscopically from the
SU(4) symmetry.  The corresponding behavior in the Zhang SO(5) model requires
that a symmetry-breaking term proportional to a chemical potential be introduced
by hand.  

2. The full SU(4) dynamics show that the SO(5) subgroup is only one of the
symmetries relevant to the cuprate problem. It is a transitional symmetry that
links AF to SC behavior, suggesting that it is most useful for the underdoped
region. The AF phases at half filling and the optimally doped superconductors
are more economically described by our SO(4) and SU(2)$\tsub p$ symmetries,
respectively.

3. SU(4) symmetry leads naturally to pseudogap behavior, with
the SO(5) subgroup being central to this property. 

4. The methodology of the SU(4) dynamical symmetry approach shows  that 
the SO(5) subgroup is an effective symmetry operating in a 
severely truncated space.  It should be interpreted, not in terms of an 
approximate symmetry of a Hubbard or $t$--$J$ Hamiltonian, but in terms of an 
exact (emergent) {\em dynamical symmetry}. Thus, its microscopic validity---as 
for that of its parent SU(4) symmetry---must be judged by the physical 
correctness of the matrix elements evaluated in that truncated model space, not 
by whether a particular Hamiltonian thought to have some relevance for the full 
space possesses such a symmetry (see the discussion in \S\ref{typesTruncation}).

For exact SO(5) symmetry, antiferromagnetic and superconducting states are
degenerate and there is no barrier between them at half filling (see the
$n/\Omega =1$ curve of \fig{eSurfaces1D}(b)). But this is inconsistent with
observed Mott insulating behavior at half-filling in the cuprates, because the
symmetric Zhang SO(5) model predicts no charge gap at half filling of the
lattice. Thus, for antiferromagnetic insulator properties to exist at half
filling, it is necessary to break SO(5) symmetry \cite{zha97}. As we have
discussed in \S\ref{weaklyBrokenSO5}, this symmetry breaking is implicit in the
SU(4) model, occurring naturally if $\sigma \ne \tfrac12$ in the Hamiltonian
\eqnoeq{eq1}. Furthermore, SU(4) symmetry implies the constraint
\begin{equation}
\langle \DdagD+\QdotQ
+\pidotpi\rangle
=\frac{1-x^2}{4}\Omega^2 .
\end{equation}
This ensures a doping dependence in the solutions that describes the transition
from AF to SC in the cuprates, as discussed in \S\ref{AF-SCcompetition}.

Hence, the SU(4) coherent state analysis indicates that the
phenomenologically-required SO(5) symmetry breaking, and the doping dependence 
in the
solutions, occur as natural consequences in the SU(4) model. They need not be
introduced empirically, as proposed in the original Zhang SO(5) model. A 
projected
Zhang SO(5) model was introduced in Ref.~\cite{zha99} that uses Gutzwiller 
projection
to satisfy the large-$U$ Hubbard (no double occupancy) constraint.  There is no 
need
to introduce such a projection if SO(5) is treated as a subgroup of
SU(4), because the SU(4) symmetry itself already implies a no double occupancy
constraint \cite{guid04} . 

We conclude that cuprate high-temperature superconductivity in the underdoped 
region
may be described by a Hamiltonian that conserves SU(4) but breaks SO(5) 
explicitly in
a manner favoring AF order over SC order, as has been discussed in
\S\ref{weaklyBrokenSO5}. The weakly-broken SO(5) symmetry acts as a critical
dynamical symmetry mediating the transition between superconducting states, 
described
by the SU(2)$\tsub p$ dynamical symmetry, and antiferromagnetic states, 
described by
the SO(4) dynamical symmetry. Thus it dominates the behavior in the underdoped
region.

\subsection{SU(4) and the Hubbard and $\bm t$--$\bm J$ models}

As discussed more extensively in \S\ref{typesTruncation}, a Hubbard or {\em t-J}
model and the dynamical symmetry approach applied here are alternative ways to
simplify a strongly-correlated electron system. In the Hubbard or {\em t-J}
models a greatly simplified Hamiltonian is chosen but no specific
configuration-space truncation is assumed (though practically a truncation is
required).  In contrast, our only approximation in the theory discussed here is
the  space truncation, since the symmetry-dictated Hamiltonian includes
all possible interactions in the truncated space, with the effect of the
excluded space absorbed into the effective interactions of the truncated space.
The validity of this approach depends entirely on validity of the choice of
truncated space and its effective interactions, which may be tested by
comparing calculated SU(4) matrix elements with data.

The Hamiltonian and wavefunctions for effective low-energy theories of the kind 
described here need not (likely should not!) resemble those of a Hubbard or 
{\em 
t-J} model. Quantum mechanically only matrix elements are related to 
observables, not operators or wavefunctions separately, as illustrated in
\fig{matrixElements}.  Thus,  a direct comparison of Hamiltonians or 
wavefunctions between two theories is valid {\em only if the two theories are 
defined within the same space.}  If they are defined in different spaces, the 
only comparison that quantum mechanics permits is that of matrix elements 
evaluated in the two spaces; a comparison of Hamiltonians or wavefunctions 
separately has no physical content.  

We may view emergent dynamical symmetries as operating in a truncated
collective subspace in which the truncation has been implemented primarily by
symmetry considerations.  Thus, if the Hubbard or {\em t-J} models and the SU(4)
model are both valid descriptions of high-temperature SC, their
physical matrix elements must be similar, making it highly unlikely that
their Hamiltonians or wavefunctions separately would be similar, since they are
defined in very different spaces.

\section{High Critical Temperatures}

Why are the critical temperatures for transition to the superconducting state 
unusually high for cuprates, iron-based superconductors, and many other 
unconventional superconductors (when measured in appropriate units for each 
case)? The unusual properties, including high $T\tsub c$, of unconventional 
superconductors are not because they have unconventional (not $s$-wave) pairing 
formfactors.  The unconventional formfactors are not causes but rather are {\em 
symptoms} of a deeper and more important issue.  The essential point is 
proximity of (one or more) other collective modes to SC in the phase diagram, 
suggesting that SC and the other mode are both possible ground states.  If the 
two competing modes are related to each other, they may compete for the same 
Hilbert space.  If they do compete for the same Hilbert space, this competition 
will tend naturally to produce unconventional formfactors but it can do 
something much more fundamental.  

\subsection{Unification of Competing Order}

If the formfactor is unconventional it is a likely sign of competing order, and 
if the competing order is {\em related to the SC} in the right way [both being 
generators of a higher symmetry like SU(4), implying that they compete for the 
same Hilbert subspace] the competing order parent state can 
``precondition'' the system for the SC phase transition.  This allows it to 
occur at a higher value of $T\tsub c$ because the competing-order ground state 
is a low-entropy state that can be rotated collectively into the SC state, as 
illustrated in \fig{collectiveArrowAlignment}.
\doublefig
{collectiveArrowAlignment}   
{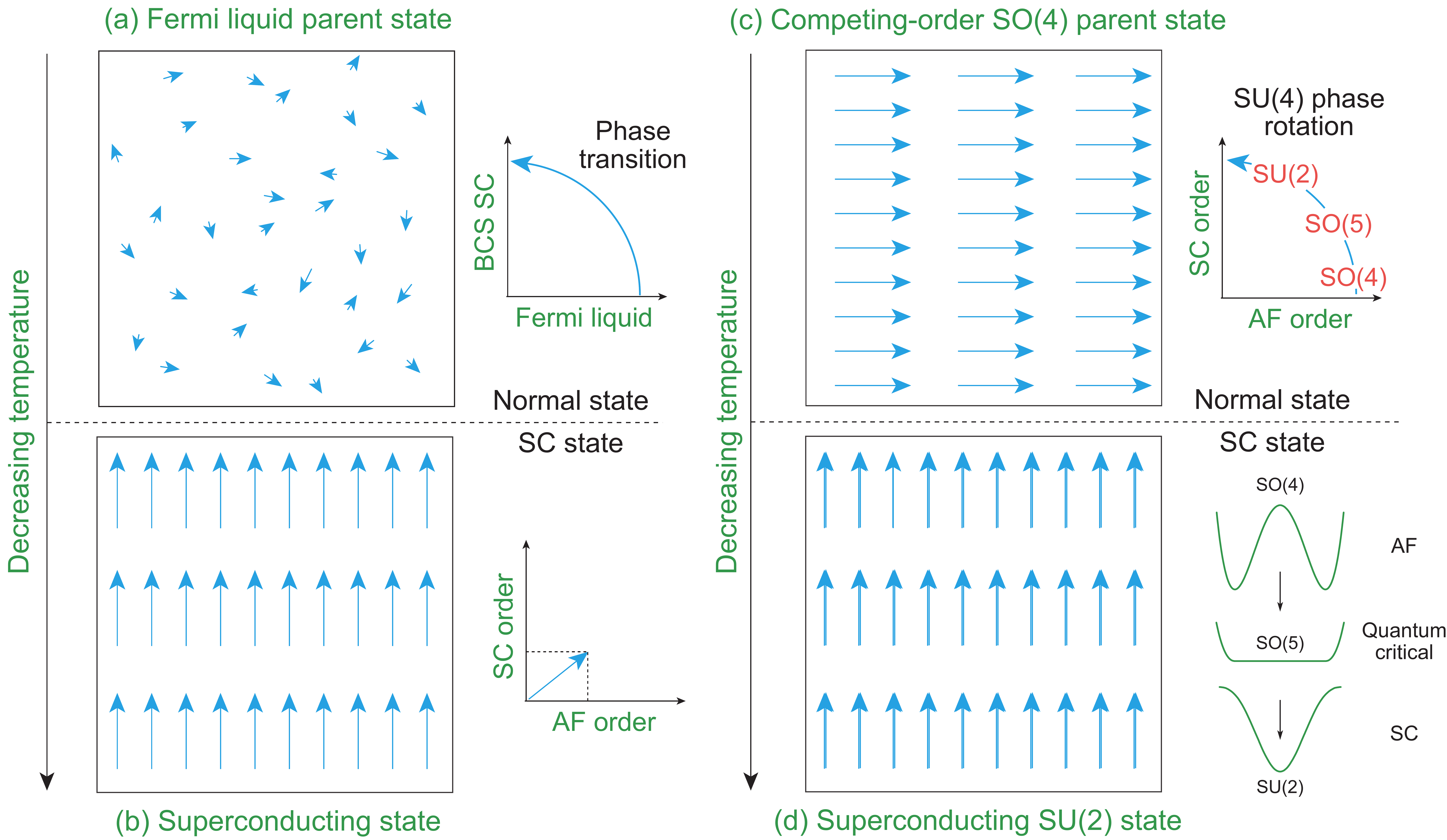}
{0pt}         
{0pt}         
{0.202}         
{(Left)~Formation of a normal (BCS) superconducting condensate. The vertical 
projection of an arrow represents the amount of pairing order; the horizontal 
projection of an arrow represents the amount of some competing order (such as 
antiferromagnetism).  (a)~Since we assume no net order in the parent state, the 
arrows are short (indicating matrix elements of non-collective strength) and 
randomly oriented (indicating no SC or competing order in the bulk).  
(b)~Producing a superconducting state requires imposing order (aligning each 
randomly oriented arrow vertically) on a high-entropy initial state, which can 
occur only a very low temperature and implies a small value of $T\tsub c$. 
(Right)~Formation of a superconducting condensate in a system having an order 
like antiferromagnetism that competes with superconductivity as its normal 
ground state. (c)~This requires imposing order on a state that is not 
superconducting, but is already highly ordered and thus of low entropy.  (d)~If 
the SC and the competing order are {\em both generators of some higher  
symmetry} like SU(4), then the transition from the competing order to the SC 
state is a collective rotation in the group space of the higher symmetry, which 
requires little change in entropy and can occur at a high value of $T\tsub c$.  
In essence the SC state already exists in the competing-order ground state; it 
only has to be pointed in the right direction by a rotational nudge in the 
group space.
For the cuprates doping can provide the required nudge, as indicated in 
\fig{cooperInstability2}(b).
}

In \fig{collectiveArrowAlignment}, superconducting order is indicated 
schematically by an up arrow and a competing order (assume it to be 
antiferromagnetism for discussion) by a right arrow.  In the normal BCS case of 
\fig{collectiveArrowAlignment}(a) there is no net SC or AF in the initial 
state, 
so all arrows must individually be lengthened  and ordered vertically in the 
superconducting phase transition, as indicated in 
\fig{collectiveArrowAlignment}(b).  This transition from a high-entropy initial 
state to a highly-ordered final state requires a correspondingly low 
temperature 
to implement. In the competing order case of \fig{collectiveArrowAlignment}(c) 
the initial state is already ordered in a way such that a simple collective 
rotation in the group space produces the SC state, as in 
\fig{collectiveArrowAlignment}(d). This is generally the case when 
superconductivity and the order competing with it are unified in a higher 
symmetry like SU(4).

\subsection{The Generalized Cooper Instability and High-$\bm T_{\rm\bf c}$}

The preceding argument is a cartoon version of the proof in 
\S \ref{pairingInstability} that the antiferromagnetic insulator ground state 
at half filling in the cuprates is inherently unstable against condensing 
Cooper pairs with doping, implying that  the AF Mott state contains 
hidden within it a superconductor that can appear spontaneously with a 
slight disturbance.  The spontaneous appearance of a significant pair gap 
$\Delta$ with infinitessimal doping in Figs.\ \ref{fig:esurfaces}(c) and 
\ref{fig:cooperInstability2}(b) is a quantitative implementation of the phase 
transition illustrated schematically in \fig{collectiveArrowAlignment}. 
However, the realistic case in \fig{cooperInstability2} is more complex than 
the schematic picture in \fig{collectiveArrowAlignment} in that a pairing 
gap appears spontaneously for doping $x \ne 0$ but AF correlations 
decrease with increased doping but remain finite
until the rotation from antiferromagnetism to pure superconductivity is 
complete at the critical doping point $x= x\tsub c$; see \fig{Gaps}.

The SC transition between Figs.\  \ref{fig:collectiveArrowAlignment}(c) and 
\ref{fig:collectiveArrowAlignment}(d) can occur spontaneously if there is no 
barrier to the SU(4) rotation. The $\sufour \supset \sofive$ critical dynamical 
symmetry limit exhibits such a property. At low doping the energy surface 
implies degenerate AF and SC ground states with effectively no energy barrier 
separating them [see the curves in \fig{eSurfaces1D}(b) for $n/\Omega = (1-x) 
\sim 1$].  This suggests that the AF and SC phases can be connected by a 
sequence of infinitesimal SU(4) rotations through intermediate SO(5) states 
having 
different mixtures of AF and SC order that are nearly degenerate in energy with 
the pure antiferromagnetic and superconducting states.

\subsection{An Information Argument}

The preceding entropy arguments may be expressed as an information argument.   
Figure \ref{fig:collectiveArrowAlignment}(d)  is obtained from 
\fig{collectiveArrowAlignment}(c) by collectively rotating all vectors. This 
can 
be specified in terms of a single rotation angle applied to all vectors, which 
requires minimal information. Conversely, in \fig{collectiveArrowAlignment}(a) 
there is no order in the parent state and each arrow must be lengthened and 
oriented separately to give \fig{collectiveArrowAlignment}(b).  This requires 
supplying a much larger amount of information.  Thus the reduction in entropy 
necessary to condense the superconducting state from the parent state is much 
greater in Figs.\ 
\ref{fig:collectiveArrowAlignment}(a)--\ref{fig:collectiveArrowAlignment}(b) 
than in Figs.\ 
\ref{fig:collectiveArrowAlignment}(c)--\ref{fig:collectiveArrowAlignment}(d).  

The information argument also highlights the fundamental distinction between 
competing collective modes that are independent and those that are related by a 
higher symmetry.  If the competing modes are independent, a large amount of 
information is required to change the competing-order state into the SC state 
because they are not fundamentally related.  Microscopically, one 
collective mode must first be broken up and then reassembled into the other 
collective mode, which will hinder onset of
superconductivity and decrease $T\tsub c$. 
In the case that  competing modes are related by a higher symmetry, the 
higher symmetry {\em already encodes the relationship between the two modes.} 
Hence only a small amount of additional information is required to produce the 
superconducting state from the competing-order state, because they arise from 
the same collective Hilbert subspace spanned by generators of the higher 
symmetry, and correspond to subgroups of the same highest symmetry.

This difference may be illustrated further by considering the competion of 
charge degrees of freedom with SC in the cuprate superconductors.  We have seen 
that the minimal symmetry that can describe the cuprate superconductors is 
U(1)$\times$SU(4), where SU(4) describes the competion of AF and SC, and U(1) 
is associated with a commensurate charge density wave. The direct product  
between U(1) and SU(4) implies physically that the commensurate charge density 
wave is independent of the SU(4) description of AF and SC. Thus it cannot lower 
$T\tsub c$ by the mechanism described above. 

A more complex charge density wave 
would require a symmetry having more generators [the smallest 
compact Lie groups with with more generators than than the 15 of SU(4) are 
SO(7) 
or Sp(6) with 21 generators, SU(5) with 24 generators, and SO(8) with 28 
generators].  Thus, for the minimal SU(4) model charge density waves would not 
be rotated into superconductivity by group generators, and by our argument they 
should not facilitate  high values of $T\tsub c$.  We conclude that 
in a minimal model charge density waves represent a competing order lacking a 
structural relationship with AF and SC that at best  has 
little effect, and at worst hinders,  the formation of a superconducting 
state.  As discussed in \S \ref{criticalInhomo}, charge density waves can 
perturb SU(4) symmetry for underdoped compounds where the energy surfaces 
may become critical.  This can enable inhomogeneities such as 
stripes in a narrow range of doping, but is not likely to  
influence $T\tsub c$ significantly.

\subsection{The Role of Microscopic Physics}

The values of effective interaction parameters (influenced by underlying 
microscopic physics) will modulate exactly how large $T\tsub c$ is in a 
specific 
compound, but the generic reason for abnormally high-$T\tsub c$ in HTSC 
compounds---and unconventional superconductors in general---is the 
preconditioning implied by the symmetry relationship between the 
competing-order 
ground state and the superconductivity that emerges from it, as illustrated in 
\fig{Tc_dependence}.  
\singlefig
{Tc_dependence}       
{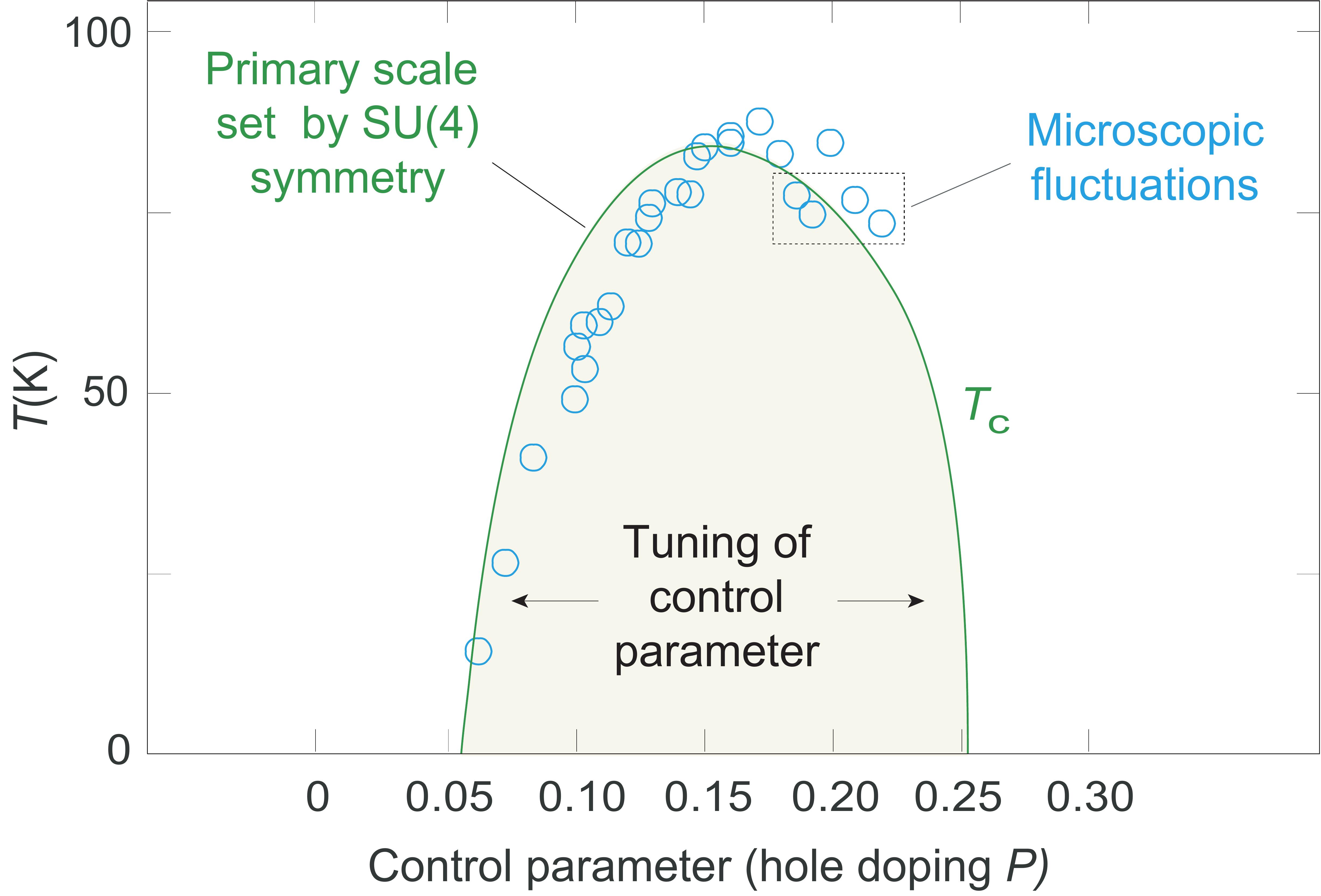}    
{0pt}         
{0pt}         
{0.172}         
{Primary factors affecting the temperature dependence of the superconducting 
transition temperature $T\tsub c$ in an unconventional superconductor. Cuprate 
data (open circles) and an SU(4) calculation (solid curve) were 
adapted from \fig{gapsResolve_k}. Error bars have been omitted but they are less 
than or equal to the sizes of the data points in most cases.  The control 
parameter in this example is hole doping. The location and shape of the curve is 
determined by SU(4) symmetry.  Its height is set by the symmetry modulated by 
the strength of the effective interaction in the truncated space, which depends 
on the average microphysics of the space excluded by the truncation. The 
fluctuations of data around the curve presumably reflect varying microphysics 
between compounds with different doping fractions that are not accounted for in 
this simple calculation. }
That is, the cuprate superconductors exhibit systematically high values of 
$T\tsub c$ because of a fundamental scale set by SU(4) symmetry modulated by 
the strength of the effective interactions, but $T\tsub c$ 
can  vary from compound to compound within 
that scale because of  microscopic physics, which influences the 
effective interaction parameters, and from variation of control parameters 
(doping in \fig{Tc_dependence}) that reflect the possibility of quantum phase 
transitions within the highest symmetry arising from competing multiple 
orders (subgroups of the highest symmetry).  Succinctly, the underlying 
microscopic physics affects the properties of superconductivity such as $T\tsub 
c$ in unconventional SC, but to first order it does so only parametrically.

\section{Universality of Superconducting and Superfluid Behavior}

Let's use the terms ``normal'' to describe standard BCS superconductivity
and  ``unusual'' to denote superconductivity that results from
Cooper pairing, but with some essential properties different from than of normal
superconductivity. (Thus unconventional superconductors fall in our class
labeled ``unusual''.)  Based on our discussion of cuprate and iron-based
superconductors in this review, on the general understanding that other forms of
SC such as that in the heavy fermion and organic superconductors
may have close resemblance to these, and the systematic application of dynamical
symmetries to superconducting and superfluid behavior in nuclear structure 
physics, we are led
to a sweeping conjecture \cite{guid13,guid2017b}.

\doTheorem{All superconductivity and superfluidity in all fields can be 
understood in terms of a generalized Cooper instability realized in terms of an 
operator algebra that is abelian in the simplest cases, but non-abelian in the 
most interesting cases.  The large and fundamental differences in underlying 
microscopic physics across these fields are important only parametrically to the 
superconducting mechanism.} 

\noindent
Normal superconductivity corresponds to a U(1) subgroup in the dynamical 
symmetry chain $\sutwo \supset \uone$, where the SU(2) group is generated by 
pseudospin and charge operators. Physically, it is realized when the pairing 
collectivity can be described largely independent of other collective modes 
and the pairing is not modfied by effects such as strong on-site 
Coulomb repulsion.  It 
corresponds to the standard BCS model. Unusual superconductivity corresponds to 
a physical situation where the pairing collectivity cannot be cleanly decoupled 
from other collective modes in the system. It corresponds to non-abelian 
subgroups of the algebra. High-temperature superconductivity in the cuprates and 
iron compounds are examples, corresponding to the dynamical symmetries of 
\eq{eq3}. Superconductivity and superfluidity in nuclear structure, which often 
occurs in the presence of collectivity associated with strong quadrupole 
deformation, is another \cite{FDSM}. 

Superficially, unusual superconductivity may not look like  BCS 
superconductivity, but that is deceiving. The complex behavior is not capricious 
but rather is related systematically to normal BCS superconductivity by 
replacement of an abelian operator algebra with a non-abelian algebra that 
couples superconductivity strongly and non-linearly to other emergent collective 
modes. This, as discussed more extensively in Refs.~\cite{guid13,guid2017b}, 
permits the Cooper instability to be realized across very different physical 
systems that may exhibit complex and varied detailed behavior but clearly 
recognizable 
general features.

\section{What is Special about SU(4) Symmetry?}

In this review we have presented a substantial amount of formalism, expressed 
often in the language of Lie algebras and Lie groups. It is 
important that this
not obscure the underlying motivation for this formalism, which is grounded
directly in the phenomenology of high-$T\tsub c$ superconductivity, and can be
given an intuitive physical meaning largely separate from the mathematics 
employed.

\subsection{The Physical Meaning of SU(4) Symmetry}

SU(4)
symmetry is concise mathematical shorthand for a minimal physical
model involving bondwise pairing and antiferromagnetism that conserves charge
and spin, and  implements no double occupancy of sites by components of the
collective pairs. It is favored under physical conditions
where superconductivity and antiferromagnetism lie near each other in the phase
diagram and strong electron--electron correlations disfavor double occupation of
the lattice sites.  These conditions are fulfilled
very well in the cuprates and at least approximately so in the iron
superconductors, and  similar ones may hold in a variety of other
unconventional superconductors such as the heavy fermion and organic
superconductors.  

To the question ``why SU(4)'', we may then give a simple answer by  asking
a slightly more precise question.  If we have a condensed matter
system with strong electron--electron correlations and a tendency toward
both magnetic and superconducting order, what is the simplest arrangement of
these complex and partially antagonistic ingredients that minimizes the energy
of the ground state?  The surprisingly concise answer is that we must arrange
the spin, charge, magnetic, and pairing operators so that they satisfy a set of
physical constraints corresponding mathematically to closure of a non-abelian
Lie algebra under commutation. The {\em simplest possibility} for those degrees 
of freedom 
is SU(4).

This relationship is no more surprising than that between
the physical observation that angular momentum is always conserved and the
statement of that physics in terms of an SU(2) Lie algebra.
In both cases we are expressing a physical observation in concise mathematical
terms using a symmetry implied by Lie algebras. But, there are 
important differences. 

(1)~The SU(4) theory represents {\em dynamical symmetries} relating the 
interactions
of different physical degrees of freedom in the Hamiltonian, while angular
momentum SU(2) symmetry implies only a {\em conservation law,} not dynamical
constraints. There is a parallel with gauge field theories, where {\em global 
gauge invariance} implies only a
conservation law, but {\em local gauge invariance} is a much deeper
statement about dynamics. SU(4) symmetry for strongly-correlated fermions
is a statement about dynamics, not just conservation laws.

(2)~As a consequence, the symmetry associated with SU(4) is more difficult to
uncover because it implies a more abstract and complex pattern
in the observables than that associated with conservation of angular momentum. 

(3)~Angular momentum SU(2) is
an exact symmetry for any closed system but the SU(4) theory is based on an
approximate symmetry that is expected to be realized only when a certain set of
physical conditions is satisfied (a strongly-correlated electron 
system with emergent
antiferromagnetism and superconductivity proximate in the phase diagram).

\subsection{Intuitively Correct Limits}

The physical validity of the SU(4) dynamical symmetry prescription is
confirmed by detailed quantitative comparison of prediction with data, but it is
also reinforced by the observation that it has intuitively-correct physical
limits. As we have demonstrated in this review: 

(1)~In the limit that the antiferromagnetic interactions of the effective
Hamiltonian may be neglected, we recover from the SU(4) coherent state
the standard BCS gap equations, but for pairs exhibiting a $g(\bm k)$ formfactor
that could be conventional $s$-wave or could be unconventional, depending on the
underlying microscopic physics.

(2)~In the limit that the pairing interactions of the effective Hamiltonian may
be neglected, we recover a theory with matrix elements corresponding to those of
a N\'eel antiferromagnetic state.

(3)~This N\'eel antiferromagnetic state has suppressed double site occupancy
because overlap of pairs breaks SU(4) symmetry. As we have explained, this is
consistent with Mott insulator character (in the cuprates) or poor metal
behavior (in the iron superconductors) for the normal state.

(4)~In the limit that both the pairing and antiferromagnetic interactions of the
effective Hamiltonian may be neglected, we still recover a state with correlated
electrons and thus a tendency to suppress onsite interactions in the ground
state, but without significant AF or SC order. 

(5)~In the limit that electron--electron repulsion is weak, bondwise pairs are
not favored energetically over onsite pairs and the SU(4) symmetry is enlarged
to its parent group SO(8), which adds to the SU(4) generators onsite pairs and
associated interactions. If physically we assume very weak electron--electron
correlations and negligible antiferromagnetism, the SO(8) states favor a dynamical
symmetry that corresponds to normal BCS superconductivity for onsite pairs. Thus the
dynamical symmetries deriving from SO(8) and $\soeight \supset \sufour$ may be
capable of describing the entire range of observed superconductivity in condensed
matter within a unified framework. 

These observations make it less surprising that our application of
non-traditional mathematical methods to the strongly-correlated electron problem
yields results that admit an interpretation of superconductivity, unconventional
or otherwise, having a highly-traditional look and feel.

\subsection{What SU(4) Is Not}

Clarifying what SU(4) symmetry is not is as important as clarifying what it is. 
It is {\em not} generally expected to be a symmetry of the underlying 
weakly-interacting microscopic system, or of a Hamiltonian appropriate for the 
microscopic system such as for the Hubbard model.  SU(4) is an {\em emergent 
symmetry} that is realized only when the dynamics of the system favor the 
emergence of the corresponding collective modes (antiferromagnetism, 
superconductivity, and their interaction in a system exhibiting strong 
electron--electron correlations). As we have emphasized, emergent dynamical 
symmetries generally would not be expected to have direct connections with  
any symmetries of the underlying microscopic system, because emergent 
phenomena generally cannot be derived perturbatively from the non-interacting 
constituent system.

\subsection{Simple Descriptions and Complex Phenomena}

Finally, let us counter a possible philosophical criticism of our dynamical
symmetry approach: the potential ground states in any strongly-correlated
electron system are so complex microscopically that we might be tempted to doubt
that a simple model like the current one would have any validity.  But this
argument ignores the quite obvious point that Nature has managed to construct a
stable ground state for high-temperature superconductors having well-defined,
collective properties that vary in a controlled manner according to a few
well-chosen parameters. Thus Nature is waving a red flag indicating that the
phenomenon in question is actually simple, if we will but change our perspective
to view it in a more natural basis. 

From experience in many fields of many-body physics, this is a clear physical
signal that the superconductor is described mathematically by a {\em small
effective subspace with renormalized interactions}, and governed by a dynamical
symmetry structure of relatively small dimensionality acting within that
subspace. Thus, if an approach like the one proposed here gives correct results
for highly non-trivial phenomenology like the doping dependence of
various observable quantities, one must take seriously the possibility that the
corresponding small symmetry-dictated subspace may have relevance to the
effective behavior of real physical systems.

\section{Summary and Conclusions}

In summary, we have developed a solvable microscopic theory of unconventional
superconductivity and antiferromagnetism on a lattice with suppressed double
site occupancy. This permits an exact many-body solution for a minimal model
having charge, spin, pairing, and antiferromagnetism for special ratios
of the coupling parameters, and an approximate generalized coherent-state
solution for arbitrary coupling strengths. Superconductivity and
antiferromagnetism enter on an equal footing. The three dynamical symmetries of
the model, SU(2), SO(4), and SO(5), yield exact solutions that correspond
respectively to states exhibiting singlet-pair superconductivity,
antiferromagnetism, and a critical dynamical symmetry interpolating between
superconductivity and antiferromagnetism. The competing AF and SC order imply an
SU(4) symmetry that embodies many essential features of high-temperature
superconductivity:

\begin{enumerate}
 \item 
The SU(4) symmetry imposes no double occupancy of lattice sites for the
electrons of the correlated pairs.
\item
If the particles near the Fermi surface all contribute uniformly to the
correlated pairs, the preceding point implies that at half filling the system
favors antiferromagnetic Mott insulator behavior (as observed in the cuprates).
\item
If the particles near the Fermi surface contribute non-uniformly to the
correlated pairs the normal state at half filling is antiferromagnetic, but
could be a poor metal in the normal state (as observed in the pnictides).
\item
Doping of the AF Mott insulator ground state at half filling leads to the 
spontaneous emergence of a singlet superconductor ground state.
\item
Pseudogap states emerge naturally in the model at intermediate doping. 
\end{enumerate}

\noindent
The ground state of this theory exhibits two fundamental instabilities:

\begin{enumerate}
 \item 
The antiferromagnetic state at half filling (which for cuprates is a Mott
insulator) is unstable against condensing singlet pairs in the presence of
infinitesimal doping unless the pairing interaction vanishes. Only competing AF
fluctuations, vanishing pair correlations, or breaking of SU(4) symmetry prevent
the immediate formation of a pure singlet-pair condensate with doping.  Thus we
have generalized the Cooper instability to strongly-correlated electron systems 
with
possible competing order.
\item
In the underdoped region, the ground state is unstable against fluctuations in
both antiferromagnetic and superconducting order.  This enables all manner of
emergent behavior as the nature of the ground state can be altered qualitatively
by small perturbations (stripes, checkerboards, \ldots) The model suggests that 
such behavior is not fundamental to the high-$T\tsub c$ mechanism but rather is 
opportunistic, exploiting a fundamental softness of the  SU(4) ground 
state that is predicted by the symmetry to occur only in a narrow range of 
doping.
\end{enumerate}

\noindent
We propose that most cuprate phenomenology may be understood in terms of these 
two fundamental instabilities: the generalized Cooper instability accounts for 
the rapid appearance of superconductivity when the Mott insulator is doped with 
holes and the antiferromagnetic instability accounts largely for the pseudogap 
and its properties.

The model introduces, as a consequence of the highest symmetry and it subgroups, 
multiple energy scales. These lead to a rich variety of gaps that could play a 
physical role.  As a result, we find a phase diagram that is in quantitative 
agreement with that observed experimentally for cuprate superconductors.  The 
basic phase diagram is a direct consequence of the symmetry, with parameter 
adjustment influencing only details.

The cuprate pseudogap state has fluctuating antiferromagnetic and pairing 
character in the SU(4) description, and terminates at a quantum phase transition 
marked by a critical doping $P \simeq 0.18$; it is distinct from the 
superconducting state but related to it by a non-abelian symmetry. The pseudogap 
may be interpreted in terms of both SC--AF competition and preformed SU(4) pairs 
that condense into a singlet $d$-wave superconductor as hole doping suppresses 
fluctuations. We account quantitatively for the doping dependence of the 
pseudogap temperature $T^*$ and, because of the fluctuating nature of the state, 
we conclude that this PG state is most likely to be observed as a crossover 
rather than a distinct phase.

The structure of the SU(4) pseudogap state leads to a simultaneous quantitative 
description of the pseudogap temperature scale $T^*$ and Fermi arcs in ARPES 
experiments, including the origin of $T^*$ scaling. We have shown that requiring 
a quantitative description simultaneously of ARPES Fermi arcs and the doping 
dependence of the $T^*$ scale upon which they depend places extremely strong 
constraints on an acceptable theory of high-temperature superconductivity. If 
instead the Fermi surface for underdoped cuprates is interpreted in terms 
of small closed pockets (as suggested by quantum oscillation experiments), 
the anisotropic pseudogap correlations implied by the SU(4) model place strong 
constraints on where those pockets could be.

We have applied the method of generalized SU(4) coherent states, which provides
a systematic procedure to relate a many-body theory to its approximate
broken-symmetry solutions.  This approach may be viewed as a standardized
technology for constructing energy surfaces of many-body theories defined in
terms of the algebra of their second-quantized operators, which provides a
microscopic connection to Ginzburg--Landau methods.  Equivalently, it may be 
viewed
as implementing  the most general Hartree-Fock-Bogoliubov theory, 
subject
to a symmetry constraint on the Hamiltonian of the system. Thus the
coherent-state solution for SU(4) allows us to express results in language
familiar in condensed matter: spontaneously-broken symmetries, gap equations 
for quasiparticles, and variational
energy surfaces.

We have shown that competing antiferromagnetism and superconductivity, 
constrained by SU(4) symmetry which imposes no double lattice occupancy, leads 
on general grounds to energy surfaces in hole underdoped cuprates corresponding 
to weakly-broken $\sufour \supset \sofive$ symmetry that may be critically 
balanced between antiferromagnetic and superconducting order. These surfaces can 
be flipped between dominance of one order or the other by small fluctuations in 
the ratio of the antiferromagnetic to pairing strength. Therefore weak 
perturbations in the underdoped region, or near vortex cores or magnetic 
impurities, can produce amplified inhomogeneity having the spatial dependence of 
the perturbation but the intrinsic character of an SU(4) symmetry. (The symmetry 
defines the possible states; the perturbation selects among them.) Our results 
show that such effects can, but need not necessarily, involve spatial modulation 
of charge. More generally, we have suggested that critical dynamical symmetry 
may be a fundamental organizing principle for emergent behavior in correlated 
fermion systems, and that it provides a natural explanation for observed rich 
inhomogeneity in underdoped compounds with (paradoxically) near-universal 
overall phase diagrams in cuprate superconductors.

These considerations suggest that stripes, checkerboards and related 
inhomogeneities are secondary issues in understanding high-$T\tsub c$ 
superconductivity. They are perturbative (around a non-perturbative vacuum) 
consequences of the superconducting physics, not its cause. It is 
important to emphasize that the pseudogap and this sensitivity to spatial 
inhomogeneity in the underdoped region derive from the same fundamental physics 
of the underlying SU(4) symmetry.

The simplest charge-density wave decouples from the pairing--AF subspace in 
lowest order, suggesting that commensurate charge-density waves are not central 
to the high-$T\tsub c$ mechanism.  However, they can exploit the underdoped 
instability described above, producing a variety of induced structure for 
underdoped compounds.

We have argued that quantitative extension of this SU(4) approach to Fe-based 
superconductivity requires only that (1)~the relevant collective degrees of 
freedom are SC and AF, and (2)~the SC involves bondwise (not onsite) pairing. 
Thus, evidence has been presented that the  Fe-based high-temperature 
superconductors represent the second example (after the cuprates) of the 
non-abelian superconductors that we proposed in 2004 \cite{guid04}. The 
identification of non-abelian superconductivity in these two classes of 
compounds permits a unified model of cuprate and Fe-based superconductors to be 
constructed based on an SU(4) group (and subgroups) generated by emergent 
degrees of freedom, despite the obvious physical and microscopic differences 
between these classes of compounds. However, consistency with neutron scattering 
and ARPES data places strong constraints on possible FeAs orbital pairing 
formfactors, and closure of the SU(4) algebra suggests generally that the 
pairing in the FeAs case could correspond to  different orbital formfactors at 
the microscopic level than is required for the cuprates.

The SU(4) model presented here has connections to a number of other theoretical
approaches that have been applied to the high-temperature superconductor 
problem:

(1)~Our results provide some support for the assumption of resonating valence
bond models that the state with AF order at half filling would really like to be
a state with many characteristics of a spin-singlet liquid.  However,  the SU(4)
variational coherent state is simpler to implement and yet contains a broader
range of physics than a spin-singlet liquid, and accounts for many cuprate
properties across the entire physical doping range without introducing spinons,
holons, or related concepts having marginal experimental support in HTSC
compounds.

(2)~The Zhang SO(5) model is recovered as one symmetry-limit approximation of
the SU(4) theory. However, the present approach differs fundamentally from that
of Zhang and derives the SO(5) subgroup by approximation from a  richer theory
with broader physical implications.

(3)~The SU(4) emergent-symmetry approach differs in spirit from that of 
approaches such as the Hubbard
or $t$---$J$ models. However, these approaches need not be antagonistic. 
Because the SU(4) theory is defined in a highly-truncated subspace, it makes no
sense to compare operators or wavefunctions directly with other theories, but 
it is legitimate to
compare matrix elements. As we have demonstrated, the SU(4) model correctly
describes the matrix elements corresponding to many fundamental properties of
high-temperature superconductors. It is less certain that Hubbard or $t$--$J$
models can make the same claim, because they are difficult to solve in spaces
large enough to give definitive results. For example, there are differing
opinions on whether these approaches actually lead to a robust superconducting
state for realistic systems.

The properties in the phase diagram discussed here for cuprate and iron-based
high-temperature superconductors have similarities with properties observed in other
materials. In heavy-fermion compounds and some organic superconductors there is
evidence for significant electron--electron correlation and a superconducting phase
appears near the boundary of an AF phase. As a second example, the manganites have
strong correlations and complex competing phases, some bearing a resemblance to those
that we have discussed. Therefore, the formalism developed here to describe
multiple competing low-temperatures phases should be applicable to a much broader
range of strongly-correlated electron systems, with doping replaced or supplemented
by additional control parameters such as pressure or strength of a magnetic 
field. 

Hence, by employing techniques that are well-established in general many-body
physics, we conclude that the properties of high-temperature superconductors,
including the rapid development of a superconductor from a Mott insulator in the
cuprates, the properties of pseudogap states, and rich disorder localized 
within an otherwise
universal phase diagram, are understandable in terms of a minimal generalization
of traditional BCS theory and the Cooper instability to include
self-consistently the role of antiferromagnetism and onsite Coulomb repulsion. 
Our results represent a minimal variational solution of competing
antiferromagnetism and singlet superconductivity on a fermionic lattice with no
double occupancy for the pair components.  Therefore, we believe that the
general gap and phase structure presented here will be a necessary consequence
of any realistic theory that takes a lattice with strongly-correlated electrons
and competing pairing and antiferromagnetism as the basis for describing
high-temperature superconductivity.

The preceding  observations, coupled with the deep algebraic analogies  noted 
between AF--SC competition in condensed matter and 
deformation--superconductivity competition in nuclear physics, suggests  a 
fundamental relationship between the forms of superconductivity 
observed in many fields of science.  The robustness and similarities of 
superconductivity across so many subfields  that deal with  matter having very 
different length and energy scales, and very 
different physical environments, indicates that the Cooper-pair superconducting 
mechanism {\em cannot} depend essentially on microscopic details in any one 
field.  Indeed, it is suggested that the opposite is true:  superconductivity 
must correspond to a mechanism that is extremely robust and compatible with a 
very broad range of underlying microscopic details.  These details influence the 
theory parametrically (for example, determining the exact value of the 
superconducting transition temperature and the range of parameters over which SC 
is found) but must have little power to determine the general 
properties of superconductivity in the system, except to favor generic 
conditions that permit it to emerge.  

A common algebraic structure for the relevant quantum operators is one of the 
few things that could be similar and independent of detailed microscopic 
structure across the superconductivity observed or expected in such diverse 
physical systems. This leads us to conjecture that all superconductivity and 
superfluidity in all fields can be understood in terms of a generalized Cooper 
instability that is realized in terms of similar abelian or non-abelian operator 
algebras.

Finally, let us draw attention to the irony that this discussion may seem at 
first blush to be unconventional because of methodology, yet it leads to the 
most conventional and conservative of conclusions. The high-temperature 
superconductors are described at all dopings by a BCS formalism generalized 
self-consistently to incorporate antiferromagnetism, pairing, and on-site 
Coulomb repulsion on an equivalent footing. Microscopic details such as 
dimensionality, gap orbital symmetry, pair binding mechanism, microscopic 
structure of the magnetism, the crystal structure, presence or absence of 
disorder, and so on are  important in their own right, but their influence on 
the superconductor is primarily to set the value of coefficients in equations 
whose form has largely been determined by emergent dynamical symmetry, 
independent of those microscopic details.  The emergent collective properties 
defining the essence of the superconducting state that are so easily recognized 
across many physical systems with very different  microscopic 
structure require only that the microscopic conditions permit realization of  
emergent dynamical symmetries of the Hamiltonian like the SU(4) symmetry 
described here, largely independent of further microscopic details.

\acknowledgments 

We would like to thank  Pengcheng Dai, Elbio Dagotto, Adriana Moreo, Takeshi 
Egami, John Quinn, Hai-Hu Wen, and Wei Ku for discussions and advice that have 
greatly enhanced our understanding of strongly correlated electron systems.  
This work was partially supported by the National Key Program for S\&T Research
and Development (Grant No. 2016YFA0400501).  L. W. acknowledges grant support 
from the Basque Government Grant No.\ IT472-10 and the Spanish MICINN Grant No.\ 
FIS2012-36673-C03-03.  This work was partially supported by LightCone 
Interactive LLC.


\vfill

\clearpage

\begin{widetext}

\setcounter{section}{0}
\setcounter{table}{0}

\appendix
\section{SU(4) Subgroups and Dynamical Symmetries}
\protect\label{appendix}

The basic properties of SU(4) and its dynamical subgroups that conserve charge and
spin are summarized in \tableref{su4Properties} and in \tableref{symmetryLims}.

\begin{table}[h]
\caption {Properties of SU(4) and its subgroups (assuming no broken pairs)}
\label{table:su4Properties}
\setlength{\tabcolsep}{6 pt}
\begin{tabular}{lcccc}
\hline\\[-6pt]
Group&
Generators&
Quantum numbers&
Casimir operator&
Casimir eigenvalue
     \\ \hline\\[-6pt]
\sufour&
$\vec S$, $\vec \AF$, $\vec\pi^\dagger$, $\vec\pi$,
$\singletPair^\dagger$, $\singletPair$, $M$&
$\sigma_1=\tfrac{\Omega}{2}$, ($\sigma_2=\sigma_3=0$) &
$\pidagpi + \DdagD + \SdotS + \QdotQ + M(M-4) $&
$\tfrac{\Omega}{2}\left(\tfrac{\Omega}{2}+4\right)$
     \\
\sofour&
$\vec \AF$, $\vec S$&
$w$, $S$&
$\QdotQ + \SdotS$&
$w(w+2) + S(S+1)$
     \\ 
$\sutwo\tsub p$ &
$\singletPair^\dagger$, $\singletPair$, $M$ &
$N$, $\nu$ &
$\DdagD + M(M-1)$ &
$\tfrac14 (N-\nu)(2\Omega -N +2)$
     \\
$\sutwo\tsub s$ &
$\vec S $ &
$S$ &
$\SdotS$ &
$S(S+1)$
\\
\sofive &
$\vec S$, $\vec \pi^\dagger$, $\vec \pi$, $M$ &
$\tau$, ($\tau_2=0$) &
$\pidagpi + \SdotS$ &
$\tau(\tau+3)$
\\ \colrule
\end{tabular}
\label{table1}
\end{table}


\begin{table}[h]
\caption {The Hamiltonian, eigenstates and spectra in three dynamical symmetry
limits of the \sufour\ model. $E\tsub{g.s.}$ is the ground state energy, $\Delta
E$ the excitation energy, $N = \tfrac12 n$ is the pair number, $x = 1-n/\Omega$,
and $\kappa\tsub{so4}=\kappa\tsub{eff}+\chi\tsub{eff}$.}
\label{table:symmetryLims}
\vspace*{-15pt}
{\small
$$
\begin{array}{lll}
\hline\\[-6pt]
 \mbox{ SU(2) limit:  }|\psi(\sutwo)\rangle=\ket{N,v,S,m_S}
&\mbox{ SO(4) limit:  }|\psi(\sofour)\rangle=\ket{N,w,S,m_S}
&\mbox{ SO(5) limit:  }|\psi(\sofive)\rangle=\ket{\tau,N,S,m_S}
\\ \hline\\[-6pt]
 \ev{C\tsub{su(2)$_p$}}=\tfrac14(\Omega-v)(\Omega-v+2)
&\ev {C\tsub{so(4)}}=w(w+2),\quad w=N-\mu
&\ev{C\tsub{so(5}}=\tau(\tau+3),\quad \Omega/2-\tau=N-\lambda \\
 H =H_0+\kappa\tsub{eff}\, \SdotS
&H =H_0+\kappa\tsub{so4}\, \SdotS -\chi_{\tsub{eff}}C\tsub{so(4)}
&H =H_0+\kappa\tsub{eff}\, \SdotS \\
\vspace{4pt}
 \hspace{18pt}-G^{(0)}\tsub{eff}\left [C\tsub{su(2)$_p$}-M(M-1)\right]
&\hspace{18pt}
&\hspace{18pt}-G^{(0)}\tsub{eff}\left [C\tsub{su(4)}+M-C\tsub{so(5)}\right ] \\
\vspace{4pt}
 E\tsub{g.s.}=H_0-\frac{1}{4}G^{(0)}\tsub{eff}\Omega^2(1-x^2)
&E\tsub{g.s.}=H_0-\frac{1}{4}\chi\tsub{eff}\Omega^2(1-x)^2
&E\tsub{g.s.}=H_0-\frac{1}{4}\chi\tsub{eff}\Omega^2(1-x)^2\\
\vspace{4pt}
 \Delta E=\nu G^{(0)}\tsub{eff}\Omega+\kappa\tsub{eff}\ S(S+1),\quad \nu=v/2
&\Delta E=\mu \chi_{\tsub{eff}}(1-x)\Omega+\kappa\tsub{so4}\ S(S+1)
&\Delta E=\lambda x G^{(0)}\tsub{eff}\Omega+\kappa\tsub{eff}\ S(S+1)\\
\vspace{4pt}
 \nu = N, N-1, \ldots 0;\hspace{5pt} S=\nu, \nu-2, \ldots 0 {\rm \ or \ } 1\ \
\hspace{10pt}
&\mu=N,N-2,\ldots 0 {\rm \ or\ } 1;\hspace{5pt} S=w, w-1, \ldots 0 \ \
\hspace{10pt}
&\lambda=N,N-1,\ldots 0 {\rm \ or\ } 1;\hspace{5pt} S=\lambda,
\lambda-2, \ldots 0 \\
\hline
\nonumber
\end{array}
$$
}

\end{table}

\clearpage

\end{widetext}



\begin{thebibliography}{000}
\bibitem{particle-hole}
The particle--hole symmetry intrinsic to these models does not mean that
hole-doped and electron-doped compounds are expected to behave in the same
manner. Although the operators and basis states of the model are particle--hole
symmetric, the interactions entering the effective Hamiltonian would not be
expected to be the same for hole-doped and particle-doped compounds.  Thus, the
physical properties of hole-doped and electron-doped compounds could differ
substantially.
\bibitem{threebodynote}
Groups generally may have more than one Casimir invariant.  We shall use the
term ``Casimir'' to refer loosely to the lowest-order such invariants (which are
generally quadratic in the group generators).  In the context of the present
discussion, quadratic Casimirs are associated with 2-body interactions at the
microscopic level.  Higher-order Casimirs are then generally associated with
3-body and higher interactions.  The restriction of our Hamiltonians to
polynomials of order 2 in the Casimirs is then a physical restriction to
consideration of only 1-body and  2-body interactions.
\bibitem{coherent-parameters} Thus the most general SU(4) coherent state depends
on eight real variables. The reduction of the coherent state parameters to only
two in \eq{eq10b} follows from requiring time reversal symmetry and assuming
conservation of spin projection $S_z$  for the wavefunction.
\bibitem{footnote2}The competition between dynamical symmetries governing the
transition between spherical and deformed nuclei is discussed in \S 4.5 (in
particular, \S 4.5.4) of Ref.~\cite{FDSM}.
\bibitem{u1note} There also is the U(1) generator of charge density waves in our 
full $\ufour \supset\uone \times \sufour$ algebra that does not appear in the 
Zhang SO(5) algebra.
\end{thebibliography}
\end{document}